\documentclass[12pt,a4paper]{article}
\pdfoutput=1

\usepackage{amsmath,amsfonts,amssymb,amsthm,mathtools}

\usepackage[english]{babel}
\usepackage[utf8]{inputenc}
\usepackage[T1]{fontenc}

\usepackage{color}
\usepackage[dvipsnames]{xcolor}

\definecolor{darkblue}{rgb}{0.1,0.1,.7}
\usepackage[colorlinks, linkcolor=darkblue, citecolor=darkblue, urlcolor=darkblue, linktocpage]{hyperref}

\usepackage{graphicx}
\usepackage{latexsym}
\usepackage{subcaption}
\usepackage{authblk}

\usepackage{slashed}
\usepackage[mathscr]{eucal}
\usepackage{bbold} 
\usepackage[margin=10pt,font=small,labelfont=bf]{caption}
\usepackage{amscd}
\usepackage{bm}
\usepackage{bbm}
\usepackage{upgreek}
\usepackage[square, comma, sort&compress,numbers]{natbib}
\usepackage[all,cmtip]{xy}
\usepackage[margin=1.5in]{geometry}
\usepackage{cleveref}
\usepackage[color=cyan!30!white,linecolor=red,textsize=footnotesize]{todonotes}
\usepackage{microtype}
\usepackage{cancel,soul,ulem}

\usepackage{tikz,tikz-3dplot,tikz-cd,circuitikz}
\usetikzlibrary{
patterns,
decorations.markings,
decorations.pathmorphing,
arrows,
calc,
knots,
hobby,
decorations.pathreplacing,
shapes.geometric,
calc,
arrows,
arrows.meta,
decorations.text
}
\tikzset{>=latex} 
\usetikzlibrary{arrows.meta}
\usepackage{pgfplots} 
\pgfplotsset{compat = newest} 

\numberwithin{equation}{section}
\definecolor{amaranth}{rgb}{0.9, 0.17, 0.31}

\hypersetup{
pdfstartview={FitH}, 
pdftitle={Gauging with LSM Constraint}, 
pdfauthor={Tiwari}, 
colorlinks=true, 
linkcolor=blue, 
citecolor=amaranth, 
filecolor=magenta, 
urlcolor=blue 
}

\makeatletter
\g@addto@macro\bfseries{\boldmath}
\makeatother

\theoremstyle{definition}
\newtheorem{defn}{Definition}

\newtheorem{thm}{Theorem}

\newtheorem{conj}{Conjecture}

\theoremstyle{remark}
\newtheorem*{remark}{Remark}

\usepackage{color}
\definecolor{amaranth}{rgb}{0.9, 0.17, 0.31}
\definecolor{dcyan}{rgb}{0.0, 0.55, 0.55}

\newcommand{\mbb}{\mathbb}

\newcommand{\ket}[1]{| #1\rangle}

\makeatletter
\if@todonotes@disabled

\else

\fi
\makeatother

\begin{document}
\definecolor{tinge}{RGB}{255, 244, 195}
\sethlcolor{tinge}
\setstcolor{red}
\vspace*{-.8in}
\thispagestyle{empty}
\vspace{.3in}
{\begin{center}
\begin{Large}
\textbf{
Lieb-Schultz-Mattis anomalies and 
web of dualities induced by gauging in quantum spin chains}  
\end{Large}
\end{center}}
\vspace{.3in}
\begin{center}
\"{O}mer M. Aksoy$^{1}$,
Christopher Mudry$^{1,2}$,
Akira Furusaki$^{3,4}$,
Apoorv Tiwari$^{5}$
\\
\vspace{.3in}
\small{
$^{1}$\textit{Condensed Matter Theory Group, Paul Scherrer Institute,
CH-5232 Villigen PSI, Switzerland}\\
$^{2}$\textit{Institut de Physique, EPF Lausanne, CH-1015 Lausanne, Switzerland}\\
$^{3}$\textit{RIKEN Center for Emergent Matter Science, Wako, Saitama, 351-0198, Japan}\\
$^{4}$\textit{Condensed Matter Theory Laboratory, RIKEN, Wako, Saitama, 351-0198, Japan}\\
$^{5}$\textit{Department of Physics, KTH Royal Institute of Technology,
Stockholm, 106 91 Sweden}\\
\vspace{.5cm}
}
\vspace{.3cm}
\vspace{.1in}
\end{center}
\vspace{.3in}

\begin{abstract}
\noindent
Lieb-Schultz-Mattis (LSM) theorems
impose non-perturbative constraints on
the zero-temperature phase diagrams
of quantum lattice Hamiltonians 
(always assumed to be local in this paper).
LSM theorems have recently been interpreted as
the lattice counterparts to mixed 't Hooft anomalies
in quantum field theories that arise from 
a combination of
crystalline and global internal symmetry groups.
Accordingly, LSM theorems have been reinterpreted as LSM anomalies.
In this work, we provide a systematic diagnostic for LSM anomalies
in one spatial dimension. We show that
gauging subgroups of the global internal symmetry group
of a quantum lattice model obeying an LSM anomaly
delivers a dual quantum lattice Hamiltonian such that its
internal and crystalline symmetries mix non-trivially
through a group extension.
This mixing of crystalline and internal
symmetries after gauging is a direct consequence of the LSM anomaly,
i.e., it can be used as a diagnostic of an LSM anomaly.
We exemplify this procedure for a quantum spin-1/2 chain
obeying an LSM anomaly resulting from combining a global internal
$\mathbb{Z}^{\,}_{2}\times\mathbb{Z}^{\,}_{2}$
symmetry with translation or reflection symmetry.
We establish a triality of models by gauging a
$\mathbb{Z}^{\,}_{2}\subset\mathbb{Z}^{\,}_{2}\times\mathbb{Z}^{\,}_{2}$
symmetry in two ways,
one of which amounts to performing a Kramers-Wannier duality,
while the other implements a Jordan-Wigner duality.
We discuss the mapping of the phase diagram
of the quantum spin-1/2 $XYZ$ chains
under such a triality. We show that the
deconfined quantum critical transitions between Neel and
dimer orders are mapped to either
topological or conventional Landau-Ginzburg transitions.
Finally, we extend our results to  
$\mathbb{Z}^{\,}_{n}$
clock models with 
$\mathbb{Z}^{\,}_{n}\times\mathbb{Z}^{\,}_{n}$ global internal 
symmetry, and provide a reinterpretation of the dual 
internal symmetries in terms of $\mathbb{Z}^{\,}_{n}$
charge and dipole symmetries.
\end{abstract}
\vskip 0.7cm \hspace{0.7cm}
\newpage
\setcounter{page}{1}
\noindent\rule{\textwidth}{.1pt}\vspace{-1.2cm}
\begingroup
\hypersetup{linkcolor=black}
\setcounter{tocdepth}{3}
\setcounter{secnumdepth}{3}
\tableofcontents
\endgroup
\noindent\rule{\textwidth}{.2pt}
\setlength{\abovedisplayskip}{15pt}
\setlength{\belowdisplayskip}{15pt}

\section{Introduction}
\label{sec:Introduction and results}

\subsection{Motivation}

The Lieb-Schultz-Mattis (LSM) Theorem \cite{Lieb61} and its
extensions~\cite{Affleck1986,Aizenman1994,Oshikawa1997,Yamanaka1997,%
Koma2000,Oshikawa2000,Hastings2004,Hastings2005,Chen2011,Roy2012,%
Parameswaran2013,Watanabe2015,Watanabe2016,Cheng2016,Qi2017,Cho2017,Po2017,%
Watanabe2018,Tasaki2018,Metlitski2018,Yang2018,Jian2018,Cheng2019,%
Kobayashi2019,Ogata2019,He2020,Else2020,Hetenyi2020,Yao2020a,Ogata2021,Bachmann2020,%
Dubinkin2021,Yao2021,Aksoy2021b,Tasaki2022,Yao2022,Gioia2022,Cheng2023,Yao2023,%
Seiberg2023}
are no-go theorems that constrain the
low-energy properties of lattice Hamiltonians with certain
combinations of internal and crystalline symmetries.
While in its original form the LSM Theorem applies to spin-1/2 chains with
$\mathrm{SO}(3)$ spin-rotation and translation symmetries, 
many generalizations for general crystalline
\cite{Roy2012,Parameswaran2013,Watanabe2015,Po2017,Watanabe2018}
and internal symmetries, for systems with bosonic and fermionic
\cite{Cheng2019,Aksoy2021b,Seiberg2023} degrees of freedom,
and for spatial dimensions greater than one
\cite{Oshikawa2000,Hastings2004,Hastings2005,Cheng2016,Po2017,Else2020,
Yao2020a,Yao2021,Aksoy2021b,Yao2022}
have been proposed.

LSM Theorems rule out
a ground state that is trivially gapped and symmetric,
i.e., a ground state that is simultaneously gapped,
non-degenerate on any closed space manifold,
and symmetric under the relevant internal and
crystalline symmetries. Conversely,
LSM Theorems predict that
ground states that are symmetric must support either gapless
excitations or topological order.

Global symmetries play a pivotal role in organizing 
various aspects of  quantum systems.
In particular, operators and states organize into 
representations of the symmetry group, while the phase diagrams 
and dynamics are constrained by symmetries.
Global symmetries of quantum systems can be anomalous.
A quantum 't Hooft anomaly \cite{tHooft1979} arises 
when the partition function coupled to a symmetry background 
gauge field is not invariant under gauge transformations of this background.
Instead, 
the partition function transforms by a $\mathrm{U}(1)$ phase factor 
that cannot be absorbed by the addition of local terms.
While quantum anomalies were initially investigated 
in continuum quantum field theories with fermions and 
Lie group symmetries \cite{Alvarez-Gaume:1984zst, Treiman:1986ep, Bertlmann:1996xk}, 
in recent years there has been much progress in 
understanding anomalies in more general contexts 
involving bosonic systems, 
finite symmetries \cite{Kapustin:2014tfa, Kapustin:2014zva, Gaiotto:2017yup}, 
and lattice quantum systems \cite{Chen:2011bcp, Kawagoe2021, Cheng2023, Moradi23}.
The low-energy dynamics and phase diagrams of systems
with 't Hooft anomalies are strongly constrained in a manner
reminiscent of LSM Theorems.
The anomaly matching condition \cite{tHooft1979}
requires that any trivially gapped ground state necessarily
breaks the full symmetry down to a subgroup that trivializes the anomaly.

This similarity between the constraints imposed at low energies
by LSM Theorems for lattice Hamiltonians, on the one hand,
and by `t Hooft anomalies on the other hand, suggests a close connection between
LSM Theorems and 't Hooft anomalies
\cite{Cheng2016,Cho2017,Komargodski2018,Jian2018,Else2020,Cheng2023,Seifnashri2023}.
More precisely, LSM Theorems can be connected to mixed
't-Hooft anomalies by showing that the long-wavelength
continuum descriptions of lattice Hamiltonians, for which
an LSM Theorem applies,
support mixed 't-Hooft anomalies between symmetries
that originate from internal and crystalline symmetries participating
in the LSM theorem.
This means that, while neither the internal nor the crystalline symmetry
are individually anomalous, their combination is.
This translates to the fact that, while there is no obstruction to
a trivially gapped and symmetric ground state under either internal 
or crystalline symmetry, any such state cannot be gapped and symmetric under 
the full symmetry group that participates in the LSM Theorem.
Equivalently, the full internal symmetry group cannot be 
gauged, while preserving the crystalline symmetries.
However, there is no obstruction to gauging a non-anomalous 
subgroup of internal symmetries for which there is no LSM Theorem.
Accordingly, LSM Theorems have been  reinterpreted as LSM anomalies,
a terminology that we will follow in this paper.

In recent years, there has been much progress towards classifying
topological phases of matter with crystalline symmetries and
understanding the corresponding quantum anomalies~
\cite{Huang2017,Thorngren2018,Khalaf2017,Guo2018,Trifunovic2018,Else2018,%
Rasmussen2018,Shiozaki2018,Song2020}. 
Such classifications have often relied on the intuition
that in the long wavelength continuum description,
some crystalline symmetries
appear as internal symmetries.
Despite this progress, the lattice understanding of anomalies
\cite{Else2014,Kawagoe2021,Moradi23},
in particular those involving crystalline symmetries,
is very much an evolving subject \cite{Cheng2023}.
Challenges arise because it is often unclear
how to probe crystalline symmetries through coupling
to crystalline backgrounds
\cite{Shiozaki2016,Thorngren2018,Else2018,Song2020},
as is routinely done with internal symmetries and gauge fields.
What is even less clear is how to dynamically gauge a crystalline
symmetry by summing over the crystalline backgrounds.
These issues make pinpointing LSM anomalies on the lattice a subtle task.

In this work, we circumvent these obstacles
in local quantum lattice models by gauging non-anomalous 
subgroups of their internal symmetries.
This approach is always viable since the chosen internal
symmetry is non-anomalous, and methods to gauge internal 
symmetries are well-known from lattice gauge theory.

Gauging global symmetries is a powerful way to 
manipulate the symmetry structure of a quantum system
\cite{Frohlich2009,Tachikawa2017,Bhardwaj2017,Gaiotto2021}.
By starting with a system with a known symmetry 
structure, like a finite group with certain anomalies, 
and gauging non-anomalous sub-symmetries, one obtains 
dual (gauged) theories with novel symmetry 
structures
\cite{Tachikawa2017,Bhardwaj2017,Delcamp2019,Borla2020,Bhardwaj2022a, Roumpedakis:2022aik, Lootens:2021tet, Gaiotto2021,
Bhardwaj2022b, Bhardwaj2022c,Bhardwaj2022d,Bartsch2022a,Bartsch2022b,Delcamp2023,Moradi23, 
Bhardwaj:2023kri, Schafer-Nameki:2023jdn,Seifert2023}.
Generalized gauging procedures 
have recently emerged as effective methods 
to study generalized symmetry structures in both 
continuous and lattice systems.
For instance, in one spatial dimension, gauging a non-anomalous finite
symmetry that participates in a mixed anomaly results in a dual
(gauged) theory with a non-anomalous global symmetry that extends the
residual symmetries left after gauging.  In higher dimensions, this
group extension becomes a higher group \cite{Tachikawa2017,Delcamp2019}. 
Interestingly, these gauging procedures have also been used to furnish
non-invertible symmetry structures~\cite{Kaidi2021}.

Another reason to study gauging of finite global symmetries is
that such gaugings are realized as dualities in quantum systems.
For example, the well-known 
Kramers-Wannier~\cite{Kramers41,Montroll42,Wannier45,
Dobson69,Frankel70,Stephenson70,Mittag71} and 
Jordan-Wigner~\cite{Jordan28}
dualities are essentially gaugings of the
$\mathbb{Z}^{\,}_{2}$
internal and
$\mathbb{Z}^{\,}_{2}$
fermion-parity symmetry in one-dimensional lattice models
\cite{McKean64,Kadanoff71,Wegner71,Balian75,Jose77,Peskin78,Fradkin78,%
Korthals-Altes78,Drouffe78,Kogut79,Bellissard79,Horn79,Drouffe79,
Ukawa80,Savit80,Druhl82,Buchstaber03,Mathur16,
Kapustin17,Radicevic18,Karch19,Thorngren2020}.
Dualities can be used to provide profound non-perturbative
insights into quantum systems and are therefore very valuable. 

In this work, we study the gauging of subgroups of internal symmetry
which participate in LSM anomalies.
More precisely, we choose a
subgroup such that neither the gauged subgroup nor the remaining
symmetries have an LSM anomaly with the crystalline symmetries, while an
LSM anomaly applies for the full internal symmetry group.
We track how the crystalline and internal symmetries organize
into the symmetry structure of the dual (gauged) theory.
We find that, as a direct consequence of an 
LSM anomaly in the pre-gauged theory, there is necessarily a
non-trivial mixing of internal and crystalline symmetries in the dual theory.
More concretely, we exemplify this procedure on a
local quantum spin-$1/2$ chain that has a global
$\mathbb{Z}^{\,}_{2}\times\mathbb{Z}^{\,}_{2}$
internal symmetry in addition to translation and reflection crystalline
symmetries~\cite{Ogata2019,Ogata2021}.
The local representatives of the internal symmetry operators satisfy
a projective representation of
$\mathbb{Z}^{\,}_{2}\times\mathbb{Z}^{\,}_{2}$
which, in turn,
implies an LSM anomaly involving either translation
or reflection symmetry. We gauge a subgroup
$\mathbb{Z}^{\,}_{2}\subset\mathbb{Z}^{\,}_{2}\times\mathbb{Z}^{\,}_{2}$  
of the global internal symmetry
$\mathbb{Z}^{\,}_{2}\times\mathbb{Z}^{\,}_{2}$
in two ways, which amounts to performing
Kramers-Wannier (KW) or Jordan-Wigner (JW) dualities, respectively.
We establish a triality of the original model and its duals under KW
or JW dualities.  After the KW duality, we find that the
dual symmetry becomes non-Abelian, more precisely a semi-direct
product of the internal and crystalline symmetries.  After the
JW transformation too, the LSM anomaly
gets traded for a symmetry structure that involves
a non-trivial fermionic group extension of the internal and
crystalline symmetry groups.

Starting with the original LSM Theorem \cite{Lieb61},
many LSM Theorems have been probed and proven
using background gauge fields (or equivalently 
twisted boundary conditions) of internal symmetries
\cite{Lieb61,Oshikawa2000,Yao2021,Aksoy2021b,Yao2022,Yao2023}. 
Our work presents a novel method for probing LSM anomalies based on 
dynamical gauging of internal sub-symmetries which 
provides an indirect yet robust way to pin-point the existence of
LSM anomalies. We, therefore, confirm that gauging non-anomalous subgroups of 
finite symmetries with LSM anomalies leads to a non-anomalous
group extension in the dual theory,
a fact known for finite internal symmetries 
with mixed anomalies \cite{Tachikawa2017}.

A deconfined quantum critical point (DQCP) describes a continuous transition
between phases with distinct symmetries. Such transitions are 
driven by deconfinement of point defects of symmetry breaking order parameters
such that the defects of the order parameter of one phase bind a non-vanishing 
expectation value of the order parameter of the other phase and vice versa
\cite{Senthil04,Senthil04Levin,Senthil04PRB,Senthil06Fisher}. 
DQCPs arise naturally in models with symmetries that carry mixed anomalies,
where the relationship between defects of the order parameters 
can be traced back to the mixed anomaly between two subgroups. 
For instance, the paradigmatic example of DQCP is
the conjectured continuous
transition between the Neel and valance-bond-solid (VBS) orders of the
Heisenberg antiferromagnet on the square lattice.
The former order preserves the crystalline $\mathrm{C}^{\,}_{4}$
rotation symmetry and breaks the internal $\mathrm{SO}(3)$ symmetry,
while the latter order breaks the $\mathrm{C}^{\,}_{4}$ symmetry
and preservs the $\mathrm{SO}(3)$ symmetry. 
Indeed, there exists an LSM anomaly between these symmetries,
which rules out a trivially gapped ground state
that is symmetric under both $\mathrm{SO}(3)$ and $\mathrm{C}^{\,}_{4}$~\cite{Po2017,Metlitski2018,Else2020}.
Under gauging a non-anomalous subgroup, a DQCP is often mapped to 
conventional Landau-Ginzburg-type transitions,
where symmetries preserved by one phase is a subgroup of the other
\cite{Moradi2022,Chatterjee2023,Zhang2023,Moradi23}. 
Motivated by the relation between mixed anomalies and DQCP, 
we study the phase diagram of the quantum spin-1/2 $XYZ$ chain
under the KW and JW dualities. This model features deconfined phase transitions
between Neel and dimer ordered phases\cite{Mudry19}. 
We show that as the crystalline and 
internal symmetries are mixed after gauging, the DQCP between Neel and Dimer
ordered phases are mapped to either (i) topological phase transitions between
two phases with same symmetries or (ii) conventional Landau-Ginzburg-type 
symmetry breaking transitions. Therein, we demonstrate how  
dualities can be utilized to recast DQCPs
and also understand the phase diagrams
of spin-1/2 and interacting Majorana chains. 

\subsection{Summary of the main results}

Common to most LSM Theorems are
quantum dynamical degrees of freedom
defined on the sites of a lattice with a quantum dynamics
that is local and invariant under
\begin{subequations}\label{eq: defs Gtot Gspa and Gint in results}
\begin{itemize}
\item
the symmetry group
\begin{equation}
\mathrm{G}^{\,}_{\mathrm{tot}}=
\mathrm{G}^{\,}_{\mathrm{spa}}
\times
\mathrm{G}^{\,}_{\mathrm{int}}
\label{eq: defs Gtot Gspa and Gint in results a}
\end{equation}
built from the direct product
of a space (crystalline) symmetry group 
$\mathrm{G}^{\,}_{\mathrm{spa}}$
with a global internal symmetry group 
$\mathrm{G}^{\,}_{\mathrm{int}}$
\item
such that 
$\mathrm{G}^{\,}_{\mathrm{int}}$
cannot be gauged in its entirety,
while preserving the symmetry under
the full space subgroup $G^{\,}_{\mathrm{spa}}$.%
~\footnote{%
~This is so whenever the internal symmetry group
is represented globally by a group homomorphism, while it is
represented  locally by a non-trivial projective representation
of the internal symmetry group. Indeed, whereas it is possible
to construct a many-body state that is a gauge singlet,
the local Hilbert space does not admit a state that is a
gauge singlet.}
\end{itemize}
In this work, by way of explicit examples, we study 
the dualities induced by gauging a subgroup of the 
internal symmetry $\mathrm{G}^{\,}_{\mathrm{int}}$
that does not participate in the LSM anomaly. 
The paradigmatic example that we shall follow 
is the case of quantum spin-1/2 degrees of freedom at 
every site of lattice $\Lambda$
(a chain of even cardinality $|\Lambda|$) with
\begin{equation}
\mathrm{G}^{\,}_{\mathrm{spa}}=\mathrm{Z}^{t}_{|\Lambda|}\rtimes\mathbb{Z}^{r}_{2}
\label{eq: defs Gtot Gspa and Gint in results b}
\end{equation}
the space symmetry generated by translation ($t$)
and site-centered reflection ($r$),
and
\begin{equation}
\mathrm{G}^{\,}_{\mathrm{int}}=\mathbb{Z}^{x}_{2}\times\mathbb{Z}^{y}_{2}
\label{eq: defs Gtot Gspa and Gint in results c}
\end{equation}
the global internal symmetry generated by $\pi$ rotations along
the $x$ and $y$ axes in internal spin-1/2 space, respectively.
It is known that both translation and reflection symmetries 
participate in LSM anomalies with the global internal 
$\mathbb{Z}^{x}_{2}\times\mathbb{Z}^{y}_{2}$ symmetry.
Our main results are as follows.
\end{subequations}

\begin{enumerate}
\item 
With the tools reviewed in Sec.\
\ref{sec:Triality through bond algebra isomorphisms},  
we show in Sec.\
\ref{sec:Triality of crystalline and internal symmetries on a chain}
that, under both KW and JW dualities, the LSM anomaly that was present
before gauging is no longer operative for the dual symmetries
after gauging, instead
the dual global symmetry group $\mathrm{G}^{\vee}_{\mathrm{tot}}$ 
is a group extension of the dual internal symmetry group $\mathrm{G}^{\vee}_{\mathrm{int}}$ 
by the dual crystalline symmetry group
$\mathrm{G}^{\vee}_{\mathrm{spa}}$. 
For instance,  the Abelian global symmetry group
$\mathbb{Z}^{r}_{2}\times\mathbb{Z}^{x}_{2}\times\mathbb{Z}^{y}_{2}$ 
that is generated by a site-centered reflection together with
$\pi$ rotations along $x$ and $y$ axes in internal spin-1/2 space
maps under KW duality to the non-Abelian group
$\mathrm{D}^{\,}_{8}$ (dihedral group of order $8$)
defined in Eq.\
(\ref{eq:total symmetry group for KW duality b}).

\item 
As a concrete application, 
we study the zero-temperature phase diagram of
the quantum spin-1/2 antiferromagnetic $XYZ$
chain with nearest- and next-nearest-neighbor couplings 
under the KW and JW dualities in Sec.\
\ref{sec:Application to quantum spin-1/2 degrees of freedom on a chain}.
For this quantum $XYZ$ chain, 
the presence of LSM anomalies precludes
a non-degenerate gapped ground state that is simultaneously symmetric under
both the crystalline (translation or reflection) symmetry group
(\ref{eq: defs Gtot Gspa and Gint in results b})
and the global internal symmetry group
(\ref{eq: defs Gtot Gspa and Gint in results c}).
As such, in the parameter space of interest,
three gapped phases are realized that spontaneously break
either the crystalline symmetry or the global internal symmetry.
The boundaries separating in parameter space these gapped
spontaneously symmetry broken (SSB) phases
are continuous phase transitions that realize
deconfined quantum criticality (DQC)~\cite{Mudry19}.
We present a triality of phase diagrams with
Figs.\ \ref{fig:comparing spin-1/2 XY phase diagram and its KW dual}
and 
\ref{fig:comparing spin-1/2 XY phase diagram and its JW dual}
such that the dual phase diagrams contain
non-degenerate symmetric and gapped ground states due to the absence of
any LSM anomaly after gauging.
We find that the continuous DQC phase transitions
prior to gauging the quantum $XYZ$ chain
are to be reinterpreted as continuous phase transitions
that are either of the conventional Landau-Ginzburg type
or of the topological type after gauging.

\item
To solidify the correspondence between LSM anomalies
prior to gauging and mixing of internal and crystalline symmetries
after gauging, we study in Sec.\ \ref{sec:Zn generalization}
the family labeled by $n=2,3,\cdots$ of
$\mathbb{Z}^{\,}_{n}$ clock models with
$\mathbb{Z}^{\,}_{n}\times\mathbb{Z}^{\,}_{n}$ global internal symmetry.
We show that,
when a $\mathbb{Z}^{\,}_{n}\subset 
\mathbb{Z}^{\,}_{n}\times\mathbb{Z}^{\,}_{n}$ subgroup is gauged,
the dual of the remaining $\mathbb{Z}^{\,}_{n}$ symmetry generator mixes with 
translation for any $n$, while a mixing occurs 
with reflection only when $n$ is even.
This result is consistent with the following conjecture. 
The mixing induced by gauging
betwen dual crystallline symmetries and dual global internal symmetries
occurs if and only if the crystalline and global internal symmetries
are subject to an LSM anomaly prior to gauging.

\item 
We unravel a connection between Hamiltonians with
spatially modulated internal symmetries, such as 
a $\mathbb{Z}^{\,}_{n}$-dipole symmetry, and Hamiltonians with global
(spatially uniform) internal symmetries,
such as $\mathbb{Z}^{\,}_{n}\times\mathbb{Z}^{\,}_{n}$ symmetry. 
We show that gauging a $\mathbb{Z}^{\,}_{n}$-charge symmetry 
induces a duality between Hamiltonians with
$\mathbb{Z}^{\,}_{n}$-charge, $\mathbb{Z}^{\,}_{n}$-dipole, and translation 
(or link-centered reflection) symmetries, and Hamiltonians with
global uniform $\mathbb{Z}^{\,}_{n}\times\mathbb{Z}^{\,}_{n}$ symmetry, 
translation or (site-centered reflection) symmetry, and an LSM anomaly.

\end{enumerate}

\subsection{Comparison with the literature}

Since 2016,
LSM Theorems have been reinterpeted as mixed 't Hooft anomalies
involving crystalline symmetries in the following loose sense.
On the one hand, it has been argued that LSM Theorems can be
understood as 't Hooft anomalies of emergent internal symmetries
arising from crystalline symmetries in the low-energy continuum limit%
~\cite{Cho2017,Metlitski2018,Kobayashi2019}.
On the other hand, Refs.%
~\cite{Cheng2016,Po2017,Jian2018,Else2020}
have made the salient observation that
LSM Theorems can be applicable to the boundaries of crystalline topological
phases in one higher dimension.

Most studies of mixed 't-Hooft anomalies have treated the cases
of internal symmetries in relativistic quantum field theories
for which the partial gauging leaves the space-time symmetries 
unaffected. In this context, the dualities induced by gauging a 
subgroup of a finite group are worked out in Refs.\ 
\cite{Tachikawa2017,Bhardwaj2017}.
For instance, the duality between the  
$\mathbb{Z}^{\,}_{2}\times\mathbb{Z}^{\,}_{2}\times\mathbb{Z}^{\,}_{2}$
global internal symmetry
with a mixed anomaly involving all three $\mathbb{Z}^{\,}_{2}$
subgroups and the $\mathrm{D}^{\,}_{8}$ internal symmetry has been
established in Ref.\ \cite{Tachikawa2017}.
Our results in Sec.\
\ref{sec:Triality of crystalline and internal symmetries on a chain}
parallels this fact for LSM anomalies
which involves crystalline symmetries. From this point of view, our 
results confirm the crystalline equivalence principle%
~\cite{Thorngren2018,Else2018}
that suggests a one-to-one correspondence
between global internal and crystalline symmetries.

The connection between
LSM anomalies with translation symmetry
and mixed 't Hooft anomalies has been studied for lattice models%
~\cite{Cheng2023,Seiberg2023,Seifnashri2023}
since late 2022. It has been shown in Refs.\ \cite{Seiberg2023,Seifnashri2023}
that a dual non-invertible translation  symmetry appears
when global internal symmetries that are participating in LSM anomalies 
are dynamically gauged. Here, we complement this picture by
studying the dualities that are induced by gauging a subgroup
that \textit{does not participate in the LSM anomaly}.

We are unaware of prior derivations or studies of
the KW and JW dual Hamiltonians
(\ref{eq:def Hamiltonian b'=0})  
and
(\ref{eq:def Hamiltonian f=1})
of the quantum spin-1/2 antiferromagnetic $XYZ$ chain
with nearest- and next-nearest-neighbor antiferromagnetic couplings 
obeying periodic boundary conditions.
This is also true of the KW dual Hamiltonian
(\ref{eq:tau-dual Hamiltonian with OPB})
that is dictated by the choice of open boundary conditions.
To the best of our knowledge,
KW dualization has been mostly applied in the literature
to the (classical) Ising limiting cases
of quantum spin-1/2 antiferromagnetic $XYZ$
chains or to the one-dimensional transverse-field Ising model,
as is reviewed in Secs.\
\ref{sec:Triality through bond algebra isomorphisms}
and
\ref{sec:Application to quantum spin-1/2 degrees of freedom on a chain}.
The derivation and discussion of Figs.\
\ref{fig:comparing spin-1/2 XY phase diagram and its KW dual}
and
\ref{fig:comparing spin-1/2 XY phase diagram and its JW dual}
are the main original results of this paper. 

The KW and JW dual Hamiltonians
(\ref{eq:def Hamiltonian b'=0})  
and
(\ref{eq:def Hamiltonian f=1})
are examples of quantum Hamiltonians with simultaneous
charge, dipole, and translation symmetries.
Gauging the charge symmetry simply brings back
Hamiltonians
(\ref{eq:def Hamiltonian b'=0})  
and
(\ref{eq:def Hamiltonian f=1})
to the quantum spin-1/2 XYZ chain with the Hamiltonian
(\ref{eq:def Hamiltonian b=0})
according to the triality encoded by Fig.\
\ref{fig:Triality diagram}.
The study of Hamiltonians with charge and multipolar symmetries
has gained popularity (see Refs.\ 
\cite{Burnell2023,Bulmash2023,Han2023,Delfino2023,Lam2023}
and references therein).
However, our observation that
gauging the charge symmetry in the presence of the additional
dipole and translation symmetries can produce a local Hamiltonian
with translation and global internal symmetry characterized by
an LSM anomaly is another original result of this paper.

\subsection{Organization}

The rest of the paper is organized as follows.

In Sec.\
\ref{sec:Triality through bond algebra isomorphisms}, we
review the implementation of the KW and JW dualities as 
bond-algebra isomorphisms due to gauging an
internal $\mathbb{Z}^{\,}_{2}$ symmetry. 
Therein, we establish the triality of three bond algebras. 

In Sec.\
\ref{sec:Triality of crystalline and internal symmetries on a chain},
we discuss how additional internal and crystalline symmetries
are modified  under gauging an internal sub-symmetry.
In particular, we show that the LSM anomaly disappears
after gauging at the cost of a group extension between crystalline and
internal symmetries.

In Sec.\
\ref{sec:Application to quantum spin-1/2 degrees of freedom on a chain},
we study the phase diagram of the quantum spin-1/2 $XYZ$ chain
and its fate under the gauging-related dualities. 

Section\ \ref{sec:Zn generalization} showcases a generalization to the 
$\mathbb{Z}^{\,}_{n}$-clock models, where we consider an LSM anomaly 
between internal $\mathbb{Z}^{\,}_{n}\times\mathbb{Z}^{\,}_{n}$ symmetry and 
translations and reflections. We conjecture that the LSM anomaly
with the reflection symmetry is present only when $n$ is even.
We confirm this conjecture by showing that
the mixing between reflection and internal symmetries only appears
when $n$ is even while mixing with translation is always present. 
We conclude in Sec.\ \ref{sec:conclusions}.

\section{Triality of $\mathbb{Z}^{\,}_{2}$-symmetric bond algebras on a chain}
\label{sec:Triality through bond algebra isomorphisms}

The first incarnation of duality was discovered by
Jordan and Wigner in 1928 \cite{Jordan28},
who showed by algebraic means
that there exists a one-to-one correspondence between
creation and annihilation operators
of hard-core bosons on the one hand and spinless fermions
on the other hand,
provided both can be labeled by an index belonging to an ordered set
(as would be the case when this label enumerates the sites of a one-dimensional
lattice for example)%
~\footnote{%
~Wigner and Jordan also introduced in Ref.\ \cite{Jordan28}
Majorana operators, i.e., Hermitian operators obeying
a Clifford algebra.}.
The second incarnation of duality was discovered by
Kramers and Wannier in 1941
\cite{Kramers41,Montroll42,Wannier45,Dobson69,Frankel70,Stephenson70,Mittag71},
who showed that the low- and high-temperature
expansions of the classical Ising model on the square lattice
with nearest-neighbor interactions
were related by a one-to-one transformation of the temperature.
Common to both incarnations of duality is the following
defining property. If there exists a correspondence between a set
of observables
$\widehat{O}^{\,}_{\iota}$
labeled by the index $\iota$
whose (quantum) statistical properties are governed
by the (quantum) partition function $Z$
and a second set of observables
$\widehat{O}^{\vee}_{\iota}$
labeled by the index $\iota^{\vee}$
whose (quantum) statistical properties are governed
by the (quantum) partition function $Z^{\vee}$
such that the equality
\begin{equation}
\left\langle
\prod_{\iota}
\widehat{O}^{\,}_{\iota}
\right\rangle^{\,}_{Z}=
\left\langle
\prod_{\iota^{\vee}}
\widehat{O}^{\vee}_{\iota^{\vee}}
\right\rangle^{\,}_{Z^{\vee}}
\end{equation}
between their correlation functions hold,
then the pairs of observables
$(\widehat{O}^{\vee}_{\iota},\widehat{O}^{\vee}_{\iota^{\vee}})$
and the pair of partition functions
$(Z,Z^{\vee})$
form dual pairs.
The Jordan-Wigner duality was used by
Lieb, Schultz, and Mattis to show that the
quantum $XY$ spin-1/2 chain with
nearest-neighbor antiferromagnetic coupling is critical
\cite{Lieb61}. Kramers and Wannier predicted the value taken
by the transition temperature in the Ising model by postulating that 
it undergoes no more than one transition between the high- and
low-temperature phases.

It was recognized by McKean in 1964 that the Kramers‐Wannier
duality can be derived by means of the Poisson summation formula
for the Abelian group $\mathbb{Z}^{\,}_{2}$
\cite{McKean64,Drouffe78,Druhl82,Buchstaber03}.
In the 1970's, in connection with lattice gauge theories
\cite{Kogut79},
the interplay between global and local symmetries
in establishing dualities took
center stage starting with Kadanoff and Ceva on the one hand
and Wegner on the other hand 
\cite{Kadanoff71,Wegner71,Balian75,Jose77,Peskin78,Fradkin78,%
Korthals-Altes78,Bellissard79,Horn79,Drouffe79,Ukawa80,Savit80,Mathur16}.
The counterpart to lattice dualities in field theory is
bosonization
\cite{Coleman75,Mandelstam75,Witten84}.
Subtle signatures of lattice dualities in massive field theories
were investigated in Refs.\
\cite{Schroer79,BenTov15}.
An influential approach to dualities was proposed by
Fr\"ohlich et al.\ in 2004 who sought to
read off the possible strong/weak-coupling
dualities leaving a given critical model
fixed solely from knowledge of its universality class
\cite{Froelich04, Frohlich2009, Ruelle05, Brunner:2013ota, 
Carqueville:2012dk, Bhardwaj2017, Lin21}.
A field-theoretical generalization of this approach has
been used to study various possible strong/weak-coupling
as well as boson/fermion dualities \cite{Kapustin17,Karch19,Thorngren2020}.

The goal of this section is to treat
the Jordan-Wigner (JW) and Kramers-Wannier (KW) dualities
on equal footing. To this end, we are going to
review the construction of
Kramers-Wannier and Jordan-Wigner dualities
obeyed by lattice bond algebras
\cite{Cobanera10,Cobanera11}
following a gauging approach \cite{Radicevic18,Moradi23}.
Equipped with theses tools,
we will present our main results in
Sec.\
\ref{sec:Triality of crystalline and internal symmetries on a chain}
in which we study the fate of crystalline transformations of the lattice 
such as translation and reflection under the dualities (triality) of
Sec.\
\ref{sec:Triality through bond algebra isomorphisms}.

Starting from
$\mathbb{Z}^{\,}_{2}$-symmetric
quantum spin-$1/2$ $XYZ$ chains defined on the lattice
\begin{subequations}
\label{eq:def Lambda and Lambdavee}
\begin{equation}
\Lambda:=
\left\{\left.\vphantom{\frac{1}{2}}
j
\ \right|\
j=1,\cdots,2N
\right\},
\label{eq:def Lambda and Lambdavee a}
\end{equation}
we are thus going to gauge the global $\mathbb{Z}^{\,}_{2}$ symmetry
in two ways. The first way
delivers a bosonic bond algebra with global
$\mathbb{Z}^{\,}_{2}$-symmetry 
that is supported on the dual lattice
\begin{equation}
\Lambda^{\star}:=
\left\{\left.
j^{\star}\equiv j+\frac{1}{2}
\ \right|\
j\in\Lambda
\right\},
\label{eq:def Lambda and Lambdavee b}
\end{equation}
\end{subequations}
i.e., the links of the 
lattice $\Lambda$.
The second way delivers a
fermionic bond algebra with global $\mathbb{Z}^{\,}_{2}$ 
fermion parity symmetry
that is supported on the lattice $\Lambda$%
~\footnote{%
~As we shall explain in Sec.\ \ref{subsec:Jordan-Wigner duality b and f},
it will be convenient to implement a unitary transformation 
which renders the dual fermionic bond algebra on the lattice $\Lambda$.}.
We will then establish a triality between all three bond algebras,
i.e., any pair of the three bond algebras form dual pairs
provided appropriate consistency conditions are imposed.

\subsection{The $\mathbb{Z}^{\,}_{2}$ symmetric bond algebra}
\label{subsec:Definition of the bond algebra cal Bb}

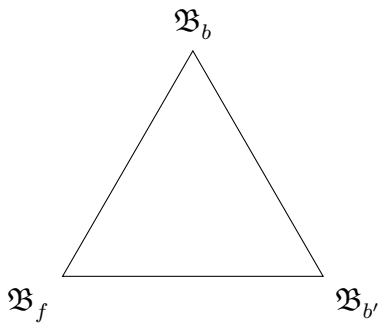
\begin{figure}[t!]
\begin{center}
\begin{tikzpicture}
\node[regular polygon, regular polygon sides=3, minimum size=4cm, draw] (T)
at (0,0) {};
\node at (T.corner 1)(A)[above]{$\mathfrak{B}^{\,}_{b}$};
\node at (T.corner 2)(A)[below left]{$\mathfrak{B}^{\,}_{f^{\vphantom{\star}}}$};
\node at (T.corner 3)(C)[below right]{$\mathfrak{B}^{\,}_{b^{\prime}}$};
\end{tikzpicture}
\end{center}
\caption{
Triality between the triplet of bond algebras
(\ref{eq:three trial bond algebras}).  
}
\label{fig:Triality diagram}
\end{figure}

To each site $j\in\Lambda$,
we assign the triplet $\hat{\bm{\sigma}}^{\,}_{j}$ of
operators whose components
$\hat{\sigma}^{\alpha}_{j}$
with $\alpha=x,y,z$ 
obey the Pauli algebra 
\begin{subequations}
\label{eq:def Pauli algebra on mathfrak{H}b}
\begin{equation}
\hat{\sigma}^{\alpha}_{j}\,
\hat{\sigma}^{\beta}_{j}=
\delta^{\alpha\beta}\,
\widehat{\mbb{1}}^{\,}_{\mathcal{H}^{\,}_{b}}
+
\mathrm{i}
\epsilon^{\alpha\beta\gamma}\,
\hat{\sigma}^{\gamma}_{j},
\qquad
\left[
\hat{\sigma}^{\alpha}_{i},
\hat{\sigma}^{\beta}_{j}
\right]=0,
\qquad
i<j\in\Lambda,
\label{eq:def Pauli algebra on mathfrak{H}b a}
\end{equation}
where $\alpha,\beta,\gamma=x,y,z$ and the summation convention over
repeated indices is implied.
We will be interested in organizing the sub-space of linear operators
that are symmetric with respect to a $\mathbb{Z}^{z}_{2}$
symmetry generated by
\begin{equation}
\widehat{U}^{\,}_{r^{z}_{\pi}}
:=
\prod_{j=1}^{2N}
\hat{\sigma}^{z}_{j}.
\label{eq:definition and symmetry of cal Bb b}
\end{equation}
Here, $\widehat{U}^{\,}_{r^{z}_{\pi}}$
implements
a global rotation by $\pi$ about the $z$ axis in internal
spin-1/2 space attached to the lattice $\Lambda$.
We consider general symmetry twisted boundary conditions labeled by
$b =0,1\in\mathbb{Z}^{z}_{2}$ as
\begin{equation}
\begin{split}
\hat{\sigma}^{x}_{j+2N}&=
\left(\widehat{U}^{\,}_{r^{z}_{\pi}}\right)^{b}\,
\hat{\sigma}^{x}_{j}\,
\left(\widehat{U}^{\dag}_{r^{z}_{\pi}}\right)^{b}
= (-1)^{b}\,\widehat{\sigma}^{x}_{j}, \\   
\hat{\sigma}^{y}_{j+2N}&=
\left(\widehat{U}^{\,}_{r^{z}_{\pi}}\right)^{b}\,
\hat{\sigma}^{y}_{j}\,
\left(\widehat{U}^{\dag}_{r^{z}_{\pi}}\right)^{b}
= (-1)^{b}\,\widehat{\sigma}^{y}_{j}, \\   
\hat{\sigma}^{z}_{j+2N}&=
\left(\widehat{U}^{\,}_{r^{z}_{\pi}}\right)^{b}\,
\hat{\sigma}^{z}_{j}\,
\left(\widehat{U}^{\dag}_{r^{z}_{\pi}}\right)^{b}
= \widehat{\sigma}^{z}_{j}.  
\end{split}
\label{eq:def Pauli algebra on mathfrak{H}b b}
\end{equation}
These operators act on the
$2^{2N}$-dimensional Hilbert space
\begin{equation}
\begin{split}
\mathcal{H}^{\,}_{b}:=&\,
\mathrm{span}\,
\left\{\left.
\bigotimes\limits_{j\in\Lambda}
\left(
\frac{
\hat{\sigma}^{x}_{j}
-
\mathrm{i}
\hat{\sigma}^{y}_{j}
}
{
2
}
\right)^{n^{\,}_{j}}
|\uparrow\rangle^{\,}_{j}
\ \right|\
n^{\,}_{j}=0,1,
\qquad
\hat{\sigma}^{z}_{j}\,|\uparrow\rangle^{\,}_{j}=|\uparrow\rangle^{\,}_{j}
\right\}
\cong
\mathbb{C}^{2^{2N}}.
\end{split}
\label{eq:def Pauli algebra on mathfrak{H}b c}
\end{equation}
\end{subequations}
We define the bond algebra
\begin{equation}
\mathfrak{B}^{\,}_{b}\equiv
\left\langle
\hat{\sigma}^{z}_{j},
\quad
\hat{\sigma}^{x}_{j}\,
\hat{\sigma}^{x}_{j+1}
\ \Big|\
j\in\Lambda
\right\rangle,
\label{eq:definition and symmetry of cal Bb}
\end{equation}
that is spanned by all complex-valued linear combinations
of products of the generators
$\hat{\sigma}^{z}_{i}$
and
$\hat{\sigma}^{x}_{j}\,
\hat{\sigma}^{x}_{j+1}$
for any $i,j\in\Lambda$.
The bond algebra $\mathfrak{B}^{\,}_{b}$
is equivalent to the algebra of all operators
acting on the Hilbert space $\mathcal{H}^{\,}_{b}$
that are symmetric under the $\mathbb{Z}^{z}_{2}$ symmetry.
Since the algebra $\mathfrak{B}^{\,}_{b}$ is $\mathbb{Z}^{z}_{2}$-symmetric, 
all operators in it can be block diagonalized into eigenspaces of 
$\widehat{U}^{\,}_{r^{z}_{\pi}}$.
Correspondingly, it is also convenient to decompose the Hilbert space
(\ref{eq:def Pauli algebra on mathfrak{H}b c})
in terms of the definite eigenvalues
of the operator $\widehat{U}^{\,}_{r^{z}_{\pi}}$,
i.e., the decomposition
\begin{align}
\label{eq:decomposition H_b}
\mathcal{H}^{\,}_{b}=
\mathcal{H}^{\,}_{b;\,+}
\oplus
\mathcal{H}^{\,}_{b;\,-},\qquad 
\mathcal{H}^{\,}_{b;\,\pm}:=
\frac{1}{2}
\left(
\widehat{\mbb{1}}^{\,}_{\mathcal{H}^{\,}_{b}}
\pm
\widehat{U}^{\,}_{r^{z}_{\pi}}
\right)
\mathcal{H}^{\,}_{b}\,.
\end{align}
In what follows, we are going to construct two additional bond algebras
$\mathfrak{B}^{\,}_{b^{\prime}}$
and
$\mathfrak{B}^{\,}_{f^{\vphantom{\star}}}$
and explain under what conditions any pair of the triplet of bond algebras
\begin{equation}
\mathfrak{B}^{\,}_{b},
\qquad
\mathfrak{B}^{\,}_{b^{\prime}},
\qquad
\mathfrak{B}^{\,}_{f^{\vphantom{\star}}}
\label{eq:three trial bond algebras}
\end{equation}
are dual to each other. 
If we place
each one of the three bond algebras at the vertices of
a triangle as is done in Fig.\ \ref{fig:Triality diagram},
we may interpret each side
of this triangle as a duality relation. 
We call this web
of dualities triality. 
The strategy that we shall use to establish each of
the dualities consists of the following three steps:
\begin{enumerate}
\item
We gauge the global $\mathbb{Z}^{z}_{2}$
symmetry by extending the Hilbert space to include gauge degrees of freedom.
This is done in two ways which correspond
to introducing bosonic or fermionic gauge degrees of freedom,
respectively. 
\item
We perform a unitary transformation on the extended Hilbert space,
which includes both the matter and gauge degrees of freedom.
This unitary effectively localizes the Gauss constraints
on the matter degrees of freedom. 
\item
We solve the Gauss constraints and project onto the gauge invariant
subspace of the extended Hilbert space. Upon doing so, the matter
degrees of freedom freeze out, thus delivering the dual bond
algebra. 
\end{enumerate}
These dualities are invertible and therefore the same procedure
can be carried out starting from either
the bond algebra
$\mathfrak{B}^{\,}_{b^{\prime}}$
or
$\mathfrak{B}^{\,}_{f^{\vphantom{\star}}}$,
as we shall detail below.

\subsection{Bosonic gauging and the Kramers-Wannier duality}
\label{subsec:Kramers-Wannier duality b to b'}

To gauge the global $\mathbb{Z}^{z}_{2}$ symmetry 
generated by the unitary operator
\eqref{eq:definition and symmetry of cal Bb b}, we introduce
$\mathbb{Z}^{\,}_{2}$-valued gauge fields on the links of the lattice
$\Lambda$.  In other words, to each site $j^{\star}\in\Lambda^{\star}$
of the dual lattice, we assign the triplet
$\hat{\bm{\tau}}^{\,}_{j^{\star}}$ of operators whose components
$\hat{\tau}^{\alpha}_{j^{\star}}$ with $\alpha=x,y,z$ obey the Pauli
algebra
\begin{subequations}
\label{eq:def Pauli algebra on mathfrak{H}b'}
\begin{align}
\hat{\tau}^{\alpha}_{j^{\star}}\,
\hat{\tau}^{\beta}_{j^{\star}}=
\delta^{\alpha\beta}\,
\widehat{\mbb{1}}^{\,}_{\mathcal{H}^{\,}_{b^{\prime}}}
+
\mathrm{i}
\epsilon^{\alpha\beta\gamma}\,
\hat{\tau}^{\gamma}_{j^{\star}},
\qquad
\left[
\hat{\tau}^{\alpha}_{i^{\star}},
\hat{\tau}^{\beta}_{j^{\star}}
\right]=0,
\qquad
i^{\star}<j^{\star}\in\Lambda^{\star}\,.
\label{eq:def Pauli algebra on mathfrak{H}b' a}
\end{align}
In analogy with \eqref{eq:def Pauli algebra on mathfrak{H}b b},
we also introduce the twisted boundary conditions
\begin{align}
\hat{\tau}^{x}_{j^{\star}+2N}=
(-1)^{b^{\prime}}
\hat{\tau}^{x}_{j^{\star}},
\quad
\hat{\tau}^{y}_{j^{\star}+2N}=
(-1)^{b^{\prime}}
\hat{\tau}^{y}_{j^{\star}},
\quad
\hat{\tau}^{z}_{j^{\star}+2N}=
\hat{\tau}^{z}_{j^{\star}},
\qquad
b^{\prime}=0,1.
\label{eq:def Pauli algebra on mathfrak{H}b' b}
\end{align}
In what follows, we will see that $b'$ plays an important role in
understanding how symmetry eigensectors map under gauging.
These link operators act on the $2^{2N}$-dimensional Hilbert space
\begin{equation}
\small
\begin{split}
\mathcal{H}^{\,}_{b^{\prime}}:=&\,
\mathrm{span}\,
\left\{\left.
\bigotimes\limits_{j^{\star}\in\Lambda^{\star}}
\left(
\frac{
\hat{\tau}^{x}_{j^{\star}}
-
\mathrm{i}
\hat{\tau}^{y}_{j^{\star}}
}
{
2
}  
\right)^{n^{\,}_{j^{\star}}}\,
|\uparrow\rangle^{\,}_{j^{\star}}
\ \right|\
n^{\,}_{j^{\star}}=0,1,
\quad
\hat{\tau}^{z}_{j^{\star}}\,|\uparrow\rangle^{\,}_{j^{\star}}=
|\uparrow\rangle^{\,}_{j^{\star}}
\right\}
\cong
\mathbb{C}^{2^{2N}}.
\end{split}
\label{eq:def Pauli algebra on mathfrak{H}b' c}
\end{equation}
\end{subequations}
We define the extended bond algebra
\begin{subequations}
\label{eq:def bond algebra cal Bbb'}
\begin{equation}
\mathfrak{B}^{\,}_{b,b^{\prime}}\equiv
\left\langle
\hat{\sigma}^{z}_{j},
\quad
\hat{\sigma}^{x}_{j}\,
\hat{\tau}^{z}_{j^{\star}}\,
\hat{\sigma}^{x}_{j+1}
\ \Big|\
j\in\Lambda
\right\rangle,
\label{eq:def bond algebra cal Bbb' a}
\end{equation}
where $j^{\star}:=j+1/2$ as defined in Eq.\ 
\eqref{eq:def Lambda and Lambdavee b}
(this is always assumed throughout the paper).
Any operator in 
$\mathfrak{B}^{\,}_{b,b^{\prime}}$
acts on the extended $2^{4N}$-dimensional Hilbert space
\begin{equation}
\mathcal{H}^{\,}_{b,b^{\prime}}:=
\mathcal{H}^{\,}_{b}
\otimes
\mathcal{H}^{\,}_{b^{\prime}}\,.
\label{eq:def bond algebra cal Bbb' b}
\end{equation}
\end{subequations}
What distinguishes
$\mathfrak{B}^{\,}_{b,b^{\prime}}$
from the set of all operators on $\mathcal{H}^{\,}_{b,b^{\prime}}$
is that any element from
$\mathfrak{B}^{\,}_{b,b^{\prime}}$
is invariant under conjugation with any one
of the $2N$ local unitary operators (i.e., Gauss operators)
\begin{subequations}
\label{eq:sym bond algebra cal Bbb'}
\begin{equation}
\widehat{G}^{\,}_{b,b^{\prime};j}:=
\hat{\tau}^{x}_{j^{\star}-1}\,
\hat{\sigma}^{z}_{j}\,
\hat{\tau}^{x}_{j^{\star}}=
\widehat{G}^{\,}_{b,b^{\prime};j+2N}
\label{eq:sym bond algebra cal Bbb' (a)}
\end{equation}
that satisfy the conditions
\begin{equation}
\left(\widehat{G}^{\,}_{b,b^{\prime};j}\right)^{2}=
\widehat{\mbb{1}}^{\,}_{\mathcal{H}^{\,}_{b,b^{\prime}}},
\qquad
\left[
\widehat{G}^{\,}_{b,b^{\prime};i},
\widehat{G}^{\,}_{b,b^{\prime};j}
\right]=0,
\qquad
\prod_{j}\widehat{G}^{\,}_{b,b^{\prime};j}
=
(-1)^{b'}\,
\widehat{U}^{\,}_{r^{z}_\pi}\,.
\label{eq:sym bond algebra cal Bbb' (a) (b)}
\end{equation}
\end{subequations} 

Products of operators $\widehat{G}^{\,}_{b,b^{\prime};j}$
over subsets of $\Lambda$
generate local $\mathbb{Z}^{\,}_{2}$ gauge transformations. 
A generic local $\mathbb{Z}^{\,}_{2}$ gauge transformation 
\begin{subequations}
\label{eq:extended bond algebra local transformations KW}
\begin{align}
&
\hat{\sigma}^{x}_{j}
\mapsto 
(-1)^{\lambda^{\,}_{j}}\,
\hat{\sigma}^{x}_{j},
\qquad
\hat{\sigma}^{y}_{j}
\mapsto 
(-1)^{\lambda^{\,}_{j}}\,
\hat{\sigma}^{y}_{j},
\qquad\qquad\,\,
\hat{\sigma}^{z}_{j}
\mapsto 
\hat{\sigma}^{z}_{j},
\\
&
\hat{\tau}^{x}_{j^{\star}}
\mapsto 
\hat{\tau}^{x}_{j^{\star}},
\qquad\qquad\,\,\,\,
\hat{\tau}^{y}_{j^{\star}}
\mapsto 
(-1)^{\lambda^{\,}_{j+1}-\lambda^{\,}_{j}}\,
\hat{\tau}^{y}_{j^{\star}},
\qquad
\hat{\tau}^{z}_{j^{\star}}
\mapsto 
(-1)^{\lambda^{\,}_{j+1}-\lambda^{\,}_{j}}\,
\hat{\tau}^{z}_{j^{\star}},
\end{align}
with $\lambda^{\,}_{j}=0,1$ is implemented by the operator
\begin{align}
\widehat{G}^{\,}_{b,b';\bm{\lambda}}:=
\prod_{j=1}^{2N}
\left(
\widehat{G}^{\,}_{b,b^{\prime};j}
\right)^{\lambda^{\,}_{j}},
\qquad
\bm{\lambda} = (\lambda^{\,}_{1},\, \cdots, \,\lambda^{\,}_{2N}),
\label{eq:sym bond algebra cal Bbb' c}
\end{align}
\end{subequations}
and specified by the string of $\mathbb{Z}^{\,}_{2}$-valued scalars
$\lambda^{\,}_{j}=0,1$ with $j=1,\cdots,2N$.
While the bond algebra
\eqref{eq:def bond algebra cal Bbb' a}
is invariant under any local transformations
\eqref{eq:extended bond algebra local transformations KW},
these are yet to be associated to \textit{gauge symmetries} or,
equivalently, redundancies in our description%
~\footnote{%
~To make the connection with gauge theories in $(1+1)$-dimensional
spacetime continuum, we observe that $\hat{\tau}^{x}_{j^{\star}}$
is reminiscent of a 
$\mathbb{Z}^{\,}_{2}$-valued electric field $\sim e^{\mathrm{i} E}$, 
while $\hat{\tau}^{z}_{j^{\star}}$ is reminiscent
of a $\mathbb{Z}^{\,}_{2}$-valued gauge field $\sim e^{\mathrm{i} A}$.
Accordingly, under the local transformation 
specified by a $\mathbb{Z}^{\,}_{2}$-valued 
scalar field $\bm{\lambda}$, the 
electric field is invariant,
while the gauge field changes by $\pi\,\mathrm{d}\bm{\lambda}$.
}.
We elevate the local transformations 
\eqref{eq:extended bond algebra local transformations KW} 
to local gauge symmetries
by requiring that any two states in the Hilbert space
$\mathcal{H}^{\,}_{b,b^{\prime}}$ are equivalent if they are 
related by a gauge transformation. In particular,
we demand that any state
$\ket{\psi^{\,}_{\mathrm{phys}}}$
is a physical one if and only if 
\begin{subequations}
\label{eq:pure gauge theory condition KW}
\begin{align}
\widehat{G}^{\,}_{b,b^{\prime};j}\,
\ket{\psi^{\,}_{\mathrm{phys}}}=
+
\ket{\psi^{\,}_{\mathrm{phys}}},
\label{eq:pure gauge theory condition KW a}
\end{align}
for any $j=1,\cdots,2N$. The choice of $+$ sign for 
any $j$ corresponds to a background with no $\mathbb{Z}^{\,}_{2}$ matter, i.e.,
we have a pure gauge theory. Observe that
Eqs.\
(\ref{eq:pure gauge theory condition KW a})
and
(\ref{eq:sym bond algebra cal Bbb' (a) (b)})
imply that
\begin{align}
\widehat{U}^{\,}_{r^{z}_\pi}\,
\ket{\psi^{\,}_{\mathrm{phys}}}=
(-1)^{b'}\,
\ket{\psi^{\,}_{\mathrm{phys}}}.
\label{eq:pure gauge theory condition KW b}
\end{align} 
\end{subequations}

We project the extended Hilbert space
(\ref{eq:def bond algebra cal Bbb' b})
into the gauge invariant sector where
condition \eqref{eq:pure gauge theory condition KW} holds. This is facilitated
by first performing the unitary transformation~\cite{Moradi23}
(see also Refs.\
\cite{Horn79,Cobanera13,Santos15,Mathur16,
Tantivasadakarn21})
\begin{subequations}
\label{eq:def extended unitary on Hilbert bb'}
\begin{equation}
\widehat{U}^{\,}_{b,b^{\prime}}:=
\prod_{j=1}^{2N}
\widehat{U}^{\,}_{b,b^{\prime};j^{\star}}, \qquad \widehat{U}^{\,}_{b,b^{\prime};j^{\star}}
:=
\widehat{P}^{\,\,+}_{b,b^{\prime};j^{\star}}
+
\hat{\tau}^{x}_{j^{\star}}\,
\widehat{P}^{\,\,-}_{b,b^{\prime};j^{\star}}
\label{eq:def extended unitary on Hilbert bb' a}
\end{equation}
with pairwise commuting projectors
\begin{equation}
\widehat{P}^{\,\,\pm}_{b,b^{\prime};j^{\star}}:=
\frac{1}{2}
\left(
\widehat{\mbb{1}}^{\,}_{\mathcal{H}^{\,}_{b,b^{\prime}}}
\pm
\hat{\sigma}^{x}_{j}\,
\hat{\sigma}^{x}_{j+1}
\right),
\qquad
j^{\star}\in \Lambda^{\star}\,.
\label{eq:def extended unitary on Hilbert bb' c}
\end{equation}
\end{subequations}
For any $j=1,\cdots,2N$, there follows the transformation laws
\begin{subequations}
\label{eq:trsf law under unitary for KW duality}
\begin{align}
&
\widehat{U}^{\,}_{b,b^{\prime}}\,
\hat{\sigma}^{x}_{j}\,
\left(
\widehat{U}^{\,}_{b,b^{\prime}}
\right)^{\dag}=
\hat{\sigma}^{x}_{j},
\qquad
&
\widehat{U}^{\,}_{b,b^{\prime}}\,
\hat{\sigma}^{z}_{j}\,
\left(
\widehat{U}^{\,}_{b,b^{\prime}}
\right)^{\dag}=
\hat{\tau}^{x}_{j^{\star}-1}\,
\hat{\sigma}^{z}_{j}\,
\hat{\tau}^{x}_{j^{\star}},
\label{eq:trsf law under unitary for KW duality a}
\\
&
\widehat{U}^{\,}_{b,b^{\prime}}\,
\hat{\tau}^{x}_{j^{\star}}\,
\left(
\widehat{U}^{\,}_{b.b'}
\right)^{\dag}=
\hat{\tau}^{x}_{j^{\star}},
\qquad
&
\widehat{U}^{\,}_{b,b^{\prime}}\,
\hat{\tau}^{z}_{j^{\star}}\,
\left(
\widehat{U}^{\,}_{b,b^{\prime}}
\right)^{\dag}=
\hat{\sigma}^{x}_{j}\,
\hat{\tau}^{z}_{j^{\star}}\,
\hat{\sigma}^{x}_{j+1},
\label{eq:trsf law under unitary for KW duality b}
\end{align}
for the generators of the Pauli algebras on the lattices $\Lambda$
and $\Lambda^{\star}$ together with the image 
\begin{equation}
\widehat{U}^{\,}_{b,b^{\prime}}\,
\widehat{G}^{\,}_{b,b^{\prime};j}\,
\left(\widehat{U}^{\,}_{b,b^{\prime}}\,\right)^{\dag}=
\hat{\sigma}^{z}_{j}
\label{eq:trsf law under unitary for KW duality c}
\end{equation}
\end{subequations}
of the local Gauss operator.

Thus, projection onto the subspace
where the condition \eqref{eq:pure gauge theory condition KW}
holds amounts to 
setting the action of $\hat{\sigma}^{z}_{j}$
on physical states to the identity ($\hat{\sigma}^{z}_{j}\equiv1$)
after the unitary transformation
\eqref{eq:trsf law under unitary for KW duality}.
More concretely, if we define the projector
\begin{subequations}
\label{eq:def gauss projector KW duality}
\begin{equation}
\widehat{P}^{\,}_{b,b^{\prime};\mathrm{G}}:=
\prod_{j\in\Lambda}
\frac{1}{2}
\left[
\widehat{\mbb{1}}^{\,}_{\mathcal{H}^{\,}_{b,b^{\prime}}}
+
\widehat{U}^{\,}_{b,b^{\prime}}\,
\widehat{G}^{\,}_{b,b';\,j}\,
\left(\widehat{U}^{\,}_{b,b^{\prime}}\,\right)^{\dag}
\right]\,
\label{eq:def gauss projector KW duality a}
\end{equation}
onto the $2^{2N}$-dimensional gauge-invariant subspace
\begin{equation}
\mathcal{H}^{\,\vee}_{b^{\prime}}:=
\widehat{P}^{\,}_{b,b^{\prime};\mathrm{G}}\,
\mathcal{H}^{\,}_{b,b^{\prime}}\subset
\mathcal{H}^{\,}_{b,b^{\prime}},
\label{eq:def gauss projector KW duality b}
\end{equation}
\end{subequations}
we find that the $2N$ triplets of projected operators
\begin{subequations}
\label{eq:dual Pauli algebra for KW duality}
\begin{align}
\hat{\tau}^{x\,\vee}_{j^{\star}}:=&\,
\widehat{P}^{\,}_{b,b^{\prime};\mathrm{G}}
\left[
\widehat{U}^{\,}_{b,b^{\prime}}\,
\hat{\tau}^{x}_{j^{\star}}\,
\left(\widehat{U}^{\,}_{b,b^{\prime}}\right)^{\dag}
\right]
\widehat{P}^{\,}_{b,b^{\prime};\mathrm{G}}
\nonumber\\
=&\,
\widehat{P}^{\,}_{b,b^{\prime};\mathrm{G}}\,
\hat{\tau}^{x}_{j^{\star}}\,
\widehat{P}^{\,}_{b,b^{\prime};\mathrm{G}},
\label{eq:dual Pauli algebra for KW duality a}
\\
\hat{\tau}^{y\,\vee}_{j^{\star}}:=&\,
\widehat{P}^{\,}_{b,b^{\prime};\mathrm{G}}
\left[
\widehat{U}^{\,}_{b,b^{\prime}}\,
\hat{\sigma}^{x}_{j}\,
\hat{\tau}^{y}_{j^{\star}}\,
\hat{\sigma}^{x}_{j+1}\,
\left(\widehat{U}^{\,}_{b,b^{\prime}}\right)^{\dag}
\right]
\widehat{P}^{\,}_{b,b^{\prime};\mathrm{G}}
\nonumber\\
=&\,
\widehat{P}^{\,}_{b,b^{\prime};\mathrm{G}}\,
\hat{\tau}^{y}_{j^{\star}}\,
\widehat{P}^{\,}_{b,b^{\prime};\mathrm{G}},
\label{eq:dual Pauli algebra for KW duality b}
\\
\hat{\tau}^{z\,\vee}_{j^{\star}}:=&\,
\widehat{P}^{\,}_{b,b^{\prime};\mathrm{G}}
\left[
\widehat{U}^{\,}_{b,b^{\prime}}\,
\hat{\sigma}^{x}_{j}\,
\hat{\tau}^{z}_{j^{\star}}\,
\hat{\sigma}^{x}_{j+1}\,
\left(\widehat{U}^{\,}_{b,b^{\prime}}\right)^{\dag}
\right]
\widehat{P}^{\,}_{b,b^{\prime};\mathrm{G}}
\nonumber\\
=&\,
\widehat{P}^{\,}_{b,b^{\prime};\mathrm{G}}\,
\hat{\tau}^{z}_{j^{\star}}\,
\widehat{P}^{\,}_{b,b^{\prime};\mathrm{G}},  
\label{eq:dual Pauli algebra for KW duality c}
\end{align}
\end{subequations}
realize a Pauli algebra on the Hilbert space
$\mathcal{H}^{\,\vee}_{b^{\prime}}$ that is isomorphic to
the Pauli algebra
(\ref{eq:def Pauli algebra on mathfrak{H}b' a})
on the Hilbert space
$\mathcal{H}^{\,}_{b^{\prime}}$.
This implies that the projection to
$\mathcal{H}^{\,\vee}_{b^{\prime}}$
of the bond algebra
$\mathfrak{B}^{\,}_{b,b^{\prime}}$ 
delivers the dual bond algebra 
\begin{align}
\mathfrak{B}^{\,}_{b^{\prime}}
:=&\,
\widehat{P}^{\,}_{b,b^{\prime};\mathrm{G}}
\left[
\widehat{U}^{\,}_{b,b^{\prime}}\,
\mathfrak{B}^{\,}_{b,b^{\prime}}\,
\left(\widehat{U}^{\,}_{b,b^{\prime}}\right)^{\dag}
\right]
\widehat{P}^{\,}_{b,b^{\prime};\mathrm{G}}\,
\nonumber\\
=&\,
\left\langle
\hat{\tau}^{x\,\vee}_{j^{\star}-1}\,
\hat{\tau}^{x\,\vee}_{j^{\star}},
\quad
\hat{\tau}^{z\,\vee}_{j^{\star}}
\ \Big|\
j^{\star}\in\Lambda^{\star}
\right\rangle.
\label{eq:definition and symmetry of cal Bb'}
\end{align}
This is the bond algebra of operators that 
are symmetric under the dual $\mathbb{Z}^{z^{\vee}}_{2}$
symmetry with the generator
\begin{equation}
\widehat{U}^{\,\vee}_{r^{z}_{\pi}}:=
\prod_{j^{\star}\in\Lambda^{\star}}
\hat{\tau}^{z\,\vee}_{j^{\star}}
\label{eq:definition and symmetry of cal Bb' b}
\end{equation}
of the global rotation by $\pi$ about the $z$ axis in internal
spin-1/2 space attached to the dual lattice $\Lambda^{\star}$.
Note that the twisted boundary conditions in
\eqref{eq:def Pauli algebra on mathfrak{H}b' b}
were nothing but symmetry twisted boundary conditions with respect to
$\mathbb{Z}^{z^{\vee}}_{2}$.
As was done for the Hilbert space $\mathcal{H}^{\,}_{b}$ in Eq.\
\eqref{eq:decomposition H_b},
it is convenient to decompose the Hilbert space $\mathcal{H}^{\vee}_{b'}$
in terms of the definite eigenvalue sectors of the operator
$\widehat{U}^{\vee}_{r^{z}_{\pi}}$,
i.e., the decomposition
\begin{subequations}
\label{eq:decomposition Hvee_b'}
\begin{align}
\mathcal{H}^{\vee}_{b'}
=
\mathcal{H}^{\vee}_{b';\,+}
\oplus
\mathcal{H}^{\vee}_{b';\,-}
\end{align}
holds, where
\begin{align}
\mathcal{H}^{\vee}_{b';\,\pm}
:=
\frac{1}{2}
\left(
\widehat{\mbb{1}}^{\,}_{\mathcal{H}^{\vee}_{b'}}
\pm
\widehat{U}^{\vee}_{r^{z}_{\pi}}
\right)
\mathcal{H}^{\vee}_{b'}.
\end{align}
\end{subequations}

The duality between the bond algebras
\eqref{eq:definition and symmetry of cal Bb}
and 
\eqref{eq:definition and symmetry of cal Bb'}
demands the following consistency conditions.
Because of the twisted boundary conditions
(\ref{eq:def Pauli algebra on mathfrak{H}b b})
and
(\ref{eq:def Pauli algebra on mathfrak{H}b' b}),
the pair of operators
\begin{subequations}
\begin{equation}
\left(
\hat{\sigma}^{z}_{j},
\qquad
\hat{\tau}^{x\,\vee}_{j^{\star}-1}\,
\hat{\tau}^{x\,\vee}_{j^{\star}}
\right)
\end{equation}
and the pair of operators
\begin{equation}
\left(
\hat{\sigma}^{x}_{j}\,
\hat{\sigma}^{x}_{j+1},
\qquad
\hat{\tau}^{z\,\vee}_{j^{\star}}
\right)
\end{equation}
\end{subequations}
each form a dual pair if and only if the pair of operators
\begin{subequations}
\label{eq:consistency duality b to b'}
\begin{align}
\left(
\prod_{j=1}^{2N}
\hat{\sigma}^{z}_{j}=
\widehat{U}^{\,}_{r^{z}_{\pi}},
\qquad
\prod_{j=1}^{2N}
\left(
\hat{\tau}^{x\,\vee}_{j^{\star}-1}\,
\hat{\tau}^{x\,\vee}_{j^{\star}}
\right)=
(-1)^{b^{\prime}}\,
\widehat{\mbb{1}}^{\,}_{\mathcal{H}^{\,\vee}_{b^{\prime}}}
\right)
\label{eq:consistency duality b to b' a}
\end{align}
and the pair of operators
\begin{align}
\left(
\prod_{j=1}^{2N}
\left(
\hat{\sigma}^{x}_{j}\,
\hat{\sigma}^{x}_{j+1}
\right)
=(-1)^{b}\,
\widehat{\mbb{1}}^{\,}_{\mathcal{H}^{\,}_{b}},
\qquad
\prod_{j=1}^{2N}
\hat{\tau}^{z\,\vee}_{j^{\star}}=:
\widehat{U}^{\,\vee}_{r^{z}_{\pi}}
\right)
\label{eq:consistency duality b to b' b}
\end{align}
\end{subequations}
each form a dual pair, respectively.
This is to say that duality between 
the bond algebras 
\eqref{eq:definition and symmetry of cal Bb}
and 
\eqref{eq:definition and symmetry of cal Bb'}
holds only on the $2^{2N-1}$-dimensional subspaces
\begin{subequations}
\label{eq:final dualities between domain defs if b to b'}
\begin{align}
&
\mathcal{H}^{\,}_{b;\,(-1)^{b'}}
=
\frac{1}{2}
\left[
\widehat{\mbb{1}}^{\,}_{\mathcal{H}^{\,}_{b}}
+
(-1)^{b^{\prime}}\,
\widehat{U}^{\,}_{r^{z}_{\pi}}
\right]
\mathcal{H}^{\,}_{b},
\label{eq:final dualities between domain defs if b to b' (a)}
\\
&
\mathcal{H}^{\vee}_{b';\,(-1)^{b}}
=
\frac{1}{2}
\left[
\widehat{\mbb{1}}^{\,}_{\mathcal{H}^{\,\vee}_{b^{\prime}}}
+
(-1)^{b}\,
\widehat{U}^{\,\vee}_{r^{z}_{\pi}}
\right]
\mathcal{H}^{\,\vee}_{b^{\prime}},
\label{eq:final dualities between domain defs if b to b' (b)}
\end{align}
\end{subequations}
of $2^{2N}$-dimensional Hilbert spaces 
$\mathcal{H}^{\,}_{b}$ and $\mathcal{H}^{\vee}_{b'}$, respectively.
This duality between the bond algebras 
\eqref{eq:definition and symmetry of cal Bb}
and 
\eqref{eq:definition and symmetry of cal Bb'}
acting on Hilbert spaces
\eqref{eq:final dualities between domain defs if b to b' (a)}
and
\eqref{eq:final dualities between domain defs if b to b' (b)},
respectively,
is nothing but the Kramers-Wannier (KW) duality. 
Under the KW duality,
the boundary conditions ($b=0,1$) of the bond algebra
\eqref{eq:definition and symmetry of cal Bb}
dictates the eigenvalue of the generator $\widehat{U}^{\vee}_{r^{z}_{\pi}}$
of the global dual symmetry,
while the eigenvalue of the generator $\widehat{U}^{\,}_{r^{z}_{\pi}}$ 
of the global symmetry that was gauged dictates the 
boundary conditions ($b'=0,1$) of the dual bond algebra
\eqref{eq:definition and symmetry of cal Bb'}.
Table \ref{Table:Kramers-Wannier dualization b to b'}
summarizes this correspondence (see
Ref.\ \cite{Kapustin17} for an alternative 
field-theoretical derivation of this mapping of the 
symmetry eigensectors under the KW duality).

\begin{table}
\caption{
\label{Table:Kramers-Wannier dualization b to b'} 
The four Kramers-Wannier dualizations that follow from the
consistency conditions
(\ref{eq:final dualities between domain defs if b to b'}).
The first column specifies the twisted boundary conditions.
The choice of twisted boundary conditions is selected by
$b=0,1$ prior to dualization.
The choice of twisted boundary conditions is selected by
$b^{\prime}=0,1$ after dualization.
The second column gives the dual subspace of the
Hilbert space prior to dualization.
The third column gives the dual subspace of the
Hilbert space after dualization.
}
\centering
\begin{tabular}{l|cccc}
\hline \hline
$(b,b^{\prime})$
&\hphantom{AAAAA}&
$\mathcal{H}^{\,}_{b;\,(-1)^{b'}}$
&\hphantom{AAAAA}&
$\mathcal{H}^{\vee}_{b';\,(-1)^{b}}$
\\
\hline
$(0,0)$
&\hphantom{AAAAA}&
$
\frac{1}{2}
\left(
\widehat{\mbb{1}}^{\,}_{\mathcal{H}^{\,}_{b}}
+
\widehat{U}^{\,}_{r^{z}_{\pi}}
\right)
\mathcal{H}^{\,}_{b}
$
&\hphantom{AAAAA}&
$
\frac{1}{2}
\left(
\widehat{\mbb{1}}^{\,}_{\mathcal{H}^{\,\vee}_{b^{\prime}}}
+
\widehat{U}^{\,\vee}_{r^{z}_{\pi}}
\right)
\mathcal{H}^{\,\vee}_{b^{\prime}}
$
\\
$(1,0)$
&\hphantom{AAAAA}&
$
\frac{1}{2}
\left(
\widehat{\mbb{1}}^{\,}_{\mathcal{H}^{\,}_{b}}
+
\widehat{U}^{\,}_{r^{z}_{\pi}}
\right)
\mathcal{H}^{\,}_{b}
$
&\hphantom{AAAAA}&
$
\frac{1}{2}
\left(
\widehat{\mbb{1}}^{\,}_{\mathcal{H}^{\,\vee}_{b^{\prime}}}
-
\widehat{U}^{\,\vee}_{r^{z}_{\pi}}
\right)
\mathcal{H}^{\,\vee}_{b^{\prime}}
$
\\
$(0,1)$
&\hphantom{AAAAA}&
$
\frac{1}{2}
\left(
\widehat{\mbb{1}}^{\,}_{\mathcal{H}^{\,}_{b}}
-
\widehat{U}^{\,}_{r^{z}_{\pi}}
\right)
\mathcal{H}^{\,}_{b}
$
&\hphantom{AAAAA}&
$
\frac{1}{2}
\left(
\widehat{\mbb{1}}^{\,}_{\mathcal{H}^{\,\vee}_{b^{\prime}}}
+
\widehat{U}^{\,\vee}_{r^{z}_{\pi}}
\right)
\mathcal{H}^{\,\vee}_{b^{\prime}}
$
\\
$(1,1)$
&\hphantom{AAAAA}&
$
\frac{1}{2}
\left(
\widehat{\mbb{1}}^{\,}_{\mathcal{H}^{\,}_{b}}
-
\widehat{U}^{\,}_{r^{z}_{\pi}}
\right)
\mathcal{H}^{\,}_{b}
$
&\hphantom{AAAAA}&
$
\frac{1}{2}
\left(
\widehat{\mbb{1}}^{\,}_{\mathcal{H}^{\,\vee}_{b^{\prime}}}
-
\widehat{U}^{\,\vee}_{r^{z}_{\pi}}
\right)
\mathcal{H}^{\,\vee}_{b^{\prime}}
$
\\
\hline \hline
\end{tabular}
\end{table}

\subsection{Fermionic gauging and the Jordan-Wigner duality}
\label{subsec:Jordan-Wigner duality b and f}

We now describe a distinct fermionic gauging of the
global $\mathbb{Z}^{z}_{2}$
symmetry with generator Eq.~\eqref{eq:definition and symmetry of cal Bb b}.
In contrast to the previous section, we introduce a pair of 
Majorana operators on every link of the lattice $\Lambda$.
These represent fermionic gauge degrees of freedom.
More precisely, to each site $j^{\star}\in\Lambda^{\star}$,
we assign the pair
$\hat{\beta}^{\,}_{j^{\star}}=\hat{\beta}^{\dag}_{j^{\star}}$
and
$\hat{\alpha}^{\,}_{j^{\star}}=\hat{\alpha}^{\dag}_{j^{\star}}$
of Majorana operators obeying the Clifford algebra
\begin{subequations}
\label{eq:def Majorana algebra on mathfrak{H}bF}
\begin{align}
\begin{split}
&
\left\{\vphantom{\hat{\beta}^{\,}_{i^{\star}}}
\hat{\alpha}^{\,}_{i^{\star}},
\hat{\alpha}^{\,}_{j^{\star}}
\right\}=
\left\{
\hat{\beta}^{\,}_{i^{\star}},
\hat{\beta}^{\,}_{j^{\star}}
\right\}=
2\delta^{\,}_{i^{\star},j^{\star}}\,
\widehat{\mbb{1}}^{\,}_{\mathcal{H}^{\,}_{f^{\vphantom{\star}}}},
\qquad
\left\{
\hat{\alpha}^{\,}_{i^{\star}},
\hat{\beta}^{\,}_{j^{\star}}
\right\}=0,
\quad
i^{\star},j^{\star}\in\Lambda^{\star}.
\end{split}
\label{eq:def Majorana algebra on mathfrak{H}bF a}
\end{align}
We also introduce the fermion parity operator 
\begin{equation}
\widehat{P}_{\mathrm{F}}:=
\prod_{j^\star\in\Lambda^{\star}}
\left(
\mathrm{i}
\hat{\beta}_{j^{\star}}\,
\hat{\alpha}_{j^{\star}}
\right),
\label{eq:def Majorana algebra on mathfrak{H}bF b}
\end{equation}
together with the cyclic group
\begin{equation}
\mathbb{Z}^{\mathrm{F}}_{2}=
\left\{
p^{\,}_{\mathrm{F}},\
\left(p^{\,}_{\mathrm{F}}\right)^{2}\equiv e
\right\}
\label{eq:def Majorana algebra on mathfrak{H}bF c}
\end{equation}
of order two with $p^{\,}_{\mathrm{F}}$ represented by
$\widehat{P}_{\mathrm{F}}$. The pair $\widehat{P}_{\mathrm{F}}$
and
$\mathbb{Z}^{\mathrm{F}}_{2}$ 
will play a central role in what follows.
We work in a Hilbert space with boundary conditions
twisted with respect to the fermion parity operator, i.e.,
\begin{equation}
\begin{split}
\hat{\alpha}_{j^{\star}+2N}&=
\left(\widehat{P}^{\,}_{\mathrm{F}}\right)^{f}
\hat{\alpha}_{j^{\star}}
\left(\widehat{P}^{\dagger}_{\mathrm{F}}\right)^{f}    
=
(-1)^{f}\,
\hat{\alpha}^{\,}_{j^{\star}},
\\
\hat{\beta}^{\,}_{j^{\star}+2N}
&=
\left(\widehat{P}^{\,}_{\mathrm{F}}\right)^{f}
\hat{\beta}^{\,}_{j^{\star}}
\left(\widehat{P}^{\dagger}_{\mathrm{F}}\right)^{f}
=
(-1)^{f}\,
\hat{\beta}^{\,}_{j^{\star}},
\end{split}
\label{eq:def Majorana algebra on mathfrak{H}bF d} 
\end{equation}
where $f=0,1$. These
$2N$ doublets of Majorana operators act on the
$2^{2N}$-dimensional Hilbert space
\begin{equation}
\begin{split}
\mathcal{H}^{\,}_{f^{\vphantom{\star}}}:=&\,
\mathrm{span}\,
\left\{\left.
\left[\,
\prod\limits_{j^{\star}\in\Lambda^{\star}}
\left(
\frac{
\hat{\beta}^{\,}_{j^{\star}}
-
\mathrm{i}
\hat{\alpha}^{\,}_{j^{\star}}
}
{
2
}
\right)^{ n^{\,}_{j^{\star}}}\,
\right]
|0\rangle
\ \right|\
n^{\,}_{j^{\star}}=0,1,
\qquad
\frac{
\hat{\beta}^{\,}_{j^{\star}}
+
\mathrm{i}
\hat{\alpha}^{\,}_{j^{\star}}
}
{
2
}\,
|0\rangle=0
\right\}
\cong
\mathbb{C}^{2^{2N}}.
\end{split}
\label{eq:def Majorana algebra on mathfrak{H}bF e}
\end{equation}
\end{subequations}

We define the extended bond algebra
\begin{subequations}
\label{eq:def bond algebra cal Bbf}
\begin{equation}
\mathfrak{B}^{\,}_{b,f^{\vphantom{\star}}}\equiv
\left\langle
\hat{\sigma}^{z}_{j},
\quad
\hat{\sigma}^{x}_{j}\,
\left(
\mathrm{i}
\hat{\beta}^{\,}_{j^{\star}}\,
\hat{\alpha}^{\,}_{j^{\star}}\,
\right)
\hat{\sigma}^{x}_{j+1}
\ \Big|\
j\in\Lambda
\right\rangle\,
\label{eq:def bond algebra cal Bbf a}
\end{equation}
that is spanned by all complex-valued linear combinations
of products of the generators
$\hat{\sigma}^{z}_{i}$
and
$\hat{\sigma}^{x}_{j}\,
\left(
\mathrm{i}
\hat{\beta}^{\,}_{j^{\star}}\,
\hat{\alpha}^{\,}_{j^{\star}}\,
\right)
\hat{\sigma}^{x}_{j+1}$
for any $i,j\in\Lambda$.
Any element of
$\mathfrak{B}^{\,}_{b,f^{\vphantom{\star}}}$
acts on the extended $2^{4N}$-dimensional Hilbert space
\begin{equation}
\mathcal{H}^{\,}_{b,f^{\vphantom{\star}}}:=
\mathcal{H}^{\,}_{b}
\otimes
\mathcal{H}^{\,}_{f^{\vphantom{\star}}}\,.
\label{eq:def bond algebra cal Bbf b}
\end{equation}
\end{subequations}
What distinguishes
$\mathfrak{B}^{\,}_{b,f}$
from the set of all operators on
$\mathcal{H}^{\,}_{b,f}$
is that any element from
$\mathfrak{B}^{\,}_{b,f}$
is invariant under conjugation with any one of the $2N$
local unitary operators (i.e., Gauss operators)
\begin{subequations}
\label{eq:sym bond algebra cal Bbf}    
\begin{equation}
\widehat{G}^{\,}_{b,f;j}:=
\mathrm{i}
\hat{\beta}^{\,}_{j^{\star}-1}\,
\hat{\sigma}^{z}_{j}\,
\hat{\alpha}^{\,}_{j^{\star}}
=
\widehat{G}^{\,}_{b,f;j+2N}
\label{eq:sym bond algebra cal Bbf (a)}    
\end{equation}
that satisfy the conditions
\begin{equation}
\left(\widehat{G}^{\,}_{b,f;j}\right)^{2}=  
\widehat{\mbb{1}}^{\,}_{\mathcal{H}^{\,}_{b,f^{\vphantom{\star}}}},
\qquad
\left[
\widehat{G}^{\,}_{b,f;i},
\widehat{G}^{\,}_{b,f;j}
\right]=0,
\qquad
\prod_{j\in \Lambda}\widehat{G}^{\,}_{b,f;j}=
(-1)^{f}\,
\widehat{P}_{\mathrm{F}}\,
\widehat{U}^{\,}_{r^{z}_{\pi}}\,.
\label{eq:sym bond algebra cal Bbf (b)}    
\end{equation}
\end{subequations}

Products of operators $\widehat{G}^{\,}_{b,f;j}$
over subsets of $\Lambda$ generate $\mathbb{Z}^{\,}_{2}$ gauge 
transformations. A generic $\mathbb{Z}^{\,}_{2}$ gauge transformation
\begin{subequations}
\label{eq:extended bond algebra local transformations JW}
\begin{align}
&
\hat{\sigma}^{x}_{j}
\mapsto 
(-1)^{\lambda^{\,}_{j}}\,
\hat{\sigma}^{x}_{j},
\qquad
\hat{\sigma}^{y}_{j}
\mapsto 
(-1)^{\lambda^{\,}_{j}}\,
\hat{\sigma}^{y}_{j},
\qquad\qquad\,\,
\hat{\sigma}^{z}_{j}
\mapsto 
\hat{\sigma}^{z}_{j},
\\
&
\hat{\alpha}^{\,}_{j^{\star}}
\mapsto 
(-1)^{\lambda^{\,}_{j}}\,
\hat{\alpha}^{\,}_{j^{\star}},
\quad\,\,
\hat{\beta}^{\,}_{j^{\star}}
\mapsto 
(-1)^{\lambda^{\,}_{j+1}}\,
\hat{\beta}^{\,}_{j^{\star}},
\end{align}
with $\lambda^{\,}_{j}=0,1$ is implemented by the operator
\begin{align}
\widehat{G}^{\,}_{b,f;\bm{\lambda}}:=
\prod_{j=1}^{2N}
\left(
\widehat{G}^{\,}_{b,f;j}
\right)^{\lambda^{\,}_{j}},
\qquad
\bm{\lambda} = (\lambda^{\,}_{1},\, \cdots, \,\lambda^{\,}_{2N}),
\label{eq:sym bond algebra cal Bbf c}
\end{align}
\end{subequations}
and specified by the string of $\mathbb{Z}^{\,}_{2}$-valued scalars
$\lambda^{\,}_{j}=0,1$ with $j=1,\cdots,2N$. 
While the bond algebra
\eqref{eq:def bond algebra cal Bbf a}
is invariant under any local transformations
\eqref{eq:extended bond algebra local transformations JW},
these are yet to be associated to \textit{gauge symmetries} or,
equivalently, redundancies in our description%
~\footnote{%
~To make the connection with gauge theories in
$(1+1)$-dimensional continuum,
we observe that both
$\mathrm{i}\hat{\beta}^{\,}_{j^{\star}-1}\,\hat{\alpha}^{\,}_{j^{\star}}$
and
$\mathrm{i}\hat{\beta}^{\,}_{j^{\star}}\,\hat{\alpha}^{\,}_{j^{\star}}$
are reminiscent of a 
$\mathbb{Z}^{\,}_{2}$-valued electric field
$\sim e^{\mathrm{i} E}$
and a $\mathbb{Z}^{\,}_{2}$-valued gauge field $\sim e^{\mathrm{i} A}$
in that they obey the same transformation laws under local
$\mathbb{Z}^{\,}_{2}$ gauge transformations, respectively. 
}.
We elevate the local transformations 
\eqref{eq:extended bond algebra local transformations JW} 
to local gauge symmetries
by requiring that any two states in the Hilbert space
$\mathcal{H}^{\,}_{b,f}$ are equivalent if they are 
related by a gauge transformation. In particular,
we demand that any state
$\ket{\psi^{\,}_{\mathrm{phys}}}$
is a physical one if and only if
\begin{subequations}
\label{eq:pure gauge theory condition JW}
\begin{align}
\widehat{G}^{\,}_{b,f;j}\,
\ket{\psi^{\,}_{\mathrm{phys}}}
=
+
\ket{\psi^{\,}_{\mathrm{phys}}}
\label{eq:pure gauge theory condition JWa}
\end{align}
for any $j=1,\cdots,2N$. The choice of $+$ sign for 
any $j$ corresponds to a background with no $\mathbb{Z}^{\,}_{2}$ matter,
i.e., we have a pure gauge theory. Observe that
Eqs.\
(\ref{eq:pure gauge theory condition JWa})
and
(\ref{eq:sym bond algebra cal Bbf (b)})
imply that (compare with Eq. (\ref{eq:pure gauge theory condition KW b}))
\begin{align}
\widehat{U}^{\,}_{r^{z}_{\pi}}\,
\ket{\psi^{\,}_{\mathrm{phys}}}
=
(-1)^{f}\, \widehat{P}_{\mathrm{F}}\,
\ket{\psi^{\,}_{\mathrm{phys}}}\,.
\label{eq:pure gauge theory condition JWb}
\end{align}  
\end{subequations}

We project the extended Hilbert space
(\ref{eq:def bond algebra cal Bbf b})
into the gauge invariant sector where condition
\eqref{eq:pure gauge theory condition JW}
holds. This is facilitated
by first performing the unitary transformation 
\begin{subequations}
\label{eq:def extended unitary on Hilbert bf}
\begin{equation}
\widehat{U}^{\,}_{b,f}:=
\prod_{j=1}^{2N}
\widehat{U}^{\,}_{b,f;j},
\label{eq:def extended unitary on Hilbert bf a}
\end{equation}
where
\begin{equation}
\begin{split}
\widehat{U}^{\,}_{b,f;j}:=&\,
\left(
\hat{\sigma}^{x}_{j}
\right)^{
\widehat{P}^{\,\,-}_{b,f;j}
}
=
\widehat{P}^{\,\,+}_{b,f;j}
+
\hat{\sigma}^{x}_{j}\,
\widehat{P}^{\,\,-}_{b,f;j}=
\widehat{U}^{\,}_{b,f;j+2N},
\end{split}
\label{eq:def extended unitary on Hilbert bf b}
\end{equation}
with pairwise commuting projectors
\begin{equation}
\widehat{P}^{\,\,\pm}_{b,f;j}:=
\frac{1}{2}
\left(
\widehat{\mbb{1}}^{\,}_{\mathcal{H}^{\,}_{b,f^{\vphantom{\star}}}}
\pm
\mathrm{i}
\hat{\beta}^{\,}_{j^{\star}-1}\,
\hat{\alpha}^{\,}_{j^{\star}}\,
\right),
\qquad
j\in\Lambda.
\label{eq:def extended unitary on Hilbert bf c}
\end{equation}
\end{subequations}
For any $j\in\Lambda$, there follows the transformation rules
\begin{subequations}
\label{eq:trsf law under unitary for JW duality}
\begin{equation}
\begin{alignedat}{2}
&
\widehat{U}^{\,}_{b,f}\,
\hat{\sigma}^{x}_{j}\,
\left(
\widehat{U}^{\,}_{b,f}
\right)^{\dag}=
\hat{\sigma}^{x}_{j},
\qquad \qquad \quad
\widehat{U}^{\,}_{b,f}\,
\hat{\sigma}^{z}_{j}\,
\left(
\widehat{U}^{\,}_{b,f}
\right)^{\dag}&&=
\mathrm{i}
\hat{\beta}^{\,}_{j^{\star}-1}\,
\hat{\sigma}^{z}_{j}\,
\hat{\alpha}^{\,}_{j^{\star}}, \\
&
\widehat{U}^{\,}_{b,f}\,
\hat{\beta}^{\,}_{j^{\star}}\,
\left(
\widehat{U}^{\,}_{b,f}
\right)^{\dag}=
\hat{\beta}^{\,}_{j^{\star}}\,
\hat{\sigma}^{x}_{j+1},
\qquad
\widehat{U}^{\,}_{b,f}\,
\hat{\alpha}^{\,}_{j^{\star}}\,
\left(
\widehat{U}^{\,}_{b,f}
\right)^{\dag}&&=
\hat{\sigma}^{x}_{j}\,
\hat{\alpha}^{\,}_{j^{\star}}
\end{alignedat}
\end{equation}
for the spin operators on the lattice
$\Lambda$
and Majorana operators on the dual lattice
$\Lambda^{\star}$
together with the image
\begin{equation}
\widehat{U}^{\,}_{b,f}\,
\widehat{G}^{\,}_{b,f;j}\,
\left(\widehat{U}^{\,}_{b,f}\,\right)^{\dag}=
\hat{\sigma}^{z}_{j}.
\label{eq:trsf law under unitary for JW duality c}
\end{equation}
\end{subequations}
Thus, projection onto the subspace where the condition
\eqref{eq:pure gauge theory condition JW}
holds amounts to setting the action of $\hat{\sigma}^{z}_{j}$
on physical states to the identity ($\hat{\sigma}^{z}_{j}\equiv1$)
after the unitary transformation
\eqref{eq:trsf law under unitary for JW duality}.
More concretely, if we define the projector
\begin{subequations}
\label{eq:def gauss projector JW duality}
\begin{equation}
\widehat{P}^{\,}_{b,f;\mathrm{G}}:=
\prod_{j\in\Lambda}
\frac{1}{2}
\left[
\widehat{\mbb{1}}^{\,}_{\mathcal{H}^{\,}_{b,f^{\vphantom{\star}}}}
+
\widehat{U}^{\,}_{b,f}\,
\widehat{G}^{\,}_{b,f;j}\,
\left(\widehat{U}^{\,}_{b,f}\,\right)^{\dag}
\right]
\label{eq:def gauss projector JW duality a}
\end{equation}
onto the $2^{2N}$-dimensional gauge-invariant subspace
\begin{equation}
\mathcal{H}^{\,\vee}_{f^{\vphantom{\star}}}:=
\widehat{P}^{\,}_{b,f;\mathrm{G}}\,
\mathcal{H}^{\,}_{b,f^{\vphantom{\star}}}\subset
\mathcal{H}^{\,}_{b,f^{\vphantom{\star}}},
\label{eq:def gauss projector JW duality b}
\end{equation}
\end{subequations}
we find that the $2N$ doublets of projected operators
\begin{subequations}
\label{eq:dual Majorana algebra for JW duality}
\begin{align}
\hat{\beta}^{\vee}_{j+1}:=&\,
\widehat{P}^{\,}_{b,f;\mathrm{G}}
\left[
\widehat{U}^{\,}_{b,f}\,
\left(
\hat{\beta}^{\,}_{j^{\star}}\,
\hat{\sigma}^{x}_{j+1}
\right)
\left(\widehat{U}^{\,}_{b,f}\right)^{\dag}
\right]
\widehat{P}^{\,}_{b,f;\mathrm{G}}
\nonumber\\
=&\,
\widehat{P}^{\,}_{b,f;\mathrm{G}}\,
\hat{\beta}^{\,}_{j^{\star}}\,
\widehat{P}^{\,}_{b,f;\mathrm{G}},
\label{eq:dual Majorana algebra for JW duality a}
\\  
\hat{\alpha}^{\vee}_{j}:=&\, \ \
\widehat{P}^{\,}_{b,f;\mathrm{G}}
\left[
\widehat{U}^{\,}_{b,f}\,
\left(
\hat{\sigma}^{x}_{j}\,
\hat{\alpha}^{\,}_{j^{\star}}
\right)
\left(\widehat{U}^{\,}_{b,f}\right)^{\dag}
\right]
\widehat{P}^{\,}_{b,f;\mathrm{G}}
\nonumber\\
=&\,
\widehat{P}^{\,}_{b,f;\mathrm{G}}\,
\hat{\alpha}^{\,}_{j^{\star}}\,
\widehat{P}^{\,}_{b,f;\mathrm{G}},
\label{eq:dual Majorana algebra for JW duality b}
\end{align}
\end{subequations}
realize a Clifford algebra on the Hilbert space
$\mathcal{H}^{\,\vee}_{f^{\vphantom{\star}}}$ that is isomorphic to
the Clifford algebra
(\ref{eq:def Majorana algebra on mathfrak{H}bF a})
on the Hilbert space
$\mathcal{H}^{\,}_{f^{\vphantom{\star}}}$.
The lattice label that we choose for
$\hat{\beta}^{\vee}_{j+1}$
and
$\hat{\alpha}^{\vee}_{j}$
is a matter of convention since the relation between 
$j$ and $j^{\star}=j+\frac{1}{2}$
is one to one%
~\footnote{%
~This choice implies a relative 
translation of the $\hat{\beta}^{\vee}_{j}$
operators compared to the $\hat{\alpha}^{\vee}_{j}$
operators. It is done
to simplify the discussion of the phase diagram
in Sec.\
\ref{sec:Application to quantum spin-1/2 degrees of freedom on a chain}.
As we shall see in Sec.\
\ref{subsec:Triality betweeb cal Bf cal Bb' cal Bb},
while such a ``half''-translation is a unitary transformation
on the fermionic bond algebras, it corresponds to implementing the 
KW duality described in Sec.\
\ref{subsec:Kramers-Wannier duality b to b'}
on the bosonic bond algebras obtained by gauging fermion parity.
}.
We also find that the projection to
$\mathcal{H}^{\,\vee}_{f^{\vphantom{\star}}}$
of the bond algebra
$\mathfrak{B}^{\,}_{b,f^{\vphantom{\star}}}$
is the bond algebra
\begin{subequations}
\label{eq:definition and symmetry of cal Bf}
\begin{align}
\mathfrak{B}^{\,}_{f^{\vphantom{\star}}}:=&\,
\left\langle
\mathrm{i}
\hat{\beta}^{\vee}_{j}\,
\hat{\alpha}^{\vee}_{j},
\qquad
\mathrm{i}
\hat{\beta}^{\vee}_{j+1}\,
\hat{\alpha}^{\vee}_{j}
\ \Big|\
j\in\Lambda
\right\rangle,
\label{eq:definition and symmetry of cal Bf a}
\end{align}
which is the algebra of operators invariant under conjugation by the generator
\begin{equation}
\widehat{P}^{\,\vee}_{\mathrm{F}}:=
\prod_{j\in\Lambda}
\left(
\mathrm{i}
\hat{\beta}^{\vee}_{j}\,
\hat{\alpha}^{\vee}_{j}
\right)
\label{eq:definition and symmetry of cal Bf b}
\end{equation}
\end{subequations}
of a global fermion-parity symmetry $\mathbb{Z}^{\mathrm{F}}_{2}$.
As was done for the Hilbert space $\mathcal{H}^{\,}_{b}$ in Eq.\
\eqref{eq:decomposition H_b},
it is convenient to decompose the Hilbert space $\mathcal{H}^{\vee}_{f}$
to the definite eigenvalue sectors of the operator
$\widehat{P}^{\vee}_{\mathrm{F}}$,
i.e., the decomposition
\begin{subequations}
\label{eq:decomposition Hvee_f}
\begin{align}
\mathcal{H}^{\vee}_{f}
=
\mathcal{H}^{\vee}_{f;\,+}
\oplus
\mathcal{H}^{\vee}_{f;\,-},
\end{align}
holds where
\begin{align}
\mathcal{H}^{\vee}_{f;\,\pm}
:=
\frac{1}{2}
\left(
\widehat{\mbb{1}}^{\,}_{\mathcal{H}^{\vee}_{f}}
\pm
\widehat{P}^{\vee}_{\mathrm{F}}
\right)
\mathcal{H}^{\vee}_{f}.
\end{align}
\end{subequations}
The duality between the bond algebras
(\ref{eq:definition and symmetry of cal Bb})
and
(\ref{eq:definition and symmetry of cal Bf})
demands certain consistency conditions.
In particular, the twisted boundary conditions
(\ref{eq:def Pauli algebra on mathfrak{H}b b})
and
(\ref{eq:def Majorana algebra on mathfrak{H}bF d}) require that
the pairs of operators
\begin{subequations}
\label{eq:consistency duality b to bF}
\begin{equation}
\left(
\hat{\sigma}^{z}_{j},\qquad
\mathrm{i}
\hat{\beta}^{\vee}_{j}\,
\hat{\alpha}^{\vee}_{j}
\right)
\end{equation}
and
\begin{equation}
\left(
\hat{\sigma}^{x}_{j}\,
\hat{\sigma}^{x}_{j+1},
\qquad
\mathrm{i}
\hat{\beta}^{\vee}_{j+1}\,
\hat{\alpha}^{\vee}_{j}
\right)
\end{equation}
each form a dual pair
if and only if  the pairs of operators
\begin{align}
\left(
\prod_{j=1}^{2N}
\hat{\sigma}^{z}_{j}=
\widehat{U}^{\,}_{r^{z}_{\pi}},
\qquad
\prod_{j=1}^{2N}
\left(
\mathrm{i}
\hat{\beta}^{\vee}_{j}\,
\hat{\alpha}^{\vee}_{j}
\right)=
\widehat{P}^{\,\vee}_{\mathrm{F}}
\right)
\label{eq:consistency duality b to bF a}
\end{align}
and 
\begin{align}
\left(
\prod_{j=1}^{2N}
\left(
\hat{\sigma}^{x}_{j}\,
\hat{\sigma}^{x}_{j+1}
\right)
=(-1)^{b}\,
\widehat{\mbb{1}}^{\,}_{\mathcal{H}^{\,}_{b}},
\qquad
\prod_{j=1}^{2N}
\left(
\mathrm{i}
\hat{\beta}^{\vee}_{j+1}\,
\hat{\alpha}^{\vee}_{j}
\right)=
(-1)^{f^{\vphantom{\star}}+ 1}\,
\widehat{P}^{\,\vee}_{\mathrm{F}}
\right)
\label{eq:consistency duality b to bF b}
\end{align}
\end{subequations}
each form a dual pair, respectively.
This is to say that the duality between 
the bond algebras 
\eqref{eq:definition and symmetry of cal Bb}
and 
\eqref{eq:definition and symmetry of cal Bf}
holds only on the $2^{2N-1}$-dimensional subspaces
\begin{subequations}
\label{eq:final dualities between domain defs if b to bF}
\begin{align}
&
\mathcal{H}^{\,}_{b;\,(-1)^{b+f+1}}
=
\frac{1}{2}
\left[
\widehat{\mbb{1}}^{\,}_{\mathcal{H}^{\,}_{b}}
+
(-1)^{b+f+1}\,
\widehat{U}^{\,}_{r^{z}_{\pi}}
\right]\mathcal{H}^{\,}_{b},
\label{eq:final dualities between domain defs if b to bF (a)}
\\
&
\mathcal{H}^{\,\vee}_{f;\,(-1)^{b+f+1}}
=
\frac{1}{2}
\left[
\widehat{\mbb{1}}^{\,}_{\mathcal{H}^{\,\vee}_{f^{\vphantom{\star}}}}
+
(-1)^{b+f+1}\,
\widehat{P}^{\,\vee}_{\mathrm{F}}
\right]
\mathcal{H}^{\,\vee}_{f^{\vphantom{\star}}},    
\label{eq:final dualities between domain defs if b to bF (b)}
\end{align}
\end{subequations}of $2^{2N}$-dimensional Hilbert spaces 
$\mathcal{H}^{\,}_{b}$ and $\mathcal{H}^{\vee}_{f}$, respectively.
This is the
Jordan-Wigner (JW) duality. 
Table \ref{Table:Jordan-Wigner dualization b to bF}
summarizes the correspondence between symmetry eigensectors
on either side of this duality
(see Refs.\ \cite{Kapustin17, Seiberg2023} for an alternative 
field-theoretical derivation of the mapping of symmetry
eigensectors under the JW duality).

\begin{table}
\caption{
\label{Table:Jordan-Wigner dualization b to bF}
The four Jordan-Wigner dualizations that follow from
the consistency conditions
(\ref{eq:final dualities between domain defs if b to bF}).
The first column specifies the twisted boundary conditions.
The choice of twisted boundary conditions is selected by
$b=0,1$ prior to dualization.
The choice of twisted boundary conditions is selected by
$f^{\vphantom{\star}}=0,1$ after dualization.
The second column gives the dual subspace of the
Hilbert space prior to dualization.
The third column gives the dual subspace of the
Hilbert space after dualization.
}
\centering
\begin{tabular}{l|cccc}
\hline \hline
$(b,f^{\vphantom{\star}})$
&\hphantom{AAAAA}&
$\mathcal{H}^{\,}_{b;\,(-1)^{b+f+1}}$
&\hphantom{AAAAA}&
$\mathcal{H}^{\,\vee}_{f;\,(-1)^{b+f+1}}$
\\
\hline
$(0,0)$
&\hphantom{AAAAA}&
$
\frac{1}{2}
\left(
\widehat{\mbb{1}}^{\,}_{\mathcal{H}^{\,}_{b}}
-
\widehat{U}^{\,}_{r^{z}_{\pi}}
\right)
\mathcal{H}^{\,}_{b}
$
&\hphantom{AAAAA}&
$
\frac{1}{2}
\left(
\widehat{\mbb{1}}^{\,}_{\mathcal{H}^{\,\vee}_{f^{\vphantom{\star}}}}
-
\widehat{P}^{\,\vee}_{\mathrm{F}}
\right)
\mathcal{H}^{\,\vee}_{f^{\vphantom{\star}}}
$
\\
$(1,0)$
&\hphantom{AAAAA}&
$
\frac{1}{2}
\left(
\widehat{\mbb{1}}^{\,}_{\mathcal{H}^{\,}_{b}}
+
\widehat{U}^{\,}_{r^{z}_{\pi}}
\right)
\mathcal{H}^{\,}_{b}
$
&\hphantom{AAAAA}&
$
\frac{1}{2}
\left(
\widehat{\mbb{1}}^{\,}_{\mathcal{H}^{\,\vee}_{f^{\vphantom{\star}}}}
+
\widehat{P}^{\,\vee}_{\mathrm{F}}
\right)
\mathcal{H}^{\,\vee}_{f^{\vphantom{\star}}}
$
\\
$(0,1)$
&\hphantom{AAAAA}&
$
\frac{1}{2}
\left(
\widehat{\mbb{1}}^{\,}_{\mathcal{H}^{\,}_{b}}
+
\widehat{U}^{\,}_{r^{z}_{\pi}}
\right)
\mathcal{H}^{\,}_{b}
$
&\hphantom{AAAAA}&
$
\frac{1}{2}
\left(
\widehat{\mbb{1}}^{\,}_{\mathcal{H}^{\,\vee}_{f^{\vphantom{\star}}}}
+
\widehat{P}^{\,\vee}_{\mathrm{F}}
\right)
\mathcal{H}^{\,\vee}_{f^{\vphantom{\star}}}
$
\\
$(1,1)$
&\hphantom{AAAAA}&
$
\frac{1}{2}
\left(
\widehat{\mbb{1}}^{\,}_{\mathcal{H}^{\,}_{b}}
-
\widehat{U}^{\,}_{r^{z}_{\pi}}
\right)
\mathcal{H}^{\,}_{b}
$
&\hphantom{AAAAA}&
$
\frac{1}{2}
\left(
\widehat{\mbb{1}}^{\,}_{\mathcal{H}^{\,\vee}_{f^{\vphantom{\star}}}}
-
\widehat{P}^{\,\vee}_{\mathrm{F}}
\right)
\mathcal{H}^{\,\vee}_{f^{\vphantom{\star}}}
$
\\
\hline \hline
\end{tabular}
\end{table}

\subsection{A triality of bond algebras}
\label{subsec:Triality betweeb cal Bf cal Bb' cal Bb}

In Sec.\
\ref{subsec:Kramers-Wannier duality b to b'},
we gauged the internal global symmetry group
$\mathbb{Z}^{z}_{2}$
of the bond algebra $\mathfrak{B}^{\,}_{b}$
defined in Eq.\ (\ref{eq:definition and symmetry of cal Bb})
by minimal coupling to the local generator
$\hat{\tau}^{z}_{j^{\star}}$
of rotation by $\pi$ about the $z$ axis in internal spin-1/2 space
of the site $j^{\star}\in\Lambda^{\star}$
with the help of the local Gauss operator defined in Eq.\
(\ref{eq:sym bond algebra cal Bbb'}).
In Sec.\
\ref{subsec:Jordan-Wigner duality b and f},
we gauged instead the internal global symmetry group
$\mathbb{Z}^{z}_{2}$
by minimal coupling to the local generator
$\mathrm{i}\hat{\beta}^{\,}_{j^{\star}}\,\hat{\alpha}^{\,}_{j^{\star}}$
of fermion-parity on the site $j^{\star} \in \Lambda^{\star}$,
with the help of the local Gauss operator defined in Eq.\
(\ref{eq:sym bond algebra cal Bbf}).

To complete the triality%
~\footnote{%
~We leave it to the reader to construct the two dualizations
$\mathfrak{B}^{\,}_{b}$
and
$\mathfrak{B}^{\,}_{f}$
by gauging the bond algebra
$\mathfrak{B}^{\,}_{b^{\prime}}$.
}, we construct the two dualizations
$\mathfrak{B}^{\,}_{b^{\prime}}$
and
$\mathfrak{B}^{\,}_{b}$
of the bond algebra
\begin{subequations}
\label{eq:def bond algebra cal Bf}
\begin{align}
&
\mathfrak{B}^{\,}_{f}\equiv
\left\langle
\mathrm{i}\hat{\beta}^{\,}_{j}\,\hat{\alpha}^{\,}_{j}, \ 
\mathrm{i}\hat{\beta}^{\,}_{j+1}\,\hat{\alpha}^{\,}_{j}
\ \Big|\
j\in\Lambda
\right\rangle,
\label{eq:def bond algebra cal Bf a}
\end{align}
where the Majorana operators
$\hat{\alpha}^{\,}_{j}=\hat{\alpha}^{\dag}_{j}$
and
$\hat{\beta}^{\,}_{j}=\hat{\beta}^{\dag}_{j}$ 
with $i,j\in\Lambda$ satisfy the Clifford algebra
\begin{align}
\left\{\vphantom{\hat{\beta}^{\,}_{j}}
\hat{\alpha}^{\,}_{j},\,
\hat{\alpha}^{\,}_{j'}
\right\}=
\left\{
\hat{\beta}^{\,}_{j},\,
\hat{\beta}^{\,}_{j'}
\right\}=
2\,\delta^{\,}_{j,j'},
\qquad
\left\{
\hat{\alpha}^{\,}_{j},\,
\hat{\beta}^{\,}_{j'}
\right\}
=
0,
\label{eq:def bond algebra cal Bf b}
\end{align}
and obey the fermion-parity twisted boundary conditions
\begin{align}
\hat{\alpha}^{\,}_{j+2N}=
(-1)^{f}\,
\hat{\alpha}^{\,}_{j},
\qquad
\hat{\beta}^{\,}_{j+2N}
=
(-1)^{f}\,
\hat{\beta}^{\,}_{j},
\qquad
f=0,1,
\label{eq:def bond algebra cal Bf c}
\end{align}
for any $j,j'\in\Lambda$.
The domain of definition of these $2N$ doublets of Majorana operators is the 
Hilbert space
\begin{equation}
\begin{split}
\mathcal{H}^{\,}_{f}:=&\,
\mathrm{span}\,
\left\{\left.
\left[
\prod\limits_{j=1}^{2N}
\left(
\frac{
\hat{\beta}^{\,}_{j}
-
\mathrm{i}
\hat{\alpha}^{\,}_{j}
}
{
2
}
\right)^{ n^{\,}_{j}}
\right]
|0\rangle
\ \right|\
n^{\,}_{j}=0,1,
\quad
\frac{
\hat{\beta}^{\,}_{j}
+
\mathrm{i}
\hat{\alpha}^{\,}_{j}
}
{
2
}\,
|0\rangle=0
\right\}
\cong
\mathbb{C}^{2^{2N}}.
\end{split}
\label{eq:def bond algebra cal Bf d}
\end{equation}
\end{subequations}
The bond algebra (\ref{eq:def bond algebra cal Bf a})
is symmetric under conjugation by the global fermion parity%
~\footnote{%
~The rational for choosing the multiplicative real-valued phase factor
on the right-hand side of Eq.\
(\ref{eq:definition and symmetry of cal Bf b})
will be given in Sec.\
\ref{subsec:G total frak H f=0}.
}
\begin{subequations}
\begin{align}
\widehat{P}^{\,}_{\mathrm{F}}:=&\,
\prod_{j=1}^{2N}
\mathrm{i}
\hat{\beta}^{\,}_{j}\,
\hat{\alpha}^{\,}_{j}
\label{eq:def fermionic parity triality a}
\\
=&\,
(-1)^{f+1}
\prod_{j=1}^{2N}
\mathrm{i}
\hat{\beta}^{\,}_{j}\,
\hat{\alpha}^{\,}_{j+1}.
\label{eq:def fermionic parity triality b}
\end{align}
\end{subequations}
We are going to show that
gauging the global fermion parity symmetry generated by
the representation
(\ref{eq:def fermionic parity triality a})
of $\widehat{P}^{\,}_{\mathrm{F}}$
delivers the bond algebra $\mathfrak{B}^{\,}_{b^{\prime}}$
on the dual lattice,
while gauging the global fermion parity symmetry generated by
the representation
(\ref{eq:def fermionic parity triality b})
of $\widehat{P}^{\,}_{\mathrm{F}}$
delivers the bond algebra $\mathfrak{B}^{\,}_{b}$
on the dual lattice.

\subsubsection{Unit-cell preserving gauging of fermion parity}
\label{subsubsec:Unit-cell preserving gauging of fermion parity}

We trade the bond algebra $\mathfrak{B}^{\,}_{f}$ defined in
Eq.\ (\ref{eq:def bond algebra cal Bf})
by the minimally coupled bond algebra
\begin{subequations}
\label{eq:def cal Bextf if unit-cell preserving gauging PF} 
\begin{align}
&
\mathfrak{B}^{\,}_{f,b^{\prime}}:=
\left\langle
\mathrm{i}
\hat{\beta}^{\,}_{j}\,
\hat{\alpha}^{\,}_{j},
\
\mathrm{i}
\hat{\beta}^{\,}_{j+1}\,
\hat{\tau}^{z}_{j^{\star}}\,
\hat{\alpha}^{\,}_{j}
\ \Big|\
j\in\Lambda
\right\rangle
\label{eq:def cal Bextf if unit-cell preserving gauging PF a}  
\end{align}
with the tensor product
\begin{equation}
\mathcal{H}^{\,}_{f,b^{\prime}}:=
\mathcal{H}^{\,}_{f}
\otimes
\mathcal{H}^{\,}_{b^{\prime}}
\label{eq:def cal Bextf if unit-cell preserving gauging PF b} 
\end{equation}
\end{subequations}
as domain of definition
[here, $\mathcal{H}^{\,}_{b^{\prime}}$ was defined in Eq.\
(\ref{eq:def Pauli algebra on mathfrak{H}b'})].
This extended  bond algebra is symmetric with respect to any one of the $2N$
pairwise-commuting Gauss operators
\begin{align}
\widehat{G}^{\,}_{f,b^{\prime};j}:=
\hat{\tau}^{x}_{j^{\star}-1}\,
\left(
\mathrm{i}
\hat{\beta}^{\,}_{j}\,
\hat{\alpha}^{\,}_{j}
\right)
\hat{\tau}^{x}_{j^{\star}}=
\widehat{G}^{\,}_{f,b^{\prime};j+2N},
\qquad
j\in\Lambda.
\label{eq:def Gauss operator if unit-cell preserving gauging PF} 
\end{align}
These are Gauss operators since they will soon be used to
define kinematic constraints on the Hilbert space as is standard
in the gauging procedure. 
We call these Gauss operators
(\ref{eq:def Gauss operator if unit-cell preserving gauging PF})
unit-cell preserving, for the local transformation it implements on the 
Majorana operators
only acts non-trivially on a single site of the direct lattice
$\Lambda$. One verifies that
\begin{align}
\prod_{j\in\Lambda}
\widehat{G}^{\,}_{f,b^{\prime};j}=
\prod_{j\in\Lambda}
\hat{\tau}^{x}_{j^{\star}-1}\,
\left(
\mathrm{i}
\hat{\beta}^{\,}_{j}\,
\hat{\alpha}^{\,}_{j}\,
\right)
\hat{\tau}^{x}_{j^{\star}}=
(-1)^{b^{\prime}}
\prod_{j\in\Lambda}
\left(
\mathrm{i}
\hat{\beta}^{\,}_{j}\,
\hat{\alpha}^{\,}_{j}
\right)\equiv
(-1)^{b^{\prime}}\,
\widehat{P}^{\,}_{\mathrm{F}}.
\end{align}
In words, the product of all local Gauss operators
equals the global fermion parity up to a sign fixed by the
twisted boundary conditions $b^{\prime}=0,1$.

We define the unitary transformation
of the Hilbert space
(\ref{eq:def cal Bextf if unit-cell preserving gauging PF b})
through
\begin{subequations}
\begin{align}
\widehat{U}^{\,}_{f,b^{\prime}}:=
\prod_{j\in\Lambda}
\widehat{U}^{\,}_{f,b^{\prime};j},
\qquad
\widehat{U}^{\,}_{f,b^{\prime};j}:=
\left(
\hat{\tau}^{x}_{j^{\star}}
\right)^{\widehat{P}^{\,\,-}_{f,b^{\prime};j}}=
\widehat{P}^{\,\,+}_{f,b^{\prime};j}
+
\hat{\tau}^{x}_{j^{\star}}\,
\widehat{P}^{\,\,-}_{f,b^{\prime};j},
\end{align}
with the $2N$ pairwise-commuting projectors
\begin{align}
\widehat{P}^{\,\,\pm}_{j}:=
\frac{1\pm\mathrm{i}\hat{\beta}^{\,}_{j+1}\,\hat{\alpha}^{\,}_{j}}{2}=
\widehat{P}^{\,\,\pm}_{j+2N}.
\end{align}
\end{subequations}
For any $j=1,\dots,2N$, the following transformation laws hold
\begin{subequations}
\label{eq:trsf law under unitary for unit cell preserving gauging}
\begin{align}
&
\widehat{U}^{\,}_{f,b^{\prime}}\,
\hat{\tau}^{x}_{j^{\star}}\,
\left(
\widehat{U}^{\,}_{f,b^{\prime}}
\right)^{\dag}=
\hat{\tau}^{x}_{j^{\star}},
\qquad
&
\widehat{U}^{\,}_{f,b^{\prime}}\,
\hat{\tau}^{z}_{j^{\star}}\,
\left(
\widehat{U}^{\,}_{f,b^{\prime}}
\right)^{\dag}=
\mathrm{i}
\hat{\beta}^{\,}_{j+1}\,
\hat{\tau}^{z}_{j^{\star}}\,
\hat{\alpha}^{\,}_{j},
\label{eq:trsf law under unitary for unit cell preserving gauging a}
\\
&
\widehat{U}^{\,}_{f,b^{\prime}}\,
\hat{\beta}^{\,}_{j}\,
\left(
\widehat{U}^{\,}_{f,b^{\prime}}
\right)^{\dag}=
\hat{\tau}^{x}_{j^{\star}-1}\,
\hat{\beta}^{\,}_{j},
\qquad
&
\widehat{U}^{\,}_{f,b^{\prime}}\,
\hat{\alpha}^{\,}_{j}\,
\left(
\widehat{U}^{\,}_{f,b^{\prime}}
\right)^{\dag}=
\hat{\alpha}^{\,}_{j}\,
\hat{\tau}^{x}_{j^{\star}},
\label{eq:trsf law under unitary for unit cell preserving gauging b}
\end{align}
for the operators on the lattices $\Lambda$
and $\Lambda^{\star}$ together with the image 
\begin{equation}
\widehat{U}^{\,}_{f,b^{\prime}}\,
\widehat{G}^{\,}_{f,b^{\prime};j}\,
\left(\widehat{U}^{\,}_{f,b^{\prime}}\,\right)^{\dag}=
\mathrm{i}
\hat{\beta}^{\,}_{j}\,
\hat{\alpha}^{\,}_{j}
\label{eq:trsf law under unitary for unit cell preserving gauging c}
\end{equation}
\end{subequations}
of the local Gauss operator.
Thus, if we define the projector
\begin{subequations}
\label{eq:def gauss projector for unit cell preserving gauging}
\begin{equation}
\widehat{P}^{\,}_{f,b^{\prime};\mathrm{G}}:=
\prod_{j\in\Lambda}
\frac{1}{2}
\left[
\widehat{\mbb{1}}^{\,}_{\mathcal{H}^{\,}_{f^{\vphantom{\star}},b^{\prime}}}
+
\widehat{U}^{\,}_{f,b^{\prime}}\,
\widehat{G}^{\,}_{f,b^{\prime};j}\,
\left(\widehat{U}^{\,}_{f,b^{\prime}}\,\right)^{\dag}
\right]
\label{eq:def gauss projector for unit cell preserving gauging a}
\end{equation}
onto the $2^{2N}$-dimensional gauge-invariant subspace
\begin{equation}
\mathcal{H}^{\,\vee}_{b^{\prime}}:=
\widehat{P}^{\,}_{f,b^{\prime};\mathrm{G}}\,
\mathcal{H}^{\,}_{f^{\vphantom{\star}},b^{\prime}}\subset
\mathcal{H}^{\,}_{f^{\vphantom{\star}},b^{\prime}},
\label{eq:def gauss projector for unit cell preserving gauging b}
\end{equation}
\end{subequations}
we find that the $2N$ triplets of projected operators
\begin{subequations}
\begin{align}
\hat{\tau}^{x\,\vee}_{j^{\star}}:=&\,
\widehat{P}^{\,}_{f,b^{\prime};\mathrm{G}}
\left[
\widehat{U}^{\,}_{f,b^{\prime}}\,
\hat{\tau}^{x}_{j^{\star}}\,
\left(\widehat{U}^{\,}_{f,b^{\prime}}\right)^{\dag}
\right]
\widehat{P}^{\,}_{f,b^{\prime};\mathrm{G}}
\nonumber\\
=&\, 
\widehat{P}^{\,}_{f,b^{\prime};\mathrm{G}}\,
\hat{\tau}^{x}_{j^{\star}}\,
\widehat{P}^{\,}_{f,b^{\prime};\mathrm{G}},
\\
\hat{\tau}^{y\,\vee}_{j^{\star}}:=&\,
\widehat{P}^{\,}_{f,b^{\prime};\mathrm{G}}
\left[
\widehat{U}^{\,}_{f,b^{\prime}}\,
\mathrm{i}\hat{\beta}^{\,}_{j+1}\, 
\hat{\tau}^{y}_{j^{\star}}\,
\hat{\alpha}^{\,}_{j}\,
\left(\widehat{U}^{\,}_{f,b^{\prime}}\right)^{\dag}
\right]
\widehat{P}^{\,}_{f,b^{\prime};\mathrm{G}}
\nonumber\\
=&\,
\widehat{P}^{\,}_{f,b^{\prime};\mathrm{G}}\,
\hat{\tau}^{y}_{j^{\star}}\,
\widehat{P}^{\,}_{f,b^{\prime};\mathrm{G}}, 
\\
\hat{\tau}^{z\,\vee}_{j^{\star}}:=&\,
\widehat{P}^{\,}_{f,b^{\prime};\mathrm{G}}
\left[
\widehat{U}^{\,}_{f,b^{\prime}}\,
\mathrm{i}\hat{\beta}^{\,}_{j+1}\, 
\hat{\tau}^{z}_{j^{\star}}\,
\hat{\alpha}^{\,}_{j}\,
\left(\widehat{U}^{\,}_{f,b^{\prime}}\right)^{\dag}
\right]
\widehat{P}^{\,}_{f,b^{\prime};\mathrm{G}}
\nonumber\\
=&\, 
\widehat{P}^{\,}_{f,b^{\prime};\mathrm{G}}\,
\hat{\tau}^{z}_{j^{\star}}\,
\widehat{P}^{\,}_{f,b^{\prime};\mathrm{G}},
\end{align}
\end{subequations}
realize the same Pauli algebra
and obey the same twisted boundary conditions
as the $2N$ triplets
$\hat{\bm{\tau}}^{\,}_{j^{\star}}$
on the dual lattice.
We also find that the projection to
$\mathcal{H}^{\,\vee}_{b^{\prime}}$
of the bond algebra
$\mathfrak{B}^{\,}_{f,b^{\prime}}$ is the bond algebra
\begin{subequations}
\label{eq:cal Bb' if unit-cell preserving gauging PF} 
\begin{align}
\mathfrak{B}^{\,}_{b^{\prime}}:=&\,
\widehat{P}^{\,}_{f,b^{\prime};\mathrm{G}}
\left[
\widehat{U}^{\,}_{f,b^{\prime}}\,
\mathfrak{B}^{\,}_{f,b^{\prime}}\,
\left(\widehat{U}^{\,}_{f,b^{\prime}}\right)^{\dag}
\right]
\widehat{P}^{\,}_{f,b^{\prime};\mathrm{G}}\,
\nonumber\\
&=
\left\langle
\hat{\tau}^{x\,\vee}_{j^{\star}-1}\,
\hat{\tau}^{x\,\vee}_{j^{\star}},
\
\hat{\tau}^{z\,\vee}_{j^{\star}}
\ \Big|\
j^{\star}\in\Lambda^{\star}
\right\rangle,
\label{eq:cal Bb' if unit-cell preserving gauging PF a} 
\end{align}
which is symmetric with respect to a $\mathbb{Z}_{2}$ symmetry generated by
\begin{equation}
\widehat{U}^{\,\vee}_{r^{z}_{\pi}}:=
\prod_{j^{\star}\in\Lambda^{\star}}
\hat{\tau}^{z\,\vee}_{j^{\star}}
\label{eq:cal Bb' if unit-cell preserving gauging PF b} 
\end{equation}
\end{subequations}
of the global rotation by $\pi$ about the $z$ axis in internal
spin-1/2 space attached to the dual lattice $\Lambda^{\star}$.
We have recovered
Eq.\ (\ref{eq:definition and symmetry of cal Bb'})
starting from the bond algebra $\mathfrak{B}^{\,}_{f}$
instead of the bond algebra $\mathfrak{B}^{\,}_{b}$.
The duality from
$\mathfrak{B}^{\,}_{f}$
to
$\mathfrak{B}^{\,}_{b^{\prime}}$
is summarized in Tables
\ref{Table:Summary I triality}
and
\ref{Table:Summary II triality}.

\begin{table}
\caption{
\label{Table:Summary I triality}
Operator dualities for the triality between the bond algebras
$\mathfrak{B}^{\,}_{f}$
defined in Eq.\ (\ref{eq:def cal Bextf if unit-cell preserving gauging PF}),
$\mathfrak{B}^{\,}_{b^{\prime}}$
defined in Eq.\ (\ref{eq:cal Bb' if unit-cell preserving gauging PF}),
and
$\mathfrak{B}^{\,}_{b}$
defined in Eq.\ (\ref{eq:cal Bb if unit-cell non-preserving gauging PF}).
Any two operators from the same column form a dual pair.
}
\centering
\begin{tabular}{c|ccc}
\hline \hline
$\vphantom{\Bigg[}$  
Symbol for bond algebra
&
Generator
&
Generator
&
Symmetry group generated by
\\
\hline
$\vphantom{\Bigg[}$  
$\mathfrak{B}^{\,}_{f}$
&
$  
\mathrm{i}
\hat{\beta}^{\,}_{j}\,
\hat{\alpha}^{\,}_{j}$
&
$
\mathrm{i}
\hat{\beta}^{\,}_{j+1}\,
\hat{\alpha}^{\,}_{j}$
&
$\widehat{P}^{\,}_{\mathrm{F}}=
\prod\limits_{j\in\Lambda}
\left(
\mathrm{i}
\hat{\beta}^{\,}_{j}\,
\hat{\alpha}^{\,}_{j}
\right)$
\\
$\vphantom{\Bigg[}$  
$\mathfrak{B}^{\,}_{b^{\prime}}$
&
$\hat{\tau}^{x\,\vee}_{j^{\star}-1}\,
\hat{\tau}^{x\,\vee}_{j^{\star}}$
&
$\hat{\tau}^{z\,\vee}_{j^{\star}}$
&
$\widehat{U}^{\,\vee}_{r^{z}_{\pi}}=
\prod\limits_{j^{\star}\in\Lambda^{\star}}\hat{\tau}^{z\,\vee}_{j^{\star}}$
\\
$\vphantom{\Bigg[}$  
$\mathfrak{B}^{\,}_{b}$
&
$\hat{\sigma}^{z\,\vee}_{j^{\star}}$
&
$\hat{\sigma}^{x\,\vee}_{j^{\star}}\,
\hat{\sigma}^{x\,\vee}_{j^{\star}+1}$
&
$\widehat{U}^{\,\vee}_{r^{z}_{\pi}}=
\prod\limits_{j^{\star}\in\Lambda^{\star}}\hat{\sigma}^{z\,\vee}_{j^{\star}}$
\\
\hline \hline
\end{tabular}
\end{table}

\subsubsection{Unit-cell non-preserving gauging of fermion parity}
\label{subsubsec:Unit-cell non-preserving gauging of fermion parity}
Next, we trade the bond algebra $\mathfrak{B}^{\,}_{f}$ defined in
Eq.\ (\ref{eq:def bond algebra cal Bf})
by the minimally coupled bond algebra
\begin{subequations}
\label{eq:def cal Bextf if unit-cell non-preserving gauging PF} 
\begin{align}
&
\mathfrak{B}^{\,}_{f,b}:=
\left\langle
\mathrm{i}
\hat{\beta}^{\,}_{j}\,
\hat{\sigma}^{z}_{j^{\star}}\,
\hat{\alpha}^{\,}_{j},
\qquad
\mathrm{i}
\hat{\beta}^{\,}_{j+1}\,
\hat{\alpha}^{\,}_{j}
\ \Big|\
j\in\Lambda
\right\rangle
\label{eq:def cal Bextf if unit-cell non-preserving gauging PF a}  
\end{align}
with the tensor product
\begin{equation}
\mathcal{H}^{\,}_{f,b}:=
\mathcal{H}^{\,}_{f}
\otimes
\mathcal{H}^{\,}_{b}
\label{eq:def cal Bextf if unit-cell non-preserving gauging PF b} 
\end{equation}
\end{subequations}
as domain of definition
[here, $\mathcal{H}^{\,}_{b}$ is defined as in Eq.\
(\ref{eq:def Pauli algebra on mathfrak{H}b c})
except for the substitution of $\Lambda$ by $\Lambda^{\star}$].
This extended  bond algebra is invariant
under conjugation by any one of the $2N$
pairwise-commuting Gauss operators
\begin{align}
\widehat{G}^{\,}_{f,b;j}:=
\hat{\sigma}^{x}_{j^{\star}+1}\,
\left(
\mathrm{i}
\hat{\beta}^{\,}_{j+1}\,
\hat{\alpha}^{\,}_{j}
\right)
\hat{\sigma}^{x}_{j^{\star}}=
\widehat{G}^{\,}_{f,b;j+2N},
\qquad
j\in\Lambda.
\label{eq:def Gauss operator if unit-cell non-preserving gauging PF} 
\end{align}
We call the Gauss operator
(\ref{eq:def Gauss operator if unit-cell non-preserving gauging PF})
unit-cell non-preserving, for the local transformation it implements on the 
Majorana operators
only acts non-trivially on two consecutive sites of the direct lattice
$\Lambda$. We observe that the Gauss operator 
\eqref{eq:def Gauss operator if unit-cell non-preserving gauging PF}
can be obtained from the Gauss operator 
\eqref{eq:def Gauss operator if unit-cell preserving gauging PF}
by translating only the $\hat{\beta}^{\,}_{j}$ operators by
one unit-cell. As we shall see, such a ``half''-translation of the bond
algebra \eqref{eq:def bond algebra cal Bf} will deliver 
the  KW dual of the bosonic bond algebra 
\eqref{eq:cal Bb' if unit-cell preserving gauging PF}. 
One verifies that
\begin{align}
\prod_{j\in\Lambda}
\widehat{G}^{\,}_{f,b;j}=
\prod_{j\in\Lambda}
\hat{\sigma}^{x}_{j^{\star}+1}\,
\left(
\mathrm{i}
\hat{\beta}^{\,}_{j+1}\,
\hat{\alpha}^{\,}_{j}\,
\right)
\hat{\sigma}^{x}_{j^{\star}}=
(-1)^{b}
\prod_{j\in\Lambda}
\left(
\mathrm{i}
\hat{\beta}^{\,}_{j}\,
\hat{\alpha}^{\,}_{j+1}
\right)\equiv
(-1)^{b+f+1}\,
\widehat{P}^{\,}_{\mathrm{F}}.
\end{align}
In words, the product of all local Gauss operators 
equals the global fermion parity up to a sign fixed by the
twisted boundary conditions $f,b=0,1$.

We define the unitary transformation
of the Hilbert space
(\ref{eq:def cal Bextf if unit-cell non-preserving gauging PF b})
through
\begin{subequations}
\begin{align}
\widehat{U}^{\,}_{f,b}:=
\prod_{j\in\Lambda}
\widehat{U}^{\,}_{f,b;j},
\qquad
\widehat{U}^{\,}_{f,b;j}:=
\left(
\hat{\sigma}^{x}_{j^{\star}}
\right)^{\widehat{P}^{\,\,-}_{f,b;j}}=
\widehat{P}^{\,\,+}_{f,b;j}
+
\hat{\sigma}^{x}_{j^{\star}}\,
\widehat{P}^{\,\,-}_{f,b;j},
\end{align}
with the $2N$ pairwise-commuting projectors
\begin{align}
\widehat{P}^{\,\,\pm}_{j}:=
\frac{1\pm \mathrm{i}\hat{\beta}^{\,}_{j}\,\hat{\alpha}^{\,}_{j}}{2}=
\widehat{P}^{\,\,\pm}_{j+2N}.
\end{align}
\end{subequations}
For any $j=1,\cdots,2N$, there follows the transformation laws
\begin{subequations}
\label{eq:trsf law under unitary for unit cell non-preserving gauging}
\begin{align}
&
\widehat{U}^{\,}_{f,b}\,
\hat{\sigma}^{x}_{j^{\star}}\,
\left(
\widehat{U}^{\,}_{f,b}
\right)^{\dag}=
\hat{\sigma}^{x}_{j^{\star}},
\qquad
&
\widehat{U}^{\,}_{f,b}\,
\hat{\sigma}^{z}_{j^{\star}}\,
\left(
\widehat{U}^{\,}_{f,b}
\right)^{\dag}=
\mathrm{i}
\hat{\beta}^{\,}_{j}\,
\hat{\sigma}^{z}_{j^{\star}}\,
\hat{\alpha}^{\,}_{j},
\label{eq:trsf law under unitary for unit cell non-preserving gauging a}
\\
&
\widehat{U}^{\,}_{f,b}\,
\hat{\beta}^{\,}_{j}\,
\left(
\widehat{U}^{\,}_{f,b}
\right)^{\dag}=
\hat{\beta}^{\,}_{j}\,
\hat{\sigma}^{x}_{j^{\star}},
\qquad
&
\widehat{U}^{\,}_{f,b}\,
\hat{\alpha}^{\,}_{j}\,
\left(
\widehat{U}^{\,}_{f,b}
\right)^{\dag}=
\hat{\alpha}^{\,}_{j}\,
\hat{\sigma}^{x}_{j^{\star}},
\label{eq:trsf law under unitary for unit cell non-preserving gauging b}
\end{align}
for the operators on the lattices $\Lambda$
and $\Lambda^{\star}$ together with the image 
\begin{equation}
\widehat{U}^{\,}_{f,b}\,
\widehat{G}^{\,}_{f,b;j}\,
\left(\widehat{U}^{\,}_{f,b}\,\right)^{\dag}=
\mathrm{i}
\hat{\beta}^{\,}_{j+1}\,
\hat{\alpha}^{\,}_{j}
\label{eq:trsf law under unitary for unit cell non-preserving gauging c}
\end{equation}
\end{subequations}
of the local Gauss operator.
Thus, if we define the projector
\begin{subequations}
\label{eq:def gauss projector for unit cell non-preserving gauging}
\begin{equation}
\widehat{P}^{\,}_{f,b;\mathrm{G}}:=
\prod_{j\in\Lambda}
\frac{1}{2}
\left[
\widehat{\mbb{1}}^{\,}_{\mathcal{H}^{\,}_{f^{\vphantom{\star}},b}}
+
\widehat{U}^{\,}_{f,b}\,
\widehat{G}^{\,}_{f,b;j}\,
\left(\widehat{U}^{\,}_{f,b}\,\right)^{\dag}
\right]
\label{eq:def gauss projector for unit cell non-preserving gauging a}
\end{equation}
onto the $2^{2N}$-dimensional gauge-invariant subspace
\begin{equation}
\mathcal{H}^{\,\vee}_{b}:=
\widehat{P}^{\,}_{f,b;\mathrm{G}}\,
\mathcal{H}^{\,}_{f^{\vphantom{\star}},b}\subset
\mathcal{H}^{\,}_{f^{\vphantom{\star}},b},
\label{eq:def gauss projector for unit cell non-preserving gauging b}
\end{equation}
\end{subequations}
we find that the $2N$ triplets of projected operators
\begin{subequations}
\begin{align}
\hat{\sigma}^{x\,\vee}_{j^{\star}}:=&\,
\widehat{P}^{\,}_{f,b;\mathrm{G}}
\left[
\widehat{U}^{\,}_{f,b}\,
\hat{\sigma}^{x}_{j^{\star}}\,
\left(\widehat{U}^{\,}_{f,b}\right)^{\dag}
\right]
\widehat{P}^{\,}_{f,b;\mathrm{G}}
\nonumber\\
=&\,
\widehat{P}^{\,}_{f,b;\mathrm{G}}\,
\hat{\sigma}^{x}_{j^{\star}}\,
\widehat{P}^{\,}_{f,b;\mathrm{G}},
\\
\hat{\sigma}^{y\,\vee}_{j^{\star}}:=&\,
\widehat{P}^{\,}_{f,b;\mathrm{G}}
\left[
\widehat{U}^{\,}_{f,b}\,
\mathrm{i}\hat{\beta}^{\,}_{j}\, 
\hat{\sigma}^{y}_{j^{\star}}\,\hat{\alpha}^{\,}_{j}\,
\left(\widehat{U}^{\,}_{f,b}\right)^{\dag}
\right]
\widehat{P}^{\,}_{f,b;\mathrm{G}}
\nonumber\\
=&\,
\widehat{P}^{\,}_{f,b;\mathrm{G}}\,
\hat{\sigma}^{y}_{j^{\star}}\,
\widehat{P}^{\,}_{f,b;\mathrm{G}}, 
\\
\hat{\sigma}^{z\,\vee}_{j^{\star}}:=&\,
\widehat{P}^{\,}_{f,b;\mathrm{G}}
\left[
\widehat{U}^{\,}_{f,b}\,
\mathrm{i}\hat{\beta}^{\,}_{j}\, 
\hat{\sigma}^{z}_{j^{\star}}\,\hat{\alpha}^{\,}_{j}\,
\left(\widehat{U}^{\,}_{f,b}\right)^{\dag}
\right]
\widehat{P}^{\,}_{f,b;\mathrm{G}}
\nonumber\\
=&\,
\widehat{P}^{\,}_{f,b;\mathrm{G}}\,
\hat{\sigma}^{z}_{j^{\star}}\,
\widehat{P}^{\,}_{f,b;\mathrm{G}},
\end{align}
\end{subequations}
realize the same Pauli algebra
and obey the same twisted boundary conditions
as the $2N$ triplets
$\hat{\bm{\sigma}}^{\,}_{j^{\star}}$
on the dual lattice.
We also find that the projection to
$\mathcal{H}^{\,\vee}_{b}$
of the bond algebra
$\mathfrak{B}^{\,}_{f,b}$ is the bond algebra
\begin{subequations}
\label{eq:cal Bb if unit-cell non-preserving gauging PF} 
\begin{align}
\mathfrak{B}^{\,}_{b}:=&\,
\widehat{P}^{\,}_{f,b;\mathrm{G}}
\left[
\widehat{U}^{\,}_{f,b}\,
\mathfrak{B}^{\,}_{f,b}\,
\left(\widehat{U}^{\,}_{f,b}\right)^{\dag}
\right]
\widehat{P}^{\,}_{f,b;\mathrm{G}}\,
\nonumber\\
&=
\left\langle
\hat{\sigma}^{z\,\vee}_{j^{\star}},
\qquad
\hat{\sigma}^{x\,\vee}_{j^{\star}}\,
\hat{\sigma}^{x\,\vee}_{j^{\star}+1}
\ \Big|\
j^{\star}\in\Lambda^{\star}
\right\rangle
\label{eq:cal Bb if unit-cell non-preserving gauging PF a} 
\end{align}
with the generator
\begin{equation}
\widehat{U}^{\,\vee}_{r^{z}_{\pi}}:=
\prod_{j^{\star}\in\Lambda^{\star}}
\hat{\sigma}^{z\,\vee}_{j^{\star}}
\label{eq:cal Bb if unit-cell non-preserving gauging PF b} 
\end{equation}
\end{subequations}
of the global rotation by $\pi$ about the $z$ axis in internal
spin-1/2 space attached to the dual lattice $\Lambda^{\star}$.
We have recovered
Eq.\ (\ref{eq:definition and symmetry of cal Bb})
starting from the bond algebra $\mathfrak{B}^{\,}_{f}$
(up to the substitution $\Lambda\to\Lambda^{\star}$).
The duality from
$\mathfrak{B}^{\,}_{f}$
to
$\mathfrak{B}^{\,}_{b}$
is summarized in Tables
\ref{Table:Summary I triality}
and
\ref{Table:Summary II triality}.

\begin{table}
\caption{
\label{Table:Summary II triality}
Compatibility conditions for the triality between the bond algebras
$\mathfrak{B}^{\,}_{f}$
defined in Eq.\ (\ref{eq:def cal Bextf if unit-cell preserving gauging PF}),
$\mathfrak{B}^{\,}_{b^{\prime}}$
defined in Eq.\ (\ref{eq:cal Bb' if unit-cell preserving gauging PF}),
and
$\mathfrak{B}^{\,}_{b}$
defined in Eq.\ (\ref{eq:cal Bb if unit-cell non-preserving gauging PF}).
The Hilbert space on which the bond algebra
$\mathfrak{B}^{\,}_{f}$
is defined is $\mathcal{H}^{\,}_{f}\cong\mathbb{C}^{2^{2N}}$
with $f=0,1$ selecting the twisted boundary conditions.
The Hilbert space on which the bond algebra
$\mathfrak{B}^{\,}_{b^{\prime}}$
is defined is $\mathcal{H}^{\,\vee}_{b^{\prime}}\cong\mathbb{C}^{2^{2N}}$
with $b^{\prime}=0,1$ selecting the twisted boundary conditions.
The Hilbert space on which the bond algebra
$\mathfrak{B}^{\,}_{b}$
is defined is $\mathcal{H}^{\,\vee}_{b}\cong\mathbb{C}^{2^{2N}}$
with $b^{\,}=0,1$ selecting the twisted boundary conditions.
Choosing two out of the triplet $(f,\, b^{\prime},\,b)$
determines the third according to the rule $f = b+b'+1$ mod 2.
Triality is defined by the fact
that duality holds between any two bond algebras
$\mathfrak{B}^{\,}_{f}$,
$\mathfrak{B}^{\,}_{b^{\prime}}$,
and
$\mathfrak{B}^{\,}_{b}$
provided their domain of definitions are restricted to
$\mathcal{H}^{\,}_{f;\,(-1)^{b+f+1}}\cong\mathbb{C}^{2^{2N-1}}$,
$\mathcal{H}^{\,\vee}_{b';\,(-1)^{b}}\cong\mathbb{C}^{2^{2N-1}}$,
and
$\mathcal{H}^{\,\vee}_{b;\,(-1)^{b'}}\cong\mathbb{C}^{2^{2N-1}}$,
respectively.
}
\centering
\begin{tabular}{c|cccccc}
\hline \hline
$\vphantom{\Bigg[}$  
$\hphantom{A}\
(f,\,b^{\prime},\,b)\ \hphantom{A}$
&\hphantom{A}&
$\mathcal{H}^{\,}_{f;\,(-1)^{b+f+1}}$
&\hphantom{A}&
$\mathcal{H}^{\,\vee}_{b';\,(-1)^{b}}$
&\hphantom{A}&
$\mathcal{H}^{\,\vee}_{b;\,(-1)^{b'}}$
\\
\hline
$\vphantom{\Bigg[}$  
$(0,1,0)$
&\hphantom{A}&
$
\frac{1}{2}
\left(
\widehat{\mbb{1}}^{\,}_{\mathcal{H}^{\,}_{f}}
-
\widehat{P}^{\,}_{\mathrm{F}}
\right)
\mathcal{H}^{\,}_{f}
$
&\hphantom{A}&
$
\frac{1}{2}
\left(
\widehat{\mbb{1}}^{\,}_{\mathcal{H}^{\,\vee}_{b^{\prime}}}
+
\widehat{U}^{\vee}_{r^{z}_{\pi}}
\right)
\mathcal{H}^{\,\vee}_{b^{\prime}}
$
&\hphantom{A}&
$
\frac{1}{2}
\left(
\widehat{\mbb{1}}^{\,}_{\mathcal{H}^{\,\vee}_{b}}
-
\widehat{U}^{\vee}_{r^{z}_{\pi}}
\right)
\mathcal{H}^{\,\vee}_{b}
$
\\
$\vphantom{\Bigg[}$  
$(0,0,1)$
&\hphantom{A}&
$
\frac{1}{2}
\left(
\widehat{\mbb{1}}^{\,}_{\mathcal{H}^{\,}_{f}}
+
\widehat{P}^{\,}_{\mathrm{F}}
\right)
\mathcal{H}^{\,}_{f}
$
&\hphantom{A}&
$
\frac{1}{2}
\left(
\widehat{\mbb{1}}^{\,}_{\mathcal{H}^{\,\vee}_{b^{\prime}}}
-
\widehat{U}^{\vee}_{r^{z}_{\pi}}
\right)
\mathcal{H}^{\,\vee}_{b^{\prime}}
$
&\hphantom{A}&
$
\frac{1}{2}
\left(
\widehat{\mbb{1}}^{\,}_{\mathcal{H}^{\,\vee}_{b}}
+
\widehat{U}^{\vee}_{r^{z}_{\pi}}
\right)
\mathcal{H}^{\,\vee}_{b}
$
\\
$\vphantom{\Bigg[}$  
$(1,0,0)$
&\hphantom{A}&
$
\frac{1}{2}
\left(
\widehat{\mbb{1}}^{\,}_{\mathcal{H}^{\,}_{f}}
+
\widehat{P}^{\,}_{\mathrm{F}}
\right)
\mathcal{H}^{\,}_{f}
$
&\hphantom{A}&
$
\frac{1}{2}
\left(
\widehat{\mbb{1}}^{\,}_{\mathcal{H}^{\,\vee}_{b^{\prime}}}
+
\widehat{U}^{\vee}_{r^{z}_{\pi}}
\right)
\mathcal{H}^{\,\vee}_{b^{\prime}}
$
&\hphantom{A}&
$
\frac{1}{2}
\left(
\widehat{\mbb{1}}^{\,}_{\mathcal{H}^{\,\vee}_{b}}
+
\widehat{U}^{\vee}_{r^{z}_{\pi}}
\right)
\mathcal{H}^{\,\vee}_{b}
$
\\
$\vphantom{\Bigg[}$  
$(1,1,1)$
&\hphantom{A}&
$
\frac{1}{2}
\left(
\widehat{\mbb{1}}^{\,}_{\mathcal{H}^{\,}_{f}}
-
\widehat{P}^{\,}_{\mathrm{F}}
\right)
\mathcal{H}^{\,}_{f}
$
&\hphantom{A}&
$
\frac{1}{2}
\left(
\widehat{\mbb{1}}^{\,}_{\mathcal{H}^{\,\vee}_{b^{\prime}}}
-
\widehat{U}^{\vee}_{r^{z}_{\pi}}
\right)
\mathcal{H}^{\,\vee}_{b^{\prime}}
$
&\hphantom{A}&
$
\frac{1}{2}
\left(
\widehat{\mbb{1}}^{\,}_{\mathcal{H}^{\,\vee}_{b}}
-
\widehat{U}^{\vee}_{r^{z}_{\pi}}
\right)
\mathcal{H}^{\,\vee}_{b}
$
\\
\hline \hline
\end{tabular}
\end{table}

Demanding the triality  of the three bond algebras 
$\mathfrak{B}^{\,}_{f}$, $\mathfrak{B}^{\,}_{b'}$,
and 
$\mathfrak{B}^{\,}_{b}$ that are defined in Eqs.\ 
\eqref{eq:def bond algebra cal Bf a}, 
\eqref{eq:cal Bb' if unit-cell preserving gauging PF a},
and \eqref{eq:cal Bb if unit-cell non-preserving gauging PF a},
respectively, puts a constraint on the possible boundary conditions
specified by the triplet $(b,\,b',\,f)$ of twisted boundary conditions.
Indeed, this triality implies
that one may start from any one of these bond algebras located at
the vertices in Fig.\ \ref{fig:Triality diagram}
and execute two successive dualities in such a way that the two remaining
vertices from Fig.\ \ref{fig:Triality diagram} are visited. 
The duality between the bond algebras 
$\mathfrak{B}^{\,}_{f}$ and $\mathfrak{B}^{\,}_{b'}$
holds on the restricted subspaces
\begin{subequations}
\label{eq:pairwise dualities}
\begin{align}
\mathcal{H}^{\,}_{f;(-1)^{b'}}
\longleftrightarrow
\mathcal{H}^{\vee}_{b';(-1)^{f+b'+1}},
\label{eq:pairwise dualities a}
\end{align}
while the duality between the 
bond algebras $\mathfrak{B}^{\,}_{f}$ and $\mathfrak{B}^{\,}_{b}$
holds on the restricted subspaces 
\begin{align}
\mathcal{H}^{\,}_{f;(-1)^{b+f+1}}
\longleftrightarrow
\mathcal{H}^{\vee}_{b;(-1)^{b+f+1}}.
\label{eq:pairwise dualities b}
\end{align}
\end{subequations}
Let us choose the corresponding subspaces of $\mathcal{H}^{\,}_{f}$
in Eqs.\ \eqref{eq:pairwise dualities a} and \eqref{eq:pairwise dualities b} 
to be identical. We then find that the
triality of all three bond algebras holds when 
\begin{subequations}
\label{eq:triality relation}
\begin{align}
\mathcal{H}^{\vee}_{b';(-1)^{f+b'+1}}
\longleftrightarrow
\mathcal{H}^{\,}_{f;(-1)^{b'}} 
\equiv
\mathcal{H}^{\,}_{f;(-1)^{b+f+1}}
\longleftrightarrow
\mathcal{H}^{\vee}_{b;(-1)^{f+b+1}},
\label{eq:triality relation a}
\end{align}
which implies the relation
\begin{align}
f = b + b' + 1 \text{ mod 2}.
\label{eq:triality relation b}
\end{align}
\end{subequations}
The duality between $\mathcal{H}^{\vee}_{b';(-1)^{f+b'+1}}$
and $\mathcal{H}^{\vee}_{b;(-1)^{f+b+1}}$ follows since one 
can first dualize the former to obtain $\mathcal{H}^{\,}_{f;(-1)^{b'}}$
and then dualize $\mathcal{H}^{\,}_{f;(-1)^{b+f+1}}$ to obtain 
$\mathcal{H}^{\vee}_{b;(-1)^{f+b+1}}$ if the condition
\eqref{eq:triality relation b} holds.

\section{LSM anomalies and triality}
\label{sec:Triality of crystalline and internal symmetries on a chain}

In Sec.\ \ref{sec:Triality through bond algebra isomorphisms},
we have established dualities between any two of the three
bond algebras
$\mathfrak{B}^{\,}_{b}$,
$\mathfrak{B}^{\,}_{b'}$,
and $\mathfrak{B}^{\,}_{f}$.
What is common to all three bond algebras is the presence of
a cyclic symmetry group of order two, namely
$\mathbb{Z}^{z}_{2}$, 
$\mathbb{Z}^{z^{\vee}}_{2}$,
and
$\mathbb{Z}^{\mathrm{F}}_{2}$, respectively.
Any Hamiltonian
$\widehat{H}^{\,}_{b}$
that is an element of the
bond algebra $\mathfrak{B}^{\,}_{b}$
has a $\mathbb{Z}^{z}_{2}$ symmetry 
generated by $\widehat{U}^{\,}_{r^{z}_{\pi}}$.
It follows that its duals
$\widehat{H}^{\vee}_{b'}$ 
and
$\widehat{H}^{\vee}_{f}$
obtained by KW and JW dualities, respectively,
are symmetric under the dual symmetries
$\mathbb{Z}^{z^{\vee}}_{2}$
and
$\mathbb{Z}^{\mathrm{F}}_{2}$,
respectively. 
The question that we address in this section
is the fate of additional crystalline and internal symmetries of 
such a Hamiltonian $\widehat{H}^{\,}_{b}$
under the dualities described in Sec.\
\ref{sec:Triality through bond algebra isomorphisms}.
In particular, we show how the presence of
an LSM anomaly
manifests itself
in the dual bond algebras $\mathfrak{B}^{\,}_{b'}$
and $\mathfrak{B}^{\,}_{f}$.

\subsection{Symmetry structure with an LSM anomaly}
\label{subsec:G total frak H b=0}

We consider the bond algebra $\mathfrak{B}^{\,}_{b}$
with $b=0$%
~\footnote{%
~This choice ensures that the total symmetry group is a direct product of 
crystalline and internal symmetries.
}
and impose two independent crystalline symmetries of
the lattice $\Lambda$, namely translation and (site-centered)
reflection 
\begin{subequations}
\label{eq:def crystalline symmetries Lambda}
\begin{equation}
\begin{split}
&
\widehat{U}^{\,}_{t}\,
\hat{\bm{\sigma}}^{\,}_{j}\,
\widehat{U}^{\dag}_{t}\,
=
\hat{\bm{\sigma}}^{\,}_{t(j)},
\qquad
t(j):=j+1\hbox{ mod }2N,
\\
&
\widehat{U}^{\,}_{r}\,
\hat{\bm{\sigma}}^{\,}_{j}\,
\widehat{U}^{\dag}_{r}\,
=
\hat{\bm{\sigma}}^{\,}_{r(j)},
\qquad
r(j):=2N-j\hbox{ mod }2N,
\end{split}
\end{equation}
implemented by the unitary operators
\begin{align}
&
\widehat{U}^{\,}_{t}:=
\prod_{j=1}^{2N-1}
\frac{1}{2}
\left(
\widehat{\mbb{1}}^{\,}_{\mathcal{H}^{\,}_{b=0}}
+
\hat{\bm{\sigma}}^{\,}_{j}
\cdot
\hat{\bm{\sigma}}^{\,}_{t(j)}
\right),
\qquad
\widehat{U}^{\,}_{r}:=
\prod_{j=1}^{N-1}
\frac{1}{2}
\left(
\widehat{\mbb{1}}^{\,}_{\mathcal{H}^{\,}_{b=0}}
+
\hat{\bm{\sigma}}^{\,}_{j}
\cdot
\hat{\bm{\sigma}}^{\,}_{r(j)}
\right),
\label{eq:reps symmetries t and r for cal Bb a} 
\end{align}
\end{subequations}
respectively.
The product on the second term in Eq.\
(\ref{eq:reps symmetries t and r for cal Bb a})
has the upper bound $N-1$ since
the reflection has two fixed points $N$ and $2N$ in $\Lambda$.
The pair of operators
$\widehat{U}^{\,}_{t}$
and
$\widehat{U}^{\,}_{r}$
generates a $2^{2N}$-dimensional representation of the space group
\begin{subequations}
\label{eq:space group Lambda}
\begin{equation}
\mathrm{G}^{\,}_{\mathrm{spa}}:=
\mathbb{Z}^{t}_{2N}
\rtimes
\mathbb{Z}^{r}_{2}
\end{equation}
with
\begin{equation}
\mathbb{Z}^{t}_{2N}\equiv
\left\{\vphantom{\Big[}
t,\ t^{2},\ \cdots,\ t^{2N-1},\ t^{2N}\equiv e
\right\},
\qquad
\mathbb{Z}^{r}_{2}\equiv
\left\{\vphantom{\Big[}
r,\ r^{2}\equiv e
\right\},
\qquad
r\,t = t^{2N-1}\,r.
\end{equation}
\end{subequations}

Next, we impose the global internal symmetries 
implemented by the unitary operators
\begin{align}
\widehat{U}^{\,}_{r^{x}_{\pi}}:=
\prod_{j=1}^{2N}
\hat{\sigma}^{x}_{j},
\qquad
\widehat{U}^{\,}_{r^{y}_{\pi}}:=
\prod_{j=1}^{2N}
\hat{\sigma}^{y}_{j},
\qquad
\widehat{U}^{\,}_{r^{z}_{\pi}}:=
(-1)^{N}\,
\widehat{U}^{\,}_{r^{x}_{\pi}}\,
\widehat{U}^{\,}_{r^{y}_{\pi}}=
\prod_{j=1}^{2N}
\hat{\sigma}^{z}_{j}.
\label{eq:reps rxpi rypi rzpi for cal Bb}
\end{align}
The pair of operators
$\widehat{U}^{\,}_{r^{x}_{\pi}}$
and
$\widehat{U}^{\,}_{r^{y}_{\pi}}$
generates a $2^{2N}$-dimensional representation of
the global internal symmetry group
\begin{subequations}
\label{eq: internal symmetry}
\begin{equation}
\mathrm{G}^{\,}_{\mathrm{int}}\equiv  
\mathbb{Z}^{x}_{2}\times\mathbb{Z}^{y}_{2}
\end{equation}
with
\begin{equation}
\mathbb{Z}^{x}_{2}\equiv
\left\{\vphantom{\Big[}
r^{x}_{\pi},\
(r^{x}_{\pi})^{2}\equiv e
\right\},
\qquad
\mathbb{Z}^{y}_{2}\equiv
\left\{\vphantom{\Big[}
r^{y}_{\pi},\
(r^{y}_{\pi})^{2}\equiv e
\right\}.
\end{equation}
\end{subequations}
Note that the $\mathbb{Z}^{z}_{2}$ symmetry of the bond algebra 
\eqref{eq:definition and symmetry of cal Bb}
corresponds to the diagonal element in the group
$\mathbb{Z}^{x}_{2}\times\mathbb{Z}^{y}_{2}$,
i.e., $r^{z}_{\pi} = r^{x}_{\pi}\,r^{y}_{\pi}$. 
Importantly, the total symmetry group has the direct product structure
\begin{equation}
\mathrm{G}^{\,}_{\mathrm{tot}}\equiv
\mathrm{G}^{\,}_{\mathrm{spa}}
\times
\mathrm{G}^{\,}_{\mathrm{int}}.
\label{eq:def Gtot for Hb}
\end{equation}
While the global representation of the
$\mathbb{Z}^{x}_{2}\times\mathbb{Z}^{y}_{2}$ 
group in Eq.\ \eqref{eq:reps rxpi rypi rzpi for cal Bb} is a
group homomorphism, it is
locally projective due to the Pauli algebra
\begin{equation}
\hat{\sigma}^{x}_{j}\,
\hat{\sigma}^{y}_{j}=
-
\hat{\sigma}^{y}_{j}\,
\hat{\sigma}^{x}_{j},
\qquad
\hat{\sigma}^{y}_{j}\,
\hat{\sigma}^{z}_{j}=
-
\hat{\sigma}^{z}_{j}\,
\hat{\sigma}^{y}_{j},
\qquad
\hat{\sigma}^{z}_{j}\,
\hat{\sigma}^{x}_{j}=
-
\hat{\sigma}^{x}_{j}\,
\hat{\sigma}^{z}_{j},
\label{eq:Pauli algebra is a non-trivial proj rep SO(3)}
\end{equation}
for any $j\in\Lambda$.
Therefore, it follows that 
the presence of the $\mathrm{G}^{\,}_{\mathrm{tot}}$
symmetry constrains the phase diagram of any symmetric 
Hamiltonian owing to the (generalized) 
Lieb-Schultz-Mattis (LSM) Theorems.
We are going to invoke two LSM Theorems~\cite{Lieb61,Ogata2019,Ogata2021}
that apply to one-dimensional spin chains with 
translation and reflection symmetries, respectively.
Importantly, the proofs of these two theorems make use of the fact that  
the total symmetry group 
\eqref{eq:def Gtot for Hb} 
is a direct product 
of the crystalline
space group with the internal group.
This is another motivation for choosing $b=0$.
In what follows, $|\Lambda|$ denotes the cardinality ($2N$)
of the set $\Lambda$.

\begin{thm}[Translation LSM]\label{thm:LSM translation}
Consider a one-dimensional lattice Hamiltonian
with the symmetry group
$\mathrm{G}^{\,}_{\mathrm{tot}}\equiv
\mathrm{Z}^{t}_{|\Lambda|}\times
\mathbb{Z}^{x}_{2}\times\mathbb{Z}^{y}_{2}$,
where the subgroup $\mathrm{Z}^{t}_{|\Lambda|}$ generates lattice 
translations and the subgroup $\mathbb{Z}^{x}_{2}\times\mathbb{Z}^{y}_{2}$
generates internal discrete spin-rotation symmetry.
If the unit cell with respect to the translation
symmetry $\mathrm{Z}^{t}_{|\Lambda|}$
hosts a half-integer spin representation
of $\mathbb{Z}^{x}_{2}\times\mathbb{Z}^{y}_{2}$,
then the ground states cannot be simultaneously gapped,
non-degenerate, and
$\mathrm{G}^{\,}_{\mathrm{tot}}$-symmetric.
\end{thm}

\begin{defn}[Translation LSM anomaly]\label{def:Translation LSM anomaly}
When Theorem  \ref{thm:LSM translation} holds,
we say that there is a translation LSM anomaly.
\end{defn}

\begin{thm}[Reflection LSM]\label{thm:LSM reflection}
Consider a one-dimensional lattice Hamiltonian
with the symmetry group
$\mathrm{G}^{\,}_{\mathrm{tot}}\equiv
\mathrm{Z}^{r}_{2}\times
\mathbb{Z}^{x}_{2}\times\mathbb{Z}^{y}_{2}$,
where the subgroup $\mathrm{Z}^{r}_{2}$ generates
site-centered reflection and the subgroup
$\mathbb{Z}^{x}_{2}\times\mathbb{Z}^{y}_{2}$
generates internal discrete spin-rotation symmetry.
If each reflection center
hosts a half-integer spin representation
of $\mathbb{Z}^{x}_{2}\times\mathbb{Z}^{y}_{2}$,
then the ground states cannot be simultaneously gapped,
non-degenerate, and
$\mathrm{G}^{\,}_{\mathrm{tot}}$-symmetric.
\end{thm}

\begin{defn}[Reflection LSM anomaly]\label{def:Reflection LSM anomaly}
When the reflection LSM Theorem \ref{thm:LSM reflection} holds,
we say that there is a reflection LSM anomaly.
\end{defn}

\begin{remark}[LSM anomaly versus mixed 't Hooft anomaly]
The translation LSM and reflection LSM
Theorems (anomalies)
have been interpreted as the presence of a
mixed 't Hooft anomaly between crystalline symmetry groups,
either $\mathbb{Z}^{t}_{2N}$ or $\mathbb{Z}^{r}_{2}$, and internal
symmetry group $\mathbb{Z}^{x}_{2}\times\mathbb{Z}^{y}_{2}$
\cite{Cheng2016, Cho2017, Jian2018, Else2020, Cheng2023}.
Accordingly, one cannot gauge the full internal symmetry group
$\mathrm{G}^{\,}_{\mathrm{int}}$, while maintaining the space group
$\mathrm{G}^{\,}_{\mathrm{spa}}$. However, a non-anomalous subgroup
$\mathrm{H}^{\,}_{\mathrm{int}}\subset \mathrm{G}^{\,}_{\mathrm{int}}$
can still be consistently gauged.
\end{remark}

In what follows, we will show that under the KW and JW dualities
introduced in Secs.\ \ref{subsec:Kramers-Wannier duality b to b'}
and
\ref{subsec:Jordan-Wigner duality b and f}, respectively, the direct
product structure of $\mathrm{G}^{\,}_{\mathrm{tot}}$ is altered
through a mixing of crystalline and internal symmetries.
As both dualities correspond to gauging the non-anomalous diagonal
subgroup $\mathbb{Z}^{z}_{2}\subset
\mathbb{Z}^{x}_{2}\times\mathbb{Z}^{y}_{2}$,
our main result can be interpreted as the
incompatibility between gauge-invariant representations of elements in
the subgroup
$(\mathbb{Z}^{x}_{2}\times\mathbb{Z}^{y}_{2})/\mathbb{Z}^{z}_{2}$ and
crystalline symmetries $\mathbb{Z}^{t}_{2N}$ and
$\mathbb{Z}^{r}_{2}$ under the KW or JW dualities.
We conjecture that an analogue of this result
holds for general space groups
$\mathrm{G}^{\,}_{\mathrm{spa}}$
and internal symmetry groups
$\mathrm{G}^{\,}_{\mathrm{int}}$
if an LSM anomaly is present.
In Sec.\ \ref{sec:Zn generalization}, we confirm that this conjecture is
true for the generalization to
$\mathrm{G}^{\,}_{\mathrm{int}}=\mathbb{Z}^{\,}_{n}\times\mathbb{Z}^{\,}_{n}$
and $\mathrm{H}^{\,}_{\mathrm{int}}=\mathbb{Z}^{\,}_{n}$.

\subsection{Kramers-Wannier dual of the LSM anomaly }
\label{subsec:G total frak H b'=0}

We are going to construct the dual total symmetry group
$\mathrm{G}^{\,\vee}_{\mathrm{tot}}$
under the KW duality introduced in Sec.\
\ref{subsec:Kramers-Wannier duality b to b'}.
To this end, we define the action of the
crystalline and internal symmetries on
the extended Hilbert space
$\mathcal{H}^{\,}_{b,b^{\prime}}$ defined in Eq.\ 
\eqref{eq:def bond algebra cal Bbb' b} 
and then project these symmetries onto the dual Hilbert space 
$\mathcal{H}^{\vee}_{b'}$.
For simplicity, we set $b'=0$.

The extension of the 
crystalline symmetries 
\eqref{eq:def crystalline symmetries Lambda}
on the Hilbert space $\mathcal{H}^{}_{b,b^{\prime}}$
are obtained by demanding the covariance of the Gauss operators
(\ref{eq:sym bond algebra cal Bbb' (a)})
under translation and reflection.
We thus define the unitary operators
\begin{subequations}
\label{eq:reps symmetries t and r for cal Bb' ext} 
\begin{align}
&
\widehat{U}^{\mathrm{ext}}_{t}:=
\left[
\prod_{j=1}^{2N-1}
\frac{1}{2}
\left(
\widehat{\mbb{1}}^{\,}_{\mathcal{H}^{\,}_{b=0,\,b'=0}}
+
\hat{\bm{\sigma}}^{\,}_{j}
\cdot
\hat{\bm{\sigma}}^{\,}_{t(j)}
\right)
\right]
\left[
\prod_{j=1}^{2N-1}
\frac{1}{2}
\left(
\widehat{\mbb{1}}^{\,}_{\mathcal{H}^{\,}_{b=0,\,b'=0}}
+
\hat{\bm{\tau}}^{\,}_{j^{\star}}
\cdot
\hat{\bm{\tau}}^{\,}_{t(j^{\star})}
\right)
\right],
\label{eq:reps symmetries t and r for cal Bb' ext a} 
\\
&
\widehat{U}^{\mathrm{ext}}_{r}:=
\left[
\prod_{j=1}^{N-1}
\frac{1}{2}
\left(
\widehat{\mbb{1}}^{\,}_{\mathcal{H}^{\,}_{b=0,\,b'=0}}
+
\hat{\bm{\sigma}}^{\,}_{j}
\cdot
\hat{\bm{\sigma}}^{\,}_{r(j)}
\right)
\right]
\left[
\prod_{j=1}^{N}
\frac{1}{2}
\left(
\widehat{\mbb{1}}^{\,}_{\mathcal{H}^{\,}_{b=0,\,b'=0}}
+
\hat{\bm{\tau}}^{\,}_{j^{\star}}
\cdot
\hat{\bm{\tau}}^{\,}_{r(j^{\star})}
\right)
\right],
\label{eq:reps symmetries t and r for cal Bb' ext b} 
\end{align}
that implement the transformation rules 
\eqref{eq:def crystalline symmetries Lambda} for the 
$\hat{\bm{\sigma}}$ operators on lattice $\Lambda$, and the 
transformation rules
\begin{align}
&
\widehat{U}^{\mathrm{ext}}_{t}\,
\hat{\bm{\tau}}^{\,}_{j^{\star}}\,
\left(\widehat{U}^{\mathrm{ext}}_{t}\right)^{\dag}\,
=
\hat{\bm{\tau}}^{\,}_{t(j^{\star})},
\qquad
t(j^{\star}):=j^{\star}+1\hbox{ mod }2N,
\label{eq:reps symmetries t and r for cal Bb' ext c} 
\\
&
\widehat{U}^{\mathrm{ext}}_{r}\,
\hat{\bm{\tau}}^{\,}_{j^{\star}}\,
\left(\widehat{U}^{\mathrm{ext}}_{r}\right)^{\dag}\,
=
\hat{\bm{\tau}}^{\,}_{r(j^{\star})},
\qquad
r(j^{\star}):=2N-j^{\star}\hbox{ mod }2N,
\label{eq:reps symmetries t and r for cal Bb' ext d}
\end{align}
\end{subequations}
for the $\hat{\bm{\tau}}$ operators on the dual lattice $\Lambda^{\star}$.
Transformation rules
(\ref{eq:reps symmetries t and r for cal Bb' ext c})
and
(\ref{eq:reps symmetries t and r for cal Bb' ext d})
correspond to two independent crystalline symmetries of the 
dual lattice $\Lambda^{\star}$, namely translation and (link-centered)
reflection symmetries. 
As promised, the operators $\widehat{U}^{\mathrm{ext}}_{t}$ and 
$\widehat{U}^{\mathrm{ext}}_{r}$ are not gauge invariant but transform
the local Gauss operators
(\ref{eq:sym bond algebra cal Bbb'})
according to the covariant rules
\begin{subequations}
\label{eq:Gauss operator covariance cyrstalline}
\begin{align}
\widehat{U}^{\mathrm{ext}}_{t}\,
\widehat{G}^{\,}_{j}\,
\left(\widehat{U}^{\mathrm{ext}}_{t}\right)^{\dag}\,
=
\widehat{G}^{\,}_{t(j)},
\label{eq:Gauss operator covariance cyrstalline a}
\\
\widehat{U}^{\mathrm{ext}}_{r}\,
\widehat{G}^{\,}_{j}\,
\left(\widehat{U}^{\mathrm{ext}}_{r}\right)^{\dag}\,
=
\widehat{G}^{\,}_{r(j)},
\label{eq:Gauss operator covariance cyrstalline b}
\end{align}
\end{subequations}
respectively, for any $j\in\Lambda$.
After the projection to the dual Hilbert space $\mathcal{H}^{\vee}_{b'=0}$,
the counterparts to the translation 
(\ref{eq:reps symmetries t and r for cal Bb' ext a})
and reflection
(\ref{eq:reps symmetries t and r for cal Bb' ext b})
are implemented by the unitary operators
\begin{subequations}
\label{eq:reps symmetries t and r for cal Bb'} 
\begin{align}
&
\widehat{U}^{\vee}_{t}:=
\prod_{j=1}^{2N-1}
\frac{1}{2}
\left(
\widehat{\mbb{1}}^{\,}_{\mathcal{H}^{\,\vee}_{b^{\prime}=0}}
+
\hat{\bm{\tau}}^{\,\vee}_{j^{\star}}
\cdot
\hat{\bm{\tau}}^{\,\vee}_{t(j^{\star})}
\right),
\label{eq:reps symmetries t and r for cal Bb' a} 
\\
&
\widehat{U}^{\,\vee}_{r}:=
\prod_{j=1}^{N}
\frac{1}{2}
\left(
\widehat{\mbb{1}}^{\,}_{\mathcal{H}^{\,\vee}_{b^{\prime}=0}}
+
\hat{\bm{\tau}}^{\,\vee}_{j^{\star}}
\cdot
\hat{\bm{\tau}}^{\,\vee}_{r(j^{\star})}
\right),
\label{eq:reps symmetries t and r for cal Bb' b} 
\end{align}
\end{subequations}
respectively. 
The product on 
the right-hand side of Eq.\
(\ref{eq:reps symmetries t and r for cal Bb' b})
has the upper bound $N$ since
the reflection has no fixed points in $\Lambda^{\star}$.
The pair of operators
$\widehat{U}^{\,\vee}_{t}$
and
$\widehat{U}^{\,\vee}_{r}$
generates a $2^{2N}$-dimensional representation of the space group
$\mathrm{G}^{\,}_{\mathrm{spa}}$ through the semi-direct product
\begin{subequations}
\label{eq:G vee spa b'}
\begin{equation}
\mathrm{G}^{\,\vee}_{\mathrm{spa}}:=
\mathbb{Z}^{t}_{2N}
\rtimes
\mathbb{Z}^{r}_{2}
\end{equation}
with
\begin{equation}
\mathbb{Z}^{t}_{2N}\equiv
\left\{\vphantom{\Big[}
t,\ t^{2},\ \cdots,\ t^{2N-1},\ t^{2N}\equiv e
\right\},
\qquad
\mathbb{Z}^{r}_{2}\equiv
\left\{\vphantom{\Big[}
r,\ r^{2}\equiv e
\right\}.
\end{equation}
\end{subequations}
We note that the dual space group
$\mathrm{G}^{\vee}_{\mathrm{spa}}$
is isomorphic to the space group
$\mathrm{G}^{\,}_{\mathrm{spa}}$
defined in Eq.\ \eqref{eq:space group Lambda}.
However, the action of the dual reflection symmetry 
\eqref{eq:reps symmetries t and r for cal Bb' b} 
differs from that of reflection symmetry 
\eqref{eq:reps symmetries t and r for cal Bb a}
in the sense that it 
acts as a link-centered reflection 
on the dual lattice $\Lambda^{\star}$ 
and  does not admit any fixed points on 
$\Lambda^{\star}$.

The duals of the internal symmetries
\eqref{eq:reps rxpi rypi rzpi for cal Bb}
are constructed
by using the isomorphism between the bond algebras 
\eqref{eq:definition and symmetry of cal Bb} and 
\eqref{eq:definition and symmetry of cal Bb'}%
~\footnote{%
~The fact that internal symmetries are tensor products over all
sites of some local symmetry is crucial to validate the use
of the dual bond algebra. For example,    
applying the isomorphism between the bond algebras
\eqref{eq:definition and symmetry of cal Bb}
and 
\eqref{eq:definition and symmetry of cal Bb'} on the
generators of the crystalline symmetries produces operators that 
are gauge invariant, i.e., they commute with the local Gauss operators.
This is quite different from 
Eq.\ \eqref{eq:Gauss operator covariance cyrstalline},
according to  which the local Gauss operators transform
non-trivially but in a covariant manner under
conjugation by
$\widehat{U}^{\mathrm{ext}}_{t}$
and
$\widehat{U}^{\mathrm{ext}}_{r}$. 
}.
However, the corresponding local representations 
$\hat{\sigma}^{x}_{j}$ and $\hat{\sigma}^{y}_{j}$
do not belong to the bond algebra 
\eqref{eq:definition and symmetry of cal Bb}. 
In other words, they are not 
invariant under the global symmetry $\widehat{U}^{\,}_{r^{z}_{\pi}}$.
Therefore, when extending
to the Hilbert space $\mathcal{H}^{\,}_{b=0,\,b'=0}$, 
the operators 
$\widehat{U}^{\,}_{r^{x}_{\pi}}$ and $\widehat{U}^{\,}_{r^{y}_{\pi}}$
must be minimally coupled by the appropriate insertions of
$\hat{\tau}^{z}_{j^{\star}}$
operators. We focus on the operator
$\widehat{U}^{\,}_{r^{x}_{\pi}}$
as the case of 
$\widehat{U}^{\,}_{r^{y}_{\pi}}$
is treated analogously. We can extend the action of 
$\widehat{U}^{\,}_{r^{x}_{\pi}}$ to 
the Hilbert space 
$\mathcal{H}^{\,}_{b=0,\,b'=0}$ either according to the definition
\begin{subequations}
\label{eq:minimal coupling of rxpi}
\begin{align}
\widehat{U}^{\mathrm{ext}}_{r^{x}_{\pi}}
:=
\prod_{j=1}^{N}
\hat{\sigma}^{x}_{2j-1}\,
\hat{\tau}^{z}_{(2j-1)^{\star}}\,
\hat{\sigma}^{x}_{2j},
\label{eq:minimal coupling of rxpi a}
\end{align}
where $\hat{\tau}^{z}_{(2j)^{\star}-1}$ are inserted 
only on odd sites of the dual lattice $\Lambda^{\star}$,
or according to the definition
\begin{align}
\widehat{U}^{\mathrm{ext}}_{r^{x}_{\pi}}
:=
\prod_{j=1}^{N}
\hat{\sigma}^{x}_{2j}\,
\hat{\tau}^{z}_{(2j)^{\star}}\,
\hat{\sigma}^{x}_{2j+1},   
\label{eq:minimal coupling of rxpi b}
\end{align}
\end{subequations}
where $\hat{\tau}^{z}_{(2j)^{\star}-1}$ are inserted 
only on even sites of the dual lattice 
$\Lambda^{\star}$.
Crucially,
neither definition
(\ref{eq:minimal coupling of rxpi a})
nor definition
(\ref{eq:minimal coupling of rxpi b}) 
are invariant under translation
(\ref{eq:reps symmetries t and r for cal Bb' ext a})
or
reflection
(\ref{eq:reps symmetries t and r for cal Bb' ext b}), 
i.e., the extended operator 
$\widehat{U}^{\mathrm{ext}}_{r^{x}_{\pi}}$ is not 
invariant under the action of translation by one unit cell or 
by reflection. This incompatibility is
rooted in the non-trivial 
local projective representation
\eqref{eq:Pauli algebra is a non-trivial proj rep SO(3)}. 
Equivalently, this is a result of the two LSM Theorems 
\ref{thm:LSM translation} and 
\ref{thm:LSM reflection} with translation and reflection
symmetries, respectively.

By projecting onto the Hilbert space $\mathcal{H}^{\,}_{b'=0}$
operators \eqref{eq:minimal coupling of rxpi a} and 
\eqref{eq:minimal coupling of rxpi b}%
~\footnote{%
~Operators
$\widehat{U}^{\,\vee}_{\mathrm{o}}$
and
$\widehat{U}^{\,\vee}_{\mathrm{e}}$
also follow from similarly dualizing 
$\widehat{U}^{\,}_{r^{y}_{\pi}}$.
Only $\hat{\tau}^{z\,\vee}_{j}$ enters in the products
making up
$\widehat{U}^{\,\vee}_{\mathrm{e}}$
and
$\widehat{U}^{\,\vee}_{\mathrm{o}}$
in Eq.\ (\ref{eq:reps ro re rz dual b'}).
Hence, these dual generators of the internal symmetries
are not realized projectively locally. 
\label{footnote:KW dual symmetries are not projective}
},
we identify the following dual internal symmetries
\begin{align}
\widehat{U}^{\,\vee}_{\mathrm{o}}=
\prod_{j=1}^{N}
\hat{\tau}^{z\,\vee}_{2j-1+\frac{1}{2}},
\qquad
\widehat{U}^{\,\vee}_{\mathrm{e}}=
\prod_{j=1}^{N}
\hat{\tau}^{z\,\vee}_{2j+\frac{1}{2}},
\qquad
\widehat{U}^{\,\vee}_{r^{z}_{\pi}}=
\widehat{U}^{\,\vee}_{\mathrm{o}}\,
\widehat{U}^{\,\vee}_{\mathrm{e}}=
\prod_{j=1}^{2N}
\hat{\tau}^{z\,\vee}_{j+\frac{1}{2}}.
\label{eq:reps ro re rz dual b'} 
\end{align}
Note that the product of $\widehat{U}^{\,\vee}_{\mathrm{o}}$ 
and $\widehat{U}^{\,\vee}_{\mathrm{e}}$ delivers the dual symmetry 
of the bond algebra $\mathfrak{B}^{\,}_{b'=0}$ defined in Eq.\ 
\eqref{eq:definition and symmetry of cal Bb' b}.
The pair of operators
$\widehat{U}^{\,\vee}_{\mathrm{o}}$
and
$\widehat{U}^{\,\vee}_{\mathrm{e}}$
generates a $2^{2N}$-dimensional representation of the symmetry group
$\mathrm{G}^{\,}_{\mathrm{int}}$ through the direct product
\begin{subequations}
\label{eq:G vee int b'}
\begin{equation}
\mathrm{G}^{\,\vee}_{\mathrm{int}}\equiv  
\mathbb{Z}^{\mathrm{o}}_{2}\times\mathbb{Z}^{\mathrm{e}}_{2}
\label{eq:G vee int b' a}
\end{equation}
with
\begin{equation}
\mathbb{Z}^{\mathrm{o}}_{2}\equiv
\left\{\vphantom{\Big[}
r^{\,}_{\mathrm{o}},\
(r^{\,}_{\mathrm{o}})^{2}\equiv e
\right\},
\qquad
\mathbb{Z}^{\mathrm{e}}_{2}\equiv
\left\{\vphantom{\Big[}
r^{\,}_{\mathrm{e}},\
(r^{\,}_{\mathrm{e}})^{2}\equiv e
\right\}.
\label{eq:G vee int b' v}
\end{equation}
\end{subequations}

Unlike the case in Sec.\ \ref{subsec:G total frak H b=0}
with the space group $\mathrm{G}^{\,}_{\mathrm{spa}}$,
the action of $\mathrm{G}^{\,\vee}_{\mathrm{spa}}$
on $\mathrm{G}^{\,\vee}_{\mathrm{int}}$
is now non-trivial as it is given by the composition rules
\begin{subequations}
\label{eq:main result Gveespace if KW}
\begin{align}
\widehat{U}^{\vee}_{t}\,
\widehat{U}^{\,\vee}_{\mathrm{o}}
\left(
\widehat{U}^{\,\vee}_{t}
\right)^{\dag}=
\widehat{U}^{\,\vee}_{\mathrm{e}},
\qquad
\widehat{U}^{\vee}_{t}\,
\widehat{U}^{\,\vee}_{\mathrm{e}}
\left(
\widehat{U}^{\,\vee}_{t}
\right)^{\dag}=
\widehat{U}^{\,\vee}_{\mathrm{o}},
\label{eq:main result Gveespace if KW a}
\end{align}
and
\begin{align}
\widehat{U}^{\vee}_{r}\,
\widehat{U}^{\,\vee}_{\mathrm{o}}
\left(
\widehat{U}^{\,\vee}_{r}
\right)^{\dag}=
\widehat{U}^{\,\vee}_{\mathrm{e}},
\qquad
\widehat{U}^{\vee}_{r}\,
\widehat{U}^{\,\vee}_{\mathrm{e}}
\left(
\widehat{U}^{\,\vee}_{r}
\right)^{\dag}=
\widehat{U}^{\,\vee}_{\mathrm{o}}.
\label{eq:main result Gveespace if KW b}
\end{align}
\end{subequations}
In other words, the dual symmetry group
\begin{equation}
\mathrm{G}^{\,\vee}_{\mathrm{tot}}\equiv
\mathrm{G}^{\,\vee}_{\mathrm{spa}}
\ltimes
\mathrm{G}^{\,\vee}_{\mathrm{int}}
\label{eq:total symmetry group for KW duality}
\end{equation}
of the Hamiltonian $\widehat{H}^{\,\vee}_{b^{\prime}=0}$
in the bond algebra
(\ref{eq:definition and symmetry of cal Bb'}) with $b'=0$
that is dual to the Hamiltonian $\widehat{H}^{\,}_{b=0}$
in the bond algebra 
\eqref{eq:definition and symmetry of cal Bb}
with $b=0$ is a semi-direct product of
crystalline symmetries
$\mathrm{G}^{\,\vee}_{\mathrm{spa}}$
and
internal symmetries
$\mathrm{G}^{\,\vee}_{\mathrm{int}}$.

One observes that the two LSM Theorems
\ref{thm:LSM translation} and \ref{thm:LSM reflection}
do not apply to the dual symmetry group $\mathrm{G}^{\,\vee}_{\mathrm{tot}}$.
This is because the local representation of $\mathrm{G}^{\,\vee}_{\mathrm{int}}$
is not projective (see footnote
\ref{footnote:KW dual symmetries are not projective})
unlike that of $\mathrm{G}^{\,}_{\mathrm{int}}$.
We further note that while being isomorphic to $\mathrm{G}^{\,}_{\mathrm{spa}}$
the dual crystalline symmetry group $\mathrm{G}^{\vee}_{\mathrm{spa}}$ 
is such that
\begin{enumerate}
\item 
the ``natural'' unit cell on which the internal symmetry group 
$\mathrm{G}^{\,\vee}_{\mathrm{int}}$ acts onsite is associated with
the generator
$\left(\widehat{U}^{\vee}_{t}\right)^{2}$
of translations, i.e., it is twice that
of the unit cell associated with the generator
$\widehat{U}^{\vee}_{t}$ of translations%
~\footnote{%
~The factor of two 
here is directly related to the fact that the 
local non-trivial projective representation
\eqref{eq:Pauli algebra is a non-trivial proj rep SO(3)}
becomes a trivial representation on doubled unit cells.
},

\item 
the operator $\widehat{U}^{\vee}_{r}$ acts as a link-centered reflection 
on lattice $\Lambda^{\star}$ such that there are no invariant unit cells.
\end{enumerate}
Both properties
can be interpreted as a trivialization of mixed anomalies
between internal and spatial symmetries
under the gauging of a subgroup
of the internal symmetries.  

There is another useful reinterpretation of the
dual internal symmetries whose generators are
defined in Eq.\ (\ref{eq:reps ro re rz dual b'}).
First, we have the identity
\begin{equation}
\widehat{U}^{\vee}_{r^{z}_{\pi}}=:
e^{\mathrm{i}\pi\,\widehat{Q}^{\vee}},
\qquad
\widehat{Q}^{\vee}:=
\sum_{j^{\star}\in\Lambda^{\star}}
\widehat{Q}^{\vee}_{j^{\star}},
\qquad
\left[
\widehat{Q}^{\vee}_{i^{\star}},
\widehat{Q}^{\vee}_{j^{\star}}  
\right]=0,
\qquad
\forall i^{\star},j^{\star}\in\Lambda^{\star},
\end{equation}
where the local Hermitean operator
$\widehat{Q}^{\vee}_{j^{\star}}$
has the $\mathbb{Z}^{\,}_{2}$-valued
local charge eigenvalue $q^{\,}_{j^{\star}}=0,1$.
Second, we have the identity
\begin{equation}
\widehat{U}^{\vee}_{\mathrm{o}}=
\left(-\mathrm{i}\right)^{2N}\,
e^{\mathrm{i}\pi\,\widehat{D}^{\vee}}=
\widehat{U}^{\vee}_{t}\,
\widehat{U}^{\,\vee}_{\mathrm{e}}
\left(
\widehat{U}^{\,\vee}_{t}
\right)^{\dag},
\qquad
\widehat{D}^{\vee}:=
\sum_{j^{\star}\in\Lambda^{\star}}
\widehat{Q}^{\vee}_{j^{\star}}\,  
j^{\star}.
\end{equation}
The symmetry generator
$\widehat{U}^{\vee}_{\mathrm{o}}$
can thus be thought of as the exponential of the conserved global
$\mathbb{Z}^{\,}_{2}$-dipole operator
$\widehat{D}^{\vee}$
associated to the conserved global $\mathbb{Z}^{\,}_{2}$-charge operator
$\widehat{Q}^{\vee}$. The punchline is now the following.
Gauging the $\mathbb{Z}^{\,}_{2}$ charge symmetry
generated by $\widehat{U}^{\vee}_{r^{z}_{\pi}}$
induces a duality between Hamiltonians 
invariant under both
$\mathbb{Z}^{\,}_{2}$-dipole and translation (or link-centered reflection) 
symmetries%
~\footnote{%
The presence of both translation 
(or link-centered reflection) and dipole symmetries imply the presence
of a charge symmetry.}
that are free from LSM anomalies
and Hamiltonians invariant under
$\mathbb{Z}^{x}_{2}\times\mathbb{Z}^{y}_{2}$ internal 
and translation (or site-centered reflection) symmetries with LSM anomalies.
In other words, spatially modulated symmetries, such as a dipole symmetry,
can be mapped to a global uniform symmetry at the cost of introducing 
an LSM anomaly.

In anticipation of the discussion of the phase diagram
of the quantum spin-1/2 $XYZ$ chain in Sec.\ 
\ref{sec:Application to quantum spin-1/2 degrees of freedom on a chain},
we close this discussion by focusing
on the reflection symmetry subgroup 
$\mathbb{Z}^{r}_{2}$ of $\mathrm{G}^{\vee}_{\mathrm{spa}}$.
As a consequence of the underlying LSM anomaly,
the Abelian group 
\begin{subequations}
\begin{align}
\mathbb{Z}^{r}_{2}\times\mathbb{Z}^{x}_{2}\times\mathbb{Z}^{y}_{2}
\end{align}
formed by the subgroup of reflection symmetry $\mathbb{Z}^{r}_{2}$
together with the group of internal symmetries
$G^{\,}_{\mathrm{int}}\equiv\mathbb{Z}^{x}_{2}\times\mathbb{Z}^{y}_{2}$ 
is mapped to the non-Abelian dihedral group of order eight
\begin{equation}
\begin{split}
\mathrm{D}^{\,}_{8}:=&\,
\left\{\vphantom{\Big[}
r,\ r^{2}\equiv e
\right\}
\ltimes
\left\{\left.\vphantom{\Big[}
e,\
r^{\,}_{\mathrm{o}},\
r^{\,}_{\mathrm{e}},\
r^{\,}_{\mathrm{o}}\,
r^{\,}_{\mathrm{e}}
\ \right|\
\left(r^{\,}_{\mathrm{o}}\right)^{2}\equiv
\left(r^{\,}_{\mathrm{e}}\right)^{2}\equiv
e,
\qquad
r^{\,}_{\mathrm{o}}\,
r^{\,}_{\mathrm{e}}=
r^{\,}_{\mathrm{e}}\,
r^{\,}_{\mathrm{o}}
\right\}
\\
=&\,
\left\{\left.\vphantom{\Big[}
e,\
a,\
a^{2},\
a^{3},\
r,\
r\,a,\
r\,a^{2},\
r\,a^{3}
\ \right|\
a\equiv r\,r^{\,}_{\mathrm{o}},
\quad
a^{4}\equiv r^{2}\equiv e,
\quad
r\,a\,r=
a^{3}
\right\}
\end{split}
\label{eq:total symmetry group for KW duality b}
\end{equation}	
\end{subequations}
after gauging the diagonal subgroup $\mathbb{Z}^{z}_{2}\subset 
\mathbb{Z}^{x}_{2}\times\mathbb{Z}^{y}_{2}$ by KW duality.

\subsection{Jordan-Wigner dual of the LSM anomaly 
}
\label{subsec:G total frak H f=0}

We are going to construct the dual total symmetry group
$\mathrm{G}^{\vee,\,\mathrm{F}}_{\mathrm{tot}}$
under the JW duality introduced in Sec.\
\ref{subsec:Jordan-Wigner duality b and f}.
To this end, we define the action of the
crystalline and internal symmetries on
the extended Hilbert space
$\mathcal{H}^{\,}_{b,f}$ defined in Eq.\ 
(\ref{eq:def bond algebra cal Bbf b})
and then project these symmetries onto the dual Hilbert space 
$\mathcal{H}^{\vee}_{f}$.
We keep the boundary condition $f$ unspecified for the time being.

The extension of the crystalline symmetries 
\eqref{eq:def crystalline symmetries Lambda}
on the Hilbert space $\mathcal{H}^{}_{b,f}$
are obtained by demanding the covariance of the Gauss operators
(\ref{eq:sym bond algebra cal Bbf (a)})
under translation and reflection.
We thus define the unitary operators 
\begin{subequations}
\label{eq:reps symmetries t and r for cal Bf} 
\begin{align}
&
\widehat{U}^{\,\vee}_{t,\,f}:=
\left(
\mathrm{i}\hat{\beta}^{\,}_{1}\,\hat{\alpha}^{\,}_{1}
\right)^{f}
\prod_{j=1}^{2N-1}
\frac{\mathrm{i}}{2}
\left[
\left(
\hat{\beta}^{\vee}_{j}
-
\hat{\beta}^{\vee}_{t(j)}
\right)
\left(
\hat{\alpha}^{\vee}_{j}
-
\hat{\alpha}^{\vee}_{t(j)}
\right)
\right],
\label{eq:reps symmetries t and r for cal Bf a}
\\
&
\widehat{U}^{\,\vee}_{r,\,f}:=
\left(
\mathrm{i}\hat{\beta}^{\,}_{2N}\,
\hat{\alpha}^{\,}_{2N}
\right)^{f}
\prod_{j=1}^{2N}
\frac{1}{\sqrt{2}}
\left(
\widehat{\mbb{1}}^{\,}_{\mathcal{H}^{\vee}_{f}}
+
\hat{\beta}^{\vee}_{r(j)}\,
\hat{\alpha}^{\vee}_{j}
\right),
\label{eq:reps symmetries t and r for cal Bf b} 
\end{align}
\end{subequations}
where the  global fermion parity $\widehat{P}^{\,\vee}_{\mathrm{F}}$
takes the form \eqref{eq:definition and symmetry of cal Bf b}. 
For any $j\in\Lambda$, conjugation of
$\hat{\alpha}^{\vee}_{j}$ and $\hat{\beta}^{\vee}_{j}$
by $\widehat{U}^{\,\vee}_{t,\,f}$ and $\widehat{U}^{\,\vee}_{r,\,f}$
implement the maps
\begin{subequations}
\begin{align}
&
\hat{\alpha}^{\vee}_{j}\mapsto
(-1)^{f\,\delta^{\,}_{j,2N}}\,
\hat{\alpha}^{\vee}_{t(j)},
\qquad
\hat{\beta}^{\vee}_{j}\mapsto
(-1)^{f\,\delta^{\,}_{j,2N}}\,
\hat{\beta}^{\vee}_{t(j)},
\\
&
\hat{\alpha}^{\vee}_{j}\mapsto
+
(-1)^{f\,\delta^{\,}_{j,2N}}\,
\hat{\beta}^{\vee}_{r(j)},
\qquad
\hat{\beta}^{\vee}_{j}\mapsto
-
(-1)^{f\,\delta^{\,}_{j,2N}}\,
\hat{\alpha}^{\vee}_{r(j)},
\end{align}
\end{subequations}
respectively.

We note that,
unlike the dual spin operators 
$\hat{\bm{\tau}}^{\,}_{j^{\star}}$
defined on the dual lattice $\Lambda^{\star}$
in Sec.\
\ref{subsec:G total frak H b'=0},
the Majorana operators are defined on the direct lattice $\Lambda$.
This is due to the fact that we applied an isomorphism implementing
an additional half lattice translation in the process of JW duality
[see Eq.\ \eqref{eq:dual Majorana algebra for JW duality}].
Due to this nuance,
the reflection symmetry acts differently
on the Majorana degrees of freedom than it did on the spins from
Sec.\ \ref{subsec:G total frak H b'=0},
since none of the sites of $\Lambda^{\star}$
are invariant under reflection,
while the sites $j=N,\,2\,N\in \Lambda$ are left fixed under reflection.
Furthermore, in the fermionic case,
reflection is not an order two operation. Instead,
one verifies that
\begin{subequations}
\begin{equation}
\left(
\widehat{U}^{\,\vee}_{r,\,f}
\right)^{2}=
-
\widehat{P}^{\,\vee}_{\mathrm{F}}
\,.
\label{eq:reps symmetries t and r for cal Bf c} 
\end{equation}
Similarly, translation is not an order $2N$ operator if $f=1$,
instead
\begin{equation}
\left(
\widehat{U}^{\,\vee}_{t,\,f}
\right)^{2N}=
\left(\widehat{P}^{\,\vee}_{\mathrm{F}}\right)^{f}.
\label{eq:reps symmetries t and r for cal Bf d}     
\end{equation}
\end{subequations}
This leads to a mixing of crystalline symmetries with the fermion parity.
We denote the crystalline group obtained after JW duality as
$\mathrm{G}^{\vee,\,\mathrm{F}}_{\mathrm{spa}}$. 
This group is obtained by
the central extension of $\mathrm{G}^{\vee}_{\mathrm{spa}}$ defined in 
Eq.\ \eqref{eq:G vee spa b'} by fermion parity 
$\mathbb{Z}^{\mathrm{F}}_{2}$ specified by the short exact sequence
\begin{align}
0 \rightarrow 
\mathbb{Z}^{\mathrm{F}}_{2}
\rightarrow
\mathrm{G}^{\vee,\,\mathrm{F}}_{\mathrm{spa}}
\rightarrow
\mathrm{G}^{\vee}_{\mathrm{spa}}
\rightarrow
0
\end{align}
with the extension class $[\gamma^{\,}_{f}]\in
H^{2}(\mathrm{G}^{\vee}_{\mathrm{spa}},\mathbb{Z}^{\mathrm{F}}_{2})$
and the extension map 
\begin{align}
\gamma^{\,}_{f}(r,r):= p^{\,}_{\mathrm{F}},
\qquad 
\gamma^{\,}_{f}(t^{a}, t^{b})=
\left(p^{\,}_{\mathrm{F}}\right)^{f\,\lfloor (a+b)/2N\rfloor},
\qquad
\gamma^{\,}_{f}(r, t)=
\left(p^{\,}_{\mathrm{F}}\right)^{f},
\end{align}
where $p^{\,}_{\mathrm{F}}$ was defined in Eq.\
(\ref{eq:def Majorana algebra on mathfrak{H}bF c})
and 
$\lfloor\cdot\rfloor$ is the lower floor function.
All other maps can be derived using these relations and
the cocycle condition for $\gamma^{\,}_{f}$.
Having defined the crystalline symmetries, 
we now turn to the internal symmetries.

After the JW duality, the internal symmetry operators are obtained by
dualizing $\widehat{U}^{\,}_{r^{x}_{\pi}}$ and
$\widehat{U}^{\,}_{r^{y}_{\pi}}$ in Eq.\
\eqref{eq:reps rxpi rypi rzpi for cal Bb}.
More precisely, under the JW duality%
~\footnote{%
~We obtain the operator $\widehat{U}^{\,\vee}_{\mathrm{o}}$
from dualizing $\widehat{U}^{\,}_{r^{x}_{\pi}}$
and multiplying with $(-1)^{N}$.
This multiplicative factor simplifies the algebra.
}
\begin{subequations}
\label{eq:reps ro re rz dual f}
\begin{align}
&
\widehat{U}^{\,}_{r^{x}_{\pi}}=
\prod_{j=1}^{N}
\hat{\sigma}^{x}_{2j-1}\,
\hat{\sigma}^{x}_{2j} \longmapsto 
\widehat{U}^{\,\vee}_{\mathrm{o}}:=
\prod_{j=1}^{N}
\left(
\mathrm{i}
\hat{\alpha}^{\vee}_{2j-1}\,
\hat{\beta}^{\vee}_{2j}
\right),
\label{eq:reps ro re rz dual f a}  
\\
&
\widehat{U}^{\,}_{r^{y}_{\pi}}=
\prod_{j=1}^{N}
\hat{\sigma}^{y}_{2j-1}\,
\hat{\sigma}^{y}_{2j}\longmapsto 
\widehat{U}^{\,\vee}_{\mathrm{e}}:=
\,
\prod_{j=1}^{N}
\left(
\mathrm{i}
\hat{\beta}^{\vee}_{2j-1}\,
\hat{\alpha}^{\vee}_{2j}
\right)\,.
\label{eq:reps ro re rz dual f b}  
\end{align}
The pair $\widehat{U}^{\,\vee}_{\mathrm{o}}$ and
$\widehat{U}^{\,\vee}_{\mathrm{e}}$
of dual internal symmetry operators
compose to the fermion parity operator,
\begin{equation}
\widehat{U}^{\,\vee}_{\mathrm{o}}\,
\widehat{U}^{\,\vee}_{\mathrm{e}}=
\widehat{P}^{\,\vee}_{\mathrm{F}}\,.
\label{eq:reps ro re rz dual f c}  
\end{equation}
\end{subequations}
The pair of operators
$\widehat{U}^{\,\vee}_{\mathrm{o}}$
and
$\widehat{U}^{\,\vee}_{\mathrm{e}}$
generates a $2^{2N}$-dimensional representation of the internal symmetry group
\begin{subequations}
\label{eq:G vee int f}
\begin{equation}
\mathrm{G}^{\,\vee,\,\mathrm{F}}_{\mathrm{int}}\equiv  
\mathbb{Z}^{\mathrm{o}}_{2}\times\mathbb{Z}^{\mathrm{e}}_{2}
\label{eq:G vee int f a}
\end{equation}
with
\begin{equation}
\mathbb{Z}^{\mathrm{o}}_{2}\equiv
\left\{\vphantom{\Big[}
r^{\,}_{\mathrm{o}},\
(r^{\,}_{\mathrm{o}})^{2}\equiv e
\right\},
\qquad
\mathbb{Z}^{\mathrm{e}}_{2}\equiv
\left\{\vphantom{\Big[}
r^{\,}_{\mathrm{e}},\
(r^{\,}_{\mathrm{e}})^{2}\equiv e
\right\}.
\label{eq:G vee int f b}
\end{equation}
\end{subequations}

The generators \eqref{eq:reps symmetries t and r for cal Bf} 
of the dual crystalline symmetries act on the 
operators 
$\widehat{U}^{\,\vee}_{\mathrm{o}}$
and
$\widehat{U}^{\,\vee}_{\mathrm{e}}$
according to the composition rules
\begin{equation}
\label{eq:main result Gveespace if JW}
\begin{alignedat}{3}
&
\widehat{U}^{\vee}_{t}\,
\widehat{U}^{\,\vee}_{\mathrm{o}}
\left(
\widehat{U}^{\,\vee}_{t}
\right)^{\dag}=
(-1)^{f+1}\,
\widehat{U}^{\,\vee}_{\mathrm{e}},
\qquad
\qquad
&&
\widehat{U}^{\vee}_{r}\,
\widehat{U}^{\,\vee}_{\mathrm{o}}
\left(
\widehat{U}^{\,\vee}_{r}
\right)^{\dag}&&=
(-1)^{f+1}\,\widehat{U}^{\,\vee}_{\mathrm{e}},
\\
&
\widehat{U}^{\vee}_{t}\,
\widehat{U}^{\,\vee}_{\mathrm{e}}
\left(
\widehat{U}^{\,\vee}_{t}
\right)^{\dag}=
(-1)^{f+1}\,\widehat{U}^{\,\vee}_{\mathrm{o}},
\qquad
\qquad
&&
\widehat{U}^{\vee}_{r}\,
\widehat{U}^{\,\vee}_{\mathrm{e}}
\left(
\widehat{U}^{\,\vee}_{r}
\right)^{\dag}&&=
(-1)^{f+1}\,\widehat{U}^{\,\vee}_{\mathrm{o}},
\\
&
\widehat{U}^{\vee}_{t}\,
\widehat{P}^{\,\vee}_{\mathrm{F}}
\left(
\widehat{U}^{\,\vee}_{t}
\right)^{\dag}=
\widehat{P}^{\,\vee}_{\mathrm{F}}, 
\qquad
\qquad
&&\widehat{U}^{\vee}_{r}\,
\widehat{P}^{\,\vee}_{\mathrm{F}}
\left(
\widehat{U}^{\,\vee}_{r}
\right)^{\dag}&&=
\widehat{P}^{\,\vee}_{\mathrm{F}}\,.
\end{alignedat}
\end{equation}
The total symmetry group $\mathrm{G}^{\vee,\,\mathrm{F}}_{\mathrm{tot}}$
is obtained by taking the semi-direct product of 
$\mathrm{G}^{\vee,\,\mathrm{F}}_{\mathrm{spa}}$
and $\mathrm{G}^{\vee,\,\mathrm{F}}_{\mathrm{int}}$ 
together with coseting
by the fermion parity group $\mathbb{Z}^{\mathrm{F}}_{2}$
defined in Eq.\ (\ref{eq:def Majorana algebra on mathfrak{H}bF c}), i.e.,
\begin{subequations}
\begin{equation}
\mathrm{G}^{\vee,\,\mathrm{F}}_{\mathrm{tot}}=
\left(
\mathrm{G}^{\vee,\,\mathrm{F}}_{\mathrm{spa}}
\ltimes
\mathrm{G}^{\vee,\,\mathrm{F}}_{\mathrm{int}}
\right)
\,\Big/\,\mathbb{Z}^{\mathrm{F}}_{2}.
\label{eq:GtotVF JW}
\end{equation} 
Here, the semi-direct product
$
\mathrm{G}^{\vee,\,\mathrm{F}}_{\mathrm{spa}}
\ltimes
\mathrm{G}^{\vee,\,\mathrm{F}}_{\mathrm{int}}
$ 
is specified by the action 
\begin{equation}
\begin{alignedat}{3}
t\,
r^{\,}_{\mathrm{o}}
t^{-1}&=
r^{\,}_{\mathrm{e}}, 
\qquad 
&&
r\,
r^{\,}_{\mathrm{o}}
r^{-1}
&&=
r^{\,}_{\mathrm{e}},
\\
t\,
r^{\,}_{\mathrm{e}}
t^{-1}&=
r^{\,}_{\mathrm{o}}, 
\qquad 
&&
r\,
r^{\,}_{\mathrm{e}}
r^{-1}
&&=
r^{\,}_{\mathrm{0}},
\\
t\,
p^{\,}_{\mathrm{F}}
t^{-1}&=
p^{\,}_{\mathrm{F}}, 
\qquad 
&&
r\,
p^{\,}_{\mathrm{F}}
r^{-1}
&&=
p^{\,}_{\mathrm{F}}.
\end{alignedat}
\end{equation}
\end{subequations}
of dual crystalline symmetry group
$\mathrm{G}^{\vee,\,\mathrm{F}}_{\mathrm{spa}}$ on the dual internal symmetry group
$\mathrm{G}^{\,\vee,\,\mathrm{F}}_{\mathrm{int}}$.
We emphasize that the structure of
$\mathrm{G}^{\vee,\,\mathrm{F}}_{\mathrm{tot}}$ is different from
$\mathrm{G}^{\vee}_{\mathrm{tot}}$ in Eq.\
\eqref{eq:total symmetry group for KW duality}
obtained via the KW duality. More precisely, under
the JW duality the resulting dual total symmetry group
$\mathrm{G}^{\,\vee,\,\mathrm{F}}_{\mathrm{tot}}$
is assembled from the
crystalline
$\mathrm{G}^{\,\vee,\,\mathrm{F}}_{\mathrm{spa}}$
and internal
$\mathrm{G}^{\,\vee,\,\mathrm{F}}_{\mathrm{int}}$
symmetry groups using a
nontrivial central extension in addition to the semi-direct product
structure.
In contrast, the dual of
$\mathrm{G}^{\,}_{\mathrm{tot}}$
under the KW duality described in Sec.\
\ref{subsec:G total frak H b'=0} is
a semi-direct product of the crystalline and internal symmetry groups.
Having set $b=b'=0$, in Secs.\
\ref{subsec:G total frak H b=0}
and
\ref{subsec:G total frak H b'=0},
triality of the bond algebras
enforces
$f=1$
as prescribed in Eq.\
\eqref{eq:triality relation}.
Finally, we observe that the two LSM Theorems
\ref{thm:LSM translation} and \ref{thm:LSM reflection}
do not apply to the dual symmetry group
$\mathrm{G}^{\,\vee,\,\mathrm{F}}_{\mathrm{tot}}$.
This is because the local representation of
$\mathrm{G}^{\,\vee,\,\mathrm{F}}_{\mathrm{int}}$
is not projective,
unlike that of $\mathrm{G}^{\,}_{\mathrm{int}}$.
As was the case with the KW dual
$\mathrm{G}^{\,\vee}_{\mathrm{int}}$
generated by the operators in Eq.\
(\ref{eq:reps ro re rz dual b'}),
the trivialization of mixed anomalies
between internal and spatial symmetries under
the JW gauging of a subgroup of the
internal symmetries can be attributed to a doubling of the natural unit cell
for the JW dual internal symmetries.

It is possible to reinterpret the dual symmetries that are
defined in Eq.\ \eqref{eq:reps ro re rz dual f}
as $\mathbb{Z}^{\,}_{2}$-charge and  $\mathbb{Z}^{\,}_{2}$-dipole symmetries,
respectively.  To see this, we employ a ``half''-translation
\begin{align}
\hat{\alpha}^{\vee}_{j}\mapsto \hat{\beta}^{\vee}_{j+1},
\qquad
\hat{\beta}^{\vee}_{j}\mapsto \hat{\alpha}^{\vee}_{j},
\end{align} 
after which the operators $\widehat{U}^{\vee}_{\mathrm{o}}$
and $\widehat{U}^{\vee}_{\mathrm{e}}$ act on only even 
and only odd sites, respectively.
If so, the dual internal symmetries satisfy
\begin{subequations}
\begin{align}
&
\widehat{P}^{\,\vee}_{\mathrm{F}}=:
e^{\mathrm{i}\pi\,\widehat{Q}^{\,\vee}},
\qquad
\widehat{Q}^{\,\vee}:=
\sum_{j^{\star}\in\Lambda^{\star}}
\widehat{Q}^{\,\vee}_{j^{\star}},
\qquad
\left[
\widehat{Q}^{\,\vee}_{i^{\star}},
\widehat{Q}^{\,\vee}_{j^{\star}}  
\right]=0,
\\
&
\widehat{U}^{\,\vee}_{\mathrm{o}}=
e^{\mathrm{i}\pi\,\widehat{D}^{\,\vee}}=
(-1)^{f+1}
\widehat{U}^{\,\vee}_{t}\,
\widehat{U}^{\,\vee}_{\mathrm{e}}
\left(
\widehat{U}^{\,\vee}_{t}
\right)^{\dag},
\qquad
\widehat{D}^{\,\vee}:=
\sum_{j^{\star}\in\Lambda^{\star}}
\widehat{Q}^{\,\vee}_{j^{\star}}\,  
j^{\star}.
\end{align}
\end{subequations}
where the local Hermitean operator
$\widehat{Q}^{\,\vee}_{j^{\star}}$
has the $\mathbb{Z}^{\,}_{2}$-valued
local charge eigenvalue $q^{\,}_{j^{\star}}=0,1$,
for any $i^{\star},j^{\star}\in\Lambda^{\star}$.
As was in Sec.\ \ref{subsec:G total frak H b'=0},
the symmetry generator
$\widehat{U}^{\,\vee}_{\mathrm{o}}$
can thus be thought of as a 
fermion dipole parity operator.
As we have discussed in Sec.\ 
\ref{subsec:Triality betweeb cal Bf cal Bb' cal Bb},
there are two ways of gauging fermion parity symmetry. 
These result in either (i) Hamiltonians with 
$\mathbb{Z}^{x}_{2}\times\mathbb{Z}^{y}_{2}$ internal 
and translation (or site-centered reflection) symmetries
with LSM anomalies, or (ii) 
$\mathbb{Z}^{\,}_{2}$-charge, 
$\mathbb{Z}^{\,}_{2}$-dipole,
and translation (or link-centered reflection) symmetries. 
Notice that both dual Hamiltonians are bosonic, however, the former 
is invariant under global uniform symmetries with LSM anomalies while 
the latter is invariant under spatially modulated symmetries. 

\section{Triality and the phase diagram of the quantum spin-1/2 $XYZ$ chain}
\label{sec:Application to quantum spin-1/2 degrees of freedom on a chain}

We apply the triality
derived in Secs.\
\ref{sec:Triality through bond algebra isomorphisms}
and
\ref{sec:Triality of crystalline and internal symmetries on a chain}
to the study of the zero-temperature phase diagram of
the quantum spin-1/2 $XYZ$ chain.  
This model is dualized to a spin-1/2 cluster model
and a model of interacting Majorana degrees of freedom under
the KW and JW dualities, respectively.
The triality allows to give three equivalent interpretations
of the zero-temperature phase diagram with an emphasis on
the symmetry structure of the bond algebras
$\mathfrak{B}^{\,}_{b}$,
$\mathfrak{B}^{\,}_{b'}$,
and
$\mathfrak{B}^{\,}_{f}$,
respectively.

\subsection{Quantum spin-1/2 $XYZ$ chain}
\label{subsec:Quantum XYZ model}

Consider the quantum spin-1/2 antiferromagnetic chain
with nearest- and next-nearest-neighbor antiferromagnetic couplings
with periodic boundary conditions described by the Hamiltonian
\begin{align}
\begin{split}
\widehat{H}^{\,}_{b=0}:=&\,
J^{\,}_{1}
\sum_{j\in\Lambda}
\left(
\Delta^{\,}_{x}\,
\hat{\sigma}^{x}_{j}\,
\hat{\sigma}^{x}_{j+1}
+
\Delta^{\,}_{y}\,
\hat{\sigma}^{y}_{j}\,
\hat{\sigma}^{y}_{j+1}
+
\Delta^{\,}_{z}\,
\hat{\sigma}^{z}_{j}\,
\hat{\sigma}^{z}_{j+1}
\right)
\\
&\,
+
J^{\,}_{2}\,
\sum_{j\in\Lambda}
\left(
\Delta^{\,}_{x}\,
\hat{\sigma}^{x}_{j}\,
\hat{\sigma}^{x}_{j+2}
+
\Delta^{\,}_{y}\,
\hat{\sigma}^{y}_{j}\,
\hat{\sigma}^{y}_{j+2}
+
\Delta^{\,}_{z}\,
\hat{\sigma}^{z}_{j}\,
\hat{\sigma}^{z}_{j+2}
\right)
\end{split}
\label{eq:def Hamiltonian b=0}
\end{align}
with the domain of definition
$\mathcal{H}^{\,}_{b=0}$
defined in Eq.\
(\ref{eq:def Pauli algebra on mathfrak{H}b}).
Both the dimensionful couplings
$J^{\,}_{1}$
and
$J^{\,}_{2}$
together with the
dimensionless couplings
$\Delta^{\,}_{x}$,
$\Delta^{\,}_{y}$,
and
$\Delta^{\,}_{z}$
are taken to be real-valued and non-negative.
For convenience, we set the cardinality of the lattice to be 
\begin{align}
|\Lambda|\equiv2N = 0 \text{ mod }4,
\end{align}
i.e.,  $N$ is an even integer.
The symmetries of
$\widehat{H}^{\,}_{b=0}$
that we shall keep track of are given in
Sec.\
\ref{subsec:G total frak H b=0}.
We choose to work with periodic boundary conditions
for the same reasons as in Sec.\
\ref{subsec:G total frak H b=0}, i.e., for $b=0$,
the total relevant symmetry group
$\mathrm{G}^{\,}_{\mathrm{tot}}$
is a direct product of a crystalline symmetry group
$\mathrm{G}^{\,}_{\mathrm{spa}}$
and of an internal symmetry group
$\mathrm{G}^{\,}_{\mathrm{int}}$.
Accordingly,
the spectrum of $\widehat{H}^{\,}_{b=0}$
is constrained by 
LSM Theorems 
\ref{thm:LSM translation}
and
\ref{thm:LSM reflection}.
Either one of
Theorems 
\ref{thm:LSM translation}
and
\ref{thm:LSM reflection} 
requires that any gapped phase in the parameter space of the model 
either breaks spontaneously the global internal symmetry
$\mathrm{G}^{\,}_{\mathrm{int}}$
or the crystalline symmetry
$\mathrm{G}^{\,}_{\mathrm{spa}}$,
or it is infinitely degenerate
in the thermodynamic limit%
~\cite{Dobson69,Stephenson70,Frankel70,Redner81,Harada83}.
The question that will be answered in Sec.\
\ref{sec:Application to quantum spin-1/2 degrees of freedom on a chain}
is that of the fate of Theorems 
\ref{thm:LSM translation}
and
\ref{thm:LSM reflection}
under the triality of
Secs.\
\ref{sec:Triality through bond algebra isomorphisms}
and
\ref{sec:Triality of crystalline and internal symmetries on a chain}.
To this end, we shall reinterpret the
zero-temperature
phase diagram of $\widehat{H}^{\,}_{b=0}$
after it has undergone a KW and JW
dualization to the Hamiltonians
$\widehat{H}^{\,\vee}_{b'}$
and
$\widehat{H}^{\,\vee}_{f}$,
respectively. 

\begin{figure}[t!]
\begin{center}
\begin{tikzpicture}[scale=0.85]
\begin{axis}[
title={},
axis line style = thick,
xlabel={$0\leq\tanh(\Delta)\le1$},
ylabel={$0\leq\tanh(J)\leq1$},
xmin=0, xmax=1,
ymin=0, ymax=1,
ticks=none,
]
\addplot[
domain=0:1,
thick,
dashed,
]{0.46};
\addplot[black, mark=*, mark size=4pt, only marks]
coordinates {(0.7,0)};
\addplot[black, mark=*, mark size=8pt, only marks]
coordinates {(0.7,1)};
\addplot[black, mark=square*, mark size=4pt, only marks]
coordinates {(0,0) (1,0)};
\addplot[black, mark=square*, mark size=8pt, only marks]
coordinates {(1,1) (0,1)};
\addplot[black, mark=diamond*, mark size=4pt, only marks]
coordinates {(0,0.46) (1,0.46)};
\node[draw] at (0.5,0.55){Majumdar-Ghosh line};
\end{axis}
\end{tikzpicture}
\end{center}
\caption{
Exactly soluble points in the 
phase diagram of Hamiltonian
(\ref{eq:def Hamiltonian b=0})
in the reduced coupling space
(\ref{eq:reduced coupling space}).
The small squares at the lower left and right corners of the phase diagram
each realize the antiferromagnetic nearest-neighbor Ising chain.
The large squares at the upper left and right corners of the phase diagram
each realize two identical and decoupled
antiferromagnetic nearest-neighbor Ising chains.
The ground states are long-range ordered,
gapped, and two-fold (four-fold) degenerate
for the lower (upper) corners.
The line $(\Delta,J=0)$ realizes non-interacting fermions
without fermion-number conservation, except when $\Delta=1$.
The small circle at $(\Delta=1,J=0)$
realizes non-interacting spinless fermions at half filling 
with a nearest-neighbor uniform hopping amplitude.
The ground state is gapless with a quantum criticality encoded by
a $c=1$ conformal field theory in $(1+1)$-dimensional spacetime
in the thermodynamic limit.
The large circle at $(\Delta=1,J=\infty)$
realizes 
two decoupled chains of non-interacting spinless fermions at half filling
with a nearest-neighbor uniform hopping amplitude.
The ground state is gapless with a quantum criticality behavior encoded by
a $c=2$ conformal field theory in $(1+1)$-dimensional spacetime
in the thermodynamic limit.
The diamonds at $(\Delta,J)=(0,1/2),(\infty,1/2)$ are first-order
boundaries between the phases governed by the Ising fixed points
(small and large squares)
at $\Delta=0$ and $\Delta=\infty$, respectively.
The open Majumdar-Ghosh (dashed) line at $(\Delta,1/2)$
with $0<\Delta<\infty$ realizes the dimer phase.
The dimer ground states are gapped and two-fold degenerate
along the Majumdar-Ghosh (dashed) line.
}
\label{fig:exactly soluble points XY phase diagram}
\end{figure}
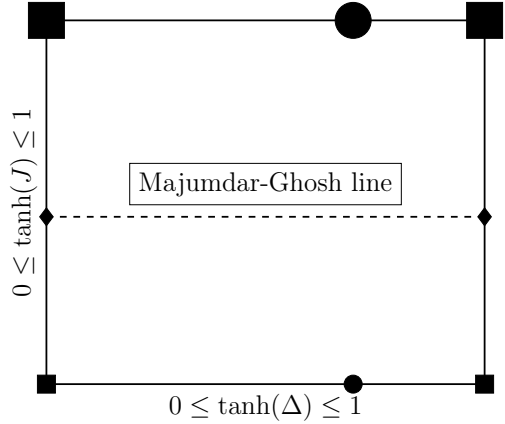

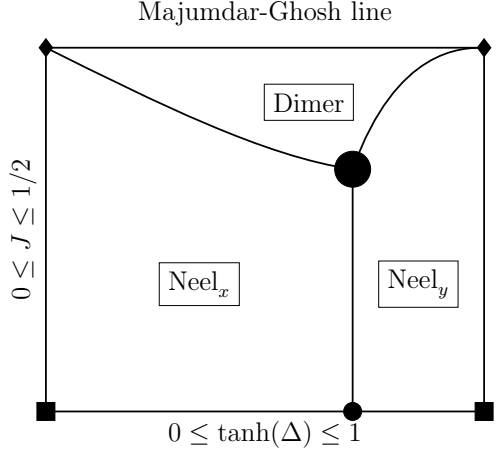
\begin{figure}[t!]
\begin{center}
\begin{tikzpicture}[scale=0.85]
\begin{axis}[
title={Majumdar-Ghosh line},
axis line style = thick,
xlabel={$0\leq\tanh(\Delta)\le1$},
ylabel={$0\leq J\leq1/2$},
xmin=0, xmax=1,
ymin=0, ymax=1,
ticks=none,
]
\addplot[black, mark=*, mark size=4pt, only marks]
coordinates {(0.7,0)};
\addplot[black, mark=square*, mark size=4pt, only marks]
coordinates {(0,0) (1,0)};
\addplot[black, mark=diamond*, mark size=4pt, only marks]
coordinates {(0,1) (1,1)};
\draw[
thick
] (0.7,0) -- (0.7,0.666);
\addplot[black, mark=, mark size=8pt, only marks]
coordinates {(0.7,0.666)};
\draw[thick] (0,1) .. controls (0.25,0.85) and (0.5,0.7) .. (0.7,0.666);
\draw[thick] (0.7,0.666)..controls (0.8,1) and (0.95,1)..(1,1);
\node[draw] at (0.35,0.35){$\hbox{Neel}^{\,}_{x}$};
\node[draw] at (0.85,0.35){$\hbox{Neel}^{\,}_{y}$};
\node[draw] at (0.6,0.85){$\hbox{Dimer}$};
\end{axis}
\end{tikzpicture}
\end{center}
\caption{
Phase diagram of Hamiltonian
(\ref{eq:def Hamiltonian b=0})
in the reduced coupling space
(\ref{eq:reduced coupling space})
with $0\leq J\leq1/2$. There are three
phases:
the $\mathrm{Neel}^{\,}_{x}$,
the $\mathrm{Neel}^{\,}_{y}$,
and the dimer phase.
Each one of these three phases
corresponds to gapped and two-fold degenerate
ground states in the thermodynamic limit.
In each phase, a non-degenerate ground state is selected by
spontaneous symmetry breaking of the
symmetry group $\mathrm{G}^{\,}_{\mathrm{tot}}$
defined in Eq.\ (\ref{eq:def Gtot for Hb}).
The dimer phase is found on both sides of
the open MG line
defined by $0<\Delta<\infty$ and $J=1/2$.
All the phase boundaries with $0<\Delta<\infty$ and $J<1/2$
are continuous quantum phase transitions that realize
deconfined quantum criticality \cite{Mudry19}. The tricritical point
(the large black circle)
where the three phases meet realizes the $\mathrm{SU}(2)^{\,}_{1}$
conformal field theory in $(1+1)$-dimensional spacetime.
}
\label{fig:spin-1/2 XY phase diagram}
\end{figure}

We define the ratios
\begin{subequations}
\label{eq:reduced coupling space} 
\begin{equation}
J:=\frac{J^{\,}_{2}}{J^{\,}_{1}},
\qquad
\Delta:=\frac{\Delta^{\,}_{y}}{\Delta^{\,}_{x}},
\label{eq:reduced coupling space a}
\end{equation}  
and consider, for simplicity, the reduced parameter space%
~\footnote{%
~The full coupling space,
if expressed in terms of dimensionless couplings only, is
$(\Delta^{\,}_{x},\Delta^{\,}_{y},\Delta^{\,}_{z},J)\in
[0,\infty[\times[0,\infty[\times[0,\infty[\times[0,\infty[$.
A detailed study of the corresponding zero-temperature phase diagram
for $0\leq J\leq1/2$ can be found in Ref.\ \cite{Mudry19}.
Knowledge of the phase diagram for the cut
$(\Delta^{\,}_{x},\Delta^{\,}_{y},0,J)\in
[0,\infty[\times[0,\infty[\times[0,\infty[$
is equivalent to knowledge of the phase diagram for the cut
$(\Delta^{\,}_{x},0,\Delta^{\,}_{z},J)\in
[0,\infty[\times[0,\infty[\times[0,\infty[$
through the application of a unitary $\mathrm{SU}(2)$-rotation
about the $x$ axis in spin-1/2 space.
}
\begin{equation}
0\leq\Delta\leq\infty,
\qquad
\Delta^{\,}_{z}=0,
\qquad
0\leq J\leq\infty.
\label{eq:reduced coupling space b}
\end{equation}
\end{subequations}
The energy eigenvalues and eigenvectors of Hamiltonian
(\ref{eq:def Hamiltonian b=0})
are known in closed forms at the four corners
\begin{subequations}
\label{eq:exactly soluble points AF J1 J2 spin1over2}
\begin{equation}
(\Delta,J)=(0,0),
\qquad
(\Delta,J)=(\infty,0),
\qquad
(\Delta,J)=(0,\infty),
\qquad
(\Delta,J)=(\infty,\infty),
\label{eq:exactly soluble points AF J1 J2 spin1over2 a}
\end{equation}
and the pair of points \cite{Lieb61}
\begin{equation}
(\Delta,J)=(1,0),
\qquad
(\Delta,J)=(1,\infty).
\label{eq:exactly soluble points AF J1 J2 spin1over2 b}
\end{equation}
The gapped ground states are also known
in closed form along the so-called Majumdar-Ghosh (MG) line
\cite{Majumdar69a,Majumdar69b,Majumdar70,Shastry81}
\begin{equation}
0\leq\Delta\leq\infty,
\qquad
J=\frac{1}{2}\,.
\label{eq:exactly soluble points AF J1 J2 spin1over2 c}
\end{equation}
The nature of the ground states of
the Hamiltonian
(\ref{eq:def Hamiltonian b=0})
at all these points in the reduced coupling space
(\ref{eq:reduced coupling space})
is summarized in
Fig.\ \ref{fig:exactly soluble points XY phase diagram}.
The ground states at the four corners
(\ref{eq:exactly soluble points AF J1 J2 spin1over2 a})
and along the open MG line $0<\Delta<\infty$, $J=1/2$
are gapped and degenerate.
Even though the exact degeneracies for any finite
cardinality $2N=|\Lambda|$
are lifted by small perturbations away from the four corners or
away from the open MG line,
these degeneracies are restored in the
thermodynamic limit $2N\to\infty$. 

Below the MG line
(\ref{eq:exactly soluble points AF J1 J2 spin1over2 c}),
there are three gapped phases \cite{Haldane82,Mudry19}, each of which
spontaneously breaks $\mathrm{G}^{\,}_{\mathrm{tot}}$ in the
thermodynamic limit through the spontaneous selection of
a ground state from two-fold degenerate ground states.
The $\mathrm{Neel}^{\,}_{x}$ phase is
adiabatically connected to the fixed-point limit at the lower left
corner $(\Delta,J)=(0,0)$ and spontaneously breaks translation
symmetry by one lattice spacing and the rotation symmetry about the
$y$-axis in spin-1/2 space. Similarly, the $\mathrm{Neel}^{\,}_{y}$
phase is adiabatically connected to the fixed-point limit
at the lower right corner
$(\Delta,J)=(\infty,0)$ and spontaneously breaks translation symmetry
and rotation symmetry about the $x$-axis in spin-1/2 space.  The dimer
phase is adiabatically connected
to the MG line and spontaneously breaks the symmetries under
translation by one lattice spacing
and the site-centered reflection, while preserving the
internal symmetries. This pattern of spontaneous symmetry breaking
precludes a continuous phase transition governed by the reduction of a
symmetry in one direction across the transition between any two of
these three phases that would follow the Landau-Ginzburg paradigm of
phase transitions.  Nevertheless, the boundaries between any two of
these three gapped phases when $0<\Delta<\infty$ realize continuous
quantum phase transitions.
In fact, they are examples of deconfined quantum critical transitions
\cite{Jiang19,Mudry19}.
A deconfined quantum critical transition is driven by
the deconfinement of point defects in one phase that nucleate locally
the local order of the phase on the other side of the transition
\cite{Senthil04,Senthil04PRB,Senthil04Levin,Senthil06Fisher}.
The two end points
\begin{equation}
(\Delta,J)=(0,1/2),
\qquad
(\Delta,J)=(\infty,1/2)
\label{eq:exactly soluble points AF J1 J2 spin1over2 d}
\end{equation}
\end{subequations}
of the MG line are gapped with a degeneracy proportional to
the cardinality $2N=|\Lambda|$.
Each becomes the phase boundary in the antiferromagnetic Ising chain
with competing nearest- and next-nearest-neighbor interactions
at which a first-order phase transition takes place in the thermodynamic limit
\cite{Dobson69,Stephenson70,Frankel70,Redner81,Harada83}.
The phase diagram of the Hamiltonian
(\ref{eq:def Hamiltonian b=0})
has been studied by numerical means both in the reduced coupling space
(\ref{eq:reduced coupling space})
\cite{Tonegawa88}
as well as without the restriction $\Delta^{\,}_{z}=0$
\cite{Tonegawa92,Nersesyan98,Hikihara01,Furukawa12,Mudry19}.
Below the MG line
(\ref{eq:exactly soluble points AF J1 J2 spin1over2 c}),
the phase diagram deduced from numerical and analytical arguments
is given in Fig.\ \ref{fig:spin-1/2 XY phase diagram}.

Since all three phases below the MG line
break translation by one lattice 
spacing, they can be distinguished by order parameters that 
break the symmetries in the subgroup
\begin{align}
\mathbb{Z}^{r}_{2}\times\mathbb{Z}^{x}_{2}\times\mathbb{Z}^{y}_{2}
\subset 
\mathrm{G}^{\,}_{\mathrm{tot}}
\label{eq:relevant subgroup}
\end{align}
defined in Eq.\ \eqref{eq:def Gtot for Hb}.
In what follows, we will limit 
the discussion to this subgroup for simplicity.
We will discuss the duals of the ground states of each gapped phase
and the duals of those operators defined in Eq.\ \
(\ref{eq:Hb ops for corr func}),
whose expectations values detect the long-range orders
that distinguish the gapped phases. 
At the two corners $(\Delta,J)=(0,0)$ and $(\Delta,J)=(\infty,0)$
and along the MG line $(\Delta,J)=(\Delta,1/2)$,
the two-fold degenerate ground states are as follows.
\begin{subequations}
\label{eq:Hb GS}
\begin{enumerate}
\item
At the lower left corner $(\Delta,J)=(0,0)$, the two degenerate ground
states are
\begin{align}
\ket{\mathrm{Neel}^{x}_{\mathrm{o}}}
:=
\ket{\rightarrow,\, \leftarrow,\,\rightarrow,\,\leftarrow,\,\cdots},
\qquad
\ket{\mathrm{Neel}^{x}_{\mathrm{e}}}
:=
\ket{\leftarrow,\, \rightarrow,\,\leftarrow,\,\rightarrow,\,\cdots},
\label{eq:Hb Neelx GS}
\end{align}
where the kets $\ket{\rightarrow}^{\,}_{j}$ and $\ket{\leftarrow}^{\,}_{j}$
denote the eigenstates of $\hat{\sigma}^{x}_{j}$ with eigenvalues
$+1$ and $-1$, respectively. 

\item 
At the lower right corner $(\Delta,J)=(\infty,0)$, the two degenerate ground
states are
\begin{align}
\ket{\mathrm{Neel}^{y}_{\mathrm{o}}}
:=
\ket{\nearrow,\, \swarrow,\,\nearrow,\,\swarrow,\,\cdots},
\qquad
\ket{\mathrm{Neel}^{y}_{\mathrm{e}}}
:=
\ket{\swarrow,\, \nearrow,\,\swarrow,\,\nearrow,\,\cdots},
\label{eq:Hb Neely GS}
\end{align}
where the kets $\ket{\nearrow}^{\,}_{j}$ and $\ket{\swarrow}^{\,}_{j}$
denote the eigenstates of $\hat{\sigma}^{y}_{j}$ with eigenvalues
$+1$ and $-1$, respectively. 

\item 
Along the MG line $(\Delta,J)=(\Delta,1/2)$, the two degenerate 
ground states are 
\begin{align}
\ket{\mathrm{Dimer}^{\,}_{\mathrm{o}}}
:=
\bigotimes_{j=1}^{N}
\ket{[2j-1,2j]},
\qquad
\ket{\mathrm{Dimer}^{\,}_{\mathrm{e}}}
:=
\bigotimes_{j=1}^{N}
\ket{[2j,2j+1]},
\label{eq:Hb dimer GS}
\end{align}
where $\ket{[j,j+1]}$ denotes the singlet 
state for two spins localized on consecutive sites $j$ 
and $j+1$. 
\end{enumerate}
\end{subequations}
These ground states are distinguished by 
the non-vanishing expectations values of the order parameters
\begin{equation}
\begin{split}
&
\widehat{O}^{\mathrm{o}}_{\mathrm{Neel}^{x}}
:=
\frac{1}{2N}
\sum_{j=1}^{2N}
(-1)^{j+1}\,
\hat{\sigma}^{x}_{j},
\\
&
\widehat{O}^{\mathrm{o}}_{\mathrm{Neel}^{y}}
:=
\frac{1}{2N}
\sum_{j=1}^{2N}
(-1)^{j+1}\,
\hat{\sigma}^{y}_{j},
\\
&
\widehat{O}^{\,}_{\mathrm{dimer}}
:=
\frac{1}{N}
\sum_{j=1}^{2N}
(-1)^{j}\,
\frac{1}{3}\,
\hat{\bm{\sigma}}^{\,}_{j}
\cdot
\hat{\bm{\sigma}}^{\,}_{j+1},
\end{split}
\label{eq:Hb order parameters}
\end{equation}
respectively. The order parameters for the
$\mathrm{Neel}^{\,}_{x}$ and $\mathrm{Neel}^{\,}_{y}$
phases are odd under $\widehat{U}^{\,}_{r^{z}_{\pi}}$ symmetry,
while the dimer order parameter is even. In other words,
the order parameter for the two Neel phases do not belong to the
bond algebra \eqref{eq:definition and symmetry of cal Bb}
and do not have an image in the dual bond algebras 
\eqref{eq:definition and symmetry of cal Bb'}
and \eqref{eq:definition and symmetry of cal Bf}.
For this reason, it is more convenient to
define the operators
\begin{subequations}
\label{eq:Hb ops for corr func}
\begin{align}
\widehat{C}^{x}_{j,j+n}:=
\hat{\sigma}^{x}_{j}\,
\hat{\sigma}^{x}_{j+n},
\label{eq:Hb ops for corr func a}
\\
\widehat{C}^{y}_{j,j+n}:=
\hat{\sigma}^{y}_{j}\,
\hat{\sigma}^{y}_{j+n},
\label{eq:Hb ops for corr func b}
\\
\widehat{D}^{\,}_{j}
:=
\frac{1}{3}\,
\hat{\bm{\sigma}}^{\,}_{j}
\cdot
\hat{\bm{\sigma}}^{\,}_{j+1},
\label{eq:Hb ops for corr func c}
\end{align}
\end{subequations}
for any $j\in\Lambda$ and any $n=1,\cdots,|\Lambda|-1$,
all of which are even under $\widehat{U}^{\,}_{r^{z}_{\pi}}$
symmetry. The first two are bilocal operators,
whose expectation values
are the two-point correlation functions
detecting the magnetic ordering in $x$- and 
$y$-directions. The last one is the local operator,
whose staggered summation over the lattice
is the order parameter of the dimer phase. 
The expectation values of the order parameters
\eqref{eq:Hb order parameters}
and operators
\eqref{eq:Hb ops for corr func} 
in the ground states \eqref{eq:Hb GS}
are given in Table\ \ref{Table:Exp values Hb}.

\def\arraystretch{1.25}
\begin{table}
\caption{
\label{Table:Exp values Hb}
The expectation values of the order parameters
\eqref{eq:Hb order parameters}
and operators
\eqref{eq:Hb ops for corr func} 
in the ground states \eqref{eq:Hb GS}
of Hamiltonian \eqref{eq:def Hamiltonian b=0}.
The states $\ket{\mathrm{Neel}^{x}}^{+}$ and
$\ket{\mathrm{Neel}^{y}}^{+}$ are defined in Eqs.\ 
\eqref{eq:def Neelx+} and \eqref{eq:def Neely+},
respectively.
}
\centering
\begin{tabular}{l|cccccc}
\hline \hline
&
$\ \widehat{O}^{\mathrm{o}}_{\mathrm{Neel}^{x}}\ $
&
$\ \widehat{O}^{\mathrm{o}}_{\mathrm{Neel}^{y}}\ $
&
$\ \widehat{O}^{\,}_{\mathrm{dimer}}\ $
&
$\ \widehat{C}^{x}_{j,j+n}\ $
&
$\ \widehat{C}^{y}_{j,j+n}\ $
&
$\widehat{D}^{\,}_{j}$
\\
\hline
$\ket{\mathrm{Neel}^{x}_{\mathrm{o}}}$
&
$+1$
&
$0$
&
$0$
&
$(-1)^{n}$
&
$0$
&
$-\frac{1}{3}$
\\
$\ket{\mathrm{Neel}^{x}_{\mathrm{e}}}$
&
$-1$
&
$0$
&
$0$
&
$(-1)^{n}$
&
$0$
&
$-\frac{1}{3}$
\\
$\ket{\mathrm{Neel}^{x}}^{+}$
&
$0$
&
$0$
&
$0$
&
$(-1)^{n}$
&
$0$
&
$-\frac{1}{3}$
\\
\hline
$\ket{\mathrm{Neel}^{y}_{\mathrm{o}}}$
&
$0$
&
$+1$
&
$0$
&
$0$
&
$(-1)^{n}$
&
$-\frac{1}{3}$
\\
$\ket{\mathrm{Neel}^{y}_{\mathrm{e}}}$
&
$0$
&
$-1$
&
$0$
&
$0$
&
$(-1)^{n}$
&
$-\frac{1}{3}$
\\
$\ket{\mathrm{Neel}^{y}}^{+}$
&
$0$
&
$0$
&
$0$
&
$0$
&
$(-1)^{n}$
&
$-\frac{1}{3}$
\\
\hline
$\ket{\mathrm{Dimer}^{\,}_{\mathrm{o}}}$
&
$0$
&
$0$
&
$+1$
&
$-\delta^{\,}_{(-1)^{j},-1}\,\delta^{\,}_{n,1}$
&
$-\delta^{\,}_{(-1)^{j},-1}\,\delta^{\,}_{n,1}$
&
$-\delta^{\,}_{(-1)^{j},-1}$
\\
$\ket{\mathrm{Dimer}^{\,}_{\mathrm{e}}}$
&
$0$
&
$0$
&
$-1$
&
$-\delta^{\,}_{(-1)^{j},+1}\,\delta^{\,}_{n,1}$
&
$-\delta^{\,}_{(-1)^{j},+1}\,\delta^{\,}_{n,1}$
&
$-\delta^{\,}_{(-1)^{j},+1}$
\\
\hline \hline
\end{tabular}
\end{table}

\subsection{Kramers-Wannier dual $\mathrm{D}^{\,}_{8}$-symmetric
spin-1/2 cluster chain}
\label{subsec:KW and D8 Spin model}

We now study the Hamiltonian dual to the Hamiltonian
\eqref{eq:def Hamiltonian b=0} under the KW duality.
As in Sec.~\ref{subsec:G total frak H b'=0},
we select
periodic boundary conditions ($b'=0$)
after the KW duality. Naive use of the dual bond algebra
(\ref{eq:definition and symmetry of cal Bb'})
delivers the Hamiltonian
\begin{align}
\begin{split}
\widehat{H}^{\,\vee}_{b^{\prime}=0}:=&\,
J^{\,}_{1}
\sum_{j^{\star}\in\Lambda^{\star}}
\left[
\Delta^{\,}_{x}\,
\hat{\tau}^{z\,\vee}_{j^{\star}}
-
\Delta^{\,}_{y}\,
\left(
\hat{\tau}^{x\,\vee}_{j^{\star}-1}\,
\hat{\tau}^{z\,\vee}_{j^{\star}}\,
\hat{\tau}^{x\,\vee}_{j^{\star}+1}
\right)
+
\Delta^{\,}_{z}
\left(
\hat{\tau}^{x\,\vee}_{j^{\star}-1}\,
\hat{\tau}^{x\,\vee}_{j^{\star}+1}
\right)
\right]
\\
&\,
+
J^{\,}_{2}  
\sum_{j^{\star}\in\Lambda^{\star}}
\left[
\Delta^{\,}_{x}\,
\hat{\tau}^{z\,\vee}_{j^{\star}}\,
\hat{\tau}^{z\,\vee}_{j^{\star}+1}
+ 
\Delta^{\,}_{y}
\left(
\hat{\tau}^{x\,\vee}_{j^{\star}-1}\,
\hat{\tau}^{z\,\vee}_{j^{\star}}\,
\hat{\tau}^{x\,\vee}_{j^{\star}+1}
\right)
\left(
\hat{\tau}^{x\,\vee}_{j^{\star}}\,
\hat{\tau}^{z\,\vee}_{j^{\star}+1}\,
\hat{\tau}^{x\,\vee}_{j^{\star}+2}
\right)
\right.
\\
&\,
\qquad\qquad\quad
\left.
+
\Delta^{\,}_{z}
\left(
\hat{\tau}^{x\,\vee}_{j^{\star}-1}\,
\hat{\tau}^{x\,\vee}_{j^{\star}+1}
\right)
\left(
\hat{\tau}^{x\,\vee}_{j^{\star}}\,
\hat{\tau}^{x\,\vee}_{j^{\star}+2}
\right)
\right]
\end{split}
\label{eq:def Hamiltonian b'=0}
\end{align}
with the domain of definition
$\mathcal{H}^{\,\vee}_{b^{\prime}=0}$
defined in Eq.\
(\ref{eq:def gauss projector KW duality b}). 
However, 
Hamiltonians
\eqref{eq:def Hamiltonian b=0}
and
\eqref{eq:def Hamiltonian b'=0}
only form a dual pair if their domains of definition
are restricted to the subspaces
$\mathcal{H}^{\,}_{b=0;+}$
and 
$\mathcal{H}^{\vee}_{b'=0;+}$, respectively 
[recall Eq.\ \eqref{eq:final dualities between domain defs if b to b'}].
With this in mind, we will first study the phase diagram
of Hamiltonian \eqref{eq:def Hamiltonian b'=0} in the full
Hilbert space $\mathcal{H}^{\vee}_{b'=0}$. 
We will then discuss the duality of phases in the restricted Hilbert 
spaces $\mathcal{H}^{\,}_{b=0;+}$ and 
$\mathcal{H}^{\vee}_{b'=0;+}$. 
Without loss of generality,
we consider only the reduced coupling space
(\ref{eq:reduced coupling space}) with $J\leq 1/2$.

The symmetries of
$\widehat{H}^{\,\vee}_{b^{\prime}=0}$
that we shall keep track of are given in
Sec.\ \ref{subsec:G total frak H b'=0}.
Because the global internal symmetry subgroup
$\mathrm{G}^{\,\vee}_{\mathrm{int}}\subset\mathrm{G}^{\,\vee}_{\mathrm{tot}}$
is represented by a trivial projective representation locally,
the LSM Theorems
\ref{thm:LSM translation}
and
\ref{thm:LSM reflection} are inoperative.
Hence, $\widehat{H}^{\,\vee}_{b^{\prime}=0}$
could exhibit a non-degenerate gapped ground state
in its phase diagram, a possibility that is indeed realized.
We restrict ourselves to the dual of subgroup
\eqref{eq:relevant subgroup},
which is the dihedral group $\mathrm{D}^{\,}_{8}$ defined in 
Eq.\ \eqref{eq:total symmetry group for KW duality b}.

By inspection, the energy 
eigenvalues and eigenvectors of Hamiltonian
(\ref{eq:def Hamiltonian b'=0})
are known in closed form at the four corners
(\ref{eq:exactly soluble points AF J1 J2 spin1over2 a}).
Along the left boundary $\Delta=0$
of the reduced coupling space
(\ref{eq:reduced coupling space}),
$\widehat{H}^{\,\vee}_{b^{\prime}=0}$
simplifies to the classical Ising model
in a uniform longitudinal magnetic field.
The same is true of the right boundary
$\Delta=\infty$, as the right boundary is unitarily equivalent
to the left boundary \cite{Santos15}%
~\footnote{%
~It is precisely this duality that was used in
Refs.\ \cite{Dobson69,Stephenson70,Frankel70}
to solve the antiferromagnetic Ising open chain with
nearest- and next-nearest-neighbor couplings.
}.
When $J=0$, the Hamiltonian
(\ref{eq:def Hamiltonian b'=0})
is a linear combination of two of 
the spin-1/2 cluster Hamiltonians that were introduced by
Suzuki in 1971 \cite{Suzuki71}, each of which is soluble in the
sense that it is a sum of pairwise commuting local Hermitian operators
that all square to the identity%
~\footnote{%
~Similarly, the upper corners
\begin{equation}
(\Delta,J)=(0,\infty),
\qquad
(\Delta,J)=(\infty,\infty)
\label{eq:upper corners reduced couplings}
\end{equation}
in the reduced coupling space (\ref{eq:reduced coupling space}) are 
exactly solvable and are gapped with two-fold degenerate ground 
states.
}.
At the lower left corner $(\Delta,J)=(0,0)$,
the ground state is the trivial paramagnet
\begin{align}
\ket{\mathrm{PM}} := \ket{\downarrow,\cdots,\downarrow},
\qquad
\hat{\tau}^{z\vee}_{j^{\star}}\,
\ket{\downarrow,\cdots,\downarrow}
=
-
\ket{\downarrow,\cdots,\downarrow},
\qquad
j^{\star}\in \Lambda^{\star},
\label{eq:H b' paramagnet}
\end{align}
which is a singlet under the
$\mathrm{D}^{\,}_{8}$
symmetry. The lower right corner $(\Delta,J)=(\infty,0)$ also 
corresponds to a non-degenerate, gapped, and $\mathrm{D}^{\,}_{8}$-symmetric
ground state $\ket{\mathrm{SPT}}$
that is defined implicitly by the eigenvalue equation
\begin{align}
\hat{\tau}^{x\,\vee}_{j^{\star}-1}\,
\hat{\tau}^{z\,\vee}_{j^{\star}}\,
\hat{\tau}^{x\,\vee}_{j^{\star}+1}\,
\ket{\mathrm{SPT}}
=
+
\ket{\mathrm{SPT}},
\qquad
j^{\star}\in \Lambda^{\star}.
\label{eq:H b' spt}
\end{align}
The ground state $\ket{\mathrm{SPT}}$ defines a
symmetry-protected topological (SPT) phase
on a closed space manifold (owing to the periodic boundary conditions).
This SPT phase  is protected by the
global internal symmetry
$\mathbb{Z}^{\mathrm{o}}_{2}\times\mathbb{Z}^{\mathrm{e}}_{2}$
in the sense that it cannot be adiabatically deformed
to the trivial paramagnetic state $\ket{\mathrm{PM}}$ without
a gap-closing phase transition or the breaking
(spontaneous or explicit) of the
$\mathbb{Z}^{\mathrm{o}}_{2}\times\mathbb{Z}^{\mathrm{e}}_{2}$ symmetry.

We emphasize that the correct KW dualization
of the Hamiltonian
(\ref{eq:def Hamiltonian b=0})
under open boundary conditions
is not the Hamiltonian
(\ref{eq:def Hamiltonian b'=0})
under open boundary conditions.
With open boundary conditions, one must modify the definition
of the local Gauss operators at the two ends of the chain
when gauging the theory.
This change is responsible
for the presence of additional terms that break the protecting
$\mathbb{Z}^{\mathrm{o}}_{2}\times\mathbb{Z}^{\mathrm{e}}_{2}$ 
symmetry at the boundaries. These additional terms lift the 
two-fold degeneracy of the SPT ground state of
the counterpart to Hamiltonian
(\ref{eq:def Hamiltonian b'=0})
corresponding to open boundary conditions.
The KW dualization
with open boundaries is explained in
Appendix \ref{appsec:Triality with open boundary conditions}.

\begin{figure}[t!]
\begin{center}
\begin{tikzpicture}[scale=0.85]
\begin{axis}[
title={Majumdar-Ghosh line},
axis line style = thick,
xlabel={$0\leq\tanh(\Delta)\le1$},
ylabel={$0\leq J\leq1/2$},
xmin=0, xmax=1,
ymin=0, ymax=1,
ticks=none,
]
\addplot[blue, mark=*, mark size=4pt, only marks]
coordinates {(0.7,0)};
\addplot[blue, mark=square*, mark size=4pt, only marks]
coordinates {(0,0) (1,0)};
\addplot[blue, mark=diamond*, mark size=4pt, only marks]
coordinates {(0,1) (1,1)};
\draw[blue,thick] (0.7,0) -- (0.7,0.666);
\addplot[blue, mark=, mark size=8pt, only marks]
coordinates {(0.7,0.666)};
\draw[red,thick] (0,1) .. controls (0.25,0.85) and (0.5,0.7) .. (0.7,0.666);
\draw[red,thick] (0.7,0.666)..controls (0.8,1) and (0.95,1)..(1,1);
\node[blue,draw] at (0.35,0.3){\hbox{PM}};
\node[blue,draw,rotate=0] at (0.85,0.3){\hbox{SPT}};
\node[blue,draw] at (0.56,0.875){\hbox{$\mathrm{D}^{\,}_{8}$ SSB}};
\end{axis}
\end{tikzpicture}
\end{center}
\caption{
Phase diagram of Hamiltonian
(\ref{eq:def Hamiltonian b'=0})
with the Hilbert space
$\mathcal{H}^{\,\vee}_{b'=0}$
as domain of definition.
The red boundaries realize a continuous quantum phase transition
that separate two phases, one of which descends from the other through
spontaneous symmetry breaking by which a symmetry-breaking
local order parameter acquires a non-vanishing expectation value
in the symmetry-broken phase, i.e.,
the Landau-Ginzburg paradigm of phase transitions.
The blue boundary realizes a continuous topological quantum phase transition
between two phases that are distinguished by a non-local order parameter.
These phases are adiabatically connected to the ground states 
\eqref{eq:H b' paramagnet} and \eqref{eq:H b' spt} for $\Delta<1$ and
$\Delta>1$, respectively.
}
\label{fig:Hamiltonian b'=0 with full Hilbert space}
\end{figure}
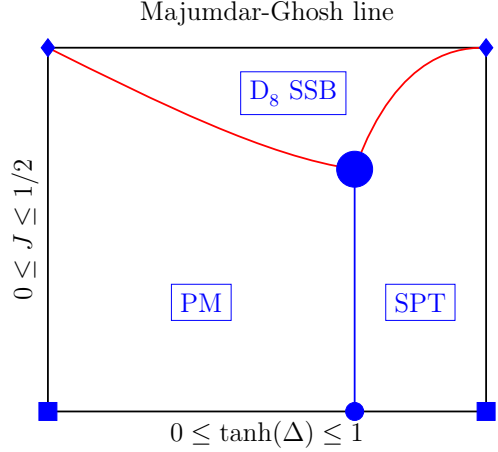

The dual ground states can also be obtained in closed analytical form
along the $J=1/2$ line in the parameter
space of the Hamiltonian \eqref{eq:def Hamiltonian b'=0}.
This is done by dualizing the
projectors onto the MG ground states of the Hamiltonian
(\ref{eq:def Hamiltonian b=0})
along the MG line
(\ref{eq:exactly soluble points AF J1 J2 spin1over2 c}),
as is detailed in Appendix
\ref{appsec:Triality of the Majumdar-Ghosh line}.
One finds that the ground
states of Hamiltonian $\widehat{H}^{\,\vee}_{b^{\prime}=0}$ are gapped
and four-fold degenerate along the open MG line. This is
confirmed by performing an exact diagonalization study of the
eigenvalue spectrum of $\widehat{H}^{\,\vee}_{b^{\prime}=0}$.  
These four ground states are
(see Appendix \ref{appsec:Triality of the Majumdar-Ghosh line})
\begin{subequations}
\label{eq:D_8 quadruplet}
\begin{align}
&
\ket{1}
=
\ket{
\downarrow,\rightarrow,\downarrow,\leftarrow,
\downarrow,\rightarrow,\downarrow,\leftarrow,
\cdots
},
\\
&
\ket{2}
=
\ket{
\downarrow,\leftarrow,\downarrow,\rightarrow,
\downarrow,\leftarrow,\downarrow,\rightarrow,
\cdots
},
\\
&
\ket{3}
=
\ket{
\rightarrow,\downarrow,\leftarrow,\downarrow,
\rightarrow,\downarrow,\leftarrow,\downarrow,
\cdots
},
\\
&
\ket{4}
=
\ket{
\leftarrow,\downarrow,\rightarrow,\downarrow,
\leftarrow,\downarrow,\rightarrow,\downarrow,
\cdots
},
\end{align}
\end{subequations}
where we chose the basis for which
$|\rightarrow\rangle^{\,}_{j}$
($|\uparrow\rangle^{\,}_{j}$)
is the eigenstate with eigenvalue $+1$
of $\hat{\tau}^{x\,\vee}_{j^{\star}}$ 
($\hat{\tau}^{z\,\vee}_{j^{\star}}$).
These four-fold degenerate ground states
spontaneously break the
dihedral group $\mathrm{D}^{\,}_{8}$ 
down to a $\mathbb{Z}^{\,}_{2}$ subgroup since
\begin{subequations}
\begin{align}
&
\widehat{U}^{\vee}_{e}\,
\widehat{U}^{\vee}_{r^{z}_{\pi}}\,
\ket{1}
=
\ket{2},
\qquad
\widehat{U}^{\vee}_{r}\,
\ket{2}
=
\ket{3},
\qquad
\widehat{U}^{\vee}_{r^{z}_{\pi}}\,
\ket{3}
=
\widehat{U}^{\vee}_{r}\,
\ket{1}
=
\ket{4},
\\
&
\widehat{U}^{\vee}_{\mathrm{o}}\,
\ket{1}
=
\ket{1},
\qquad
\widehat{U}^{\vee}_{\mathrm{o}}\,
\ket{2}
=
\ket{2},
\qquad
\widehat{U}^{\vee}_{\mathrm{e}}\,
\ket{3}
=
\ket{3},
\qquad
\widehat{U}^{\vee}_{\mathrm{e}}\,
\ket{4}
=
\ket{4},
\end{align}
\end{subequations}
i.e., the states $\ket{1}$ and $\ket{2}$ are invariant 
only under the $\mathbb{Z}^{\,}_{2}$ subgroup generated by 
$\widehat{U}^{\vee}_{\mathrm{o}}$ while the states $\ket{3}$ and 
$\ket{4}$ are invariant only under the $\mathbb{Z}^{\,}_{2}$ 
subgroup generated by 
$\widehat{U}^{\vee}_{\mathrm{e}}$. 
The phase diagram of Hamiltonian 
\eqref{eq:def Hamiltonian b'=0} on the full Hilbert space 
$\mathcal{H}^{\vee}_{b'=0}$ is shown in Fig.\ 
\ref{fig:Hamiltonian b'=0 with full Hilbert space}.
The phase boundaries in Fig.\ \ref{fig:spin-1/2 XY phase diagram}
carry over to Fig.\ \ref{fig:Hamiltonian b'=0 with full Hilbert space}
owing to the duality. As opposed to the deconfined
quantum critical lines in Fig.\ \ref{fig:spin-1/2 XY phase diagram},
the phase diagram in Fig.\
\ref{fig:Hamiltonian b'=0 with full Hilbert space}
features (i) a topological transition (blue line) between the two 
$\mathrm{D}^{\,}_{8}$-singlet states that are adiabatically connected 
to states \eqref{eq:H b' paramagnet} and \eqref{eq:H b' spt},
respectively, and (ii) two conventional symmetry breaking transitions 
(red lines) between a doublet of states that break completely
the symmetry group $\mathrm{D}^{\,}_{8}$ and the
$\mathrm{D}^{\,}_{8}$-singlet states that are adiabatically connected 
to states \eqref{eq:H b' paramagnet} and \eqref{eq:H b' spt},
respectively.

\begin{figure}[t!]
\begin{center}
\begin{minipage}{0.45\textwidth}  
\begin{tikzpicture}[scale=0.85]
\begin{axis}[
title={Majumdar-Ghosh line},
axis line style = thick,
xlabel={$0\leq\tanh(\Delta)\le1$},
ylabel={$0\leq J\leq1/2$},
xmin=0, xmax=1,
ymin=0, ymax=1,
ticks=none,
] 
\addplot[blue, mark=*, mark size=4pt, only marks]
coordinates {(0.7,0)};
\addplot[blue, mark=square*, mark size=4pt, only marks]
coordinates {(0,0) (1,0)};
\addplot[blue, mark=diamond*, mark size=4pt, only marks]
coordinates {(0,1) (1,1)};
\draw[blue,thick] (0.7,0) -- (0.7,0.666);
\addplot[blue, mark=, mark size=8pt, only marks]
coordinates {(0.7,0.666)};
\draw[blue,thick] (0,1) .. controls (0.25,0.85) and (0.5,0.7) .. (0.7,0.666);
\draw[blue,thick] (0.7,0.666)..controls (0.8,1) and (0.95,1)..(1,1);
\node[blue,draw] at (0.35,0.3){$\ket{\mathrm{Neel}^{x}}^{+}$};
\node[blue,draw, rotate=0] at (0.85,0.3){$\ket{\mathrm{Neel}^{y}}^{+}$};
\node[blue,draw] at (0.55,0.89){Dimer doublet};
\end{axis}
\end{tikzpicture}
\end{minipage}
\begin{Large}
$\rightleftharpoons$
\end{Large}
\begin{minipage}{0.45\textwidth}  
\begin{tikzpicture}[scale=0.85]
\begin{axis}[
title={Majumdar-Ghosh line},
axis line style = thick,
xlabel={$0\leq\tanh(\Delta)\le1$},
ylabel={$0\leq J\leq1/2$},
xmin=0, xmax=1,
ymin=0, ymax=1,
ticks=none,
]
\addplot[blue, mark=*, mark size=4pt, only marks]
coordinates {(0.7,0)};
\addplot[blue, mark=square*, mark size=4pt, only marks]
coordinates {(0,0) (1,0)};
\addplot[blue, mark=diamond*, mark size=4pt, only marks]
coordinates {(0,1) (1,1)};
\draw[blue,thick] (0.7,0) -- (0.7,0.666);
\addplot[blue, mark=, mark size=8pt, only marks]
coordinates {(0.7,0.666)};
\draw[red,thick] (0,1) .. controls (0.25,0.85) and (0.5,0.7) .. (0.7,0.666);
\draw[red,thick] (0.7,0.666)..controls (0.8,1) and (0.95,1)..(1,1);
\node[blue,draw] at (0.35,0.3){$\ket{\mathrm{PM}}$};
\node[blue,draw, rotate=0] at (0.85,0.3){$\ket{\mathrm{SPT}}$};
\node[blue,draw] at (0.575,0.875){\hbox{$\mathrm{D}^{+}_{8}$ doublet}};
\end{axis}
\end{tikzpicture}
\end{minipage} 
\end{center}
\caption{
The phase diagram of Hamiltonian
(\ref{eq:def Hamiltonian b=0})
restricted to the subspace
$\mathcal{H}^{\,}_{b=0;+}$
of the Hilbert space
$\mathcal{H}^{\,}_{b=0}$
is dual to the phase diagram of Hamiltonian
(\ref{eq:def Hamiltonian b'=0})
restricted to the subspace
$\mathcal{H}^{\vee}_{b'=0;+}$
of the Hilbert space
$\mathcal{H}^{\,\vee}_{b'=0}$.
The pair of dual subspaces are to be found in
the first line of Table
\ref{Table:Kramers-Wannier dualization b to b'}.
The two states $\ket{\mathrm{Neel}^{x}}^{+}$ and
$\ket{\mathrm{Neel}^{y}}^{+}$ are defined in Eqs.\ 
\eqref{eq:def Neelx+} and \eqref{eq:def Neely+}, while
dimer doublet refers to the ground states \eqref{eq:Hb dimer GS}.
On the dual side, the paramagnetic states $\ket{\mathrm{PM}}$,
and $\ket{\mathrm{SPT}}$ state are defined in Eqs.\ 
\eqref{eq:H b' paramagnet} and \eqref{eq:H b' spt}, respectively. 
By $\mathrm{D}^{+}_{8}$ doublet, we refer to the states 
$\ket{\mathrm{Dimer}^{\vee}_{\mathrm{o}}}$
and
$\ket{\mathrm{Dimer}^{\vee}_{\mathrm{e}}}$,
both of which are 
in the subspace $\mathcal{H}^{\vee}_{b'=0;+}$ and defined in 
Eq.\ \eqref{eq:D_8 doublet}.
The symbols
$\rightharpoonup$ and $\leftharpoondown$
denote gauging the diagonal subgroups 
generated by $\widehat{U}^{\,}_{r^{z}_{\pi}}$ 
and by its dual $\widehat{U}^{\vee}_{r^{z}_{\pi}}$, respectively. 
}
\label{fig:comparing spin-1/2 XY phase diagram and its KW dual}
\end{figure}

The KW duality implies that
the expectation value of any operator
from the bond algebra
\eqref{eq:definition and symmetry of cal Bb}
restricted to the Hilbert space $\mathcal{H}^{\,}_{b=0;+}$
has the same expectation value as its
dual in the bond algebra \eqref{eq:definition and symmetry of cal Bb'}
restricted to the Hilbert space $\mathcal{H}^{\vee}_{b'=0;+}$.
Under the isomorphism between the bond algebras
\eqref{eq:definition and symmetry of cal Bb} 
and
\eqref{eq:definition and symmetry of cal Bb'},
operators
\eqref{eq:Hb ops for corr func} 
dualize to
\begin{subequations}
\label{eq:Hb' ops for corr func}
\begin{align}
&
\widehat{C}^{x\,\vee}_{j^{\star},j^{\star}+n-1}
:=
\prod_{\ell=j^{\star}}^{j^{\star}+n-1}
\hat{\tau}^{z\,\vee}_{\ell^{\star}},
\label{eq:Hb' ops for corr func a}
\\
&
\widehat{C}^{y\,\vee}_{j^{\star},j^{\star}+n-1}
:=
\hat{\tau}^{x\,\vee}_{j^{\star}-1}\,
\hat{\tau}^{x\,\vee}_{j^{\star}}\,
\left(
\prod_{\ell=j^{\star}}^{j^{\star}+n-1}
\hat{\tau}^{z\,\vee}_{\ell^{\star}}
\right)
\,
\hat{\tau}^{x\,\vee}_{j^{\star}+n-1}\,
\hat{\tau}^{x\,\vee}_{j^{\star}+n},
\label{eq:Hb' ops for corr func b}
\\
&
\widehat{D}^{\vee}_{j^{\star}}
:= 
\frac{1}{3}
\left(
\hat{\tau}^{z\,\vee}_{j^{\star}}
-
\hat{\tau}^{x\,\vee}_{j^{\star}-1}\,
\hat{\tau}^{z\,\vee}_{j^{\star}}\,
\hat{\tau}^{x\,\vee}_{j^{\star}+1}
+
\hat{\tau}^{x\,\vee}_{j^{\star}-1}\,
\hat{\tau}^{x\,\vee}_{j^{\star}+1}
\right),
\label{eq:Hb' ops for corr func c}
\end{align}
\end{subequations}
for any $j^{\star}\in\Lambda^{\star}$ and any $n=1,\cdots,|\Lambda|-1$.
We observe that operators
$\widehat{C}^{x}_{j,j+n}$
and 
$\widehat{C}^{y}_{j,j+n}$
defined in Eqs.\
\eqref{eq:Hb ops for corr func a}
and
\eqref{eq:Hb ops for corr func b},
respectively,
dualize to non-local string operators,
while the local operator
$\widehat{D}^{\,}_{j}$
defined in Eq.\
\eqref{eq:Hb ops for corr func c}
remains local after dualization. 

At the lower left corner $(\Delta,J)=(0,0)$ of the phase diagram, 
only the bonding linear combination of the two Neel states
\begin{align}
\ket{\mathrm{Neel}^{x}}^{+}
:=
\frac{1}{\sqrt{2}}
\left(
\ket{\mathrm{Neel}^{x}_{\mathrm{o}}}
+
\ket{\mathrm{Neel}^{x}_{\mathrm{e}}}
\right),
\qquad
\widehat{U}^{\,}_{r^{z}_{\pi}}\,
\ket{\mathrm{Neel}^{x}}^{+}
=
+
\ket{\mathrm{Neel}^{x}}^{+},
\label{eq:def Neelx+}
\end{align}    
belongs to the subspace $\mathcal{H}^{\,}_{b=0;+}$.
Under the KW duality, this state is mapped to
the paramagnetic ground state 
$\ket{\mathrm{PM}}\in\mathcal{H}^{\vee}_{b'=0;+}$ 
defined in Eq.\  \eqref{eq:H b' paramagnet}. 
The expectation values of the dual operators
\eqref{eq:Hb' ops for corr func} in the ground state $\ket{\mathrm{PM}}$
are given in Table\ \ref{Table:Exp values under KW duality}. 
The non-vanishing expectation value of the bilocal
operator $\widehat{C}^{x}_{j,j+n}$
translates to the non-vanishing expectation value of the string operator
$\widehat{C}^{x\,\vee}_{j^{\star},j^{\star}+n-1}$ for any $j^{\star}$ and $n$. 
This is the so-called disorder operator,
whose non-vanishing expectation value detects the disordered paramagnetic 
phase~\cite{Kadanoff71}. 

At the lower right corner $(\Delta,J)=(\infty,0)$,
only the bonding linear combination of the two Neel states
\begin{align}
\ket{\mathrm{Neel}^{y}}^{+}
:=
\frac{1}{\sqrt{2}}
\left(
\ket{\mathrm{Neel}^{y}_{\mathrm{o}}}
+
\ket{\mathrm{Neel}^{y}_{\mathrm{e}}}
\right),
\qquad
\widehat{U}^{\,}_{r^{z}_{\pi}}\,
\ket{\mathrm{Neel}^{y}}^{+}
=
+
\ket{\mathrm{Neel}^{y}}^{+},
\label{eq:def Neely+}
\end{align}    
belongs to the subspace $\mathcal{H}^{\,}_{b=0;+}$.
Under the KW duality, this state is mapped to
the SPT ground state with periodic boundary conditions
$\ket{\mathrm{SPT}}\in\mathcal{H}^{\vee}_{b'=0;+}$ 
defined in Eq.\  \eqref{eq:H b' spt}. 
The expectation values of the dual operators
\eqref{eq:Hb' ops for corr func}
in the ground state $\ket{\mathrm{SPT}}$
are given in Table\ \ref{Table:Exp values under KW duality}. 
The non-vanishing expectation value of the bilocal 
operator $\widehat{C}^{y}_{j,j+n}$
translates to the non-vanishing expectation value of the string operator
$\widehat{C}^{y\,\vee}_{j^{\star},j^{\star}+n-1}$
operator for any $j^{\star}$ and $n$.
The string operator
$\widehat{C}^{y\,\vee}_{j^{\star},j^{\star}+n-1}$
is invariant under the dual internal symmetries
$\widehat{U}^{\vee}_{\mathrm{o}}$ and $\widehat{U}^{\vee}_{\mathrm{e}}$ 
defined in Eq.\ \eqref{eq:reps ro re rz dual b'},
owing to the presence of
$\hat{\tau}^{x\,\vee}_{j^{\star}-1}\,
\hat{\tau}^{x\,\vee}_{j^{\star}}$
and
$\hat{\tau}^{x\,\vee}_{j^{\star}+n-1}\,
\hat{\tau}^{x\,\vee}_{j^{\star}+n}$  
to the left and to right of the string of
(\ref{eq:Hb' ops for corr func a}),
respectively,
on the right-hand side of Eq.\
(\ref{eq:Hb' ops for corr func b}).
The operator
$\widehat{C}^{y\,\vee}_{j^{\star},j^{\star}+n-1}$ is
the so-called string order parameter that detects the SPT ground state
~\cite{PerezGarcia2008,Pollman2012,Moradi2022},
while having vanishing expectation value in the trivial ground state 
\eqref{eq:H b' paramagnet}. 

\def\arraystretch{1.25}
\begin{table}
\caption{
\label{Table:Exp values under KW duality}
The expectation values the operators 
\eqref{eq:Hb' ops for corr func} in
the dual ground states \eqref{eq:H b' paramagnet},
\eqref{eq:H b' spt}, and \eqref{eq:D_8 doublet}.
}
\centering
\begin{tabular}{l|ccc}
\hline \hline
&
$\ \widehat{C}^{x\,\vee}_{j^{\star},j^{\star}+n-1}\ $
&
$\ \widehat{C}^{y\,\vee}_{j^{\star},j^{\star}+n-1}\ $
&
$\widehat{D}^{\vee}_{j^{\star}}$
\\
\hline
$\ket{\mathrm{PM}}$
&
$(-1)^{n}$
&
$0$
&
$-\frac{1}{3}$
\\
\hline
$\ket{\mathrm{SPT}}$
&
$0$
&
$(-1)^{n}$
&
$-\frac{1}{3}$
\\
\hline
$\ket{\mathrm{Dimer}^{\vee}_{\mathrm{o}}}$
&
$-\delta^{\,}_{(-1)^{j},-1}\,\delta^{\,}_{n,1}$
&
$-\delta^{\,}_{(-1)^{j},-1}\,\delta^{\,}_{n,1}$
&
$-\delta^{\,}_{(-1)^{j},-1}$
\\
$\ket{\mathrm{Dimer}^{\vee}_{\mathrm{e}}}$
&
$-\delta^{\,}_{(-1)^{j},+1}\,\delta^{\,}_{n,1}$
&
$-\delta^{\,}_{(-1)^{j},+1}\,\delta^{\,}_{n,1}$
&
$-\delta^{\,}_{(-1)^{j},+1}$
\\
\hline \hline
\end{tabular}
\end{table}

Finally, along the MG line $(\Delta,J=1/2)$, both dimer ground states 
\eqref{eq:Hb dimer GS} belong to the subspace $\mathcal{H}^{\,}_{b=0;+}$. 
However, out of the four ground states \eqref{eq:D_8 quadruplet} 
of Hamiltonian \eqref{eq:def Hamiltonian b'=0}, only the two linear 
combinations
\begin{subequations}
\label{eq:D_8 doublet}
\begin{align}
&
\ket{\mathrm{Dimer}^{\vee}_{\mathrm{o}}}
:=
\frac{1}{\sqrt{2}}
\left(\ket{1} + \ket{2}\right),
\qquad
\widehat{U}^{\vee}_{r^{z}_{\pi}}\,
\ket{\mathrm{Dimer}^{\vee}_{\mathrm{0}}}
=
+
\ket{\mathrm{Dimer}^{\vee}_{\mathrm{o}}},
\\
&
\ket{\mathrm{Dimer}^{\vee}_{\mathrm{e}}}
:=
\frac{1}{\sqrt{2}}
\left(\ket{3} + \ket{4}\right),
\qquad
\widehat{U}^{\vee}_{r^{z}_{\pi}}\,
\ket{\mathrm{Dimer}^{\vee}_{\mathrm{e}}}
=
+
\ket{\mathrm{Dimer}^{\vee}_{\mathrm{e}}},
\end{align}
\end{subequations}
belong to the subspace $\mathcal{H}^{\vee}_{b'=0;+}$.
These two states are dual to the dimer states \eqref{eq:Hb dimer GS},
respectively.
We refer to this twofold degenerate ground state manifold as
the $\mathrm{D}^{+}_{8}$ doublet. 
The ground states \eqref{eq:D_8 doublet} break the reflection symmetry 
spontaneously,
while they are both singlets under the internal symmetry group 
$\mathbb{Z}^{\mathrm{o}}_{2}\times\mathbb{Z}^{\mathrm{e}}_{2}$.
The expectation values of the dual operators
\eqref{eq:Hb' ops for corr func} in these ground states
are given in Table\ \ref{Table:Exp values under KW duality}. 
The phase diagrams of Hamiltonians \eqref{eq:def Hamiltonian b=0} 
and \eqref{eq:def Hamiltonian b'=0} in the restricted subspaces
$\mathcal{H}^{\,}_{b=0;+}$
and $\mathcal{H}^{\vee}_{b'=0;+}$ are compared in Fig.\
\ref{fig:comparing spin-1/2 XY phase diagram and its KW dual}.

\subsection{Jordan-Wigner dual interacting Majorana chain}
\label{subsec:JW and interacting Majorana model}

We now study the Hamiltonian dual to the Hamiltonian
\eqref{eq:def Hamiltonian b=0} under the JW duality.
As in Sec.~\ref{subsec:G total frak H f=0},
we select anti-periodic boundary conditions ($f=1$)
after the JW duality. Naive use of the dual bond algebra
(\ref{eq:cal Bb if unit-cell non-preserving gauging PF a})
delivers the Hamiltonian
\begin{align}
\begin{split}
\widehat{H}^{\,\vee}_{f=1}:
=&\,
J^{\,}_{1}\,
\sum_{j\in\Lambda}
\left(
\Delta^{\,}_{x}\,
\mathrm{i}
\hat{\beta}^{\vee}_{j+1}\,
\hat{\alpha}^{\vee}_{j}
+
\Delta^{\,}_{y}\,
\mathrm{i}
\hat{\beta}^{\vee}_{j}\,
\hat{\alpha}^{\vee}_{j+1}
+
\Delta^{\,}_{z}\,
\hat{\beta}^{\vee}_{j}\,
\hat{\beta}^{\vee}_{j+1}\,
\hat{\alpha}^{\vee}_{j}\,
\hat{\alpha}^{\vee}_{j+1}
\right)
\\
&\,
+
J^{\,}_{2}\,
\sum_{j=1}^{2N}
\left(
\Delta^{\,}_{x}\,
\hat{\beta}^{\vee}_{j+1}\,
\hat{\beta}^{\vee}_{j+2}\,
\hat{\alpha}^{\vee}_{j}\,
\hat{\alpha}^{\vee}_{j+1}
+
\Delta^{\,}_{y}\,
\hat{\alpha}^{\vee}_{j+1}\,
\hat{\alpha}^{\vee}_{j+2}\,
\hat{\beta}^{\vee}_{j}\,
\hat{\beta}^{\vee}_{j+1}
+
\Delta^{\,}_{z}\,
\hat{\beta}^{\vee}_{j}\,
\hat{\beta}^{\vee}_{j+2}\,
\hat{\alpha}^{\vee}_{j}\,
\hat{\alpha}^{\vee}_{j+2}
\right)
\end{split}
\label{eq:def Hamiltonian f=1}
\end{align}
with the domain of definition
$\mathcal{H}^{\,\vee}_{f=1}$
defined in Eq.\
(\ref{eq:def Majorana algebra on mathfrak{H}bF}).
However, 
Hamiltonians
\eqref{eq:def Hamiltonian b=0}
and
\eqref{eq:def Hamiltonian f=1}
only form a dual pair if their domains of definition
are restricted to the subspaces
$\mathcal{H}^{\,}_{b=0;+}$
and 
$\mathcal{H}^{\vee}_{f=1;+}$, respectively 
[recall Eq.\
\eqref{eq:final dualities between domain defs if b to bF}].
With this in mind, we will first study the phase diagram
of Hamiltonian \eqref{eq:def Hamiltonian f=1} in the full
Hilbert space $\mathcal{H}^{\vee}_{f=1}$. 
We will then discuss the duality of phases in the restricted Hilbert 
spaces $\mathcal{H}^{\,}_{b=0;+}$ and 
$\mathcal{H}^{\vee}_{f=1;+}$. 
Without loss of generality,
we consider only the reduced coupling space
(\ref{eq:reduced coupling space}) with $J\leq 1/2$. 

The symmetries of
$\widehat{H}^{\,\vee}_{f=1}$
that we shall keep track of are given in
Sec.\ \ref{subsec:G total frak H f=0}.
Because the global internal symmetry subgroup
$\mathrm{G}^{\,\vee,\,\mathrm{F}}_{\mathrm{int}}\subset
\mathrm{G}^{\,\vee,\,\mathrm{F}}_{\mathrm{tot}}$
is represented by a trivial projective representation locally,
the LSM Theorems \ref{thm:LSM translation}
and
\ref{thm:LSM reflection}
are inoperative.
Hence, $\widehat{H}^{\,\vee}_{f=1}$
could exhibit a non-degenerate gapped ground state
in its phase diagram, a possibility that is indeed realized.
We restrict ourselves to the dual of the subgroup
\eqref{eq:relevant subgroup},
which is the subgroup
\begin{subequations}
\label{eq:fermionic dual to ref + internal}
\begin{align}
\mathrm{Z}^{\mathrm{F}r}_{4}\ltimes
\mathbb{Z}^{\mathrm{o}}_{2}\times\mathbb{Z}^{\mathrm{e}}_{2}/\mathbb{Z}^{\mathrm{F}}_{2}
\subset
\mathrm{G}^{\vee,\,\mathrm{F}}_{\mathrm{tot}},
\label{eq:fermionic dual to ref + internal a}
\end{align}
where
\begin{equation}
\mathrm{Z}^{\mathrm{F}r}_{4}:=
\left\{r,\ r^{2}=p^{\,}_{\mathrm{F}},\ r^{3},\ r^{4}=e\right\}.
\label{eq:fermionic dual to ref + internal b}
\end{equation}
\end{subequations}

By inspection, Hamiltonian
(\ref{eq:def Hamiltonian f=1})
is quadratic along the $(\Delta,J)=(\Delta,0)$ line.
At either one of the two lower corners $(\Delta,J)=(0,0)$
or $(\Delta,J)=(\infty,0)$,
$\widehat{H}^{\,\vee}_{f=1}$ simplifies to a Kitaev chain
\cite{Kitaev01}.
We denote the ground states at the points $(\Delta,J)=(0,0)$ and 
$(\Delta,J)=(\infty,0)$ by
$\ket{\mathrm{Kitaev}}$ 
and
$\ket{\overline{\mathrm{Kitaev}}}$,
respectively, such that%
~\footnote{%
~Terms $\mathrm{i}\hat{\beta}^{\vee}_{2N+1}\,
\hat{\alpha}^{\vee}_{2N}
=
-
\mathrm{i}\hat{\beta}^{\vee}_{1}\,
\hat{\alpha}^{\vee}_{2N}$ 
and 
$\mathrm{i}\hat{\beta}^{\vee}_{2N}\,
\hat{\alpha}^{\vee}_{2N+1}
=
-
\mathrm{i}\hat{\beta}^{\vee}_{2N}\,
\hat{\alpha}^{\vee}_{1}$
come with an additional minus sign because of the
anti-periodic boundary conditions
($f=1$).
}
\begin{subequations}
\label{eq:Kitaev ground states}
\begin{align}
&
\mathrm{i}\hat{\beta}^{\vee}_{1}\,
\hat{\alpha}^{\vee}_{2N}\,
\ket{\mathrm{Kitaev}}
=
+
\ket{\mathrm{Kitaev}},
\quad
\mathrm{i}\hat{\beta}^{\vee}_{j+1}\,
\hat{\alpha}^{\vee}_{j}\,
\ket{\mathrm{Kitaev}}
=
-
\ket{\mathrm{Kitaev}},
\quad
j=1,\cdots,2N-1,
\\
&
\mathrm{i}\hat{\beta}^{\vee}_{2N}\,
\hat{\alpha}^{\vee}_{1}\,
\ket{\overline{\mathrm{Kitaev}}}
=
+
\ket{\overline{\mathrm{Kitaev}}},
\quad
\mathrm{i}\hat{\beta}^{\vee}_{j}\,
\hat{\alpha}^{\vee}_{j+1}\,
\ket{\overline{\mathrm{Kitaev}}}
=
-
\ket{\overline{\mathrm{Kitaev}}},
\quad
j=1,\cdots,2N-1.
\end{align}
\end{subequations}
These two ground states are both symmetric under
the subgroup
\eqref{eq:fermionic dual to ref + internal}.
They are the ground states of two distinct and non-trivial 
invertible fermionic phases of matter%
~\footnote{%
~In fact, they are inverse of each other
under the fermionic stacking operation
\cite{Turzillo2019,Bourne2021,Aksoy2022}.
}. 
When appropriate open boundary conditions are imposed,
the counterpart to Hamiltonian
$\widehat{H}^{\,}_{f=1}$
has two-fold degenerate ground states.
This degeneracy arises owing to the existence of
two Majorana zero modes, one of which is localized a the left end
while the other is localized at the right end, of the open chain.
At the point $(\Delta,J)=(1,0)$,
$\widehat{H}^{\,\vee}_{f=1}$ simplifies to free spinless fermions
hopping with a uniform nearest-neighbor hopping amplitude along the chain%
~\footnote{%
~The upper corners
$(\Delta,J)=(0,\infty)$ and $(\Delta,J)=(\infty,\infty)$
are gapped and two-fold degenerate nder antiperiodic boundary conditions as a consequence of the
dualization from Sec.\
\ref{subsubsec:Unit-cell preserving gauging of fermion parity}.
The image under $J\to1/J$ 
of the point $(\Delta,J)=(1,0)$,
is gapless, as a consequence of the
dualization from Sec.\
\ref{subsubsec:Unit-cell preserving gauging of fermion parity}.
}.

By dualization of the
projectors onto the MG ground states of Hamiltonian
(\ref{eq:def Hamiltonian b=0})
along the MG line
(\ref{eq:exactly soluble points AF J1 J2 spin1over2 c}),
it is shown in Appendix
\ref{appsec:Triality of the Majumdar-Ghosh line}
that the ground states of Hamiltonian
$\widehat{H}^{\,\vee}_{f=1}$
are two-fold degenerate along the open MG line.
We can always choose an orthonormal basis of ground states
such that the basis elements are the dual to the dimer states
(\ref{eq:Hb dimer GS}). This dual basis is given by  
\begin{subequations}
\label{eq:BDO states}
\begin{equation}
\begin{split}
&
|\mathrm{Bonding}^{\vee}_{\mathrm{o}}\rangle:=
\left[
\prod_{j=1}^{N}
\frac{1}{\sqrt{2}}\,
\left(
\hat{c}^{\vee\,\dag}_{2j-1}
+
\hat{c}^{\vee\,\dag}_{2j}
\right)
\right]
|0\rangle,
\\
&
|\mathrm{Bonding}^{\vee}_{\mathrm{e}}\rangle:=
\left[
\prod_{j=1}^{N}
\frac{1}{\sqrt{2}}\,
\left(
\hat{c}^{\vee\,\dag}_{2j}
+
\hat{c}^{\vee\,\dag}_{2j+1}
\right)
\right]
|0\rangle,
\end{split}
\end{equation}
where the complex fermion operators are defined as 
\begin{equation}
\hat{c}^{\vee\,\dag}_{j}
:=
\frac{1}{2}
(
\hat{\alpha}^{\vee}_{j}
-
\mathrm{i}
\hat{\beta}^{\vee}_{j}
),
\qquad
\hat{c}^{\vee}_{j}
:=
\frac{1}{2}
(
\hat{\alpha}^{\vee}_{j}
+
\mathrm{i}
\hat{\beta}^{\vee}_{j}
).
\end{equation}
\end{subequations}
These states spontaneously break the reflection symmetry while
preserving the internal $\mathbb{Z}^{\mathrm{o}}_{2}\times
\mathbb{Z}^{\mathrm{e}}_{2}$ symmetry.

\def\arraystretch{1.25}
\begin{table}
\caption{
\label{Table:Exp values under JW duality}
The expectation values the operators 
\eqref{eq:Hb' ops for corr func} in
the dual ground states \eqref{eq:H b' paramagnet},
\eqref{eq:H b' spt}, and \eqref{eq:D_8 doublet}.
}
\centering
\begin{tabular}{l|ccc}
\hline \hline
&
$\ \widehat{C}^{x\,\vee}_{j,j+n}\ $
&
$\ \widehat{C}^{y\,\vee}_{j,j+n}\ $
&
$\widehat{D}^{\vee}_{j}$
\\
\hline
$\ket{\mathrm{Kitaev}}$
&
$(-1)^{n}$
&
$0$
&
$-\frac{1}{3}$
\\
\hline
$\ket{\overline{\mathrm{Kitaev}}}$
&
$0$
&
$(-1)^{n}$
&
$-\frac{1}{3}$
\\
\hline
$|\mathrm{Bonding}^{\vee}_{\mathrm{o}}\rangle$
&
$-\delta^{\,}_{(-1)^{j},-1}\,\delta^{\,}_{n,1}$
&
$-\delta^{\,}_{(-1)^{j},-1}\,\delta^{\,}_{n,1}$
&
$-\delta^{\,}_{(-1)^{j},-1}$
\\
$|\mathrm{Bonding}^{\vee}_{\mathrm{e}}\rangle$
&
$-\delta^{\,}_{(-1)^{j},+1}\,\delta^{\,}_{n,1}$
&
$-\delta^{\,}_{(-1)^{j},+1}\,\delta^{\,}_{n,1}$
&
$-\delta^{\,}_{(-1)^{j},+1}$
\\
\hline \hline
\end{tabular}
\end{table}

The JW duality implies that
the expectation value of any operator
from the bond algebra
\eqref{eq:definition and symmetry of cal Bb}
restricted to the Hilbert space $\mathcal{H}^{\,}_{b=0;+}$
has the same expectation value as its
dual in the bond algebra \eqref{eq:definition and symmetry of cal Bf}
restricted to the Hilbert space $\mathcal{H}^{\vee}_{f=1;+}$. 
Under the isomorphism between the bond algebras
\eqref{eq:definition and symmetry of cal Bb} 
and
\eqref{eq:definition and symmetry of cal Bf},
operators \eqref{eq:Hb ops for corr func} 
dualize to
\begin{subequations}
\label{eq:Hf ops for corr func}
\begin{align}
&
\widehat{C}^{x\,\vee}_{j,j+n}
:=
\prod_{\ell=j}^{j+n-1}
\mathrm{i}
\hat{\beta}^{\vee}_{\ell+1}\,
\hat{\alpha}^{\vee}_{\ell},
\label{eq:Hf ops for corr func a}
\\
&
\widehat{C}^{y\,\vee}_{j,j+n}
:=
\prod_{\ell=j}^{j+n-1}
\mathrm{i}
\hat{\beta}^{\vee}_{\ell}\,
\hat{\alpha}^{\vee}_{\ell+1},
\label{eq:Hf ops for corr func b}
\\
&
\widehat{D}^{\vee}_{j}
:= 
\frac{1}{3}
\left(
\mathrm{i}
\hat{\beta}^{\vee}_{j+1}\,
\hat{\alpha}^{\vee}_{j}
+
\mathrm{i}
\hat{\beta}^{\vee}_{j}\,
\hat{\alpha}^{\vee}_{j+1}
+
\hat{\beta}^{\vee}_{j}\,
\hat{\beta}^{\vee}_{j+1}\,
\hat{\alpha}^{\vee}_{j}\,
\hat{\alpha}^{\vee}_{j+1}
\right).
\label{eq:Hf ops for corr func c}
\end{align}
\end{subequations}
As was the case in Sec.\ \ref{subsec:KW and D8 Spin model},
we observe that operators
$\widehat{C}^{x}_{j,j+n}$
and 
$\widehat{C}^{y}_{j,j+n}$
defined in Eqs.\
\eqref{eq:Hb ops for corr func a}
and
\eqref{eq:Hb ops for corr func b},
respectively,
dualize to non-local string operators,
while the local operator
$\widehat{D}^{\,}_{j}$
defined in Eq.\
\eqref{eq:Hf ops for corr func c}
remains local after dualization. 

Under the JW duality, the bonding linear combinations
$\ket{\mathrm{Neel}^{x}}^{+}$
defined in Eq.\ \eqref{eq:def Neelx+} and 
$\ket{\mathrm{Neel}^{y}}^{+}$
defined in Eq.\ \eqref{eq:def Neely+} of
$\mathrm{Neel}^{\,}_{x}$ and $\mathrm{Neel}^{\,}_{y}$
states dualize to the two topologically nontrivial ground states
$\ket{\mathrm{Kitaev}}$ and $\ket{\overline{\mathrm{Kitaev}}}$ defined
in Eq.\ \eqref{eq:Kitaev ground states}, respectively. 
These states can be distinguished by the expectation values of the 
string order parameters $\widehat{C}^{x\,\vee}_{j,j+n}$
and $\widehat{C}^{y\,\vee}_{j,j+n}$.

\begin{figure}[t!]
\begin{center}
\begin{minipage}{0.45\textwidth}  
\begin{tikzpicture}[scale=0.85]
\begin{axis}[
title={Majumdar-Ghosh line},
axis line style = thick,
xlabel={$0\leq\tanh(\Delta)\le1$},
ylabel={$0\leq J\leq1/2$},
xmin=0, xmax=1,
ymin=0, ymax=1,
ticks=none,
] 
\addplot[blue, mark=*, mark size=4pt, only marks]
coordinates {(0.7,0)};
\addplot[blue, mark=square*, mark size=4pt, only marks]
coordinates {(0,0) (1,0)};
\addplot[blue, mark=diamond*, mark size=4pt, only marks]
coordinates {(0,1) (1,1)};
\draw[blue,thick] (0.7,0) -- (0.7,0.666);
\addplot[blue, mark=, mark size=8pt, only marks]
coordinates {(0.7,0.666)};
\draw[blue,thick] (0,1) .. controls (0.25,0.85) and (0.5,0.7) .. (0.7,0.666);
\draw[blue,thick] (0.7,0.666)..controls (0.8,1) and (0.95,1)..(1,1);
\node[blue,draw] at (0.35,0.3){$\ket{\mathrm{Neel}^{x}}^{+}$};
\node[blue,draw, rotate=0] at (0.85,0.3){$\ket{\mathrm{Neel}^{y}}^{+}$};
\node[blue,draw] at (0.55,0.89){Dimer doublet};
\end{axis}
\end{tikzpicture}
\end{minipage}
\begin{Large}
$\rightleftharpoons$
\end{Large}
\begin{minipage}{0.45\textwidth}  
\begin{tikzpicture}[scale=0.85]
\begin{axis}[
title={Majumdar-Ghosh line},
axis line style = thick,
xlabel={$0\leq\tanh(\Delta)\le1$},
ylabel={$0\leq J\leq1/2$},
xmin=0, xmax=1,
ymin=0, ymax=1,
ticks=none,
]
\addplot[blue, mark=*, mark size=4pt, only marks]
coordinates {(0.7,0)};
\addplot[blue, mark=square*, mark size=4pt, only marks]
coordinates {(0,0) (1,0)};
\addplot[blue, mark=diamond*, mark size=4pt, only marks]
coordinates {(0,1) (1,1)};
\draw[blue,thick] (0.7,0) -- (0.7,0.666);
\addplot[blue, mark=, mark size=8pt, only marks]
coordinates {(0.7,0.666)};
\draw[red,thick] (0,1) .. controls (0.25,0.85) and (0.5,0.7) .. (0.7,0.666);
\draw[red,thick] (0.7,0.666)..controls (0.8,1) and (0.95,1)..(1,1);
\node[blue,draw] at (0.35,0.3){$\ket{\mathrm{Kitaev}}$};
\node[blue,draw, rotate=0] at (0.85,0.3){$\ket{\overline{\mathrm{Kitaev}}}$};
\node[blue,draw] at (0.575,0.875){BDO};
\end{axis}
\end{tikzpicture}
\end{minipage} 
\end{center}
\caption{
The phase diagram of Hamiltonian
(\ref{eq:def Hamiltonian b=0})
restricted to the subspace
$\mathcal{H}^{\,}_{b=0;+}$
of the Hilbert space
$\mathcal{H}^{\,}_{b=0}$
is dual to the phase diagram of Hamiltonian
(\ref{eq:def Hamiltonian f=1})
restricted to the subspace
$\mathcal{H}^{\,\vee}_{f=1;+}$
of
$\mathcal{H}^{\,\vee}_{f=1}$.
The pair of dual subspaces are to be found in
the third line of Table
\ref{Table:Jordan-Wigner dualization b to bF}.
The two states
$\ket{\mathrm{Neel}^{x}}^{+}$
and
$\ket{\mathrm{Neel}^{y}}^{+}$ are defined in Eqs.\ 
\eqref{eq:def Neelx+}
and
\eqref{eq:def Neely+},
respectively,
while the box ``Dimer doublet''
refers to the ground states
\eqref{eq:Hb dimer GS}.
On the dual side,
the two non-trivial and distinct invertible topological states
$\ket{\mathrm{Kitaev}}$
and
$\ket{\overline{\mathrm{Kitaev}}}$
are defined in Eq.\ \eqref{eq:Kitaev ground states}.
The box ``BDO'' stands for the bond-density ordered phase
described by the two-fold degenerate
ground states \eqref{eq:BDO states}
along the MG line. The symbols
$\rightharpoonup$
and
$\leftharpoondown$
denote gauging the diagonal subgroups 
generated
by
$\widehat{U}^{\,}_{r^{z}_{\pi}}$ 
and
by its dual $\widehat{P}^{\vee}_{\mathrm{F}}$
defined in Eq.\
(\ref{eq:reps symmetries t and r for cal Bf c}), respectively.
}
\label{fig:comparing spin-1/2 XY phase diagram and its JW dual}
\end{figure}
As opposed to the 
KW duality, the Hamiltonian \eqref{eq:def Hamiltonian f=1}
that obeys open boundary conditions
is equivalent,
up to a unitary transformation,
to the dual of the Hamiltonian \eqref{eq:def Hamiltonian b=0}
that obeys open boundary conditions.
It is shown in
Appendix \ref{appsec:Triality with open boundary conditions}
that selecting open boundary conditions removes the consistency conditions
on the bond algebra that require the projections of the Hilbert spaces
$\mathcal{H}^{\,}_{b=0}$
and
$\mathcal{H}^{\vee}_{f=1}$
onto their subspaces
$\mathcal{H}^{\,}_{b=0,+}$
and
$\mathcal{H}^{\vee}_{f=1,+}$
for duality to hold.
The two-fold degeneracy of the
$\mathrm{Neel}^{\,}_{x}$
and
$\mathrm{Neel}^{\,}_{y}$
ground states dualizes to the two-fold 
degeneracy of the non-trivial invertible topological phases with
open boundary conditions. 
Finally, along the MG line, the two-fold degenerate dimer ground states
\eqref{eq:Hb dimer GS} of Hamiltonian \eqref{eq:def Hamiltonian b=0}
dualize to the two-fold degenerate bond-density order 
ground states \eqref{eq:BDO states} of 
Hamiltonian \eqref{eq:def Hamiltonian f=1}.
The expectations values of the operators 
\eqref{eq:Hf ops for corr func}
in the ground states of Hamiltonian 
\eqref{eq:def Hamiltonian f=1}
are given in Table \ref{Table:Exp values under JW duality}.

The phase diagrams below the MG line of Hamiltonians
(\ref{eq:def Hamiltonian b=0})
and
(\ref{eq:def Hamiltonian f=1})
defined on their domain of definitions
$\mathcal{H}^{}_{b=0;+}$
and
$\mathcal{H}^{\vee}_{f=1;+}$,
respectively, are compared in
Fig.\
\ref{fig:comparing spin-1/2 XY phase diagram and its JW dual}.
Whereas the vertical phase boundary at
$\Delta=1$ remains a line of quantum critical points (blue line)
outside of the Landau-Ginzburg paradigm,
the boundaries separating the two topologically nontrivial
singlet phases from the bond-density ordered
phase are ordinary
Landau-Ginzburg phase transitions (red lines).

\section{Quantum $\mathbb{Z}^{\,}_{n}$ clock models with
$n\text{ mod }2$ LSM anomalies}
\label{sec:Zn generalization}

We are going to generalize
the spin-$1/2$ chains with global $\mathbb{Z}^{\,}_{2}$ symmetry
that we have studied in Secs.\
\ref{sec:Triality through bond algebra isomorphisms}--%
\ref{sec:Application to quantum spin-1/2 degrees of freedom on a chain}
to clock models with global $\mathbb{Z}^{\,}_{n}$ symmetry
whereby $n=2,3,\cdots$.
Our aim is to establish how, as a consequence of an LSM anomaly,
the crystalline and internal symmetries become intertwined
under dualities obtained by gauging the global
$\mathbb{Z}^{\,}_{n}$ symmetry. We are going to show that the
non-trivial mixing of the crystalline and internal symmetries
is sensitive to the parity of $n=2,3,\cdots$.

\subsection{A generalized LSM anomaly}

Consider a one-dimensional lattice $\Lambda$ of cardinality $2Nn$
with the integers $N=1,2,\cdots$ and $n=2,3,\cdots$.
To each site $j$ of the lattice,
we assign an $n$-dimensional complex Hilbert space $\mathbb{C}^{n}$
on which we may represent the clock operator $\widehat{Z}^{\,}_{j}$
and the shift operator $\widehat{X}^{\,}_{j}$ obeying the algebra
\begin{subequations}
\label{eq:def Zn clock model on Lambda}
\begin{align}
\widehat{X}^{\,}_{i}\,
\widehat{Z}^{\,}_{j}=
\left(\omega^{\,}_{n}\right)^{\delta^{\,}_{i,j}}\,
\widehat{Z}_{j}\,
\widehat{X}_{i},
\qquad
\left(\widehat{X}^{\,}_{j}\right)^{n}=
\left(\widehat{Z}^{\,}_{j}\right)^{n}=
\widehat{\mathbb{1}}^{\,}_{\mathcal{H}^{\,}_{b}},
\qquad
\omega^{\,}_{n}
:=
e^{\mathrm{i}\frac{2\pi}{n}},
\qquad
i,j\in\Lambda,
\label{eq:def Zn clock model on Lambda b}
\end{align}
by $n$-dimensional complex-valued unitary matrices%
~\footnote{%
~The shift operator $\widehat{X}^{\,}_{j}$ and clock operator
$\widehat{Z}^{\,}_{j}$
are unitary for any $j\in\Lambda$, i.e.,
$\left(\widehat{X}^{\,}_{j}\right)^{-1}=
\left(\widehat{X}^{\,}_{j}\right)^{\dag}$
and
$\left(\widehat{Z}^{\,}_{j}\right)^{-1}=
\left(\widehat{Z}^{\,}_{j}\right)^{\dag}$.
}.
As we will impose the global internal symmetry
$\mathbb{Z}^{z}_{n}$ that is generated by the unitary operator 
\begin{align}
\widehat{U}^{\,}_{z}:=
\prod_{j\in\Lambda}
\widehat{Z}^{\,}_{j},
\label{eq:Zn clock bond algebra b}
\end{align}
we impose the twisted boundary conditions
\begin{equation}
\widehat{X}^{\,}_{j+2Nn}=
\left(\widehat{U}^{\,}_{z}\right)^{b}\,
\widehat{X}^{\,}_{j}\,
\left(\widehat{U}^{\,}_{z}\right)^{-b}=
\left(\omega^{\,}_{n}\right)^{b}\,
\widehat{X}^{\,}_{j},
\qquad
\widehat{Z}^{\,}_{j+2Nn}=
\left(\widehat{U}^{\,}_{z}\right)^{b}\,
\widehat{Z}^{\,}_{j}\,
\left(\widehat{U}^{\,}_{z}\right)^{-b}=
\widehat{Z}^{\,}_{j},
\label{eq:def Zn clock model on Lambda c}    
\end{equation}
for any $b\in\mathbb{Z}^{\,}_{n}$ on the Hilbert space
\begin{align}
\mathcal{H}^{\,}_{b}:=
\bigotimes_{j\in\Lambda}
\mathbb{C}^{n}\,.
\label{eq:def Zn clock model on Lambda d}
\end{align}
\end{subequations}

We define the bond algebra
\begin{subequations}
\label{eq:Zn clock bond algebra}
\begin{align}
\mathfrak{B}^{\,}_{b}:=
\left\langle\left. 
\widehat{Z}^{\,}_{j},
\quad 
\widehat{X}^{\,}_{j}\,
\left(\widehat{X}^{\,}_{j+1}\right)^{-1}
\ \right|\
j\in\Lambda
\right\rangle,
\label{eq:Zn clock bond algebra a}
\end{align}
\end{subequations} 
of operators that are symmetric under the $\mathbb{Z}^{z}_{n}$ symmetry generated by
Eq.\ \eqref{eq:Zn clock bond algebra b}.
We decompose the Hilbert space into definite eigenvalue sectors
of $\widehat{U}^{\,}_{z}$
\begin{align}
\mathcal{H}^{\,}_{b}
=
\bigoplus_{\alpha=0}^{n-1}
\mathcal{H}^{\,}_{b;\,\alpha},
\qquad
\mathcal{H}^{\,}_{b;\,\alpha}
:=
\widehat{P}^{\,}_{b;\,\alpha}\,
\mathcal{H}^{\,}_{b},
\qquad
\alpha 
=
0,\cdots,\,n-1,
\label{eq:Zn clock model splitting into sectors}
\end{align}
where $\widehat{P}^{\,}_{b;\,\alpha}$ 
is the projector to the subspace with
definite eigenvalue 
$\left(\omega^{\,}_{n}\right)^{\alpha}$ of $\widehat{U}^{\,}_{z}$.

In addition to the global internal $\mathbb{Z}^{z}_{n}$
symmetry of the bond algebra
\eqref{eq:Zn clock bond algebra},
we presume additional 
crystalline and internal symmetries.
To accommodate translation symmetry in a simple way,
we select periodic boundary conditions
by choosing $b=0$.
First, we shall impose two crystalline symmetries,
namely, translations and site-centered reflection of lattice
$\Lambda$
which are implemented by the unitary operators
\begin{subequations}
\label{eq:Zn space symmetries b}
\begin{align}
&
\widehat{U}^{\,}_{t}\,
\widehat{X}^{\,}_{j}\,
\widehat{U}^{\dagger}_{t}
=
\widehat{X}^{\,}_{t(j)},
\quad
\widehat{U}^{\,}_{t}\,
\widehat{Z}^{\,}_{j}\,
\widehat{U}^{\dagger}_{t}
=
\widehat{Z}^{\,}_{t(j)},
\quad
t(j)
=
j+1\hbox{ mod }2Nn,
\\
&
\widehat{U}^{\,}_{r}\,
\widehat{X}^{\,}_{j}\,
\widehat{U}^{\dagger}_{r}
=
\widehat{X}^{\,}_{r(j)},
\quad
\widehat{U}^{\,}_{r}\,
\widehat{Z}^{\,}_{j}\,
\widehat{U}^{\dagger}_{r}
=
\widehat{Z}^{\,}_{r(j)},
\quad
r(j)
=
2Nn-j\hbox{ mod }2Nn,
\end{align}
respectively. 
Choosing the lattice $\Lambda$ to be made of an even number of
sites ensures that site-centered reflection exists for any $n$
and has the two fixed points $j=Nn$ and $j=2Nn$.
The operators $\widehat{U}^{\,}_{t}$ and $\widehat{U}^{\,}_{r}$
generate the representation of the space group
\begin{align}
\mathrm{G}^{\,}_{\mathrm{spa}}
\equiv 
\mathbb{Z}^{t}_{2Nn}\rtimes \mathbb{Z}^{r}_{2}.
\end{align}
\end{subequations}
Next, we impose an additional 
global internal symmetry $\mathbb{Z}^{x}_{n}$ that is
implemented by the unitary operator
\begin{subequations}
\label{eq:Zn internal symmetries b}
\begin{align}
\widehat{U}^{\,}_{x}:=
\prod_{j\in\Lambda}
\widehat{X}^{\,}_{j},
\label{eq:Zn clock symmetry global rep}
\end{align}
i.e., the product of all local shift operators.
Together, $\widehat{U}^{\,}_{x}$ and $\widehat{U}^{\,}_{z}$
generate a global representation of 
the Abelian group
$\mathbb{Z}^{x}_{n}\times\mathbb{Z}^{z}_{n}$.  Thus, the total
symmetry group is the direct product
\begin{align}
\mathrm{G}^{\,}_{\mathrm{tot}}
\equiv
\mathrm{G}^{\,}_{\mathrm{spa}}
\times
\mathrm{G}^{\,}_{\mathrm{int}},
\qquad
\mathrm{G}^{\,}_{\mathrm{int}}
\equiv
\mathbb{Z}^{x}_{n}\times \mathbb{Z}^{z}_{n}.
\label{eq:Zn clock total sym group}
\end{align}
\end{subequations}
While the global representation of $\mathrm{G}^{\,}_{\mathrm{int}}$
is a group homomorphism,
it is locally projective due to the algebra
\begin{subequations}
\begin{align}
\widehat{X}^{\,}_{j}\,
\widehat{Z}^{\,}_{j}
=
\omega^{\,}_{n}\,
\widehat{Z}^{\,}_{j}\,
\widehat{X}^{\,}_{j},
\qquad
j\in\Lambda.
\label{eq:ZnxZn proj rep}
\end{align}
More precisely, distinct projective representations of the group 
$\mathbb{Z}^{\,}_{n}\times\mathbb{Z}^{\,}_{n}$ are labeled 
by the equivalence classes $[\omega]=0,\,1,\cdots,n-1$
taking values in the second cohomology group
\begin{align}
[\omega]\in
H^{2}
\bm{\big(}
\mathbb{Z}^{\,}_{n}\times\mathbb{Z}^{\,}_{n},\,
\mathrm{U}(1)
\bm{\big)}=
\mathbb{Z}^{\,}_{n}.    
\label{eq:ZnxZn 2nd cohomology}
\end{align}
\end{subequations}
The algebra \eqref{eq:ZnxZn proj rep} is a representative of the generator 
$[\omega]=1$ of the cohomology group \eqref{eq:ZnxZn 2nd cohomology}.
Because of the projective algebra \eqref{eq:ZnxZn proj rep},
the following LSM Theorem with translation symmetry applies.

\begin{thm}[Generalized translation LSM]\label{thm:LSM translation Zn}
Consider a one-dimensional lattice Hamiltonian
with the symmetry group
$\mathrm{G}^{\,}_{\mathrm{tot}}\equiv
\mathrm{Z}^{t}_{|\Lambda|}\times
\mathbb{Z}^{x}_{n}\times\mathbb{Z}^{y}_{n}$,
where the subgroup $\mathrm{Z}^{t}_{|\Lambda|}$ generates lattice 
translations and the subgroup $\mathbb{Z}^{x}_{n}\times\mathbb{Z}^{y}_{n}$
with $n=2,3,\cdots$
generates global internal discrete clock-rotation symmetry.
Let
$[\omega]\in
H^{2}
\bm{\big(}
\mathbb{Z}^{\,}_{n}\times\mathbb{Z}^{\,}_{n},\,
\mathrm{U}(1)
\bm{\big)}=
\mathbb{Z}{\,}_{n}$ 
denote the second cohomology class associated with
the local representation of $\mathbb{Z}^{x}_{n}\times\mathbb{Z}^{z}_{n}$
at any site of $\Lambda$.
If $[\omega] \neq 0 \text{ mod } n$, 
then the ground states cannot be simultaneously
gapped,
non-degenerate,
and
$\mathrm{G}^{\,}_{\mathrm{tot}}$-symmetric.
\end{thm}

\begin{defn}[Generalized translation LSM anomaly]
When LSM Theorem \ref{thm:LSM translation Zn}
applies,
we say that there is a translation LSM anomaly.
\end{defn}

\begin{remark}
Since LSM Theorem \ref{thm:LSM translation Zn}
holds for any integer $n=2,3,\cdots$,
the dual of $\widehat{U}^{\,}_{x}$ is not translationally invariant
for any value of $n=2,3,\cdots$.
\end{remark}

Unlike with Theorems
\ref{thm:LSM translation}--%
\ref{thm:LSM translation Zn},
we are not aware of  a rigorous proof of
the Conjecture \ref{thm:LSM reflection Zn}
that follows
(see Refs.\ \cite{Po2017,Weicheng2022}).
Conjecture \ref{thm:LSM reflection Zn} 
is expected to hold based
on the lattice homotopy arguments
introduced in Ref.\ \cite{Po2017}
and crystalline equivalence 
principle introduced in Refs.\ \cite{Thorngren2018,Else2020}. 
\addtocounter{conj}{3}
\begin{conj}[Generalized reflection LSM]\label{thm:LSM reflection Zn}
Consider a one-dimensional lattice Hamiltonian
with the symmetry group
$\mathrm{G}^{\,}_{\mathrm{tot}}\equiv
\mathrm{Z}^{r}_{2}\times
\mathbb{Z}^{x}_{n}\times\mathbb{Z}^{z}_{n}$,
where the subgroup $\mathrm{Z}^{r}_{2}$ is generated by
a site-centered reflection, while the subgroup
$\mathbb{Z}^{x}_{n}\times\mathbb{Z}^{y}_{n}$
with $n=2,3,\cdots$ is generated by two global 
internal discrete clock-rotation symmetries.
Let
$[\omega]\in
H^{2}
\bm{\big(}
\mathbb{Z}^{\,}_{n}\times\mathbb{Z}^{\,}_{n},\,
\mathrm{U}(1)
\bm{\big)}=
\mathbb{Z}{\,}_{n}$ 
denote the second cohomology class associated with
the local representation of $\mathbb{Z}^{x}_{n}\times\mathbb{Z}^{z}_{n}$
at any one of the fixed points of the reflection. 
If $[\omega] \neq 2k \text{ mod } n$ for some integer $k$,
then the ground states cannot be simultaneously gapped,
non-degenerate, and
$\mathrm{G}^{\,}_{\mathrm{tot}}$-symmetric.
\end{conj}

\begin{defn}[Reflection LSM anomaly]
When \textbf{Conjecture} \ref{thm:LSM reflection Zn} applies,
we say that there is a reflection LSM anomaly.
\end{defn}

\begin{remark}
Conjecture \ref{thm:LSM reflection Zn} reduces to 
Theorem \ref{thm:LSM reflection} for $n=2$ and $[\omega]=1$, which is 
the only non-trivial projective representation realized by 
half-integer spins. Furthermore, when $n$ is odd,
the condition 
$[\omega] = 2k \text{ mod } n$
is always satisfied for some integer $k$.
Hence, 
there is no generalized reflection LSM anomaly
when $n$ is odd.
For the algebra \eqref{eq:ZnxZn proj rep},
we have $[\omega]=1$ which implies that a non-degenerate, gapped,
and
$\mathrm{G}^{\,}_{\mathrm{tot}}$-symmetric ground state is
possible only when $n=3,5,\cdots$, while it is ruled out by 
Conjecture \ref{thm:LSM reflection Zn} when $n=2,4,\cdots$.
In what follows, we are going to confirm
this claim by showing that the
operator $\widehat{U}^{\,}_{x}$ cannot be
dualized and remain invariant under reflection 
when $n$ is even, while it can be when $n$ is odd. 
\end{remark}

\subsection{Kramers-Wannier dual of the generalized LSM anomaly}

Starting from the bond algebra
$\mathfrak{B}^{\,}_{b}$ in Eq.\ \eqref{eq:Zn clock bond algebra a},
we are going to perform a gauging of $\widehat{U}^{\,}_{z}$.
This gauging furnishes
a dual bond algebra, where the duality is nothing but a
$\mathbb{Z}^{\,}_{n}$ generalization of KW duality
described in Sec.\ \ref{subsec:Kramers-Wannier duality b to b'}.
We are then going to invoke an additional
$\mathbb{Z}^{x}_{n}$ symmetry
that is generated by $\widehat{U}^{\,}_{x}$
defined in
Eq.\ \eqref{eq:Zn clock symmetry global rep}
and construct its dual
$\widehat{U}^{\vee}_{x}$
under the $\mathbb{Z}^{\,}_{n}$
generalization of KW duality.
Our main result will be that
the action of reflection on
$\widehat{U}^{\vee}_{x}$
turns out to be non-trivial (trivial)
if $n=0\, \,\rm{mod}\, 2$
($n=1\,\, \rm{mod}\, 2$).
This result is aligned with the LSM anomaly conjecture
\ref{thm:LSM reflection Zn}.

In order to gauge $\widehat{U}^{\,}_{z}$
defined in Eq.\ \eqref{eq:def Zn clock model on Lambda c}, 
we introduce
$\mathbb{Z}^{\,}_{n}$-valued gauge degrees of freedom
on the dual lattice $\Lambda^{\star}$.
To each site $j^{\star}$ of the dual lattice $\Lambda^{\star}$,
we therefore associate an $n$-dimensional Hilbert space.
The operator algebra attached to
$j^{\star}\in\Lambda^{\star}$
is generated by the clock operator
$\widehat{Z}^{\star}_{j^{\star}}$
and the shift operator
$\widehat{X}^{\star}_{j^{\star}}$
that satisfy the same algebra
\eqref{eq:def Zn clock model on Lambda b}
except for substituting $\Lambda$ by $\Lambda^{\star}$.
To gauge the global symmetry
\eqref{eq:Zn clock bond algebra b},
we define the unitary local Gauss operator
\begin{align}
\widehat{G}^{\,}_{j}:=
\left(\widehat{X}^{\star}_{j^{\star}-1}\right)^{-1}\,
\widehat{Z}^{\,}_{j}\,
\widehat{X}^{\star}_{j^{\star}}
\label{eq:def ext Zn clock model c}
\end{align}
for any site $j\in\Lambda$.
These local Gauss operators commute pairwise on distinct sites.
By analogy to Eq.\
\eqref{eq:def Zn clock model on Lambda c},
the twisted boundary conditions 
\begin{align}
\widehat{X}^{\,\star}_{j^{\star}+2Nn}=
\left(\omega^{\,}_{n}\right)^{b^{\prime}}\,
\widehat{X}^{\,\star}_{j^{\star}},
\qquad \quad
\widehat{Z}^{\,\star}_{j^{\star}+2Nn}=
\widehat{Z}^{\,\star}_{j^{\star}}
\label{eq:def ext Zn clock model g}
\end{align}
are imposed for the
$\mathbb{Z}^{\,}_{n}$-valued gauge degrees of freedom.
The extended Hilbert space including the original (matter)
and $\mathbb{Z}^{\,}_{n}$-valued gauge degrees of freedom
defined on
$\Lambda$ and $\Lambda^{\star}$, respectively,
admits the tensor decomposition 
\begin{align}
\mathcal{H}^{\,}_{b,b^{\prime}}:=
\mathcal{H}^{\,}_{b}
\otimes
\mathcal{H}^{\,}_{b^{\prime}},
\qquad
\mathcal{H}^{\,}_{b}:=
\bigotimes_{j\in\Lambda}
\mathbb{C}^{n},
\qquad
\mathcal{H}^{\,}_{b^{\prime}}:=
\bigotimes_{j^{\star}\in\Lambda^{\star}}
\mathbb{C}^{n}.
\label{eq:def ext Zn clock model h}
\end{align}
A generic gauge transformation
\begin{equation}
\begin{alignedat}{3}
&
\widehat{Z}^{\,}_{j}\longmapsto
\widehat{Z}^{\,}_{j},
\qquad \qquad \quad
&&\widehat{X}^{\,}_{j}&&\longmapsto
\omega^{-\lambda_{j}}_{n}\,
\widehat{X}^{\,}_{j},
\\
&
\widehat{X}^{\star}_{j^{\star}}\longmapsto
\widehat{X}^{\star}_{j^{\star}},
\qquad
&&\widehat{Z}^{\star}_{j^{\star}}&&\longmapsto
\omega^{(\lambda^{\,}_{j}-\lambda^{\,}_{j+1})}_{n}\,
\widehat{Z}^{\star}_{j^{\star}},    
\end{alignedat}    
\end{equation}
is specified by the variables 
$\lambda^{\,}_{j}\in\mathbb{Z}^{\,}_{n}$ 
with $j\in\Lambda$
and  implemented by conjugation with the
unitary operator
\begin{align}
\widehat{G}^{\,}_{\bm{\lambda}}:=
\prod_{j\in\Lambda}
\left(\widehat{G}^{\,}_{j}\right)^{\lambda^{\,}_{j}},
\qquad
\bm{\lambda}:=
\left(\lambda^{\,}_{1},\cdots,\lambda^{\,}_{2Nn}\right).
\label{eq:Gauss operator Zn}
\end{align}

By minimally coupling the bond algebra 
\eqref{eq:Zn clock bond algebra a},
we obtain the extended bond algebra
\begin{align}
\mathfrak{B}^{\,}_{b,b^{\prime}}:=
\left\langle\left. 
\widehat{Z}^{\,}_{j},
\qquad
\widehat{X}^{\,}_{j}\,
\widehat{Z}^{\star}_{j^{\star}}\,
\left(\widehat{X}^{\,}_{j+1}\right)^{-1}
\ \right|\
j\in\Lambda
\right \rangle
\label{eq:def extended bond algebra Zn}
\end{align}
of gauge-invariant operators.
As was done in Sec.\ \ref{sec:Triality through bond algebra isomorphisms},
there exists a unitary operator $\widehat{U}^{\,}_{b,b^{\prime}}$
that implements the transformation \cite{Moradi23}
\begin{subequations}
\label{eq:def unitary trsf for Zn}
\begin{equation}
\begin{alignedat}{3}
\widehat{U}^{\,}_{b,b^{\prime}}\,
\widehat{X}^{\,}_{j}\,
\widehat{U}^{\dag}_{b,b^{\prime}}&=
\widehat{X}^{\,}_{j},
\qquad\qquad
&&\widehat{U}^{\,}_{b,b^{\prime}}\,
\widehat{Z}^{\,}_{j}\,
\widehat{U}^{\dag}_{b,b^{\prime}}&&=
\widehat{X}^{\star}_{j^{\star}-1}
\widehat{Z}^{\,}_{j}\,
\left(\widehat{X}^{\star}_{j^{\star}}\right)^{-1},
\\
\widehat{U}^{\,}_{b,b^{\prime}}\,
\widehat{X}^{\star}_{j^{\star}}\,
\widehat{U}^{\dag}_{b,b^{\prime}}&=
\widehat{X}^{\star}_{j^{\star}},
\qquad
&&\widehat{U}^{\,}_{b,b^{\prime}}\,
\widehat{Z}^{\star}_{j^{\star}}\,
\widehat{U}^{\dagger}_{b,b^{\prime}}&&=
\left(\widehat{X}^{\,}_{j}\right)^{-1}\,
\widehat{Z}^{\star}_{j^{\star}}\,
\widehat{X}^{\,}_{j+1},
\end{alignedat}
\end{equation}
for any $j\in\Lambda$ and $j^{\star}\in\Lambda^{\star}$.
In particular, under this transformation,
the local Gauss operator
\eqref{eq:Gauss operator Zn}
becomes
\begin{align}
\widehat{U}^{\,}_{b,b^{\prime}}\,
\widehat{G}^{\,}_{j}\,
\widehat{U}^{\dag}_{b,b^{\prime}}=
\widehat{Z}^{\,}_{j}
\end{align}
\end{subequations}
for any $j\in\Lambda$.
The subspace of physical states is defined to be the one
for which the action of all local Gauss operators
reduces to the identity. Hence, after the unitary transformation
(\ref{eq:def unitary trsf for Zn}),
the subspace of physical states is defined to be the one
for which the action of $\widehat{Z}^{\,}_{j}$
for any $j\in\Lambda$
reduces to the identity. It
is the $n^{2Nn}$-dimensional
gauge-invariant subspace $\mathcal{H}^{\,\vee}_{b^{\prime}}$
of $\mathcal{H}^{\,}_{b,b^{\prime}}$.
The projection of the bond algebra
(\ref{eq:def extended bond algebra Zn})
to the subspace $\mathcal{H}^{\,\vee}_{b^{\prime}}$
delivers the dual bond algebra
\begin{subequations}
\begin{align}
\mathfrak{B}^{\,}_{b^{\prime}}:=
\left\langle\left. 
\widehat{X}^{\star\,\vee}_{j^{\star}-1}\,
\left(\widehat{X}^{\star\,\vee}_{j^{\star}}\right)^{-1},
\qquad
\widehat{Z}^{\star\,\vee}_{j^{\star}}\,
\ \right|\
j^{\star}\in\Lambda^{\star}
\right \rangle,
\label{eq:Zn clock bond algebra b'}
\end{align}
which is symmetric under the 
dual $\mathbb{Z}^{z^{\vee}}_{n}$-symmetry generated by the unitary
operator
\begin{equation}
\widehat{U}^{\vee}_{z^\vee}:=
\prod_{j^{\star}\in\Lambda^{\star}}
\widehat{Z}^{\star\,\vee}_{j^{\star}}.
\label{eq:Zn Uz KW dual}
\end{equation}
\end{subequations}
The projected Hilbert space $\mathcal{H}^{\vee}_{b'}$
is isomorphic to the Hilbert space
\eqref{eq:def Zn clock model on Lambda d}. It
can be decomposed into subspaces
with definite eigenvalue of 
$\widehat{U}^{\vee}_{z^{\,}_{n}}$
[as was done in Eq.\ 
\eqref{eq:Zn clock model splitting into sectors}],
\begin{align}
\mathcal{H}^{\vee}_{b'}
=
\bigoplus_{\alpha=0}^{n-1}
\mathcal{H}^{\vee}_{b';\,\alpha},
\qquad
\mathcal{H}^{\vee}_{b';\,\alpha}
:=
\widehat{P}^{\vee}_{b';\,\alpha}\,
\mathcal{H}^{\vee}_{b'},
\qquad
\alpha 
=
0,\cdots,\,n-1,
\label{eq:Zn clock model splitting into sectors b'}
\end{align}
where $\widehat{P}^{\vee}_{b';\,\alpha}$ 
is the projector to subspace with
definite eigenvalue 
$\omega^{\alpha}_{n}$ of $\widehat{U}^{\vee}_{r^{z\,\vee}_{n}}$.

Consistency with the pair of twisted boundary conditions 
\eqref{eq:def Zn clock model on Lambda c}
and
\eqref{eq:def ext Zn clock model g}
requires the identification of the pair of operators
\begin{subequations}
\begin{align}
\left(\,
\prod_{j\in \Lambda}
\widehat{Z}^{\,}_{j}\equiv
\widehat{U}^{\,}_{r^{z}_{n}},
\qquad
\left(\omega^{\,}_{n}\right)^{b'}\,
\widehat{\mathbb{1}}^{\,}_{\mathcal{H}^{\,\vee}_{b^{\prime}}}
\right)
\end{align}
on the one hand and the pair of operators
\begin{align}
\left(
\left(\omega^{\,}_{n}\right)^{b}\,
\widehat{\mathbb{1}}^{\,}_{\mathcal{H}^{\,}_{b}},
\qquad
\prod_{j^{\star}\in \Lambda^{\star}}
\widehat{Z}^{\star\,\vee}_{j^{\star}}\equiv
\widehat{U}^{\vee}_{r^{z\,\vee}_{n}}
\right)
\end{align}
\end{subequations}
on the other hand. This pair of consistency conditions can only be met
if the domains of definition of all dual pairs of operators are restricted to
the dual pair of subspaces 
\begin{equation}
\mathcal{H}^{\,}_{b;\,b^{\prime}}:=
\widehat{P}^{\,}_{b;\,b'}\,
\mathcal{H}^{\,}_{b},
\qquad
\mathcal{H}^{\vee}_{b';\,b}:=
\widehat{P}^{\vee}_{b';\,b}\,
\mathcal{H}^{\,\vee}_{b^{\prime}}.
\end{equation}
The boundary conditions $b$ on the Hilbert space
$\mathcal{H}^{\,}_{b}$
dictates the definite eigenvalue subspace of the dual Hilbert space 
$\mathcal{H}^{\vee}_{b'}$ and vice versa.
This is the $\mathbb{Z}^{\,}_{n}$ generalization of the KW duality 
from Sec.\ \ref{subsec:Kramers-Wannier duality b to b'}.

We now turn to obtaining the duals of the crystalline
and internal symmetries
\eqref{eq:Zn space symmetries b}
and
\eqref{eq:Zn internal symmetries b}, respectively. 
We impose periodic boundary conditions for both
bond algebras
\eqref{eq:Zn clock bond algebra}
and 
\eqref{eq:Zn clock bond algebra b'}
by choosing $b=b'=0$.

As described in Sec.\ \ref{subsec:G total frak H b'=0}, the dual
crystalline symmetries are obtained by first extending the
operators to the Hilbert space
\eqref{eq:def ext Zn clock model h}
by demanding covariance of the Gauss operators
\eqref{eq:def ext Zn clock model c}.
We obtain the duals of operators
$\widehat{U}^{\,}_{t}$
and 
$\widehat{U}^{\,}_{r}$
defined by their actions on the Hilbert space
$\mathcal{H}^{\vee}_{b'=0}$
\begin{subequations}
\label{eq:Zn KW spatial symmetries}
\begin{align}
&
\widehat{U}^{\vee}_{t}\,
\widehat{X}^{\star\,\vee}_{j^{\star}}\,
\left(\widehat{U}^{\vee}_{t}\right)^{\dagger}
=
\widehat{X}^{\star\,\vee}_{t(j^{\star})},
\qquad\quad
\widehat{U}^{\vee}_{t}\,
\widehat{Z}^{\star\,\vee}_{j^{\star}}\,
\left(\widehat{U}^{\vee}_{t}\right)^{\dagger}
=
\widehat{Z}^{\star\,\vee}_{t(j^{\star})},
\qquad\quad
t(j^{\star})
=
j^{\star}+1,
\label{eq:Zn KW spatial symmetries a}
\\
&
\widehat{U}^{\vee}_{r}\,
\widehat{X}^{\star\,\vee}_{j^{\star}}\,
\left(\widehat{U}^{\vee}_{r}\right)^{\dagger}
=
\left(\widehat{X}^{\star\,\vee}_{r(j^{\star})}\right)^{-1},
\quad
\widehat{U}^{\vee}_{r}\,
\widehat{Z}^{\star\,\vee}_{j^{\star}}\,
\left(\widehat{U}^{\vee}_{r}\right)^{\dagger}
=
\left(\widehat{Z}^{\star\,\vee}_{r(j^{\star})}\right)^{-1},
\quad
r(j^{\star})
=
2Nn-j^{\star}.
\label{eq:Zn KW spatial symmetries b}
\end{align}
\end{subequations}
as it should be,
these transformation rules reduce to 
those in Eqs.\
(\ref{eq:reps symmetries t and r for cal Bb' ext c})
and
(\ref{eq:reps symmetries t and r for cal Bb' ext d})
when $n=2$. 
Since
$\widehat{X}^{\star\,\vee}_{j^{\star}}$
and 
$\widehat{Z}^{\star\,\vee}_{j^{\star}}\,$
are akin to electric 
field $e^{\mathrm{i}E}$ and 
gauge field $e^{\mathrm{i}A}$, respectively, 
they are to be Hermitian conjugated
under reflection in 
Eq.\ \eqref{eq:Zn KW spatial symmetries b}.

Due to the projective algebra \eqref{eq:ZnxZn proj rep},
the global symmetry operator $\widehat{U}^{\,}_{x}$
is not gauge invariant. 
We therefore dualize it by expressing it in terms
of products of local operators from the bond algebra 
\eqref{eq:Zn clock bond algebra a}.
This allows us to use the isomorphism between the dual bond algebras 
\eqref{eq:Zn clock bond algebra a} and \eqref{eq:Zn clock bond algebra b'}.
We treat the cases of $n$ even and $n$ odd separately.

\paragraph{Case of $n$ even.} 
First, we rewrite the unitary operator
$\widehat{U}^{\,}_{x}$
defined in Eq.~\eqref{eq:Zn clock symmetry global rep}
as
\begin{subequations}
\begin{equation}
\begin{split}
\widehat{U}^{\,}_{x}
=&
\left[
\widehat{X}^{\,}_{1}\,
\left(\widehat{X}^{\,}_{2}\right)^{-1}
\right]\,
\left[
\widehat{X}^{\,}_{2}\,
\left(\widehat{X}^{\,}_{3}\right)^{-1}
\right]^{2}\,
\left[
\widehat{X}^{\,}_{3}\,
\left(\widehat{X}^{\,}_{4}\right)^{-1}
\right]^{3}\,
\cdots\,
\left[
\widehat{X}^{\,}_{n-1}\,
\left(\widehat{X}^{\,}_{n}\right)^{-1}
\right]^{n-1}
\\
&
\times
\left[
\widehat{X}^{\,}_{n+1}\,
\left(\widehat{X}^{\,}_{n+2}\right)^{-1}
\right]\,
\left[
\widehat{X}^{\,}_{n+2}\,
\left(\widehat{X}^{\,}_{n+3}\right)^{-1}
\right]^{2}\,
\cdots\,
\left[
\widehat{X}^{\,}_{2Nn-1}\,
\left(\widehat{X}^{\,}_{2Nn}\right)^{-1}
\right]^{n-1},    
\end{split}
\label{eq:Zn KW duality Ux reexpress n even}
\end{equation}
where each term inside the square brackets is
a generator of the bond algebra
\eqref{eq:Zn clock bond algebra a}.
The $j$th square bracket on the right-hand side of Eq.\
(\ref{eq:Zn KW duality Ux reexpress n even})
becomes gauge invariant upon insertion of
$\widehat{Z}^{\star}_{j^{\star}}$ between the pair of
shift operators on the sites $j$ and $j+1$ of $\Lambda$.
We may then use the isomorphism between dual bond algebras 
\eqref{eq:Zn clock bond algebra a} and 
\eqref{eq:Zn clock bond algebra b'}
to obtain the dual symmetry generator
\begin{align}
\widehat{U}^{\vee}_{x^\vee}
=
\prod_{j^{\star}\in \Lambda^{\star}}
\left(\widehat{Z}^{\star\,\vee}_{j^{\star}}\right)^{j^{\star}-1/2}.
\label{eq:Zn Ux KW dual n even}
\end{align}
\end{subequations}
Note that the dual operator
\eqref{eq:Zn Ux KW dual n even}
is neither invariant under translations 
\eqref{eq:Zn KW spatial symmetries a}
nor under
reflection
\eqref{eq:Zn KW spatial symmetries b}.
Instead, one verifies the algebra 
\begin{align}
\widehat{U}^{\vee}_{t}\,
\widehat{U}^{\vee}_{x^{\vee}}\,
\left(\widehat{U}^{\vee}_{t}\right)^{\dag}
=
\left(\widehat{U}^{\vee}_{z^{\vee}}\right)^{\dag}\,
\widehat{U}^{\vee}_{x^{\vee}},
\qquad
\widehat{U}^{\vee}_{r}\,
\widehat{U}^{\vee}_{x^{\vee}}\,
\left(\widehat{U}^{\vee}_{r}\right)^{\dag}
=
\widehat{U}^{\vee}_{z^{\vee}}\,
\widehat{U}^{\vee}_{x^{\vee}}.
\label{eq:Zn KW duality algebra n even}
\end{align}
Hence, we find that the total symmetry group 
\eqref{eq:Zn clock total sym group} dualizes to the 
symmetry group
\begin{subequations}
\begin{align}
\mathrm{G}^{\vee}_{\mathrm{tot}}
\equiv
\mathrm{G}^{\vee}_{\mathrm{spa}}
\ltimes
\mathrm{G}^{\vee}_{\mathrm{int}}
\label{eq:Zn clock KW dual total sym group n even} 
\end{align}
with the internal symmetry group 
\begin{align}
\mathrm{G}^{\vee}_{\mathrm{int}}
\equiv
\mathbb{Z}^{z\,\vee}_{n}
\times
\mathbb{Z}^{x\,\vee}_{n}
\end{align}
generated by $z^\vee$ and $x^\vee$
of order $n=2,3,\cdots$ that are represented by operators 
\eqref{eq:Zn Uz KW dual} and
\eqref{eq:Zn Ux KW dual n even},
respectively. 
The spatial symmetry group 
\begin{align}
\mathrm{G}^{\vee}_{\mathrm{spa}}
\equiv
\mathbb{Z}^{t}_{2Nn}\rtimes \mathbb{Z}^{r}_{2},
\end{align}
has two generators $t$ and $r$ that are order $2Nn$ and $2$;
and represented by operators
\eqref{eq:Zn KW spatial symmetries a}
and
\eqref{eq:Zn KW spatial symmetries b}, respectively. 
The semi-direct product structure
in the dual total symmetry group 
\eqref{eq:Zn clock KW dual total sym group n even}
is due to the non-trivial group action
\begin{align}
t\,
x^{\vee}\,
t^{-1}
=
(z^{\vee})^{-1}\,
x^{\vee},
\qquad
r\,
x^{\vee}\,
r
=
z^{\vee}\,
x^{\vee},
\qquad
r\,
z^{\vee}\,
r
=
(z^{\vee})^{-1}
\end{align}
\end{subequations}
of crystalline symmetries
$\mathrm{G}^{\vee}_{\mathrm{spa}}$
on the internal symmetries
$\mathrm{G}^{\vee}_{\mathrm{int}}$.

\paragraph{Case of $n$ odd.}
For the case of $n$ odd, Eq.\
\eqref{eq:Zn KW duality Ux reexpress n even}
can be used to reexpress the operator
$\widehat{U}^{\,}_{x}$. 
However, there is an alternative expression for
$\widehat{U}^{\,}_{x}$
which was not available when $n$ is even. 
We may write
\begin{equation}
\begin{split}
\widehat{U}^{\,}_{x}
&=
\left[
\widehat{X}^{\,}_{2Nn}\,
\left(\widehat{X}^{\,}_{1}\right)^{-1}
\right]^{\frac{n+1}{2}}\,
\left[
\widehat{X}^{\,}_{1}\,
\left(\widehat{X}^{\,}_{2}\right)^{-1}
\right]^{\frac{n+3}{2}}\,
\cdots\,
\left[
\widehat{X}^{\,}_{N\,n-1}\,
\left(\widehat{X}^{\,}_{N\,n}\right)^{-1}
\right]^{\frac{n-1}{2}}
\nonumber\\
&
\times
\left[
\widehat{X}^{\,}_{2Nn}\,
\left(\widehat{X}^{\,}_{2Nn-1}\right)^{-1}
\right]^{\frac{n+1}{2}}\,
\left[
\widehat{X}^{\,}_{2Nn-1}\,
\left(\widehat{X}^{\,}_{2Nn-2}\right)^{-1}
\right]^{\frac{n+3}{2}}\,
\cdots\,
\left[
\widehat{X}^{\,}_{N\,n+1}\,
\left(\widehat{X}^{\,}_{N\,n}\right)^{-1}
\right]^{\frac{n-1}{2}},    
\end{split}
\label{eq:Zn KW duality Ux reexpress n odd}
\end{equation}
where we have utilized the fact that
$n$ is an odd integer when writing the exponents.
The $j$th square bracket on the right-hand side of Eq.\
(\ref{eq:Zn KW duality Ux reexpress n even})
becomes gauge invariant upon insertion of
$\widehat{Z}^{\star}_{j^{\star}}$ between the pair of
shift operators on the sites $j$ and $j+1$ of $\Lambda$.
We may then use
the isomorphism between dual bond algebras 
\eqref{eq:Zn clock bond algebra a}
and 
\eqref{eq:Zn clock bond algebra b'}
to obtain the dual symmetry generator
\begin{align}
\widehat{U}^{\vee}_{x^{\vee}}
=
\left[
\prod_{j=0}^{Nn-1}
\left(\widehat{Z}^{\star\,\vee}_{j^{\star}}\right)^{\frac{n+1}{2}+j}
\right]
\left[
\prod_{j=0}^{Nn-1}
\left(\widehat{Z}^{\star\,\vee}_{2Nn-j^{\star}}\right)^{\frac{n-1}{2}-j}
\right]\,.
\label{eq:Zn Ux KW dual n odd}
\end{align}
While still not invariant under translation symmetry 
\eqref{eq:Zn KW spatial symmetries a},
the unitary operator
\eqref{eq:Zn KW duality Ux reexpress n odd}
is manifestly invariant
under the reflection symmetry
\eqref{eq:Zn KW spatial symmetries b}.
Therefore, we find the algebra
\begin{subequations}
\begin{align}
\widehat{U}^{\vee}_{t}\,
\widehat{U}^{\vee}_{x^{\vee}}\,
\left(\widehat{U}^{\vee}_{t}\right)^{\dag}
=
\left(\widehat{U}^{\vee}_{z^{\vee}}\right)^{\dag}\,
\widehat{U}^{\vee}_{x^{\vee}},
\qquad
\widehat{U}^{\vee}_{r}\,
\widehat{U}^{\vee}_{x^\vee}\,
\left(\widehat{U}^{\vee}_{r}\right)^{\dag}
=
\widehat{U}^{\vee}_{x^\vee}.
\label{eq:Zn KW duality algebra n odd}
\end{align}
As opposed to the algebra
\eqref{eq:Zn KW duality algebra n even},
reflection commutes with the dual symmetry
\eqref{eq:Zn Ux KW dual n odd}.
The total symmetry group 
\begin{align}
\mathrm{G}^{\mathrm{o}\,\vee}_{\mathrm{tot}}
\equiv
\mathrm{G}^{\vee}_{\mathrm{spa}}
\ltimes
\mathrm{G}^{\vee}_{\mathrm{int}}.
\label{eq:Zn clock KW dual total sym group n odd} 
\end{align}
differs from the group \eqref{eq:Zn clock KW dual total sym group n even}
by the group action
\begin{align}
t\,
x^{\vee}\,
t^{-1}
=
(z^{\vee})^{-1}\,
x^{\vee},
\qquad
r\,
x^{\vee}\,
r
=
x^{\vee},
\qquad
r\,
z^{\vee}\,
r
=
(z^{\vee})^{-1}
\end{align}
\end{subequations}
of crystalline symmetries $\mathrm{G}^{\vee}_{\mathrm{spa}}$
on the internal symmetries $\mathrm{G}^{\vee}_{\mathrm{int}}$.

The fact that the reflection symmetric decomposition
\eqref{eq:Zn KW duality Ux reexpress n odd}
is only possible for $n$ odd is rooted in the
LSM anomaly 
\ref{thm:LSM reflection Zn},
which only applies when $n$ is an even integer%
~\footnote{%
~Our starting point is the projective algebra 
\eqref{eq:ZnxZn proj rep} which corresponds to $[\omega]=1$. 
It can be verified for general $[\omega]$ that a reflection 
symmetric decomposition is
possible only when $[\omega]= 2k$ mod $n$ for some integer $k$.
}.
Importantly, while the reflection symmetry has a nontrivial group 
action on the generator $z^{\vee}$
of the dual symmetry group $\mathbb{Z}^{z\,\vee}_{n}$,
LSM anomalies \ref{thm:LSM translation Zn} and \ref{thm:LSM reflection Zn}
appear as the incompatibility between the image
$\mathbb{Z}^{x\,\vee}_{n}$
of the ungauged internal symmetry group
$\mathbb{Z}^{x}_{n}$
and crystalline symmetries.

We observe that, for any $n$, operators 
$\widehat{U}^{\,\vee}_{z^{\vee}}$
and
$\widehat{U}^{\,\vee}_{x^{\vee}}$
implement
$\mathbb{Z}^{\,}_{n}$-charge
and
$\mathbb{Z}^{\,}_{n}$-dipole
symmetries, respectively.
As was the case with $n=2$ in Sec.\ 
\ref{sec:Triality of crystalline and internal symmetries on a chain},
we find that $\mathbb{Z}^{\,}_{n}$ clock Hamiltonians
with charge, dipole, and translation symmetries
are free from an LSM anomaly for any $n$ and
are dual to Hamitonians with global internal 
$\mathbb{Z}^{\,}_{n}\times\mathbb{Z}^{\,}_{n}$ symmetry
with an LSM anomaly.
A more detailed investigation of dualities between models with 
multipolar symmetries and models with uniform symmetries and LSM anomalies 
is left for future works.


\section{Conclusions}
\label{sec:conclusions}

In this paper, we studied the dualization
induced by the gauging of global internal sub-symmetries
of one-dimensional quantum spin chains with LSM anomalies.
We found that when the pre-gauged
theory had a non-trivial LSM anomaly,
the dual theory was free from an LSM anomaly but
had a symmetry structure wherein the crystalline and
internal symmetries combined together through non-trivial group
extensions.  Therefore, the symmetry structure of the gauged theory
was shown to serve as a diagnostic for
LSM anomalies.
Similar phenomena (restricted to only internal
symmetries) have been studied extensively in the context of quantum
field theory (see for example \cite{Tachikawa2017}), where gauging a
non-anomalous symmetry participating in a mixed anomaly delivers a
dual theory with a symmetry structure involving a group extension
controlled by the anomaly of the pre-gauged theory.
We exemplified our procedure for a
$\mathbb{Z}^{\,}_{2}\times\mathbb{Z}^{\,}_{2}$-symmetric
quantum spin-1/2 $XYZ$ chain with
LSM anomalies due to translation and reflection.
We established a triality of models by gauging a
$\mathbb{Z}^{\,}_{2}\subset\mathbb{Z}^{\,}_{2}\times\mathbb{Z}^{\,}_{2}$
symmetry in two ways,
which amount to performing Kramers-Wannier or Jordan-Wigner duality,
respectively.  We detailed the mapping of the phase diagram of the
quantum spin-1/2 $XYZ$ chain under the triality and showed that the
deconfined quantum critical transitions between Neel and
valence-bond-solid orders of the chain map to either topological
transitions or conventional Landau-Ginzburg-type transitions.

There are several future directions that could be pursued.
One avenue is the generalization of the approach developed
in this work to quantum lattice Hamiltonians with LSM anomalies
for higher-dimensional lattices.
We expect that in higher dimensions, gauging non-anomalous subgroups of
internal symmetries participating in LSM anomalies delivers
dual theories with novel symmetry structures
that may involve higher groups or even non-invertible symmetries~\cite{Seiberg2023}
mixing the dual crystalline and dual internal symmetries.
Furthermore, higher-dimensional space accommodates dualities
between phase diagrams that support phases
that are not allowed when space is one dimensional,
namely phases supporting symmetry-enriched (anomalous) topological order
or ordered phases with local order parameters that break
spontaneously a continuous symmetry group.
\cite{Moradi23}.
Another avenue is to construct fermionic models
that support novel deconfined phase transitions
by gauging sub-symmetries.


\section*{Acknowledgements}

We thank Lakshya Bhardwaj, Lea Bottini, Heidar Moradi and Sakura
Schafer Nameki for several useful discussions. \"OMA is supported by
the Swiss National Science Foundation (SNSF) under Grant No.\ 200021
184637. AT is supported by the Swedish Research Council (VR) through
grants number 2019-04736 and 2020-00214.
AF is supported by JSPS KAKENHI (Grant No.\ JP19K03680),
JST CREST (Grant No.\ JPMJCR19T2), 
and the National Science Foundation (Grant No. NSF PHY-1748958).

\begin{appendix}

\section{Triality of the Majumdar-Ghosh line}
\label{appsec:Triality of the Majumdar-Ghosh line}

\subsection{Definition and properties of the Majumdar-Ghosh line}

To study the Majumdar-Ghosh (MG) line
(\ref{eq:exactly soluble points AF J1 J2 spin1over2 c}),
we start from the fully $\mathrm{SU}(2)$-symmetric Hamiltonian
\begin{equation}
\begin{split}
\widehat{H}^{\,}_{\mathrm{SU}(2)}:=&\,
\sum_{j=1}^{2N}
\left(
\hat{\bm{\sigma}}^{\,}_{j}
\cdot
\hat{\bm{\sigma}}^{\,}_{j+1}
+
\frac{3}{2}\,
\widehat{\mbb{1}}^{\,}_{\mathcal{H}^{\,}_{b=0}}
\right)
+
\frac{1}{2}
\sum_{j=1}^{2N}
\hat{\bm{\sigma}}^{\,}_{j}
\cdot
\hat{\bm{\sigma}}^{\,}_{j+2}
\\
=&\,
\sum_{j=1}^{2N}
\left[
\frac{1}{4}
\left(
\hat{\bm{\sigma}}^{\,}_{j}
+
\hat{\bm{\sigma}}^{\,}_{j+1}
+
\hat{\bm{\sigma}}^{\,}_{j+2}
\right)^{2}
-
\frac{3}{4}\,
\widehat{\mbb{1}}^{\,}_{\mathcal{H}^{\,}_{b=0}}
\right],
\end{split}
\label{eq:def Majumdar-Ghosh Hamiltonian}
\end{equation}
where periodic boundary conditions ($b=0$) have been imposed.

We shall denote with
\begin{equation}
|[j,j+1]\rangle:=
\frac{1}{\sqrt{2}}
\left(
|\rightarrow\rangle^{\,}_{j}
\otimes
|\leftarrow\rangle^{\,}_{j+1}
-
|\leftarrow\rangle^{\,}_{j}
\otimes
|\rightarrow\rangle^{\,}_{j+1}
\right)
\label{eq:singlet}
\end{equation}
the singlet state for two spin-1/2 localized on two consecutive
sites $j$ and $j+1$ of $\Lambda$ in the basis for which
$|\rightarrow\rangle^{\,}_{j}$
($|\uparrow\rangle^{\,}_{j}$)
is the eigenstate with eigenvalue $+1$
of $\hat{\sigma}^{x}_{j}$ ($\hat{\sigma}^{z}_{j}$).
By inspection, the states
\begin{subequations}
\label{eq:def dimer two states}
\begin{equation}
|\mathrm{Dimer}^{\,}_{\mathrm{o}}\rangle:=
\bigotimes_{j=1}^{N}
|[2j-1,2j]\rangle
\label{eq:def dimer two states a}
\end{equation}
and
\begin{equation}
|\mathrm{Dimer}^{\,}_{\mathrm{e}}\rangle:=
\bigotimes_{j=1}^{N}
|[2j,2j+1]\rangle
\label{eq:def dimer two states b}
\end{equation}
are orthonormal eigenstates of $\widehat{H}^{\,}_{\mathrm{SU}(2)}$
with the degenerate eigenvalue
\begin{equation}
E^{\,}_{\mathrm{Dimer}}=0.
\label{eq:def dimer two states c}
\end{equation}
Their bonding and anti-bonding linear combinations are defined by
\begin{equation}
|\mathrm{Dimer}\rangle^{\pm}:=
\frac{1}{\sqrt{2}}
\left(
|\mathrm{Dimer}^{\,}_{\mathrm{o}}\rangle
\pm
|\mathrm{Dimer}^{\,}_{\mathrm{e}}\rangle
\right).
\label{eq:def dimer two states d}
\end{equation}
\end{subequations}
Since the square bracket on the right-hand side
of the second equality in Eq.\ (\ref{eq:def Majumdar-Ghosh Hamiltonian})
is positive definite, this energy is that of the ground state.
Shastry and Sutherland
have shown that these are the gapped ground states of Hamiltonian
(\ref{eq:def Hamiltonian b=0})
along the MG line
(\ref{eq:exactly soluble points AF J1 J2 spin1over2 c}).

The projectors onto the two dimer states are
\begin{subequations}
\begin{equation}
\widehat{P}^{\,\mathrm{o}}_{\mathrm{Dimer}}:=
\prod_{j=1}^{N}
\widehat{P}^{\,}_{[2j-1,2j]}
\end{equation}
and
\begin{equation}
\widehat{P}^{\,\mathrm{e}}_{\mathrm{Dimer}}:=
\prod_{j=1}^{N}
\widehat{P}^{\,}_{[2j,2j+1]},
\end{equation}
where
\begin{equation}
\begin{split}
\widehat{P}^{\,}_{[j,j+1]}:=&\,
\frac{1}{4}
\left(
\widehat{\mbb{1}}^{\,}_{\mathcal{H}^{\,}_{b=0}}
-
\hat{\bm{\sigma}}^{\,}_{j}
\cdot
\hat{\bm{\sigma}}^{\,}_{j+1}
\right)
\\
=&\,
\frac{1}{4}
\left\{
\widehat{\mbb{1}}^{\,}_{\mathcal{H}^{\,}_{b=0}}
-
\frac{1}{2}
\left[
\left(
\hat{\bm{\sigma}}^{\,}_{j}
+
\hat{\bm{\sigma}}^{\,}_{j+1}
\right)^{2}
-
6\,
\widehat{\mbb{1}}^{\,}_{\mathcal{H}^{\,}_{b=0}}
\right]
\right\}.
\end{split}
\end{equation}
\end{subequations}

We have the transformation laws
\begin{subequations}
\begin{align}
& 
\widehat{U}^{\,}_{t}\,
\widehat{P}^{\,\mathrm{o}}_{\mathrm{Dimer}}\,
\left(\widehat{U}^{\,}_{t}\right)^{\dag}=
\prod_{j=1}^{N}
\left[
\widehat{U}^{\,}_{t}\,
\widehat{P}^{\,}_{[2j-1,2j]}\,
\left(\widehat{U}^{\,}_{t}\right)^{\dag}
\right]=
\widehat{P}^{\,\mathrm{e}}_{\mathrm{Dimer}}, 
\\
& 
\widehat{U}^{\,}_{r}\,
\widehat{P}^{\,\mathrm{o}}_{\mathrm{Dimer}}\,
\widehat{U}^{\,}_{r}=
\prod_{j=1}^{N}
\left(
\widehat{U}^{\,}_{r}\,
\widehat{P}^{\,}_{[2j-1,2j]}\,
\widehat{U}^{\,}_{r}
\right)=
\widehat{P}^{\,\mathrm{e}}_{\mathrm{Dimer}}, 
\\
& 
\widehat{U}^{\,}_{r^{x}_{\pi}}\,
\widehat{P}^{\,\mathrm{o}}_{\mathrm{Dimer}}\,
\widehat{U}^{\,}_{r^{x}_{\pi}}=
\prod_{j=1}^{N}
\left(
\widehat{U}^{\,}_{r^{x}_{\pi}}\,
\widehat{P}^{\,}_{[2j-1,2j]}\,
\widehat{U}^{\,}_{r^{x}_{\pi}}
\right)=
\widehat{P}^{\,\mathrm{o}}_{\mathrm{Dimer}}, 
\\
&
\widehat{U}^{\,}_{r^{z}_{\pi}}\,
\widehat{P}^{\,\mathrm{o}}_{\mathrm{Dimer}}\,
\widehat{U}^{\,}_{r^{z}_{\pi}}=
\prod_{j=1}^{N}
\left(
\widehat{U}^{\,}_{r^{z}_{\pi}}\,
\widehat{P}^{\,}_{[2j-1,2j]}\,
\widehat{U}^{\,}_{r^{z}_{\pi}}
\right)=
\widehat{P}^{\,\mathrm{o}}_{\mathrm{Dimer}},
\end{align}
and
\begin{align}
& 
\widehat{U}^{\,}_{t}\,
\widehat{P}^{\,\mathrm{e}}_{\mathrm{Dimer}}\,
\left(\widehat{U}^{\,}_{t}\right)^{\dag}=
\prod_{j=1}^{N}
\left[
\widehat{U}^{\,}_{t}\,
\widehat{P}^{\,}_{[2j,2j+1]}\,
\left(\widehat{U}^{\,}_{t}\right)^{\dag}
\right]=
\widehat{P}^{\,\mathrm{o}}_{\mathrm{Dimer}}, 
\\
& 
\widehat{U}^{\,}_{r}\,
\widehat{P}^{\,\mathrm{e}}_{\mathrm{Dimer}}\,
\widehat{U}^{\,}_{r}=
\prod_{j=1}^{N}
\left(
\widehat{U}^{\,}_{r}\,
\widehat{P}^{\,}_{[2j,2j+1]}\,
\widehat{U}^{\,}_{r}
\right)=
\widehat{P}^{\,\mathrm{o}}_{\mathrm{Dimer}}, 
\\
& 
\widehat{U}^{\,}_{r^{x}_{\pi}}\,
\widehat{P}^{\,\mathrm{e}}_{\mathrm{Dimer}}\,
\widehat{U}^{\,}_{r^{x}_{\pi}}=
\prod_{j=1}^{2N}
\left(
\widehat{U}^{\,}_{r^{x}_{\pi}}\,
\widehat{P}^{\,}_{[2j,2j+1]}\,
\widehat{U}^{\,}_{r^{x}_{\pi}}
\right)=
\widehat{P}^{\,\mathrm{e}}_{\mathrm{Dimer}}, 
\\
&
\widehat{U}^{\,}_{r^{z}_{\pi}}\,
\widehat{P}^{\,\mathrm{e}}_{\mathrm{Dimer}}\,
\widehat{U}^{\,}_{r^{z}_{\pi}}=
\prod_{j=1}^{2N}
\left(
\widehat{U}^{\,}_{r^{z}_{\pi}}\,
\widehat{P}^{\,}_{[2j,2j+1]}\,
\widehat{U}^{\,}_{r^{z}_{\pi}}
\right)=
\widehat{P}^{\,\mathrm{e}}_{\mathrm{Dimer}}.
\end{align}
\end{subequations}
In words, both translation $t$ and
reflection $r$ with the fixed points $N$ and $2N$ interchange
$|\mathrm{Dimer}^{\,}_{\mathrm{o}}\rangle$
and
$|\mathrm{Dimer}^{\,}_{\mathrm{e}}\rangle$.
Observe that both rotations $r^{x}_{\pi}$ and $r^{z}_{\pi}$
map
$|\mathrm{Dimer}^{\,}_{\mathrm{o}}\rangle$ to
$(-1)^{N}\,|\mathrm{Dimer}^{\,}_{\mathrm{o}}\rangle$
(they do the same for
$|\mathrm{Dimer}^{\,}_{\mathrm{e}}\rangle$),
\begin{subequations}
\begin{align}
&
\widehat{U}^{\,}_{r^{x}_{\pi}}\,
|\mathrm{Dimer}^{\,}_{\mathrm{o}}\rangle=
(-1)^{N}\,
|\mathrm{Dimer}^{\,}_{\mathrm{o}}\rangle,
\qquad
\widehat{U}^{\,}_{r^{z}_{\pi}}\,
|\mathrm{Dimer}^{\,}_{\mathrm{o}}\rangle=
(-1)^{N}\,
|\mathrm{Dimer}^{\,}_{\mathrm{o}}\rangle,
\\
&
\widehat{U}^{\,}_{r^{x}_{\pi}}\,
|\mathrm{Dimer}^{\,}_{\mathrm{e}}\rangle=
(-1)^{N}\,
|\mathrm{Dimer}^{\,}_{\mathrm{e}}\rangle,
\qquad
\widehat{U}^{\,}_{r^{z}_{\pi}}\,
|\mathrm{Dimer}^{\,}_{\mathrm{e}}\rangle=
(-1)^{N}\,
|\mathrm{Dimer}^{\,}_{\mathrm{e}}\rangle.
\end{align}
\end{subequations}
The multiplicative phase factor $(-1)^{N}$
cancels in either one of the projectors
$\widehat{P}^{\,\mathrm{o}}_{\mathrm{Dimer}}$
and
$\widehat{P}^{\,\mathrm{e}}_{\mathrm{Dimer}}$.

It is instructive to compare the
dimer states (\ref{eq:def dimer two states})
with the Neel states
\begin{subequations}
\label{eq:def Neel e/o states in preliminaries MG line}
\begin{equation}
\begin{split}
|\mathrm{Neel}^{x}_{\mathrm{o}}\rangle:=&\,
|\rightarrow\rangle^{\,}_{1}
\otimes
|\leftarrow\rangle^{\,}_{2}
\otimes\cdots\otimes
|\rightarrow\rangle^{\,}_{2N-1}
\otimes
|\leftarrow\rangle^{\,}_{2N}
\\
\equiv&\,
|\rightarrow,\leftarrow,\cdots,\rightarrow,\leftarrow\rangle
\end{split}
\label{eq:def Neel e/o states in preliminaries MG line a}
\end{equation}
and
\begin{equation}
\begin{split}
|\mathrm{Neel}^{x}_{\mathrm{e}}\rangle:=&\,
\widehat{U}^{\,}_{r^{z}_{\pi}}\,  
|\mathrm{Neel}^{x}_{\mathrm{o}}\rangle
\\
\equiv&\,
|\rightarrow,\leftarrow,\cdots,\rightarrow,\leftarrow\rangle.
\end{split}
\label{eq:def Neel e/o states in preliminaries MG line b}
\end{equation}
\end{subequations}
The pair of Neel states
(\ref{eq:def Neel e/o states in preliminaries MG line})
are the two-fold degenerate gapped ground states of
Hamiltonian (\ref{eq:def Hamiltonian b=0})
at the lower left corner in the reduced coupling space
(\ref{eq:reduced coupling space}).

The projectors onto the Neel states are
\begin{subequations}
\label{eq:Projectors Neel e/o states in preliminaries MG line}
\begin{equation}
\begin{split}    
\widehat{P}^{\,\mathrm{o}}_{\mathrm{Neel}^{x}}=&\,
\prod_{j=1}^{2N}
\frac{1}{2}
\left[
\widehat{\mbb{1}}^{\,}_{\mathcal{H}^{\,}_{b=0}}
+
(-1)^{j+1}\,
\hat{\sigma}^{x}_{j}
\right]
\end{split}
\end{equation}
and
\begin{equation}
\begin{split}
\widehat{P}^{\,\mathrm{e}}_{\mathrm{Neel}^{x}}=&\,
\widehat{U}^{\,}_{r^{z}_{\pi}}\,
\widehat{P}^{\,\mathrm{o}}_{\mathrm{Neel}^{x}}\,
\widehat{U}^{\,}_{r^{z}_{\pi}}
\\
=&\,
\prod_{j=1}^{2N}
\frac{1}{2}
\left[
\widehat{\mbb{1}}^{\,}_{\mathcal{H}^{\,}_{b=0}}
-
(-1)^{j+1}\,
\hat{\sigma}^{x}_{j}
\right],
\end{split}
\end{equation}
\end{subequations}
respectively.
The Neel projectors corresponding to the
orthonormal pair of bonding and anti-bonding linear combinations
\begin{subequations}
\label{eq:def Neel pm states in preliminaries MG line}
\begin{equation}
|\mathrm{Neel}^{x}\rangle^{\pm}:=
\frac{1}{\sqrt{2}}
\left(
|\mathrm{Neel}^{x}_{\mathrm{o}}\rangle
\pm
|\mathrm{Neel}^{x}_{\mathrm{e}}\rangle
\right)
\end{equation}
are
\begin{equation}
\widehat{P}^{\pm}_{\mathrm{Neel}^{x}}:=
\frac{1\pm \widehat{U}^{\,}_{r^{z}_{\pi}}}{2}
\left(
\widehat{P}^{\,\mathrm{o}}_{\mathrm{Neel}^{x}}
+
\widehat{P}^{\,\mathrm{e}}_{\mathrm{Neel}^{x}}
\right).
\end{equation}
\end{subequations}
The projector
$\widehat{P}^{+}_{\mathrm{Neel}^{x}}$
is a linear combination of string of $\hat{\sigma}^{x}$'s of even length.
The projector
$\widehat{P}^{-}_{\mathrm{Neel}^{x}}$
is a linear combination of string of $\hat{\sigma}^{x}$'s of odd length.

We have the transformation laws
\begin{subequations}
\begin{align}
& 
\widehat{U}^{\,}_{t}\,
\widehat{P}^{\,\mathrm{o}}_{\mathrm{Neel}^{x}}\,
\left(\widehat{U}^{\,}_{t}\right)^{\dag}=
\widehat{P}^{\,\mathrm{e}}_{\mathrm{Neel}^{x}}, 
\\
& 
\widehat{U}^{\,}_{r}\,
\widehat{P}^{\,\mathrm{o}}_{\mathrm{Neel}^{x}}\,
\widehat{U}^{\,}_{r}=
\widehat{P}^{\,\mathrm{o}}_{\mathrm{Neel}^{x}}, 
\\
& 
\widehat{U}^{\,}_{r^{x}_{\pi}}\,
\widehat{P}^{\,\mathrm{o}}_{\mathrm{Neel}^{x}}\,
\widehat{U}^{\,}_{r^{x}_{\pi}}=
\widehat{P}^{\,\mathrm{o}}_{\mathrm{Neel}^{x}}, 
\\
&
\widehat{U}^{\,}_{r^{z}_{\pi}}\,
\widehat{P}^{\,\mathrm{o}}_{\mathrm{Neel}^{x}}\,
\widehat{U}^{\,}_{r^{z}_{\pi}}=
\widehat{P}^{\,\mathrm{o}}_{\mathrm{Neel}^{x}},
\end{align}
and
\begin{align}
& 
\widehat{U}^{\,}_{t}\,
\widehat{P}^{\,\mathrm{e}}_{\mathrm{Neel}^{x}}\,
\left(\widehat{U}^{\,}_{t}\right)^{\dag}=
\widehat{P}^{\,\mathrm{o}}_{\mathrm{Neel}^{x}}, 
\\
& 
\widehat{U}^{\,}_{r}\,
\widehat{P}^{\,\mathrm{e}}_{\mathrm{Neel}^{x}}\,
\widehat{U}^{\,}_{r}=
\widehat{P}^{\,\mathrm{e}}_{\mathrm{Neel}^{x}}, 
\\
& 
\widehat{U}^{\,}_{r^{x}_{\pi}}\,
\widehat{P}^{\,\mathrm{e}}_{\mathrm{Neel}^{x}}\,
\widehat{U}^{\,}_{r^{x}_{\pi}}=
\widehat{P}^{\,\mathrm{e}}_{\mathrm{Neel}^{x}}, 
\\
&
\widehat{U}^{\,}_{r^{z}_{\pi}}\,
\widehat{P}^{\,\mathrm{e}}_{\mathrm{Neel}^{x}}\,
\widehat{U}^{\,}_{r^{z}_{\pi}}=
\widehat{P}^{\,\mathrm{e}}_{\mathrm{Neel}^{x}}.
\end{align}
\end{subequations}
In words, translation $t$ interchanges
$|\mathrm{Neel}^{x}_{\mathrm{o}}\rangle$
and
$|\mathrm{Neel}^{x}_{\mathrm{e}}\rangle$.
Reflection $r$ with the fixed points $N$ and $2N$ in $\Lambda$
leaves each Neel state unchanged. The same is true of rotation
$r^{x}_{\pi}$.
Observe that rotation $r^{z}_{\pi}$
maps
$|\mathrm{Neel}^{x}_{\mathrm{o}}\rangle$ to
$|\mathrm{Neel}^{x}_{\mathrm{o}}\rangle$
(it does the same for
$|\mathrm{Neel}^{x}_{\mathrm{e}}\rangle$),
\begin{subequations}
\begin{align}
&
\widehat{U}^{\,}_{r^{x}_{\pi}}\,
|\mathrm{Neel}^{x}_{\mathrm{o}}\rangle=
(-1)^{N}\,
|\mathrm{Neel}^{x}_{\mathrm{o}}\rangle,
\qquad
\widehat{U}^{\,}_{r^{z}_{\pi}}\,
|\mathrm{Neel}^{x}_{\mathrm{o}}\rangle=
|\mathrm{Neel}^{x}_{\mathrm{o}}\rangle,
\\
&
\widehat{U}^{\,}_{r^{x}_{\pi}}\,
|\mathrm{Neel}^{x}_{\mathrm{e}}\rangle=
(-1)^{N}\,
|\mathrm{Neel}^{x}_{\mathrm{e}}\rangle,
\qquad
\widehat{U}^{\,}_{r^{z}_{\pi}}\,
|\mathrm{Neel}^{x}_{\mathrm{e}}\rangle=
|\mathrm{Neel}^{x}_{\mathrm{e}}\rangle.
\end{align}
\end{subequations}
The multiplicative phase factor $(-1)^{N}$
cancels in either one of the projectors
$\widehat{P}^{\,\mathrm{o}}_{\mathrm{Neel}^{x}}$
and
$\widehat{P}^{\,\mathrm{e}}_{\mathrm{Neel}^{x}}$.

Define the local order parameters
\begin{subequations}
\label{eq:def local order parameters anisotropic XY chain}
\begin{align}
&
\widehat{O}^{\,\mathrm{o}}_{\mathrm{Neel}^{\alpha}}:=
\frac{1}{2N}
\sum_{j=1}^{2N}
(-1)^{j+1}\,
\hat{\sigma}^{\alpha}_{j},
\qquad
\alpha=x,y,z,
\label{eq:def local order parameters anisotropic XY chain a}
\\
&
\widehat{O}^{\,\mathrm{o}}_{\mathrm{dimer}}:=
\frac{1}{N}
\sum_{j=1}^{2N}
(-1)^{j}\,
\frac{1}{3}\,
\hat{\bm{\sigma}}^{\,}_{j}
\cdot
\hat{\bm{\sigma}}^{\,}_{j+1}.
\label{eq:def local order parameters anisotropic XY chain b}
\end{align}
\end{subequations}
Define the two-point operator
\begin{equation}
\widehat{C}^{\alpha}_{j,j+n}:=
\hat{\sigma}^{\alpha}_{j}\,
\hat{\sigma}^{\alpha}_{j+n},
\qquad
\alpha=x,y,z.
\label{eq:def two-point product Neel XY chain}
\end{equation}
We will replace the staggered magnetization
(\ref{eq:def local order parameters anisotropic XY chain a})
with the two-point operator
(\ref{eq:def two-point product Neel XY chain})
to detect Neel order as the former cannot be dualized
when $\alpha=x,y$.
We recall the definition of the unitary operator
\begin{subequations}
\label{eq:def string operator XY chain}
\begin{align}
&
\widehat{U}^{\,}_{r}:=
\prod_{j=1}^{N-1}
\frac{1}{2}
\left(
\widehat{\mbb{1}}^{\,}_{\mathcal{H}^{\,}_{b=0}}
+
\hat{\bm{\sigma}}^{\,}_{j}
\cdot
\hat{\bm{\sigma}}^{\,}_{r(j)}
\right)
\end{align}
that implements reflection with the fixed points $N$ and $2N$
(the upper bound is $N-1$ in the product because
the two fixed points $N$ and $2N$ must be removed from the product)
and the unitary operators
\begin{align}
&
\widehat{U}^{\,}_{r^{\alpha}_{\pi}}:=
\prod_{j=1}^{2N}\hat{\sigma}^{\alpha}_{j},
\qquad
\alpha=x,y,z,
\\
&
\widehat{U}^{\,\mathrm{o}\,(2n)}_{r^{\alpha}_{\pi}}:=
\prod_{k=2j-1}^{2j-1+2n-1}\hat{\sigma}^{\alpha}_{k},
\qquad
\alpha=x,y,z,
\qquad
j=1,\cdots,2N,
\qquad
n=1,\cdots,N,
\\
&
\widehat{U}^{\,\mathrm{e}\,(2n)}_{r^{\alpha}_{\pi}}:=
\prod_{k=2j}^{2j+2n-1}\hat{\sigma}^{\alpha}_{k},
\qquad
\alpha=x,y,z,
\qquad
j=1,\cdots,2N,
\qquad
n=1,\cdots,N,
\end{align}
\end{subequations}
that implement rotations by $\pi$ around the $\alpha$ axis in
the Bloch spheres labeled by the lattice sites
$j=1,\cdots,2N$
on strings of consecutive $2n$ lattice sites.
Their expectation values in the four Neel and four Dimer states are
given in Table \ref{Table:Table Majumdar-Ghosh line}.
Observe that of the eight states
in Table \ref{Table:Table Majumdar-Ghosh line},
only six are eigenstates
of $\widehat{U}^{\,}_{r^{z}_{\pi}}$,
namely
\begin{equation}
|\mathrm{Neel}^{x}\rangle^{+},
\qquad
|\mathrm{Neel}^{x}\rangle^{-},
\qquad
|\mathrm{Dimer}^{\,}_{\mathrm{o}}\rangle,
\qquad
|\mathrm{Dimer}^{\,}_{\mathrm{e}}\rangle,
\qquad
|\mathrm{Dimer}\rangle^{+},
\qquad
|\mathrm{Dimer}\rangle^{-}.
\end{equation}
Moreover, we can distinguish
$
|\mathrm{Dimer}^{\,}_{\mathrm{o}}\rangle
$
and
$
|\mathrm{Dimer}^{\,}_{\mathrm{e}}\rangle
$
from
$
|\mathrm{Dimer}\rangle^{+}
$
and
$
|\mathrm{Dimer}\rangle^{-}
$
by using the fact that the two elements of the
first pair are interchanged by reflection
about the lattice site $N$,
while the two elements of the second pair transform
like the eigenstates of reflection
about the lattice site $N$. 

\def\arraystretch{1.25}
\begin{table}
\caption{
\label{Table:Table Majumdar-Ghosh line}
The expectation values of nine operators in eight states.
The domain of definition of all nine operators is  
$\mathcal{H}^{\,}_{b=0}$.
The first four Neel states are defined in
Eqs.\
(\ref{eq:def Neel e/o states in preliminaries MG line})
and
(\ref{eq:def Neel pm states in preliminaries MG line}).
The next four dimer states are defined in
Eqs.\
(\ref{eq:def dimer two states})
All nine operators are defined in Eqs.\
(\ref{eq:def local order parameters anisotropic XY chain}),
(\ref{eq:def two-point product Neel XY chain}),
and
(\ref{eq:def string operator XY chain}).
}
\centering
\begin{scriptsize}
\begin{tabular}{l|ccccccccc}
\hline \hline
&
$\ \widehat{C}^{x}_{j,j+n}\ $
&
$\ \widehat{O}^{\,\mathrm{o}}_{\mathrm{dimer}}\ $
&
$\ \widehat{U}^{\,}_{r}\ $
&
$\ \widehat{U}^{\,}_{r^{x}_{\pi}}\ $
&
$\ \widehat{U}^{\,}_{r^{z}_{\pi}}\ $
&
$\ \widehat{U}^{\,\mathrm{o}\,(2n)}_{r^{x}_{\pi}}\ $
&
$\ \widehat{U}^{\,\mathrm{e}\,(2n)}_{r^{x}_{\pi}}\ $
&
$\ \widehat{U}^{\,\mathrm{o}\,(2n)}_{r^{z}_{\pi}}\ $
&
$\ \widehat{U}^{\,\mathrm{e}\,(2n)}_{r^{z}_{\pi}}\ $
\\
\hline
$|\mathrm{Neel}^{x}_{\mathrm{o}}\rangle$
&
$(-1)^{n}$
&
$0$
&
$+1$
&
$(-1)^{N}$
&
$0$
&
$(-1)^{n}$
&
$(-1)^{n}$
&
$0$
&
$0$
\\
$|\mathrm{Neel}^{x}_{\mathrm{e}}\rangle$
&
$(-1)^{n}$
&
$0$
&
$+1$
&
$(-1)^{N}$
&
$0$
&
$(-1)^{n}$
&
$(-1)^{n}$
&
$0$
&
$0$
\\
$|\mathrm{Neel}^{x}\rangle^{+}$
&
$(-1)^{n}$
&
$0$
&
$+1$
&
$(-1)^{N}$
&
$+1$
&
$(-1)^{n}$
&
$(-1)^{n}$
&
$0$
&
$0$
\\
$|\mathrm{Neel}^{x}\rangle^{-}$
&
$(-1)^{n}$
&
$0$
&
$+1$
&
$(-1)^{N}$
&
$-1$
&
$(-1)^{n}$
&
$(-1)^{n}$
&
$0$
&
$0$
\\
\hline
$|\mathrm{Dimer}^{\,}_{\mathrm{o}}\rangle$
&
$-\delta^{\,}_{(-1)^{j},-1}\,\delta^{\,}_{n,1}$
&
$+1$
&
0
&
$(-1)^{N}$
&
$(-1)^{N}$
&
$(-1)^{n}$
&
$0$
&
$(-1)^{n}$
&
$0$
\\
$|\mathrm{Dimer}^{\,}_{\mathrm{e}}\rangle$
&
$-\delta^{\,}_{(-1)^{j},+1}\,\delta^{\,}_{n,1}$
&
$-1$
&
0
&
$(-1)^{N}$
&
$(-1)^{N}$
&
$0$
&
$(-1)^{n}$
&
$0$
&
$(-1)^{n}$
\\
$|\mathrm{Dimer}\rangle^{+}$
&
$-\frac{\delta^{\,}_{n,1}}{2}$
&
$0$
&
$+(-1)^{N}$
&
$(-1)^{N}$
&
$(-1)^{N}$
&
$\frac{(-1)^{n}}{2}$
&
$\frac{(-1)^{n}}{2}$
&
$\frac{(-1)^{n}}{2}$
&
$\frac{(-1)^{n}}{2}$
\\
$|\mathrm{Dimer}\rangle^{-}$
&
$-\frac{\delta^{\,}_{n,1}}{2}$
&
$0$
&
$-(-1)^{N}$
&
$(-1)^{N}$
&
$(-1)^{N}$
&
$\frac{(-1)^{n}}{2}$
&
$\frac{(-1)^{n}}{2}$
&
$\frac{(-1)^{n}}{2}$
&
$\frac{(-1)^{n}}{2}$
\\
\hline \hline
\end{tabular}
\end{scriptsize}
\end{table}

\subsection{Kramers-Wannier dualization of the Majumdar-Ghosh line}

The projectors that are the Kramers-Wannier
dual to those for the pair of dimer states are built out of
\begin{subequations}
\label{eq:Kramers-Wannier projectors dimer states}
\begin{equation}
\widehat{P}^{\,\mathrm{o}\,\vee}_{\mathrm{Dimer}}:=
\prod_{j=1}^{2N}
\widehat{P}^{\,\vee}_{[(2j-1)^{\star},(2j)^{\star}]}
\label{eq:Kramers-Wannier projectors dimer states a}
\end{equation}
and
\begin{equation}
\widehat{P}^{\,\mathrm{e}\,\vee}_{\mathrm{Dimer}}:=
\prod_{j=1}^{2N}
\widehat{P}^{\,\vee}_{[(2j)^{\star},(2j+1)^{\star}]},
\label{eq:Kramers-Wannier projectors dimer states b}
\end{equation}
where
\begin{align}
\widehat{P}^{\,\vee}_{[j^{\star},j^{\star}+1]}=&\,
\frac{1}{4}
\left(
\widehat{\mbb{1}}^{\,}_{\mathcal{H}^{\,\vee}_{b^{\prime}=0}}
-
\hat{\tau}^{z\,\vee}_{j^{\star}}
+
\hat{\tau}^{x\,\vee}_{j^{\star}-1}\,
\hat{\tau}^{z\,\vee}_{j^{\star}}\,
\hat{\tau}^{x\,\vee}_{j^{\star}+1}
-
\hat{\tau}^{x\,\vee}_{j^{\star}-1}\,
\hat{\tau}^{x\,\vee}_{j^{\star}+1}
\right)
\nonumber\\
=&\,
\frac{1}{2}
\left(
\widehat{\mbb{1}}^{\,}_{\mathcal{H}^{\,\vee}_{b^{\prime}=0}}
-
\hat{\tau}^{z\,\vee}_{j^{\star}}
\right)\,
\frac{1}{2}
\left(
\widehat{\mbb{1}}^{\,}_{\mathcal{H}^{\,\vee}_{b^{\prime}=0}}
-
\hat{\tau}^{x\,\vee}_{j^{\star}-1}\,
\hat{\tau}^{x\,\vee}_{j^{\star}+1}
\right),
\label{eq:Kramers-Wannier projectors dimer states c}
\end{align}
\end{subequations}
by restriction to the subspace $\mathcal{H}^{\,\vee}_{b^{\prime}=0}$
from Table \ref{Table:Kramers-Wannier dualization b to b'}.
As a consequence of the fact that each of
$\widehat{P}^{\,\vee}_{[2j-1,2j]}$
and
$\widehat{P}^{\,\vee}_{[2j,2j+1]}$
acts non-trivially on three consecutive sites,
we are going to show that each of projectors
$\widehat{P}^{\,\mathrm{o}\,\vee}_{\mathrm{Dimer}}$
and
$\widehat{P}^{\,\mathrm{e}\,\vee}_{\mathrm{Dimer}}$
has two degenerate orthonormal eigenstates with eigenvalue one.

The projector
\begin{subequations}
\begin{equation}
\widehat{P}^{\,\mathrm{o}\,\vee}_{\mathrm{Dimer}}:=
\prod_{j=1}^{2N}
\widehat{P}^{\,\vee}_{[(2j-1)^{\star},(2j)^{\star}]}
\end{equation}
has the degenerate pair 
\begin{equation}
|1\rangle=
|
\downarrow,\rightarrow,\downarrow,\leftarrow;
\cdots;  
\downarrow,\rightarrow,\downarrow,\leftarrow;
\downarrow,\rightarrow,\downarrow,\leftarrow;
\cdots   
\downarrow,\rightarrow,\downarrow,\leftarrow
\rangle
\end{equation}
and
\begin{equation}
\begin{split}
|2\rangle=&\,
|
\downarrow,\leftarrow,\downarrow,\rightarrow;
\cdots;   
\downarrow,\leftarrow,\downarrow,\rightarrow;
\downarrow,\leftarrow,\downarrow,\rightarrow;
\cdots; 
\downarrow,\leftarrow,\downarrow,\rightarrow
\rangle
\\
=&\,
\widehat{U}^{\,\vee}_{r^{z}_{\pi}}\,
|1\rangle
\end{split}
\end{equation}
\end{subequations}
of orthonormal eigenstates with eigenvalue one.
The projector
\begin{subequations}
\begin{equation}
\widehat{P}^{\,\mathrm{e}\,\vee}_{\mathrm{Dimer}}:=
\prod_{j=1}^{2N}
\widehat{P}^{\,\vee}_{[(2j)^{\star},(2j+1)^{\star}]}
\end{equation}
has the degenerate pair 
\begin{equation}
\begin{split}
|3\rangle=&\,
|
\rightarrow,\downarrow,\leftarrow,\downarrow;
\cdots;
\rightarrow,\downarrow,\leftarrow,\downarrow;
\rightarrow,\downarrow,\leftarrow,\downarrow;
\cdots;  
\rightarrow,\downarrow,\leftarrow,\downarrow
\rangle
\\
=&\,
\widehat{U}^{\,\vee}_{r}\,|2\rangle
\end{split}
\end{equation}
and
\begin{equation}
\begin{split}
|4\rangle=&\,
|
\leftarrow,\downarrow,\rightarrow,\downarrow;
\cdots;  
\leftarrow,\downarrow,\rightarrow,\downarrow;
\leftarrow,\downarrow,\rightarrow,\downarrow;
\cdots; 
\leftarrow,\downarrow,\rightarrow,\downarrow
\rangle
\\
=&\,
\widehat{U}^{\,\vee}_{r^{z}_{\pi}}\,
|3\rangle
\\
=&\,
\widehat{U}^{\,\vee}_{r}\,|1\rangle
\end{split}
\end{equation}
\end{subequations}
of orthonormal eigenstates with eigenvalue one.
Hence, the dual of the dimer phase when periodic boundary conditions
($b=0$)
apply has the two degenerate and orthonormal ground states
along the MG line
\begin{subequations}
\label{eq:Dimer vee o+ and Dimer vee e+ groundstates}
\begin{equation}
|\mathrm{Dimer}^{\vee}_{\mathrm{o}}\rangle^{+}=
\frac{1}{\sqrt{2}}\,
\left(
|1\rangle
+
|2\rangle
\right)
\end{equation}
and
\begin{equation}
|\mathrm{Dimer}^{\vee}_{\mathrm{e}}\rangle^{+}=
\frac{1}{\sqrt{2}}\,
\left(
|3\rangle
+
|4\rangle
\right).
\end{equation}
\end{subequations}
The two degenerate and orthonormal states
\begin{subequations}
\label{eq:Dimer vee o- and Dimer vee e- groundstates}
\begin{equation}
|\mathrm{Dimer}^{\vee}_{\mathrm{o}}\rangle^{-}=
\frac{1}{\sqrt{2}}\,
\left(
|1\rangle
-
|2\rangle
\right)
\end{equation}
and
\begin{equation}
|\mathrm{Dimer}^{\vee}_{\mathrm{e}}\rangle^{-}=
\frac{1}{\sqrt{2}}\,
\left(
|3\rangle
-
|4\rangle
\right)
\end{equation}
\end{subequations}
are the ground states along the MG line
when twisted boundary conditions ($b^{\prime}=1$) apply.
Observe that
\begin{subequations}
\begin{align}
&
\widehat{U}^{\,\vee}_{r}\,
|\mathrm{Dimer}^{\vee}_{\mathrm{o}}\rangle^{+}=
+
|\mathrm{Dimer}^{\vee}_{\mathrm{e}}\rangle^{+},
\\
&
\widehat{U}^{\,\vee}_{r}\,
|\mathrm{Dimer}^{\vee}_{\mathrm{o}}\rangle^{-}=
-
|\mathrm{Dimer}^{\vee}_{\mathrm{e}}\rangle^{-}.
\end{align}
\end{subequations}

We can dualize all operators entering
Eqs.\
(\ref{eq:def local order parameters anisotropic XY chain}),
(\ref{eq:def two-point product Neel XY chain}),
and
(\ref{eq:def string operator XY chain})
except for
$\hat{\sigma}^{x}_{j}$
and
$\hat{\sigma}^{y}_{j}$.
The dimer order parameter
(\ref{eq:def local order parameters anisotropic XY chain b})
dualizes to
\begin{equation}
\widehat{O}^{\,\mathrm{o}\,\vee}_{\mathrm{dimer}}=
\frac{1}{N}
\sum_{j^{\star}\in \Lambda^{\star}}
(-1)^{j^{\star}-1/2}\,
\frac{1}{3}
\left(
\hat{\tau}^{z\,\vee}_{j^{\star}}
-
\hat{\tau}^{x\,\vee}_{j^{\star}-1}\,
\hat{\tau}^{z\,\vee}_{j^{\star}}\,
\hat{\tau}^{x\,\vee}_{j^{\star}+1}
+
\hat{\tau}^{x\,\vee}_{j^{\star}-1}\,
\hat{\tau}^{x\,\vee}_{j^{\star}+1}
\right).
\label{eq:dual dimer order parameter}
\end{equation}
The $xx$ two-point operator
(\ref{eq:def two-point product Neel XY chain})
dualizes to the string operator made of $n$ consecutive sites from
the dual lattice given by
\begin{equation}
\widehat{C}^{x\,\vee}_{j^{\star},j^{\star}+n}=
\prod_{k=1}^{n}
\hat{\tau}^{z\,\vee}_{j^{\star}+(k-1)}.
\label{eq:dual two point xx fct}
\end{equation}
The reflection with no fixed point on the dual lattice dualizes to
\begin{align}
\widehat{U}^{\,\vee}_{r}=&\,
\prod\limits_{j=1}^{N}
\frac{1}{2}
\left(
\widehat{\mbb{1}}^{\,}_{\mathcal{H}^{\,\vee}_{b^{\prime}=0}}
+
\hat{\tau}^{x\,\vee}_{j^{\star}}\,
\hat{\tau}^{x\,\vee}_{r(j^{\star})}
+
\hat{\tau}^{y\,\vee}_{j^{\star}}\,
\hat{\tau}^{y\,\vee}_{r(j^{\star})}
+
\hat{\tau}^{z\,\vee}_{j^{\star}}\,
\hat{\tau}^{z\,\vee}_{r(j^{\star})}
\right)
\label{eq:dual U r}
\end{align}
(the upper bound is now $N$ in the product instead of $N-1$ because
there are no invariant dual lattice points under reflection, i.e.,
we need not remove the invariant fixed points).
We choose to dualize the
global rotation by $\pi$ around the $x$ and $y$ axis of the Bloch spheres
labeled by $j=1,\cdots,2N$ to the rotation by $\pi$ around
the $z$ axis of the Bloch spheres labeled by
$j^{\star}=1+\frac{1}{2},3+\frac{1}{2},\cdots,2N-3+\frac{1}{2},2N-1+\frac{1}{2}$
and
$j^{\star}=2+\frac{1}{2},4+\frac{1}{2},\cdots,2N-2+\frac{1}{2},2N+\frac{1}{2}$,
respectively,
i.e., by
\begin{equation}
\widehat{U}^{\,\vee}_{r^{x}_{\pi}}=
\prod_{j=1}^{N}
\hat{\tau}^{z\,\vee}_{2j-1+\frac{1}{2}}\equiv
\widehat{U}^{\,\vee}_{\mathrm{o}}
\label{eq:dual U rxpi global}
\end{equation}
and
\begin{equation}
\widehat{U}^{\,\vee}_{r^{y}_{\pi}}=
\prod_{j=1}^{N}
\hat{\tau}^{z\,\vee}_{2j+\frac{1}{2}}\equiv
\widehat{U}^{\,\vee}_{\mathrm{e}},
\label{eq:dual U rypi global}
\end{equation}
respectively.

The global rotation by $\pi$ around the $z$ axis of the Bloch spheres
labeled by $j=1,\cdots,2N$ dualizes to the identity
\begin{equation}
\widehat{U}^{\,\vee}_{r^{z}_{\pi}}=
\widehat{\mbb{1}}^{\,}_{\mathcal{H}^{\,\vee}_{b^{\prime}=0}}.
\label{eq:dual U rzpi global}
\end{equation}
The rotation by $\pi$ around the $x$ axis of the Bloch spheres
labeled by $j=1,\cdots,2N$ on a string of $2n$ consecutive sites
from the lattice $\Lambda$ starting from an odd site
dualizes to the rotation by $\pi$ around
the $z$ axis of the Bloch spheres labeled by
$j^{\star}=1+\frac{1}{2},\cdots,2N+\frac{1}{2}$
on a string of $n$ consecutive odd sites from the dual lattice
starting from an odd dual site $\Lambda^{\star}$, i.e., by
\begin{equation}
\widehat{U}^{\,\mathrm{o}\,(2n)\,\vee}_{r^{x}_{\pi}}=
\prod_{k=1}^{n}
\hat{\tau}^{z\,\vee}_{2j-1+2(k-1)+\frac{1}{2}}\equiv
\widehat{U}^{\,\mathrm{o}\,\vee}_{n}, 
\qquad
j=1,\cdots,2N,
\qquad
n=1,\cdots,N.
\label{eq:dual U 2n rxpi o}
\end{equation}
The rotation by $\pi$ around the $x$ axis of the Bloch spheres
labeled by $j=1,\cdots,2N$ on a string of $2n$ consecutive sites
from the lattice $\Lambda$ starting from an even site
dualizes to the rotation by $\pi$ around
the $z$ axis of the Bloch spheres labeled by
$j^{\star}=1+\frac{1}{2},\cdots,2N+\frac{1}{2}$
on a string of $n$ consecutive even sites from the dual lattice
$\Lambda^{\star}$ starting from an even dual site, i.e., by
\begin{equation}
\widehat{U}^{\,\mathrm{e}\,(2n)\,\vee}_{r^{x}_{\pi}}=
\prod_{k=1}^{n}
\hat{\tau}^{z\,\vee}_{2j+2(k-1)+\frac{1}{2}}\equiv
\widehat{U}^{\,\mathrm{e}\,\vee}_{n}, 
\qquad
j=1,\cdots,2N,
\qquad
n=1,\cdots,N.
\label{eq:dual U 2n rxpi e}
\end{equation}
The rotation by $\pi$ around the $z$ axis of the Bloch spheres
labeled by $j=1,\cdots,2N$ on a string of $2n$ consecutive sites
from the lattice $\Lambda$ starting from an odd site
dualizes to the rotation by $\pi$ around
the $x$ axis of the Bloch spheres labeled by
$j^{\star}=1+\frac{1}{2},\cdots,2N+\frac{1}{2}$
on the two end points of a string of $2n+1$ consecutive sites
from the dual lattice $\Lambda^{\star}$
starting from an even dual site, i.e., by
\begin{equation}
\widehat{U}^{\,\mathrm{o}\,(2n)\,\vee}_{r^{z}_{\pi}}=
\hat{\tau}^{x\,\vee}_{2j-1-1+\frac{1}{2}}\,
\hat{\tau}^{x\,\vee}_{2j-1+2n-1+\frac{1}{2}}\,
\qquad 
j=1,\cdots,2N,
\qquad
n=1,\cdots,N.
\label{eq:dual U 2n rzpi o}
\end{equation}
The rotation by $\pi$ around the $x$ axis of the Bloch spheres
labeled by $j=1,\cdots,2N$ on a string of $2n$ consecutive sites
from the lattice $\Lambda$ starting from an even site
dualize to the rotation by $\pi$ around
the $z$ axis of the Bloch spheres labeled by
$j^{\star}=1+\frac{1}{2},\cdots,2N+\frac{1}{2}$
on the two end points of a string of $2n+1$ consecutive sites
from the dual lattice $\Lambda^{\star}$
starting from an odd dual site, i.e., by
\begin{equation}
\widehat{U}^{\,\mathrm{e}\,(2n)\,\vee}_{r^{z}_{\pi}}=
\hat{\tau}^{x\,\vee}_{2j-1+\frac{1}{2}}\,
\hat{\tau}^{x\,\vee}_{2j+2n-1+\frac{1}{2}},
\qquad
j=1,\cdots,2N,
\qquad
n=1,\cdots,N.
\label{eq:dual U 2n rzpi e}
\end{equation}

We seek the duals of the states
\begin{equation}
|\mathrm{Neel}^{x}\rangle^{+},
\qquad
|\mathrm{Dimer}^{\,}_{\mathrm{o}}\rangle,
\qquad
|\mathrm{Dimer}^{\,}_{\mathrm{e}}\rangle,
\end{equation}
that all have the eigenvalue $+1$ under the global $\pi$ rotation
about the $z$ axis of the Bloch spheres
labeled by $j=1,\cdots,2N$ and are annihilated by
either
$\widehat{U}^{\,\mathrm{o}\,(2n)}_{r^{x}_{\pi}}$
or
$\widehat{U}^{\,\mathrm{e}\,(2n)}_{r^{x}_{\pi}}$
for the dimer states.
These are the ground states of the dual Hamiltonian
$\widehat{H}^{\,\vee}_{b^{\prime}=0}$
defined in Eq.\
(\ref{eq:def Hamiltonian b'=0})
with the domain of definition
$\mathcal{H}^{\,\vee}_{b^{\prime}=0;+}$
with either $J=\Delta=0$ for the dual to $|\mathrm{Neel}^{x}\rangle^{+}$
or $J=1/2$ for the dual to the dimer states
$
|\mathrm{Dimer}^{\,}_{\mathrm{o}}\rangle
$
and
$
|\mathrm{Dimer}^{\,}_{\mathrm{e}}\rangle
$
in the reduced coupling space
(\ref{eq:reduced coupling space}).
The ground state of the dual Hamiltonian
$\widehat{H}^{\,\vee}_{b^{\prime}=0}$
with the domain of definition
$\mathcal{H}^{\,\vee}_{b^{\prime}=0;+}$
when $J=\Delta=0$
in the reduced coupling space
(\ref{eq:reduced coupling space})
is non-degenerate and given by
\begin{subequations}
\label{eq:dual of neel x+ state}
\begin{equation}
\begin{split}
|\mathrm{Neel}^{x\,\vee}\rangle^{+}=&\,
|\downarrow\rangle^{\,}_{1}\otimes\cdots\otimes|\downarrow\rangle^{\,}_{2N}  
\\
\equiv&\,
|\downarrow,\cdots,\downarrow\rangle,
\end{split}
\end{equation}
where
\begin{equation}
\hat{\tau}^{z\,\vee}_{j+\frac{1}{2}}\,|\downarrow\rangle^{\,}_{j+\frac{1}{2}}=
-|\downarrow\rangle^{\,}_{j+\frac{1}{2}},
\qquad
\hat{\tau}^{z\,\vee}_{j+\frac{1}{2}}\,|\uparrow\rangle^{\,}_{j+\frac{1}{2}}=
+|\uparrow\rangle^{\,}_{j+\frac{1}{2}},
\quad
j=1,\cdots,2N.
\end{equation}
\end{subequations}
The ground state of the dual Hamiltonian
$\widehat{H}^{\,\vee}_{b^{\prime}=0}$
with the domain of definition
$\mathcal{H}^{\,\vee}_{b^{\prime}=0;+}$
when $J=1/2$
in the reduced coupling space
(\ref{eq:reduced coupling space})
is two-fold degenerate with the eigenstates
$|\mathrm{Dimer}^{\,\vee}_{\mathrm{o}}\rangle^{+}$
and
$|\mathrm{Dimer}^{\,\vee}_{\mathrm{e}}\rangle^{+}$
defined in Eq.\ (\ref{eq:Dimer vee o+ and Dimer vee e+ groundstates}).
The expectation values in the Neel and two dimer states are
tabulated in Table \ref{Table:Table Majumdar-Ghosh line dual}.
These entries agree with the corresponding ones in
Table \ref{Table:Table Majumdar-Ghosh line}
(lines three, five, and six).

\begin{table}
\caption{
\label{Table:Table Majumdar-Ghosh line dual}
The expectation values of nine operators in three states.
The domain of definition of all nine operators is  
$\mathcal{H}^{\,\vee}_{b^{\prime}=0}$.
The Neel state is defined in
Eq.\
(\ref{eq:dual of neel x+ state}).
The next two dimer states are defined in
Eq.\
(\ref{eq:Dimer vee o+ and Dimer vee e+ groundstates}).
All nine operators are defined in Eqs.\
(\ref{eq:dual dimer order parameter})-(\ref{eq:dual U 2n rzpi e}).
}
\centering
\begin{scriptsize}
\begin{tabular}{l|ccccccccc}
\hline \hline
&
$\ \widehat{C}^{x\,\vee}_{j,j+n}\ $
&
$\ \widehat{O}^{\,\mathrm{o}\,\vee}_{\mathrm{dimer}}\ $
&
$\ \widehat{U}^{\,\vee}_{r}\ $
&
$\ \widehat{U}^{\,\vee}_{r^{x}_{\pi}}\ $
&
$\ \widehat{U}^{\,\vee}_{r^{z}_{\pi}}\ $
&
$\widehat{U}^{\,\mathrm{o}\,(2n)\,\vee}_{r^{x}_{\pi}}$
&
$\widehat{U}^{\,\mathrm{e}\,(2n)\,\vee}_{r^{x}_{\pi}}$
&
$\widehat{U}^{\,\mathrm{o}\,(2n)\,\vee}_{r^{z}_{\pi}}$
&
$\widehat{U}^{\,\mathrm{e}\,(2n)\,\vee}_{r^{z}_{\pi}}$
\\
\hline
$|\mathrm{Neel}^{x\,\vee}\rangle^{+}$
&
$(-1)^{n}$
&
$0$
&
$+1$
&
$(-1)^{N}$
&
$+1$
&
$(-1)^{n}$
&
$(-1)^{n}$
&
$0$
&
$0$
\\
\hline
$|\mathrm{Dimer}^{\,\vee}_{\mathrm{o}}\rangle$
&
$-\delta^{\,}_{(-1)^{j},-1}\,\delta^{\,}_{n,1}$
&
$+1$
&
$0$
&
$(-1)^{N}$
&
$(-1)^{N}$
&
$(-1)^{n}$
&
$0$
&
$(-1)^{n}$
&
$0$
\\
$|\mathrm{Dimer}^{\,\vee}_{\mathrm{e}}\rangle$
&
$-\delta^{\,}_{(-1)^{j},+1}\,\delta^{\,}_{n,1}$
&
$-1$
&
$0$
&
$(-1)^{N}$
&
$(-1)^{N}$
&
$0$
&
$(-1)^{n}$
&
$0$
&
$(-1)^{n}$
\\
\hline \hline
\end{tabular}
\end{scriptsize}
\end{table}

\subsection{Jordan-Wigner dualization of the Majumdar-Ghosh line}
\label{subsec:Jordan-Wigner dualization of the Majumdar-Ghosh line}

The projectors that are the Jordan-Wigner
dual to those for the pair of dimer states are built out of
\begin{subequations}
\label{eq:Jordan-Wigner projectors dimer states}
\begin{equation}
\widehat{P}^{\,\mathrm{o}\,\vee}_{\mathrm{Dimer}}:=
\prod_{j=1}^{N}
\widehat{P}^{\,\vee}_{[2j-1,2j]}
\end{equation}
and
\begin{equation}
\widehat{P}^{\,\mathrm{e}\,\vee}_{\mathrm{Dimer}}:=
\prod_{j=1}^{N}
\widehat{P}^{\,\vee}_{[2j,2j+1]},
\end{equation}
where
\begin{equation}
\widehat{P}^{\,\vee}_{[j,j+1]}:=
\frac{1}{4}
\left(
\widehat{\mbb{1}}^{\,}_{\mathcal{H}^{\,\vee}_{f=1}}
-
\mathrm{i}
\hat{\beta}^{\vee}_{j}\,
\hat{\alpha}^{\vee}_{j+1}
-
\mathrm{i}
\hat{\beta}^{\vee}_{j+1}\,
\hat{\alpha}^{\vee}_{j}
-
\hat{\beta}^{\vee}_{j}\,
\hat{\beta}^{\vee}_{j+1}\,
\hat{\alpha}^{\vee}_{j}\,
\hat{\alpha}^{\vee}_{j+1}
\right),
\end{equation}
\end{subequations}
by restriction to the subspace $\mathcal{H}^{\,\vee}_{f=1}$
from Table \ref{Table:Jordan-Wigner dualization b to bF}.
Unlike in the case of
Eq.\ (\ref{eq:Kramers-Wannier projectors dimer states}),
each of
$\widehat{P}^{\,\vee}_{[2j-1,2j]}$
and
$\widehat{P}^{\,\vee}_{[2j,2j+1]}$
acts non-trivially on two consecutive sites.
This is why each of the projectors
$\widehat{P}^{\,\mathrm{o}\,\vee}_{\mathrm{Dimer}}$
and
$\widehat{P}^{\,\mathrm{e}\,\vee}_{\mathrm{Dimer}}$
has a non-degenerate eigenstate with eigenvalue one.

It is instructive to trade the Majorana operators for
fermionic ones. To this end, define for any $j=1,\cdots,2N$
\begin{subequations}
\begin{equation}
\hat{c}^{\vee\,\dag}_{j}:=
\frac{1}{2}
\left(
\hat{\alpha}^{\vee}_{j}
-
\mathrm{i}
\hat{\beta}^{\vee}_{j}
\right),
\qquad
\hat{c}^{\vee}_{j}:=
\frac{1}{2}
\left(
\hat{\alpha}^{\vee}_{j}
+
\mathrm{i}
\hat{\beta}^{\vee}_{j}
\right),
\end{equation}
i.e.,
\begin{equation}
\hat{\alpha}^{\vee}_{j}=
\hat{c}^{\vee}_{j}
+
\hat{c}^{\vee\,\dag}_{j},
\qquad
\hat{\beta}^{\vee}_{j}=
-\mathrm{i}
\left(
\hat{c}^{\vee}_{j}
-
\hat{c}^{\vee\,\dag}_{j}
\right).
\end{equation}
\end{subequations}
There follows the identities
\begin{subequations}
\begin{align}
\mathrm{i}    
\hat{\beta}^{\vee}_{j}\,
\hat{\alpha}^{\vee}_{j}=&\,
\left(
\hat{c}^{\vee}_{j}
-
\hat{c}^{\vee\,\dag}_{j}
\right)
\left(
\hat{c}^{\vee}_{j}
+
\hat{c}^{\vee\,\dag}_{j}
\right)
\nonumber\\
=&\,
\hat{c}^{\vee}_{j}\,
\hat{c}^{\vee\,\dag}_{j}
-
\hat{c}^{\vee\,\dag}_{j}\,
\hat{c}^{\vee}_{j}
\nonumber\\
=&\,
1
-
2\,
\hat{c}^{\vee\,\dag}_{j}\,
\hat{c}^{\vee}_{j}
\nonumber\\
\equiv&\,
1
-
2\,
\hat{n}^{\vee}_{j},
\qquad
\hat{n}^{\vee}_{j}:=
\hat{c}^{\vee\,\dag}_{j}\,
\hat{c}^{\vee}_{j},
\\ 
\hat{\beta}^{\vee}_{j}\,
\hat{\beta}^{\vee}_{j+1}\,
\hat{\alpha}^{\vee}_{j}\,
\hat{\alpha}^{\vee}_{j+1}=&\,
\left(
\mathrm{i}
\hat{\beta}^{\vee}_{j}\,
\hat{\alpha}^{\vee}_{j}\,
\right)
\left(
\mathrm{i}
\hat{\beta}^{\vee}_{j+1}\,
\hat{\alpha}^{\vee}_{j+1}
\right)
\nonumber\\
=&\,
\left(
1
-
2\,
\hat{n}^{\vee}_{j}
\right)
\left(
1
-
2\,
\hat{n}^{\vee}_{j+1}
\right),
\\ 
\mathrm{i}
\hat{\beta}^{\vee}_{j}\,
\hat{\alpha}^{\vee}_{j+1}=&\,
\left(
\hat{c}^{\vee}_{j}
-
\hat{c}^{\vee\,\dag}_{j}
\right)
\left(
\hat{c}^{\vee}_{j+1}
+
\hat{c}^{\vee\,\dag}_{j+1}
\right)
\nonumber\\
=&\,
\hat{c}^{\vee}_{j}\,
\hat{c}^{\vee}_{j+1}
+
\hat{c}^{\vee}_{j}\,
\hat{c}^{\vee\,\dag}_{j+1}
-
\hat{c}^{\vee\,\dag}_{j}\,
\hat{c}^{\vee}_{j+1}
-
\hat{c}^{\vee\,\dag}_{j}\,
\hat{c}^{\vee\,\dag}_{j+1},
\\ 
\mathrm{i}
\hat{\beta}^{\vee}_{j}\,
\hat{\alpha}^{\vee}_{j+1}
+
\mathrm{i}
\hat{\beta}^{\vee}_{j+1}\,
\hat{\alpha}^{\vee}_{j}=&\,
\hat{c}^{\vee}_{j}\,
\hat{c}^{\vee}_{j+1}
+
\hat{c}^{\vee}_{j}\,
\hat{c}^{\vee\,\dag}_{j+1}
-
\hat{c}^{\vee\,\dag}_{j}\,
\hat{c}^{\vee}_{j+1}
-
\hat{c}^{\vee\,\dag}_{j}\,
\hat{c}^{\vee\,\dag}_{j+1}
\nonumber\\
&\,
+
\hat{c}^{\vee}_{j+1}\,
\hat{c}^{\vee}_{j}
+
\hat{c}^{\vee}_{j+1}\,
\hat{c}^{\vee\,\dag}_{j}
-
\hat{c}^{\vee\,\dag}_{j+1}\,
\hat{c}^{\vee}_{j}
-
\hat{c}^{\vee\,\dag}_{j+1}\,
\hat{c}^{\vee\,\dag}_{j}
\nonumber\\
=&\,
-
2
\left(
\hat{c}^{\vee\,\dag}_{j+1}\,
\hat{c}^{\vee}_{j}
+
\hat{c}^{\vee\,\dag}_{j}\,
\hat{c}^{\vee}_{j+1}
\right),
\end{align}
and
\begin{equation}
\widehat{P}^{\,\vee}_{[j,j+1]}:=
\frac{1}{4}
\left[
\widehat{\mbb{1}}^{\,}_{\mathcal{H}^{\,\vee}_{f=1}}
+
2
\left(
\hat{c}^{\vee\,\dag}_{j+1}\,
\hat{c}^{\vee}_{j}
+
\hat{c}^{\vee\,\dag}_{j}\,
\hat{c}^{\vee}_{j+1}
\right)
-
\left(
1
-
2\,
\hat{n}^{\vee}_{j}
\right)
\left(
1
-
2\,
\hat{n}^{\vee}_{j+1}
\right)
\right].
\end{equation}
\end{subequations}
On the Hilbert space
\begin{subequations}
\begin{equation}
\begin{split}
\mathcal{H}^{\,}_{j+\frac{1}{2},j+1+\frac{1}{2}}:=
\mathrm{span}\,
\Bigg\{&
\left(
\hat{c}^{\vee\,\dag}_{j}
\right)^{ n^{\,}_{j}}\,
\left(
\hat{c}^{\vee\,\dag}_{j+1}
\right)^{n^{\,}_{j+1}}\,
|0,0\rangle
\ \Bigg|\ 
n^{\,}_{j},n^{\,}_{j+1}=0,1,
\\
&\,
\hat{c}^{\vee}_{j}\,
|0,0\rangle=
\hat{c}^{\vee}_{j+1}\,
|0,0\rangle=0
\Bigg\},
\end{split}
\end{equation}
$
\mathrm{i}
\hat{\beta}^{\vee}_{j}\,
\hat{\alpha}^{\vee}_{j+1}
$
is represented by the matrix
\begin{equation}
\begin{pmatrix}
0&0&0&-1\\
0&0&-1&0\\
0&-1&0&0\\
+1&0&0&0\\
\end{pmatrix},
\end{equation}
while $\widehat{P}^{\,\vee}_{[j,j+1]}$ is represented by the matrix
\begin{equation}
\frac{1}{4}
\left[
\begin{pmatrix}
+1&0&0&0\\
0&+1&0&0\\
0&0&+1&0\\
0&0&0&+1\\
\end{pmatrix}
+
2
\begin{pmatrix}
0&0&0&0\\
0&0&1&0\\
0&1&0&0\\
0&0&0&0\\
\end{pmatrix}
-
\begin{pmatrix}
+1&0&0&0\\
0&-1&0&0\\
0&0&-1&0\\
0&0&0&+1\\
\end{pmatrix}
\right]=
\frac{1}{2}
\begin{pmatrix}
0&0&0&0\\
0&1&1&0\\
0&1&1&0\\
0&0&0&0\\
\end{pmatrix}
\end{equation}
that annihilates the three orthonormal eigenstates
\begin{equation}
|0,0\rangle,
\qquad
\frac{1}{\sqrt{2}}
\left(
|1,0\rangle
-
|0,1\rangle
\right),
\qquad
|1,1\rangle,
\end{equation}
and projects onto the eigenstate
\begin{equation}
\frac{1}{\sqrt{2}}
\left(
|1,0\rangle
+
|0,1\rangle
\right).
\end{equation}
\end{subequations}
Hence, the two orthonormal states
\begin{equation}
|\mathrm{Bonding}^{\vee}_{\mathrm{o}}\rangle:=
\left[
\prod_{j=1}^{N}
\frac{1}{\sqrt{2}}\,
\left(
\hat{c}^{\vee\,\dag}_{2j-1}
+
\hat{c}^{\vee\,\dag}_{2j}
\right)
\right]
|0\rangle
\end{equation}
and
\begin{equation}
|\mathrm{Bonding}^{\vee}_{\mathrm{e}}\rangle:=
\left[
\prod_{j=1}^{N}
\frac{1}{\sqrt{2}}\,
\left(
\hat{c}^{\vee\,\dag}_{2j}
+
\hat{c}^{\vee\,\dag}_{2j+1}
\right)
\right]
|0\rangle
\end{equation}
are ground states of $\widehat{H}^{\,\vee}_{f=1}$ along the MG line.
We set $N$ to be an even integer, so that these states have even fermion parity
and belong to the subspace $\mathcal{H}^{\vee}_{f=1;+}$.
Their transformation laws under the symmetry group
(\ref{eq:GtotVF JW})
are
\begin{subequations}
\begin{align}
&
\widehat{U}^{\vee}_{t}\,
|\mathrm{Bonding}^{\vee}_{\mathrm{o}}\rangle=
|\mathrm{Bonding}^{\vee}_{\mathrm{e}}\rangle,
\qquad
\widehat{U}^{\vee}_{t}\,
|\mathrm{Bonding}^{\vee}_{\mathrm{e}}\rangle=
|\mathrm{Bonding}^{\vee}_{\mathrm{o}}\rangle,
\\
&
\widehat{U}^{\vee}_{r}\,
|\mathrm{Bonding}^{\vee}_{\mathrm{o}}\rangle=
|\mathrm{Bonding}^{\vee}_{\mathrm{e}}\rangle,
\qquad
\widehat{U}^{\vee}_{r}\,
|\mathrm{Bonding}^{\vee}_{\mathrm{e}}\rangle=
|\mathrm{Bonding}^{\vee}_{\mathrm{o}}\rangle,
\\
&
\widehat{U}^{\vee}_{\mathrm{o}}\,
|\mathrm{Bonding}^{\vee}_{\mathrm{o}}\rangle=
|\mathrm{Bonding}^{\vee}_{\mathrm{o}}\rangle,
\qquad
\widehat{U}^{\vee}_{\mathrm{o}}\,
|\mathrm{Bonding}^{\vee}_{\mathrm{e}}\rangle=
|\mathrm{Bonding}^{\vee}_{\mathrm{e}}\rangle,
\\
&
\widehat{U}^{\vee}_{\mathrm{e}}\,
|\mathrm{Bonding}^{\vee}_{\mathrm{o}}\rangle=
|\mathrm{Bonding}^{\vee}_{\mathrm{o}}\rangle,
\qquad
\widehat{U}^{\vee}_{\mathrm{e}}\,
|\mathrm{Bonding}^{\vee}_{\mathrm{e}}\rangle=
|\mathrm{Bonding}^{\vee}_{\mathrm{e}}\rangle.
\end{align}
\end{subequations}

\begin{proof}
Without loss of generality, we consider the case of $N=2$.
We do the substitutions
\begin{align}
\hat{c}^{\vee\,\dag}_{1}\to\hat{a}^{\dag},
\qquad
\hat{c}^{\vee\,\dag}_{2}\to\hat{b}^{\dag},
\qquad
\hat{c}^{\vee\,\dag}_{3}\to\hat{c}^{\dag},
\qquad
\hat{c}^{\vee\,\dag}_{4}\to\hat{d}^{\dag},
\end{align}
to simplify the notation. The basis of the Hilbert space is chosen to be
\begin{equation}
\left.
\begin{array}{ll}  
|a,b,c,d\rangle=
\left(\hat{a}^{\dag}\vphantom{\hat{d}^{\dag}}\right)^{a}\,
\left(\hat{b}^{\dag}\vphantom{\hat{d}^{\dag}}\right)^{b}\,
\left(\hat{c}^{\dag}\vphantom{\hat{d}^{\dag}}\right)^{c}\,
\left(\hat{d}^{\dag}\vphantom{\hat{d}^{\dag}}\right)^{d}\,
|0\rangle\
\\
\hat{a}\,|0\rangle=
\hat{b}\,|0\rangle=
\hat{c}\,|0\rangle=
\hat{d}\,|0\rangle=0\
\end{array}
\right\}
a,b,c,d=0,1.
\end{equation}
In this basis,
\begin{subequations}
\begin{align}
&
|\mathrm{Bonding}^{\vee}_{\mathrm{o}}\rangle=
\frac{1}{2}
\left(
|1,0,1,0\rangle
+
|1,0,0,1\rangle
+
|0,1,1,0\rangle
+
|0,1,0,1\rangle
\right),
\\
&
|\mathrm{Bonding}^{\vee}_{\mathrm{e}}\rangle=
\frac{1}{2}
\left(
|0,1,0,1\rangle
+
|1,1,0,0\rangle
+
|0,0,1,1\rangle
+
|1,0,1,0\rangle
\right).
\end{align}
\end{subequations}
Translation $j\mapsto j+1\hbox{ mod }4$
by one lattice spacing corresponds to
\begin{subequations}
\begin{equation}
\hat{a}\mapsto\hat{b},
\qquad
\hat{b}\mapsto\hat{c},
\qquad
\hat{c}\mapsto\hat{d},
\qquad
\hat{d}\mapsto(-1)\,\hat{a},
\end{equation}
under which
\begin{equation}
|\mathrm{Bonding}^{\vee}_{\mathrm{o}}\rangle\mapsto
|\mathrm{Bonding}^{\vee}_{\mathrm{e}}\rangle,
\qquad
|\mathrm{Bonding}^{\vee}_{\mathrm{e}}\rangle\mapsto
|\mathrm{Bonding}^{\vee}_{\mathrm{o}}\rangle.
\end{equation}
\end{subequations}
Reflection $j\mapsto 2N-j\hbox{ mod }4$
corresponds to
\begin{subequations}
\begin{equation}
\hat{a}\mapsto -\mathrm{i}\hat{c},
\qquad
\hat{b}\mapsto -\mathrm{i}\hat{b},
\qquad
\hat{c}\mapsto -\mathrm{i}\hat{a},
\qquad
\hat{d}\mapsto +\mathrm{i}\hat{d},
\end{equation}
(and not
$\hat{a}\mapsto\hat{b}$,
$\hat{b}\mapsto\hat{a}$,
$\hat{c}\mapsto\hat{d}$,
$\hat{d}\mapsto\hat{c}$)
under which
\begin{equation}
|\mathrm{Bonding}^{\vee}_{\mathrm{o}}\rangle\mapsto
|\mathrm{Bonding}^{\vee}_{\mathrm{e}}\rangle,
\qquad
|\mathrm{Bonding}^{\vee}_{\mathrm{e}}\rangle\mapsto
|\mathrm{Bonding}^{\vee}_{\mathrm{o}}\rangle.
\end{equation}
\end{subequations}

The representation of
\begin{subequations}
\begin{equation}
\widehat{U}^{\vee}_{\mathrm{o}}=
\left(\widehat{U}^{\vee}_{\mathrm{o}}\right)^{\dag}
\end{equation}
after normal ordering is
\begin{align}
\widehat{U}^{\vee}_{\mathrm{o}}=&\,
\left(
\hat{a}\,
\hat{b}
+
\hat{b}^{\dag}\,
\hat{a}^{\dag}
-
\hat{b}^{\dag}\,
\hat{a}
-
\hat{a}^{\dag}\,
\hat{b}
\right)
\left(
\hat{c}\,
\hat{d}
+
\hat{d}^{\dag}\,
\hat{c}^{\dag}
-
\hat{d}^{\dag}\,
\hat{c}
-
\hat{c}^{\dag}\,
\hat{d}
\right).
\end{align}
\end{subequations}
On the one hand,
\begin{align}
\widehat{U}^{\vee}_{\mathrm{o}}\,
|\mathrm{Bonding}^{\vee}_{\mathrm{o}}\rangle=&\,
\frac{1}{2}
\left(
\hat{a}\,
\hat{b}
+
\hat{b}^{\dag}\,
\hat{a}^{\dag}
-
\hat{b}^{\dag}\,
\hat{a}
-
\hat{a}^{\dag}\,
\hat{b}
\right)
\left(
\hat{c}\,
\hat{d}
+
\hat{d}^{\dag}\,
\hat{c}^{\dag}
-
\hat{d}^{\dag}\,
\hat{c}
-
\hat{c}^{\dag}\,
\hat{d}
\right)
\nonumber\\
&\,
\times
\left(
|1,0,1,0\rangle
+
|1,0,0,1\rangle
+
|0,1,1,0\rangle
+
|0,1,0,1\rangle
\right)
\nonumber\\
=&\,
\frac{1}{2}
\left(
\hat{a}\,
\hat{b}
+
\hat{b}^{\dag}\,
\hat{a}^{\dag}
-
\hat{b}^{\dag}\,
\hat{a}
-
\hat{a}^{\dag}\,
\hat{b}
\right)
\nonumber\\
&\,
\times
\left[
-
\hat{d}^{\dag}\,
\hat{c}\,
\left(
|1,0,1,0\rangle
+
|0,1,1,0\rangle
\right)
-
\hat{c}^{\dag}\,
\hat{d}
\left(
|1,0,0,1\rangle
+
|0,1,0,1\rangle
\right)
\right]
\nonumber\\
=&\,
\frac{1}{2}
\left(
\hat{a}\,
\hat{b}
+
\hat{b}^{\dag}\,
\hat{a}^{\dag}
-
\hat{b}^{\dag}\,
\hat{a}
-
\hat{a}^{\dag}\,
\hat{b}
\right)
\nonumber\\
&\,
\times(-1)
\left(
|1,0,0,1\rangle
+
|0,1,0,1\rangle
+
|1,0,1,0\rangle
+
|0,1,1,0\rangle
\right)
\nonumber\\
=&\,
\frac{1}{2}
\left[
\hat{b}^{\dag}\,
\hat{a}
\left(
|1,0,0,1\rangle
+
|1,0,1,0\rangle
\right)
+
\hat{a}^{\dag}\,
\hat{b}
\left(
|0,1,0,1\rangle
+
|0,1,1,0\rangle
\right)
\right]
\nonumber\\
=&\,
\frac{1}{2}
\left(
|0,1,0,1\rangle
+
|0,1,1,0\rangle
+
|1,0,0,1\rangle
+
|1,0,1,0\rangle
\right)
\nonumber\\
=&\,
|\mathrm{Bonding}^{\vee}_{\mathrm{o}}\rangle.
\end{align}
On the other hand,
\begin{align}
\widehat{U}^{\vee}_{\mathrm{o}}\,
|\mathrm{Bonding}^{\vee}_{\mathrm{e}}\rangle=&\,
\frac{1}{2}
\left(
\hat{a}\,
\hat{b}
+
\hat{b}^{\dag}\,
\hat{a}^{\dag}
-
\hat{b}^{\dag}\,
\hat{a}
-
\hat{a}^{\dag}\,
\hat{b}
\right)
\left(
\hat{c}\,
\hat{d}
+
\hat{d}^{\dag}\,
\hat{c}^{\dag}
-
\hat{d}^{\dag}\,
\hat{c}
-
\hat{c}^{\dag}\,
\hat{d}
\right)
\nonumber\\
&\,
\times
\left(
|0,1,0,1\rangle
+
|1,1,0,0\rangle
+
|0,0,1,1\rangle
+
|1,0,1,0\rangle
\right)
\nonumber\\
=&\,
\frac{1}{2}
\left(
\hat{a}\,
\hat{b}
+
\hat{b}^{\dag}\,
\hat{a}^{\dag}
-
\hat{b}^{\dag}\,
\hat{a}
-
\hat{a}^{\dag}\,
\hat{b}
\right)
\nonumber\\
&\,
\times
\left(
\hat{c}\,
\hat{d}\,
|0,0,1,1\rangle
+
\hat{d}^{\dag}\,
\hat{c}^{\dag}\,
|1,1,0,0\rangle
-
\hat{d}^{\dag}\,
\hat{c}\,
|1,0,1,0\rangle
-
\hat{c}^{\dag}\,
\hat{d}\,
|0,1,0,1\rangle
\right)
\nonumber\\
=&\,
\frac{1}{2}
\left(
\hat{a}\,
\hat{b}
+
\hat{b}^{\dag}\,
\hat{a}^{\dag}
-
\hat{b}^{\dag}\,
\hat{a}
-
\hat{a}^{\dag}\,
\hat{b}
\right)
\nonumber\\
&\,
\times
(-1)
\left(
|0,0,0,0\rangle
+
|1,1,1,1\rangle
+
|1,0,0,1\rangle
+
|0,1,1,0\rangle
\right)
\nonumber\\
=&\,
\frac{1}{2}
\nonumber\\
&\,
\times
(-1)\,
\left(
\hat{a}\
\hat{b}\,
|1,1,1,1\rangle
+
\hat{b}^{\dag}\
\hat{a}^{\dag}\,
|0,0,0,0\rangle
-
\hat{b}^{\dag}\,
\hat{a}\,
|1,0,0,1\rangle
-
\hat{a}^{\dag}\,
\hat{b}\,
|0,1,1,0\rangle
\right)
\nonumber\\
=&\,
\frac{1}{2}
\left(
|0,0,1,1\rangle
+
|1,1,0,0\rangle
+
|0,1,0,1\rangle
+
|1,0,1,0\rangle
\right)
\nonumber\\
=&\,
|\mathrm{Bonding}^{\vee}_{\mathrm{e}}\rangle.
\end{align}

We can deduce the action of
$\widehat{U}^{\vee}_{\mathrm{e}}$
on
$|\mathrm{Bonding}^{\vee}_{\mathrm{o}}\rangle$
and
$|\mathrm{Bonding}^{\vee}_{\mathrm{e}}\rangle$
from the facts that
\begin{subequations}
\begin{align}
&
\left[
\widehat{U}^{\vee}_{\mathrm{o}},
\widehat{U}^{\vee}_{\mathrm{e}}  
\right]=0,
\\
&
\widehat{U}^{\vee}_{\mathrm{o}}\,
\widehat{U}^{\vee}_{\mathrm{e}}=
\widehat{U}^{\vee}_{\mathrm{e}}\,
\widehat{U}^{\vee}_{\mathrm{o}}=
\widehat{P}^{\vee}_{\mathrm{F}},
\\
&
\left(\widehat{U}^{\vee}_{\mathrm{o}}\right)^{2}=
\left(\widehat{U}^{\vee}_{\mathrm{e}}\right)^{2}=
\widehat{\mbb{1}}^{\,}_{\mathcal{H}^{\,\vee}_{f}},
\\
&
\widehat{P}^{\vee}_{\mathrm{F}}\,
|\mathrm{Bonding}^{\vee}_{\mathrm{o}}\rangle=
|\mathrm{Bonding}^{\vee}_{\mathrm{o}}\rangle,
\\
&
\widehat{P}^{\vee}_{\mathrm{F}}\,
|\mathrm{Bonding}^{\vee}_{\mathrm{e}}\rangle=
|\mathrm{Bonding}^{\vee}_{\mathrm{e}}\rangle.
\end{align}
\end{subequations}
We then infer that
\begin{align}
\widehat{U}^{\vee}_{\mathrm{e}}\,
|\mathrm{Bonding}^{\vee}_{\mathrm{o}}\rangle=&\,
\left(
\widehat{U}^{\vee}_{\mathrm{e}}\,
\widehat{U}^{\vee}_{\mathrm{o}}
\right)\,
\left(
\widehat{U}^{\vee}_{\mathrm{o}}\,
|\mathrm{Bonding}^{\vee}_{\mathrm{o}}\rangle
\right)
\nonumber\\
=&\,
\left(
\widehat{P}^{\vee}_{\mathrm{F}}
\right)
|\mathrm{Bonding}^{\vee}_{\mathrm{o}}\rangle
\nonumber\\
=&\,
|\mathrm{Bonding}^{\vee}_{\mathrm{o}}\rangle
\end{align}
and
\begin{align}
\widehat{U}^{\vee}_{\mathrm{e}}\,
|\mathrm{Bonding}^{\vee}_{\mathrm{e}}\rangle=&\,
\left(
\widehat{U}^{\vee}_{\mathrm{e}}\,
\widehat{U}^{\vee}_{\mathrm{o}}
\right)\,
\left(
\widehat{U}^{\vee}_{\mathrm{o}}\,
|\mathrm{Bonding}^{\vee}_{\mathrm{e}}\rangle
\right)
\nonumber\\
=&\,
\left(
\widehat{P}^{\vee}_{\mathrm{F}}
\right)
|\mathrm{Bonding}^{\vee}_{\mathrm{e}}\rangle
\nonumber\\
=&\,
|\mathrm{Bonding}^{\vee}_{\mathrm{e}}\rangle.
\end{align}

\end{proof}

\section{Triality with open boundary conditions}
\label{appsec:Triality with open boundary conditions}

To treat the case of open boundary conditions,
we need to modify the bond algebras
\begin{equation}
\mathfrak{B}^{\,}_{b},
\qquad
\mathfrak{B}^{\,}_{b^{\prime}},
\qquad
\mathfrak{B}^{\,}_{f},
\label{appeq:list bond algebras}
\end{equation}
defined in Secs.\
\ref{subsec:Definition of the bond algebra cal Bb},
\ref{subsec:Kramers-Wannier duality b to b'},
and
\ref{subsec:Jordan-Wigner duality b and f},
respectively.
The following changes must be done as one repeats all steps of
Sec.\
\ref{sec:Triality through bond algebra isomorphisms}.

One must remove the term
$\hat{\sigma}^{x}_{2N}\,\hat{\sigma}^{x}_{2N+1}$
from $\mathfrak{B}^{\,}_{b}$
as the dual lattice
\begin{subequations}
\begin{equation}
\Lambda^{\star}=
\left\{\left.\vphantom{\Big[}
j^{\star}=j+\frac{1}{2}
\ \right|\  
j=1,\cdots,2N-1
\right\}
\end{equation}
has one less site than the direct lattice
\begin{equation}
\Lambda=
\left\{\vphantom{\Big[}
j=1,\cdots,2N
\right\}
\end{equation}
\end{subequations}
when open boundary conditions are imposed.

One must modify the bond algebras (\ref{appeq:list bond algebras})
according to
\begin{subequations}
\begin{align}
&
\mathfrak{B}^{\,}_{b}\to
\mathfrak{B}^{\,}_{\sigma}:=
\left\langle\vphantom{\Big[}
\hat{\sigma}^{z}_{i},
\qquad
\hat{\sigma}^{x}_{j}\,
\hat{\sigma}^{x}_{j+1}
\ \Big|\
i\in\Lambda,
\qquad
j\in\Lambda\setminus\{{2N\}}
\right\rangle,
\\
&
\mathfrak{B}^{\,}_{b,b^{\prime}}\to
\mathfrak{B}^{\,}_{\sigma,\tau}:=
\left\langle\vphantom{\Big[}
\hat{\sigma}^{z}_{i},
\qquad
\hat{\sigma}^{x}_{j}\,
\hat{\tau}^{z}_{j^{\star}}\,
\hat{\sigma}^{x}_{j+1}
\ \Big|\
i\in\Lambda,
\qquad
j\in\Lambda\setminus\{{2N\}}
\right\rangle,
\\
&
\mathfrak{B}^{\,}_{b,f}\to
\mathfrak{B}^{\,}_{\sigma,\beta\alpha}:=
\left\langle\vphantom{\Big[}
\hat{\sigma}^{z}_{i},
\qquad
\hat{\sigma}^{x}_{j}\,
\left(
\mathrm{i}
\hat{\beta}^{\,}_{j}\,
\hat{\alpha}^{\,}_{j+1}
\right)
\hat{\sigma}^{x}_{j+1}
\ \Big|\
i\in\Lambda,
\qquad
j\in\Lambda\setminus\{{2N\}}
\right\rangle,
\end{align}
with the Hilbert space 
\begin{equation}
\begin{split}
\mathcal{H}^{\,}_{b,b^{\prime}}:=
\mathcal{H}^{\,}_{b}
\otimes
\mathcal{H}^{\,}_{b^{\prime}}
\to
\mathcal{H}^{\,}_{\sigma,\tau}:=&\,
\mathcal{H}^{\,}_{\sigma}
\otimes
\mathcal{H}^{\,}_{\tau}
\\
\cong&\,
\left(
\bigotimes\limits_{j\in\Lambda}\mathbb{C}^{2}
\right)
\otimes
\left(
\bigotimes\limits_{j^{\star}\in\Lambda^{\star}}\mathbb{C}^{2}
\right)=
\mathbb{C}^{2^{2N}}
\otimes
\mathbb{C}^{2^{2N-1}}=
\mathbb{C}^{2^{4N-1}}
\end{split}
\end{equation}
of dimension $2^{4N-1}$  and that 
\begin{equation}
\begin{split}
\mathcal{H}^{\,}_{b,f}:=
\mathcal{H}^{\,}_{b}
\otimes
\mathcal{H}^{\,}_{f}
\to
\mathcal{H}^{\,}_{\sigma,\beta\alpha}:=&\,
\mathcal{H}^{\,}_{\sigma}
\otimes
\mathcal{H}^{\,}_{\beta\alpha}
\\
\cong&\,
\left(
\bigotimes\limits_{j\in\Lambda}\mathbb{C}^{2}
\right)
\otimes
\left(
\bigotimes\limits_{j\in\Lambda}\mathbb{C}^{2}
\right)=
\mathbb{C}^{2^{2N}}
\otimes
\mathbb{C}^{2^{2N}}=
\mathbb{C}^{2^{4N}}
\end{split}
\end{equation}
\end{subequations}
of dimension $2^{4N}$ as domain of definition, respectively.

One must modify the local Gauss operators
(\ref{eq:sym bond algebra cal Bbb'})
and
(\ref{eq:sym bond algebra cal Bbf})
according to
\begin{subequations}
\begin{align}
\widehat{G}^{}_{b,b^{\prime};j}\to&\,
\widehat{G}^{}_{\sigma,\tau;j}:=
\begin{cases}
\hat{\sigma}^{z}_{1}\,
\hat{\tau}^{x}_{1+\frac{1}{2}},
&
j=1,
\\
\hat{\tau}^{z}_{j-1+\frac{1}{2}}\,
\hat{\sigma}^{z}_{j}\,
\hat{\tau}^{z}_{j+\frac{1}{2}},
&
j=2,\cdots,2N-1,
\\
\hat{\tau}^{z}_{2N-1+\frac{1}{2}}\,
\hat{\sigma}^{z}_{2N},
&
j=2N,
\end{cases}
\end{align}
and
\begin{align}    
\widehat{G}^{}_{b,f;j}\to&\,
\widehat{G}^{}_{\sigma,\beta\alpha;j}:=
\mathrm{i}
\hat{\beta}^{\,}_{j}\,
\hat{\sigma}^{z}_{j}\,
\hat{\alpha}^{\,}_{j},
\qquad
j\in\Lambda,
\end{align}
\end{subequations}
respectively.

One must modify the dualized bond algebras
(\ref{eq:definition and symmetry of cal Bb'})
and
(\ref{eq:definition and symmetry of cal Bf})
according to
\begin{equation}
\mathfrak{B}^{\,}_{b'}\to
\mathfrak{B}^{\,}_{\tau}:=
\left\langle\left.
\left(
\hat{\tau}^{x\,\vee}_{i^{\star}-1}
\right)^{1-\delta^{\,}_{i,1}}
\left(
\hat{\tau}^{x\,\vee}_{i^{\star}}
\right)^{1-\delta^{\,}_{i,2N}},
\qquad
\hat{\tau}^{z\,\vee}_{j^{\star}}
\ \right|\
i\in\Lambda,
\qquad
j\in\Lambda\setminus\{2N\} 
\right\rangle
\end{equation}
and
\begin{equation}
\mathfrak{B}^{\,}_{f}\to
\mathfrak{B}^{\,}_{\beta\alpha}:=
\left\langle\left.
\mathrm{i}
\hat{\beta}^{\vee}_{j}\,
\hat{\alpha}^{\vee}_{j},
\qquad
\mathrm{i}
\hat{\beta}^{\vee}_{j}\,
\hat{\alpha}^{\vee}_{j+1}
\ \right|\
i\in\Lambda,
\quad
j\in\Lambda\setminus\{2N\} 
\right\rangle,
\end{equation}
respectively.

One must modify the consistency conditions
(\ref{eq:consistency duality b to b' a})
and
(\ref{eq:consistency duality b to bF a})
by identifying the dual pairs
\begin{subequations}
\begin{equation}
\left(\,
\prod_{j\in\Lambda}
\hat{\sigma}^{z}_{j}=
\widehat{U}^{\,}_{r^{z}_{\pi}},
\qquad
\widehat{\mbb{1}}^{\,}_{\mathcal{H}^{\,\vee}_{\tau}}
\right)
\end{equation}
and
\begin{equation}
\left(\,
\prod_{j\in\Lambda}
\hat{\sigma}^{z}_{j}=
\widehat{U}^{\,}_{r^{z}_{\pi}},
\qquad
\prod_{j\in\Lambda}
\left(
\mathrm{i}
\hat{\beta}^{\,}_{j}\,
\hat{\alpha}^{\,}_{j}
\right)=
-
\widehat{P}^{\,\vee}_{\mathrm{F}}
\right),
\end{equation}
\end{subequations}
respectively. Here,
$\mathcal{H}^{\,\vee}_{\tau}$
($\mathcal{H}^{\,\vee}_{\beta\alpha}$)
is the projection of
$\mathcal{H}^{\,}_{\sigma,\tau}$
($\mathcal{H}^{\,}_{\sigma,\beta\alpha}$)
to the subspace on which all local Gauss operators reduce to
the identity. Correspondingly, the dual pairs of Hilbert subspaces are
\begin{subequations}
\begin{equation}
\left(
\mathcal{H}^{\,\mathrm{dual}}_{\sigma},  
\qquad
\mathcal{H}^{\,\vee\,\mathrm{dual}}_{\tau}  
\right),
\qquad
\mathcal{H}^{\,\vee\,\mathrm{dual}}_{\tau}:=
\mathcal{H}^{\,\vee}_{\tau}, 
\end{equation}
where
$\mathcal{H}^{\,\mathrm{dual}}_{\sigma}$
is the $2^{2N}-1$-dimensional subspace of
$\mathcal{H}^{\,}_{\sigma}$
on which $\widehat{U}^{\,}_{r^{z}_{\pi}}$
reduces to the identity,
and
\begin{equation}
\left(
\mathcal{H}^{\,\mathrm{dual}}_{\sigma},  
\qquad
\mathcal{H}^{\,\vee\,\mathrm{dual}}_{\beta\alpha}  
\right),
\qquad
\mathcal{H}^{\,\mathrm{dual}}_{\sigma}:=
\mathcal{H}^{\,}_{\sigma},
\qquad
\mathcal{H}^{\,\vee\,\mathrm{dual}}_{\beta\alpha}:=
\mathcal{H}^{\,\vee}_{\beta\alpha},
\end{equation}
\end{subequations}
respectively.

The Kramers-Wannier dual of the Hamiltonian
(\ref{eq:def Hamiltonian b=0})
when open boundary conditions are imposed
in the reduced coupling space
(\ref{eq:reduced coupling space})
is%
~\footnote{%
~We emphasize that
the dual of the Hamiltonian
(\ref{eq:def Hamiltonian b=0})
when open boundary conditions are imposed
is not the Hamiltonian
(\ref{eq:def Hamiltonian b'=0})
when open boundary conditions are imposed.
}
\begin{equation}
\begin{split}
\widehat{H}^{\,\vee}_{\tau}=&\,
\sum_{j=1}^{2N-1}
\hat{\tau}^{z\,\vee}_{j+\frac{1}{2}}
-
\Delta\,
\left(
\underline{
\hat{\tau}^{z\,\vee}_{1+\frac{1}{2}}\,
\hat{\tau}^{x\,\vee}_{2+\frac{1}{2}}
}
+
\sum_{j=2}^{2N-2}
\hat{\tau}^{x\,\vee}_{j-1+\frac{1}{2}}\,
\hat{\tau}^{z\,\vee}_{j+\frac{1}{2}}\,
\hat{\tau}^{x\,\vee}_{j+1+\frac{1}{2}}
+
\underline{
\underline{
\hat{\tau}^{x\,\vee}_{2N-2+\frac{1}{2}}\,
\hat{\tau}^{z\,\vee}_{2N-1+\frac{1}{2}}
}
}
\right)
\\
&\,
+
J\,
\Bigg\{
\sum_{j=1}^{2N-2}
\hat{\tau}^{z\,\vee}_{j+\frac{1}{2}}\,
\hat{\tau}^{z\,\vee}_{j+1+\frac{1}{2}}
+
\Delta\,
\Bigg[
\underline{
\left(
\hat{\tau}^{z\,\vee}_{1+\frac{1}{2}}\,
\hat{\tau}^{x\,\vee}_{2+\frac{1}{2}}
\right)\,
\left(
\hat{\tau}^{x\,\vee}_{1+\frac{1}{2}}\,
\hat{\tau}^{z\,\vee}_{2+\frac{1}{2}}\,
\hat{\tau}^{x\,\vee}_{3+\frac{1}{2}}
\right)
}
\\
&\,
+
\sum_{j=2}^{2N-3}
\left(
\hat{\tau}^{x\,\vee}_{j-\frac{1}{2}}\,
\hat{\tau}^{z\,\vee}_{j+\frac{1}{2}}\,
\hat{\tau}^{x\,\vee}_{j+\frac{3}{2}}
\right)
\left(
\hat{\tau}^{x\,\vee}_{j+\frac{1}{2}}\,
\hat{\tau}^{z\,\vee}_{j+\frac{3}{2}}\,
\hat{\tau}^{x\,\vee}_{j+\frac{5}{2}}
\right)
\\
&\,
+
\underline{
\underline{
\left(
\hat{\tau}^{x\,\vee}_{2N-3+\frac{1}{2}}\,
\hat{\tau}^{z\,\vee}_{2N-2+\frac{1}{2}}\,
\hat{\tau}^{x\,\vee}_{2N-1+\frac{1}{2}}
\right)\,
\left(
\hat{\tau}^{x\,\vee}_{2N-2+\frac{1}{2}}\,
\hat{\tau}^{z\,\vee}_{2N-1+\frac{1}{2}}
\right)
}
}
\Bigg]
\Bigg\}.
\end{split}
\label{eq:tau-dual Hamiltonian with OPB}
\end{equation}
The terms that are underlined once only act non trivially on the
left boundary.
The terms that are underlined twice only act non trivially on the
right boundary. These boundary terms break explicitly the global
internal symmetry
(\ref{eq:reps ro re rz dual b'}).
These boundary terms gap the zero modes,
if present, of the bulk contributions (all terms that are not underlined)
to the Hamiltonian.

The Jordan-Wigner dual of the Hamiltonian
(\ref{eq:def Hamiltonian b=0})
when open boundary conditions are imposed
in the reduced coupling space
(\ref{eq:reduced coupling space})
is
\begin{equation}
\begin{split}
\widehat{H}^{\,\vee}_{\beta\alpha}=&\,
\sum_{j=1}^{2N-1}
\left(
\mathrm{i}
\hat{\beta}^{\vee}_{j}\,
\hat{\alpha}^{\vee}_{j+1}\,
+
\Delta\,
\mathrm{i}
\hat{\alpha}^{\vee}_{j}\,
\hat{\beta}^{\vee}_{j+1}\,
\right)
\\
&\,
+
J
\sum_{j=1}^{2N-2}
\left(
\mathrm{i}
\hat{\beta}^{\vee}_{j}\,
\hat{\beta}^{\vee}_{j+1}\,
\hat{\alpha}^{\vee}_{j+1}\,
\hat{\alpha}^{\vee}_{j+2}
+
\Delta\,
\mathrm{i}
\hat{\alpha}^{\vee}_{j}\,
\hat{\alpha}^{\vee}_{j+1}\,
\hat{\beta}^{\vee}_{j+1}\,
\hat{\beta}^{\vee}_{j+2}
\right).
\end{split}
\end{equation}
If we do the unitary transformation
\begin{equation}
\hat{\beta}^{\vee}_{j}\mapsto
+\hat{\alpha}^{\vee}_{j},
\qquad
\hat{\alpha}^{\vee}_{j}\mapsto
-\hat{\beta}^{\vee}_{j},
\end{equation}
we recover Hamiltonian
(\ref{eq:def Hamiltonian f=1})
in the reduced coupling space
(\ref{eq:reduced coupling space})
with open boundary conditions.
The two-fold degeneracy of the
$\mathrm{Neel}^{\,}_{x}$
or
$\mathrm{Neel}^{\,}_{y}$
phases is now interpreted by the
existence of a single Majorana zero mode localized at the
left and right ends of the open chain.

\end{appendix}

\newpage


\begin{small}
\bibliography{ref}

\providecommand{\href}[2]{#2}\begingroup\begin{thebibliography}{100}

\bibitem{Lieb61}
E.~{Lieb}, T.~{Schultz}, and D.~{Mattis}, {\it {Two soluble models of an
  antiferromagnetic chain}},
  \href{http://dx.doi.org/10.1016/0003-4916(61)90115-4}{{\sf Annals of Physics}
  {\sf {16} }{\sf no.~3, }{\sf (Dec., 1961) }{\sf 407--466}}.

\bibitem{Affleck1986}
I.~Affleck and E.~H. Lieb, {\it A proof of part of haldane's conjecture on spin
  chains},  \href{http://dx.doi.org/10.1007/BF00400304}{{\sf Letters in
  Mathematical Physics} {\sf {12} }{\sf no.~1, }{\sf (Jul, 1986) }{\sf
  57--69}}. \url{https://doi.org/10.1007/BF00400304}.

\bibitem{Aizenman1994}
M.~Aizenman and B.~Nachtergaele, {\it Geometric aspects of quantum spin
  states},  \href{http://dx.doi.org/10.1007/BF02108805}{{\sf Communications in
  Mathematical Physics} {\sf {164} }{\sf no.~1, }{\sf (Jul, 1994) }{\sf
  17--63}}. \url{https://doi.org/10.1007/BF02108805}.

\bibitem{Oshikawa1997}
M.~Oshikawa, M.~Yamanaka, and I.~Affleck, {\it Magnetization plateaus in spin
  chains: ``haldane gap'' for half-integer spins},
  \href{http://dx.doi.org/10.1103/PhysRevLett.78.1984}{{\sf Phys. Rev. Lett.}
  {\sf {78} }{\sf (Mar, 1997) }{\sf 1984--1987}}.
  \url{https://link.aps.org/doi/10.1103/PhysRevLett.78.1984}.

\bibitem{Yamanaka1997}
M.~Yamanaka, M.~Oshikawa, and I.~Affleck, {\it Nonperturbative approach to
  luttinger's theorem in one dimension},
  \href{http://dx.doi.org/10.1103/PhysRevLett.79.1110}{{\sf Phys. Rev. Lett.}
  {\sf {79} }{\sf (Aug, 1997) }{\sf 1110--1113}}.
  \url{https://link.aps.org/doi/10.1103/PhysRevLett.79.1110}.

\bibitem{Koma2000}
T.~Koma, {\it Spectral gaps of quantum hall systems with interactions},
  \href{http://dx.doi.org/10.1023/A:1018604925491}{{\sf Journal of Statistical
  Physics} {\sf {99} }{\sf no.~1, }{\sf (Apr, 2000) }{\sf 313--381}}.
  \url{https://doi.org/10.1023/A:1018604925491}.

\bibitem{Oshikawa2000}
M.~Oshikawa, {\it Commensurability, excitation gap, and topology in quantum
  many-particle systems on a periodic lattice},
  \href{http://dx.doi.org/10.1103/PhysRevLett.84.1535}{{\sf Phys. Rev. Lett.}
  {\sf {84} }{\sf (Feb, 2000) }{\sf 1535--1538}}.
  \url{https://link.aps.org/doi/10.1103/PhysRevLett.84.1535}.

\bibitem{Hastings2004}
M.~B. Hastings, {\it -schultz-mattis in higher dimensions},
  \href{http://dx.doi.org/10.1103/PhysRevB.69.104431}{{\sf Phys. Rev. B} {\sf
  {69} }{\sf (Mar, 2004) }{\sf 104431}}.
  \url{https://link.aps.org/doi/10.1103/PhysRevB.69.104431}.

\bibitem{Hastings2005}
M.~B. Hastings, {\it Sufficient conditions for topological order in
  insulators},  \href{http://dx.doi.org/10.1209/epl/i2005-10046-x}{{\sf
  Europhysics Letters ({EPL})} {\sf {70} }{\sf no.~6, }{\sf (Jun, 2005) }{\sf
  824--830}}. \url{https://doi.org/10.1209%2Fepl%2Fi2005-10046-x}.

\bibitem{Chen2011}
X.~Chen, Z.-C. Gu, and X.-G. Wen, {\it Complete classification of
  one-dimensional gapped quantum phases in interacting spin systems},
  \href{http://dx.doi.org/10.1103/PhysRevB.84.235128}{{\sf Phys. Rev. B} {\sf
  {84} }{\sf (Dec, 2011) }{\sf 235128}}.
  \url{https://link.aps.org/doi/10.1103/PhysRevB.84.235128}.

\bibitem{Roy2012}
R.~Roy, {\it Space group symmetries and low lying excitations of many-body
  systems at integer fillings}, . \url{https://arxiv.org/abs/1212.2944}.

\bibitem{Parameswaran2013}
S.~A. Parameswaran, A.~M. Turner, D.~P. Arovas, and A.~Vishwanath, {\it
  Topological order and absence of band insulators at integer filling in
  non-symmorphic crystals},  \href{http://dx.doi.org/10.1038/nphys2600}{{\sf
  Nature Physics} {\sf {9} }{\sf no.~5, }{\sf (Apr, 2013) }{\sf 299–303}}.
  \url{http://dx.doi.org/10.1038/nphys2600}.

\bibitem{Watanabe2015}
H.~Watanabe, H.~C. Po, A.~Vishwanath, and M.~Zaletel, {\it Filling constraints
  for spin-orbit coupled insulators in symmorphic and nonsymmorphic crystals},
  \href{http://dx.doi.org/10.1073/pnas.1514665112}{{\sf Proceedings of the
  National Academy of Sciences} {\sf {112} }{\sf no.~47, }{\sf (2015) }{\sf
  14551--14556}}. \url{https://www.pnas.org/content/112/47/14551}.

\bibitem{Watanabe2016}
H.~Watanabe, H.~C. Po, M.~P. Zaletel, and A.~Vishwanath, {\it Filling-enforced
  gaplessness in band structures of the 230 space groups},
  \href{http://dx.doi.org/10.1103/PhysRevLett.117.096404}{{\sf Phys. Rev.
  Lett.} {\sf {117} }{\sf (Aug, 2016) }{\sf 096404}}.
  \url{https://link.aps.org/doi/10.1103/PhysRevLett.117.096404}.

\bibitem{Cheng2016}
M.~Cheng, M.~Zaletel, M.~Barkeshli, A.~Vishwanath, and P.~Bonderson, {\it
  Translational symmetry and microscopic constraints on symmetry-enriched
  topological phases: A view from the surface},
  \href{http://dx.doi.org/10.1103/PhysRevX.6.041068}{{\sf Phys. Rev. X} {\sf
  {6} }{\sf (Dec, 2016) }{\sf 041068}}.
  \url{https://link.aps.org/doi/10.1103/PhysRevX.6.041068}.

\bibitem{Qi2017}
Y.~Qi, C.~Fang, and L.~Fu, {\it {Ground state degeneracy in quantum spin
  systems protected by crystal symmetries}},
  \href{http://dx.doi.org/10.48550/arXiv.1705.09190}{{\sf arXiv e-prints} {\sf
  (May, 2017) }{\sf arXiv:1705.09190}}.

\bibitem{Cho2017}
G.~Y. Cho, C.-T. Hsieh, and S.~Ryu, {\it Anomaly manifestation of
  lieb-schultz-mattis theorem and topological phases},
  \href{http://dx.doi.org/10.1103/PhysRevB.96.195105}{{\sf Phys. Rev. B} {\sf
  {96} }{\sf (Nov, 2017) }{\sf 195105}}.
  \url{https://link.aps.org/doi/10.1103/PhysRevB.96.195105}.

\bibitem{Po2017}
H.~C. Po, H.~Watanabe, C.-M. Jian, and M.~P. Zaletel, {\it Lattice homotopy
  constraints on phases of quantum magnets},
  \href{http://dx.doi.org/10.1103/PhysRevLett.119.127202}{{\sf Phys. Rev.
  Lett.} {\sf {119} }{\sf (Sep, 2017) }{\sf 127202}}.
  \url{https://link.aps.org/doi/10.1103/PhysRevLett.119.127202}.

\bibitem{Watanabe2018}
H.~Watanabe, {\it Lieb-schultz-mattis-type filling constraints in the 1651
  magnetic space groups},
  \href{http://dx.doi.org/10.1103/PhysRevB.97.165117}{{\sf Phys. Rev. B} {\sf
  {97} }{\sf (Apr, 2018) }{\sf 165117}}.
  \url{https://link.aps.org/doi/10.1103/PhysRevB.97.165117}.

\bibitem{Tasaki2018}
H.~Tasaki, {\it Lieb--schultz--mattis theorem with a local twist for general
  one-dimensional quantum systems},
  \href{http://dx.doi.org/10.1007/s10955-017-1946-0}{{\sf Journal of
  Statistical Physics} {\sf {170} }{\sf no.~4, }{\sf (2018) }{\sf 653--671}}.

\bibitem{Metlitski2018}
M.~A. Metlitski and R.~Thorngren, {\it Intrinsic and emergent anomalies at
  deconfined critical points},
  \href{http://dx.doi.org/10.1103/PhysRevB.98.085140}{{\sf Phys. Rev. B} {\sf
  {98} }{\sf (Aug, 2018) }{\sf 085140}}.
  \url{https://link.aps.org/doi/10.1103/PhysRevB.98.085140}.

\bibitem{Yang2018}
X.~Yang, S.~Jiang, A.~Vishwanath, and Y.~Ran, {\it Dyonic lieb-schultz-mattis
  theorem and symmetry protected topological phases in decorated dimer models},
   \href{http://dx.doi.org/10.1103/PhysRevB.98.125120}{{\sf Phys. Rev. B} {\sf
  {98} }{\sf (Sep, 2018) }{\sf 125120}}.
  \url{https://link.aps.org/doi/10.1103/PhysRevB.98.125120}.

\bibitem{Jian2018}
C.-M. Jian, Z.~Bi, and C.~Xu, {\it Lieb-schultz-mattis theorem and its
  generalizations from the perspective of the symmetry-protected topological
  phase},  \href{http://dx.doi.org/10.1103/PhysRevB.97.054412}{{\sf Phys. Rev.
  B} {\sf {97} }{\sf (Feb, 2018) }{\sf 054412}}.
  \url{https://link.aps.org/doi/10.1103/PhysRevB.97.054412}.

\bibitem{Cheng2019}
M.~Cheng, {\it Fermionic lieb-schultz-mattis theorems and weak
  symmetry-protected phases},
  \href{http://dx.doi.org/10.1103/PhysRevB.99.075143}{{\sf Phys. Rev. B} {\sf
  {99} }{\sf (Feb, 2019) }{\sf 075143}}.
  \url{https://link.aps.org/doi/10.1103/PhysRevB.99.075143}.

\bibitem{Kobayashi2019}
R.~Kobayashi, K.~Shiozaki, Y.~Kikuchi, and S.~Ryu, {\it Lieb-schultz-mattis
  type theorem with higher-form symmetry and the quantum dimer models},
  \href{http://dx.doi.org/10.1103/PhysRevB.99.014402}{{\sf Phys. Rev. B} {\sf
  {99} }{\sf (Jan, 2019) }{\sf 014402}}.
  \url{https://link.aps.org/doi/10.1103/PhysRevB.99.014402}.

\bibitem{Ogata2019}
Y.~Ogata and H.~Tasaki, {\it Lieb--schultz--mattis type theorems for quantum
  spin chains without continuous symmetry},
  \href{http://dx.doi.org/10.1007/s00220-019-03343-5}{{\sf Communications in
  Mathematical Physics} {\sf {372} }{\sf no.~3, }{\sf (2019) }{\sf 951--962}}.
  \url{https://doi.org/10.1007/s00220-019-03343-5}.

\bibitem{He2020}
H.~He, Y.~You, and A.~Prem, {\it Lieb-schultz-mattis--type constraints on
  fractonic matter},  \href{http://dx.doi.org/10.1103/PhysRevB.101.165145}{{\sf
  Phys. Rev. B} {\sf {101} }{\sf (Apr, 2020) }{\sf 165145}}.
  \url{https://link.aps.org/doi/10.1103/PhysRevB.101.165145}.

\bibitem{Else2020}
D.~V. Else and R.~Thorngren, {\it Topological theory of lieb-schultz-mattis
  theorems in quantum spin systems},
  \href{http://dx.doi.org/10.1103/PhysRevB.101.224437}{{\sf Phys. Rev. B} {\sf
  {101} }{\sf (Jun, 2020) }{\sf 224437}}.
  \url{https://link.aps.org/doi/10.1103/PhysRevB.101.224437}.

\bibitem{Hetenyi2020}
B.~Het\'enyi, {\it Interaction-driven polarization shift in the
  $t\text{\ensuremath{-}}v\text{\ensuremath{-}}{V}^{\ensuremath{'}}$ lattice
  fermion model at half filling: Emergent haldane phase},
  \href{http://dx.doi.org/10.1103/PhysRevResearch.2.023277}{{\sf Phys. Rev.
  Res.} {\sf {2} }{\sf (Jun, 2020) }{\sf 023277}}.
  \url{https://link.aps.org/doi/10.1103/PhysRevResearch.2.023277}.

\bibitem{Yao2020a}
Y.~Yao and M.~Oshikawa, {\it Generalized boundary condition applied to
  lieb-schultz-mattis-type ingappabilities and many-body chern numbers},
  \href{http://dx.doi.org/10.1103/PhysRevX.10.031008}{{\sf Phys. Rev. X} {\sf
  {10} }{\sf (Jul, 2020) }{\sf 031008}}.
  \url{https://link.aps.org/doi/10.1103/PhysRevX.10.031008}.

\bibitem{Ogata2021}
Y.~Ogata, Y.~Tachikawa, and H.~Tasaki, {\it General lieb--schultz--mattis type
  theorems for quantum spin chains},
  \href{http://dx.doi.org/10.1007/s00220-021-04116-9}{{\sf Communications in
  Mathematical Physics} {\sf {385} }{\sf no.~1, }{\sf (Jul, 2021) }{\sf
  79--99}}. \url{https://doi.org/10.1007/s00220-021-04116-9}.

\bibitem{Bachmann2020}
S.~Bachmann, A.~Bols, W.~De~Roeck, and M.~Fraas, {\it A many-body index for
  quantum charge transport},
  \href{http://dx.doi.org/10.1007/s00220-019-03537-x}{{\sf Communications in
  Mathematical Physics} {\sf {375} }{\sf no.~2, }{\sf (Apr, 2020) }{\sf
  1249--1272}}. \url{https://doi.org/10.1007/s00220-019-03537-x}.

\bibitem{Dubinkin2021}
O.~Dubinkin, J.~May-Mann, and T.~L. Hughes, {\it Lieb-schultz-mattis-type
  theorems and other nonperturbative results for strongly correlated systems
  with conserved dipole moments},
  \href{http://dx.doi.org/10.1103/PhysRevB.103.125133}{{\sf Phys. Rev. B} {\sf
  {103} }{\sf (Mar, 2021) }{\sf 125133}}.
  \url{https://link.aps.org/doi/10.1103/PhysRevB.103.125133}.

\bibitem{Yao2021}
Y.~Yao and M.~Oshikawa, {\it Twisted boundary condition and lieb-schultz-mattis
  ingappability for discrete symmetries},
  \href{http://dx.doi.org/10.1103/PhysRevLett.126.217201}{{\sf Phys. Rev.
  Lett.} {\sf {126} }{\sf (May, 2021) }{\sf 217201}}.
  \url{https://link.aps.org/doi/10.1103/PhysRevLett.126.217201}.

\bibitem{Aksoy2021b}
{\"O}.~M. Aksoy, A.~Tiwari, and C.~Mudry, {\it Lieb-schultz-mattis type
  theorems for majorana models with discrete symmetries},
  \href{http://dx.doi.org/10.1103/PhysRevB.104.075146}{{\sf Phys. Rev. B} {\sf
  {104} }{\sf (Aug, 2021) }{\sf 075146}}.
  \url{https://link.aps.org/doi/10.1103/PhysRevB.104.075146}.

\bibitem{Tasaki2022}
H.~Tasaki, {\it {The Lieb-Schultz-Mattis Theorem: A Topological Point of
  View}},  \href{http://dx.doi.org/10.48550/arXiv.2202.06243}{{\sf arXiv
  e-prints} {\sf (Feb., 2022) }{\sf arXiv:2202.06243}}.

\bibitem{Yao2022}
Y.~Yao and A.~Furusaki, {\it Geometric approach to lieb-schultz-mattis theorem
  without translation symmetry under inversion or rotation symmetry},
  \href{http://dx.doi.org/10.1103/PhysRevB.106.045125}{{\sf Phys. Rev. B} {\sf
  {106} }{\sf (Jul, 2022) }{\sf 045125}}.
  \url{https://link.aps.org/doi/10.1103/PhysRevB.106.045125}.

\bibitem{Gioia2022}
L.~Gioia and C.~Wang, {\it Nonzero momentum requires long-range entanglement},
  \href{http://dx.doi.org/10.1103/PhysRevX.12.031007}{{\sf Phys. Rev. X} {\sf
  {12} }{\sf (Jul, 2022) }{\sf 031007}}.
  \url{https://link.aps.org/doi/10.1103/PhysRevX.12.031007}.

\bibitem{Cheng2023}
M.~{Cheng} and N.~{Seiberg}, {\it {Lieb-Schultz-Mattis, Luttinger, and 't Hooft
  -- anomaly matching in lattice systems}},
  \href{http://dx.doi.org/10.48550/arXiv.2211.12543}{{\sf arXiv e-prints} {\sf
  (Nov., 2022) }{\sf arXiv:2211.12543}}.

\bibitem{Yao2023}
Y.~{Yao}, L.~{Li}, M.~{Oshikawa}, and C.-T. {Hsieh}, {\it {Lieb-Schultz-Mattis
  theorem for 1d quantum magnets with antiunitary translation and inversion
  symmetries}},  \href{http://dx.doi.org/10.48550/arXiv.2307.09843}{{\sf arXiv
  e-prints} {\sf (July, 2023) }{\sf arXiv:2307.09843}},
  \href{http://arxiv.org/abs/2307.09843}{{\ttfamily arXiv:2307.09843
  [cond-mat.str-el]}}.

\bibitem{Seiberg2023}
N.~{Seiberg} and S.-H. {Shao}, {\it {Majorana chain and Ising model --
  (non-invertible) translations, anomalies, and emanant symmetries}},
  \href{http://dx.doi.org/10.48550/arXiv.2307.02534}{{\sf arXiv e-prints} {\sf
  (July, 2023) }{\sf arXiv:2307.02534}},
  \href{http://arxiv.org/abs/2307.02534}{{\ttfamily arXiv:2307.02534
  [cond-mat.str-el]}}.

\bibitem{tHooft1979}
G.~'t~Hooft, {\it {Naturalness, chiral symmetry, and spontaneous chiral
  symmetry breaking}},
  \href{http://dx.doi.org/10.1007/978-1-4684-7571-5_9}{{\sf NATO Sci. Ser. B}
  {\sf {59} }{\sf (1980) }{\sf 135--157}}.

\bibitem{Alvarez-Gaume:1984zst}
L.~Alvarez-Gaume, S.~Della~Pietra, and G.~W. Moore, {\it {Anomalies and Odd
  Dimensions}},  \href{http://dx.doi.org/10.1016/0003-4916(85)90383-5}{{\sf
  Annals Phys.} {\sf {163} }{\sf (1985) }{\sf 288}}.

\bibitem{Treiman:1986ep}
S.~B. Treiman, R.~Jackiw, B.~Zumino, and E.~Witten, {\it Current algebra and
  anomalies}, .
\newblock Princeton University Press, 1985.
\newblock \url{http://www.jstor.org/stable/j.ctt7ztmc8}.

\bibitem{Bertlmann:1996xk}
R.~Bertlmann, {\it Anomalies in quantum field theory}, .
\newblock International Series of Monographs on Physics. Clarendon Press, 2000.
\newblock \url{https://books.google.ch/books?id=FC\_DRRUHFXEC}.

\bibitem{Kapustin:2014tfa}
A.~{Kapustin}, {\it {Symmetry Protected Topological Phases, Anomalies, and
  Cobordisms: Beyond Group Cohomology}},
  \href{http://dx.doi.org/10.48550/arXiv.1403.1467}{{\sf arXiv e-prints} {\sf
  (Mar., 2014) }{\sf arXiv:1403.1467}}.

\bibitem{Kapustin:2014zva}
A.~{Kapustin} and R.~{Thorngren}, {\it {Anomalies of discrete symmetries in
  various dimensions and group cohomology}},
  \href{http://dx.doi.org/10.48550/arXiv.1404.3230}{{\sf arXiv e-prints} {\sf
  (Apr., 2014) }{\sf arXiv:1404.3230}}.

\bibitem{Gaiotto:2017yup}
D.~Gaiotto, A.~Kapustin, Z.~Komargodski, and N.~Seiberg, {\it {Theta, time
  Reversal, and temperature}},
  \href{http://dx.doi.org/10.1007/JHEP05(2017)091}{{\sf JHEP} {\sf {05} }{\sf
  (2017) }{\sf 091}}, \href{http://arxiv.org/abs/1703.00501}{{\ttfamily
  arXiv:1703.00501 [hep-th]}}.

\bibitem{Chen:2011bcp}
X.~Chen, Z.-X. Liu, and X.-G. Wen, {\it {Two-dimensional symmetry-protected
  topological orders and their protected gapless edge excitations}},
  \href{http://dx.doi.org/10.1103/PhysRevB.84.235141}{{\sf Phys. Rev. B} {\sf
  {84} }{\sf no.~23, }{\sf (2011) }{\sf 235141}}.

\bibitem{Kawagoe2021}
K.~Kawagoe and M.~Levin, {\it {Anomalies in bosonic symmetry-protected
  topological edge theories: Connection to F symbols and a method of
  calculation}},  \href{http://dx.doi.org/10.1103/PhysRevB.104.115156}{{\sf
  Phys. Rev. B} {\sf {104} }{\sf no.~11, }{\sf (2021) }{\sf 115156}}.

\bibitem{Moradi23}
H.~{Moradi}, {\"O}.~M. {Aksoy}, J.~H. {Bardarson}, and A.~{Tiwari}, {\it
  {Symmetry fractionalization, mixed-anomalies and dualities in quantum spin
  models with generalized symmetries}},
  \href{http://dx.doi.org/10.48550/arXiv.2307.01266}{{\sf arXiv e-prints} {\sf
  (July, 2023) }{\sf arXiv:2307.01266}}.

\bibitem{Komargodski2018}
Z.~Komargodski, T.~Sulejmanpasic, and M.~\"Unsal, {\it Walls, anomalies, and
  deconfinement in quantum antiferromagnets},
  \href{http://dx.doi.org/10.1103/PhysRevB.97.054418}{{\sf Phys. Rev. B} {\sf
  {97} }{\sf (Feb, 2018) }{\sf 054418}}.
  \url{https://link.aps.org/doi/10.1103/PhysRevB.97.054418}.

\bibitem{Seifnashri2023}
S.~{Seifnashri}, {\it {Lieb-Schultz-Mattis anomalies as obstructions to gauging
  (non-on-site) symmetries}},
  \href{http://dx.doi.org/10.48550/arXiv.2308.05151}{{\sf arXiv e-prints} {\sf
  (Aug., 2023) }{\sf arXiv:2308.05151}}.

\bibitem{Huang2017}
S.-J. Huang, H.~Song, Y.-P. Huang, and M.~Hermele, {\it Building crystalline
  topological phases from lower-dimensional states},
  \href{http://dx.doi.org/10.1103/PhysRevB.96.205106}{{\sf Phys. Rev. B} {\sf
  {96} }{\sf (Nov, 2017) }{\sf 205106}}.
  \url{https://link.aps.org/doi/10.1103/PhysRevB.96.205106}.

\bibitem{Thorngren2018}
R.~Thorngren and D.~V. Else, {\it Gauging spatial symmetries and the
  classification of topological crystalline phases},
  \href{http://dx.doi.org/10.1103/PhysRevX.8.011040}{{\sf Phys. Rev. X} {\sf
  {8} }{\sf (Mar, 2018) }{\sf 011040}}.
  \url{https://link.aps.org/doi/10.1103/PhysRevX.8.011040}.

\bibitem{Khalaf2017}
E.~Khalaf, H.~C. Po, A.~Vishwanath, and H.~Watanabe, {\it {Symmetry indicators
  and anomalous surface states of topological crystalline insulators}},
  \href{http://dx.doi.org/10.1103/PhysRevX.8.031070}{{\sf Phys. Rev. X} {\sf
  {8} }{\sf no.~3, }{\sf (2018) }{\sf 031070}},
  \href{http://arxiv.org/abs/1711.11589}{{\ttfamily arXiv:1711.11589
  [cond-mat.str-el]}}.

\bibitem{Guo2018}
M.~Guo, K.~Ohmori, P.~Putrov, Z.~Wan, and J.~Wang, {\it {Fermionic Finite-Group
  Gauge Theories and Interacting Symmetric/Crystalline Orders via Cobordisms}},
   \href{http://dx.doi.org/10.1007/s00220-019-03671-6}{{\sf Commun. Math.
  Phys.} {\sf {376} }{\sf no.~2, }{\sf (2020) }{\sf 1073--1154}}.

\bibitem{Trifunovic2018}
L.~{Trifunovic} and P.~W. {Brouwer}, {\it {Higher-order bulk-boundary
  correspondence for topological crystalline phases}},
  \href{http://dx.doi.org/10.48550/arXiv.1805.02598}{{\sf arXiv e-prints} {\sf
  (May, 2018) }{\sf arXiv:1805.02598}}.

\bibitem{Else2018}
D.~V. Else and R.~Thorngren, {\it {Crystalline topological phases as defect
  networks}},  \href{http://dx.doi.org/10.1103/PhysRevB.99.115116}{{\sf Phys.
  Rev. B} {\sf {99} }{\sf no.~11, }{\sf (2019) }{\sf 115116}}.

\bibitem{Rasmussen2018}
A.~Rasmussen and Y.-M. Lu, {\it {Classification and construction of
  higher-order symmetry protected topological phases of interacting bosons}},
  \href{http://dx.doi.org/10.1103/PhysRevB.101.085137}{{\sf Phys. Rev. B} {\sf
  {101} }{\sf no.~8, }{\sf (2020) }{\sf 085137}}.

\bibitem{Shiozaki2018}
K.~Shiozaki, C.~Z. Xiong, and K.~Gomi, {\it {Generalized homology and
  Atiyah–Hirzebruch spectral sequence in crystalline symmetry protected
  topological phenomena}},  \href{http://dx.doi.org/10.1093/ptep/ptad086}{{\sf
  Progress of Theoretical and Experimental Physics} {\sf {2023} }{\sf no.~8,
  }{\sf (07, 2023) }{\sf 083I01}}. \url{https://doi.org/10.1093/ptep/ptad086}.

\bibitem{Song2020}
Z.~Song, C.~Fang, and Y.~Qi, {\it {Real-space recipes for general topological
  crystalline states}},
  \href{http://dx.doi.org/10.1038/s41467-020-17685-5}{{\sf Nature Commun.} {\sf
  {11} }{\sf (2020) }{\sf 4197}}.

\bibitem{Else2014}
D.~V. Else and C.~Nayak, {\it {Classifying symmetry-protected topological
  phases through the anomalous action of the symmetry on the edge}},
  \href{http://dx.doi.org/10.1103/PhysRevB.90.235137}{{\sf Phys. Rev. B} {\sf
  {90} }{\sf no.~23, }{\sf (2014) }{\sf 235137}}.

\bibitem{Shiozaki2016}
K.~Shiozaki, H.~Shapourian, and S.~Ryu, {\it {Many-body topological invariants
  in fermionic symmetry-protected topological phases}: {Cases of point group
  symmetries}},  \href{http://dx.doi.org/10.1103/PhysRevB.95.205139}{{\sf Phys.
  Rev. B} {\sf {95} }{\sf no.~20, }{\sf (2017) }{\sf 205139}}.

\bibitem{Frohlich2009}
J.~Frohlich, J.~Fuchs, I.~Runkel, and C.~Schweigert, {\it Defect lines,
  dualities and generalised orbifolds}, .

\bibitem{Tachikawa2017}
Y.~Tachikawa, {\it {On gauging finite subgroups}},
  \href{http://dx.doi.org/10.21468/SciPostPhys.8.1.015}{{\sf SciPost Phys.}
  {\sf {8} }{\sf no.~1, }{\sf (2020) }{\sf 015}},
  \href{http://arxiv.org/abs/1712.09542}{{\ttfamily arXiv:1712.09542
  [hep-th]}}.

\bibitem{Bhardwaj2017}
L.~Bhardwaj and Y.~Tachikawa, {\it {On finite symmetries and their gauging in
  two dimensions}},  \href{http://dx.doi.org/10.1007/JHEP03(2018)189}{{\sf
  JHEP} {\sf {03} }{\sf (2018) }{\sf 189}}.

\bibitem{Gaiotto2021}
D.~Gaiotto and J.~Kulp, {\it Orbifold groupoids},
  \href{http://dx.doi.org/10.1007/JHEP02(2021)132}{{\sf Journal of High Energy
  Physics} {\sf {2021} }{\sf no.~2, }{\sf (Feb, 2021) }{\sf 132}}.
  \url{https://doi.org/10.1007/JHEP02(2021)132}.

\bibitem{Delcamp2019}
C.~Delcamp and A.~Tiwari, {\it {On 2-form gauge models of topological phases}},
   \href{http://dx.doi.org/10.1007/JHEP05(2019)064}{{\sf JHEP} {\sf {05} }{\sf
  (2019) }{\sf 064}}.

\bibitem{Borla2020}
U.~Borla, R.~Verresen, J.~Shah, and S.~Moroz, {\it {Gauging the Kitaev chain}},
   \href{http://dx.doi.org/10.21468/SciPostPhys.10.6.148}{{\sf SciPost Phys.}
  {\sf {10} }{\sf no.~6, }{\sf (2021) }{\sf 148}}.

\bibitem{Bhardwaj2022a}
L.~Bhardwaj, L.~E. Bottini, S.~Schafer-Nameki, and A.~Tiwari
  \href{http://dx.doi.org/10.21468/SciPostPhys.15.4.160}{{\sf SciPost Phys.}
  {\sf {15} }{\sf (2023) }{\sf 160}}.
  \url{https://scipost.org/10.21468/SciPostPhys.15.4.160}.

\bibitem{Roumpedakis:2022aik}
K.~Roumpedakis, S.~Seifnashri, and S.-H. Shao, {\it {Higher Gauging and
  Non-invertible Condensation Defects}},
  \href{http://dx.doi.org/10.1007/s00220-023-04706-9}{{\sf Commun. Math. Phys.}
  {\sf {401} }{\sf no.~3, }{\sf (2023) }{\sf 3043--3107}}.

\bibitem{Lootens:2021tet}
L.~Lootens, C.~Delcamp, G.~Ortiz, and F.~Verstraete, {\it {Dualities in
  One-Dimensional Quantum Lattice Models: Symmetric Hamiltonians and Matrix
  Product Operator Intertwiners}},
  \href{http://dx.doi.org/10.1103/PRXQuantum.4.020357}{{\sf PRX Quantum} {\sf
  {4} }{\sf no.~2, }{\sf (2023) }{\sf 020357}},
  \href{http://arxiv.org/abs/2112.09091}{{\ttfamily arXiv:2112.09091
  [quant-ph]}}.

\bibitem{Bhardwaj2022b}
L.~Bhardwaj, S.~Schafer-Nameki, and A.~Tiwari
  \href{http://dx.doi.org/10.21468/SciPostPhys.15.3.122}{{\sf SciPost Phys.}
  {\sf {15} }{\sf (2023) }{\sf 122}}.
  \url{https://scipost.org/10.21468/SciPostPhys.15.3.122}.

\bibitem{Bhardwaj2022c}
L.~Bhardwaj, L.~E. Bottini, S.~Schafer-Nameki, and A.~Tiwari, {\it
  {Non-invertible higher-categorical symmetries}},
  \href{http://dx.doi.org/10.21468/SciPostPhys.14.1.007}{{\sf SciPost Phys.}
  {\sf {14} }{\sf no.~1, }{\sf (2023) }{\sf 007}}.

\bibitem{Bhardwaj2022d}
L.~Bhardwaj, S.~Schafer-Nameki, and J.~Wu, {\it {Universal Non-Invertible
  Symmetries}},  \href{http://dx.doi.org/10.1002/prop.202200143}{{\sf Fortsch.
  Phys.} {\sf {70} }{\sf no.~11, }{\sf (2022) }{\sf 2200143}}.

\bibitem{Bartsch2022a}
T.~Bartsch, M.~Bullimore, A.~E.~V. Ferrari, and J.~Pearson, {\it
  {Non-invertible Symmetries and Higher Representation Theory I}},
  \href{http://dx.doi.org/10.48550/arXiv.2208.05993}{{\sf arXiv e-prints} {\sf
  (Aug., 2022) }{\sf arXiv:2208.05993}}.

\bibitem{Bartsch2022b}
T.~Bartsch, M.~Bullimore, A.~E.~V. Ferrari, and J.~Pearson, {\it
  {Non-invertible Symmetries and Higher Representation Theory II}},
  \href{http://dx.doi.org/10.48550/arXiv.2212.07393}{{\sf arXiv e-prints} {\sf
  (Dec., 2022) }{\sf arXiv:2212.07393}}.

\bibitem{Delcamp2023}
C.~Delcamp and A.~Tiwari, {\it {Higher categorical symmetries and gauging in
  two-dimensional spin systems}},
  \href{http://dx.doi.org/10.48550/arXiv.2301.01259}{{\sf arXiv e-prints} {\sf
  (Jan., 2023) }{\sf arXiv:2301.01259}}.

\bibitem{Bhardwaj:2023kri}
L.~Bhardwaj, L.~E. Bottini, L.~Fraser-Taliente, L.~Gladden, D.~S.~W. Gould,
  A.~Platschorre, and H.~Tillim, {\it {Lectures on generalized symmetries}},
  \href{http://dx.doi.org/10.1016/j.physrep.2023.11.002}{{\sf arXiv e-prints}
  {\sf {1051} }{\sf (Feb., 2024) }{\sf 1--87}},
  \href{http://arxiv.org/abs/2307.07547}{{\ttfamily arXiv:2307.07547
  [hep-th]}}.

\bibitem{Schafer-Nameki:2023jdn}
S.~Schafer-Nameki, {\it {ICTP Lectures on (Non-)Invertible Generalized
  Symmetries}},  \href{http://dx.doi.org/10.48550/arXiv.2305.18296}{{\sf arXiv
  e-prints} {\sf (May, 2023) }{\sf arXiv:2305.18296}}.

\bibitem{Seifert2023}
U.~F.~P. {Seifert} and S.~{Moroz}, {\it {Wegner's Ising gauge spins versus
  Kitaev's Majorana partons: Mapping and application to anisotropic confinement
  in spin-orbital liquids}},
  \href{http://dx.doi.org/10.48550/arXiv.2306.09405}{{\sf arXiv e-prints} {\sf
  (June, 2023) }{\sf arXiv:2306.09405}}.

\bibitem{Kaidi2021}
J.~Kaidi, K.~Ohmori, and Y.~Zheng, {\it {Kramers-Wannier-like Duality Defects
  in (3+1)D Gauge Theories}},
  \href{http://dx.doi.org/10.1103/PhysRevLett.128.111601}{{\sf Phys. Rev.
  Lett.} {\sf {128} }{\sf no.~11, }{\sf (2022) }{\sf 111601}}.

\bibitem{Kramers41}
H.~A. Kramers and G.~H. Wannier, {\it Statistics of the two-dimensional
  ferromagnet. part i},  \href{http://dx.doi.org/10.1103/PhysRev.60.252}{{\sf
  Phys. Rev.} {\sf {60} }{\sf (Aug, 1941) }{\sf 252--262}}.
  \url{https://link.aps.org/doi/10.1103/PhysRev.60.252}.

\bibitem{Montroll42}
E.~W. Montroll, {\it {Statistical Mechanics of Nearest Neighbor Systems II.
  General Theory and Application to Two‐Dimensional Ferromagnets}},
  \href{http://dx.doi.org/10.1063/1.1723622}{{\sf The Journal of Chemical
  Physics} {\sf {10} }{\sf no.~1, }{\sf (12, 2004) }{\sf 61--77}}.
  \url{https://doi.org/10.1063/1.1723622}.

\bibitem{Wannier45}
G.~H. Wannier, {\it The statistical problem in cooperative phenomena},
  \href{http://dx.doi.org/10.1103/RevModPhys.17.50}{{\sf Rev. Mod. Phys.} {\sf
  {17} }{\sf (Jan, 1945) }{\sf 50--60}}.
  \url{https://link.aps.org/doi/10.1103/RevModPhys.17.50}.

\bibitem{Dobson69}
J.~F. Dobson, {\it {Many‐Neighbored Ising Chain}},
  \href{http://dx.doi.org/10.1063/1.1664757}{{\sf Journal of Mathematical
  Physics} {\sf {10} }{\sf no.~1, }{\sf (11, 2003) }{\sf 40--45}}.
  \url{https://doi.org/10.1063/1.1664757}.

\bibitem{Frankel70}
N.~E. Frankel and D.~C. Rapaport, {\it {Domain Structure in a Second-Neighbour
  Ising Chain}},  \href{http://dx.doi.org/10.1143/PTP.43.1170}{{\sf Progress of
  Theoretical Physics} {\sf {43} }{\sf no.~5, }{\sf (05, 1970) }{\sf
  1170--1185}}. \url{https://doi.org/10.1143/PTP.43.1170}.

\bibitem{Stephenson70}
J.~Stephenson, {\it Two one-dimensional ising models with disorder points},
  \href{http://dx.doi.org/10.1139/p70-217}{{\sf Canadian Journal of Physics}
  {\sf {48} }{\sf no.~14, }{\sf (1970) }{\sf 1724--1734}}.
  \url{https://doi.org/10.1139/p70-217}.

\bibitem{Mittag71}
L.~Mittag and M.~J. Stephen, {\it {Dual Transformations in Many‐Component
  Ising Models}},  \href{http://dx.doi.org/10.1063/1.1665606}{{\sf Journal of
  Mathematical Physics} {\sf {12} }{\sf no.~3, }{\sf (10, 2003) }{\sf
  441--450}}. \url{https://doi.org/10.1063/1.1665606}.

\bibitem{Jordan28}
E.~Wigner and P.~Jordan, {\it {\"U}ber das paulische {\"a}quivalenzverbot},
  {\sf Z. Phys} {\sf {47} }{\sf (1928) }{\sf 631}.

\bibitem{McKean64}
J.~McKean, H.~P., {\it {Kramers‐Wannier Duality for the 2‐Dimensional Ising
  Model as an Instance of Poisson's Summation Formula}},
  \href{http://dx.doi.org/10.1063/1.1704178}{{\sf Journal of Mathematical
  Physics} {\sf {5} }{\sf no.~6, }{\sf (12, 1964) }{\sf 775--776}}.
  \url{https://doi.org/10.1063/1.1704178}.

\bibitem{Kadanoff71}
L.~P. Kadanoff and H.~Ceva, {\it Determination of an operator algebra for the
  two-dimensional ising model},
  \href{http://dx.doi.org/10.1103/PhysRevB.3.3918}{{\sf Phys. Rev. B} {\sf {3}
  }{\sf (Jun, 1971) }{\sf 3918--3939}}.
  \url{https://link.aps.org/doi/10.1103/PhysRevB.3.3918}.

\bibitem{Wegner71}
F.~J. Wegner, {\it {Duality in Generalized Ising Models and Phase Transitions
  without Local Order Parameters}},
  \href{http://dx.doi.org/10.1063/1.1665530}{{\sf Journal of Mathematical
  Physics} {\sf {12} }{\sf no.~10, }{\sf (10, 2003) }{\sf 2259--2272}}.
  \url{https://doi.org/10.1063/1.1665530}.

\bibitem{Balian75}
R.~Balian, J.~M. Drouffe, and C.~Itzykson, {\it Gauge fields on a lattice. ii.
  gauge-invariant ising model},
  \href{http://dx.doi.org/10.1103/PhysRevD.11.2098}{{\sf Phys. Rev. D} {\sf
  {11} }{\sf (Apr, 1975) }{\sf 2098--2103}}.
  \url{https://link.aps.org/doi/10.1103/PhysRevD.11.2098}.

\bibitem{Jose77}
J.~V. Jos\'e, L.~P. Kadanoff, S.~Kirkpatrick, and D.~R. Nelson, {\it
  Renormalization, vortices, and symmetry-breaking perturbations in the
  two-dimensional planar model},
  \href{http://dx.doi.org/10.1103/PhysRevB.16.1217}{{\sf Phys. Rev. B} {\sf
  {16} }{\sf (Aug, 1977) }{\sf 1217--1241}}.
  \url{https://link.aps.org/doi/10.1103/PhysRevB.16.1217}.

\bibitem{Peskin78}
M.~E. Peskin, {\it Mandelstam-'t hooft duality in abelian lattice models},
  \href{http://dx.doi.org/10.1016/0003-4916(78)90252-X}{{\sf Annals of Physics}
  {\sf {113} }{\sf no.~1, }{\sf (1978) }{\sf 122--152}}.
  \url{https://www.sciencedirect.com/science/article/pii/000349167890252X}.

\bibitem{Fradkin78}
E.~Fradkin and L.~Susskind, {\it Order and disorder in gauge systems and
  magnets},  \href{http://dx.doi.org/10.1103/PhysRevD.17.2637}{{\sf Phys. Rev.
  D} {\sf {17} }{\sf (May, 1978) }{\sf 2637--2658}}.
  \url{https://link.aps.org/doi/10.1103/PhysRevD.17.2637}.

\bibitem{Korthals-Altes78}
C.~P. Korthals~Altes, {\it Duality for z(n) gauge theories},
  \href{http://dx.doi.org/10.1016/0550-3213(78)90207-9}{{\sf Nuclear Physics B}
  {\sf {142} }{\sf no.~3, }{\sf (1978) }{\sf 315--326}}.
  \url{https://www.sciencedirect.com/science/article/pii/0550321378902079}.

\bibitem{Drouffe78}
J.~M. Drouffe, {\it Transitions and duality in gauge lattice systems},
  \href{http://dx.doi.org/10.1103/PhysRevD.18.1174}{{\sf Phys. Rev. D} {\sf
  {18} }{\sf (Aug, 1978) }{\sf 1174--1182}}.
  \url{https://link.aps.org/doi/10.1103/PhysRevD.18.1174}.

\bibitem{Kogut79}
J.~B. Kogut, {\it An introduction to lattice gauge theory and spin systems},
  \href{http://dx.doi.org/10.1103/RevModPhys.51.659}{{\sf Rev. Mod. Phys.} {\sf
  {51} }{\sf (Oct, 1979) }{\sf 659--713}}.
  \url{https://link.aps.org/doi/10.1103/RevModPhys.51.659}.

\bibitem{Bellissard79}
J.~Bellissard, {\it {A remark about the duality for non‐Abelian lattice
  fields}},  \href{http://dx.doi.org/10.1063/1.524206}{{\sf Journal of
  Mathematical Physics} {\sf {20} }{\sf no.~7, }{\sf (07, 2008) }{\sf
  1490--1493}}. \url{https://doi.org/10.1063/1.524206}.

\bibitem{Horn79}
D.~Horn, M.~Weinstein, and S.~Yankielowicz, {\it Hamiltonian approach to
  $\mathbb{Z}(n)$ lattice gauge theories},
  \href{http://dx.doi.org/10.1103/PhysRevD.19.3715}{{\sf Phys. Rev. D} {\sf
  {19} }{\sf (Jun, 1979) }{\sf 3715--3731}}.
  \url{https://link.aps.org/doi/10.1103/PhysRevD.19.3715}.

\bibitem{Drouffe79}
{\it Lattice models with a solvable symmetry group},
  \href{http://dx.doi.org/10.1016/0550-3213(79)90418-8}{{\sf Nuclear Physics B}
  {\sf {147} }{\sf no.~1, }{\sf (1979) }{\sf 132--134}}.
  \url{https://www.sciencedirect.com/science/article/pii/0550321379904188}.

\bibitem{Ukawa80}
A.~Ukawa, P.~Windey, and A.~H. Guth, {\it Dual variables for lattice gauge
  theories and the phase structure of $\mathbb{Z}(n)$ systems},
  \href{http://dx.doi.org/10.1103/PhysRevD.21.1013}{{\sf Phys. Rev. D} {\sf
  {21} }{\sf (Feb, 1980) }{\sf 1013--1036}}.
  \url{https://link.aps.org/doi/10.1103/PhysRevD.21.1013}.

\bibitem{Savit80}
R.~Savit, {\it Duality in field theory and statistical systems},
  \href{http://dx.doi.org/10.1103/RevModPhys.52.453}{{\sf Rev. Mod. Phys.} {\sf
  {52} }{\sf (Apr, 1980) }{\sf 453--487}}.
  \url{https://link.aps.org/doi/10.1103/RevModPhys.52.453}.

\bibitem{Druhl82}
K.~Dr\"uhl and H.~Wagner, {\it Algebraic formulation of duality transformations
  for abelian lattice models},
  \href{http://dx.doi.org/10.1016/0003-4916(82)90286-X}{{\sf Annals of Physics}
  {\sf {141} }{\sf no.~2, }{\sf (1982) }{\sf 225--253}}.
  \url{https://www.sciencedirect.com/science/article/pii/000349168290286X}.

\bibitem{Buchstaber03}
V.~M. Buchstaber and M.~I. Monastyrsky, {\it Generalized kramers–wannier
  duality for spin systems with non-commutative symmetry},
  \href{http://dx.doi.org/10.1088/0305-4470/36/28/301}{{\sf Journal of Physics
  A: Mathematical and General} {\sf {36} }{\sf no.~28, }{\sf (Jul, 2003) }{\sf
  7679}}. \url{https://dx.doi.org/10.1088/0305-4470/36/28/301}.

\bibitem{Mathur16}
M.~Mathur and T.~P. Sreeraj, {\it Lattice gauge theories and spin models},
  \href{http://dx.doi.org/10.1103/PhysRevD.94.085029}{{\sf Phys. Rev. D} {\sf
  {94} }{\sf (Oct, 2016) }{\sf 085029}}.
  \url{https://link.aps.org/doi/10.1103/PhysRevD.94.085029}.

\bibitem{Kapustin17}
A.~{Kapustin} and R.~{Thorngren}, {\it {Fermionic SPT phases in higher
  dimensions and bosonization}},
  \href{http://dx.doi.org/10.1007/JHEP10(2017)080}{{\sf Journal of High Energy
  Physics} {\sf {2017} }{\sf no.~10, }{\sf (Oct., 2017) }{\sf 80}}.

\bibitem{Radicevic18}
D.~Radicevic, {\it {Spin Structures and Exact Dualities in Low Dimensions}},
  \href{http://dx.doi.org/10.48550/arXiv.1809.07757}{{\sf arXiv e-prints} {\sf
  (Sept., 2018) }{\sf arXiv:1809.07757}}.

\bibitem{Karch19}
A.~Karch, D.~Tong, and C.~Turner, {\it {A web of 2d dualities: ${\bf Z}_2$
  gauge fields and Arf invariants}},
  \href{http://dx.doi.org/10.21468/SciPostPhys.7.1.007}{{\sf SciPost Phys.}
  {\sf {7} }{\sf (2019) }{\sf 007}}.
  \url{https://scipost.org/10.21468/SciPostPhys.7.1.007}.

\bibitem{Thorngren2020}
R.~Thorngren, {\it Anomalies and bosonization},
  \href{http://dx.doi.org/10.1007/s00220-020-03830-0}{{\sf COMMUNICATIONS IN
  MATHEMATICAL PHYSICS} {\sf {378} }{\sf no.~3, }{\sf (Sep, 2020) }{\sf
  1775--1816}}.

\bibitem{Senthil04}
T.~Senthil, A.~Vishwanath, L.~Balents, S.~Sachdev, and M.~P.~A. Fisher, {\it
  Deconfined quantum critical points},
  \href{http://dx.doi.org/10.1126/science.1091806}{{\sf Science} {\sf {303}
  }{\sf no.~5663, }{\sf (2004) }{\sf 1490--1494}}.
  \url{http://science.sciencemag.org/content/303/5663/1490}.

\bibitem{Senthil04Levin}
M.~Levin and T.~Senthil, {\it Deconfined quantum criticality and n\'eel order
  via dimer disorder},
  \href{http://dx.doi.org/10.1103/PhysRevB.70.220403}{{\sf Phys. Rev. B} {\sf
  {70} }{\sf (Dec, 2004) }{\sf 220403}}.
  \url{https://link.aps.org/doi/10.1103/PhysRevB.70.220403}.

\bibitem{Senthil04PRB}
T.~Senthil, L.~Balents, S.~Sachdev, A.~Vishwanath, and M.~P.~A. Fisher, {\it
  Quantum criticality beyond the landau-ginzburg-wilson paradigm},
  \href{http://dx.doi.org/10.1103/PhysRevB.70.144407}{{\sf Phys. Rev. B} {\sf
  {70} }{\sf (Oct, 2004) }{\sf 144407}}.
  \url{https://link.aps.org/doi/10.1103/PhysRevB.70.144407}.

\bibitem{Senthil06Fisher}
T.~Senthil and M.~P.~A. Fisher, {\it Competing orders, nonlinear sigma models,
  and topological terms in quantum magnets},
  \href{http://dx.doi.org/10.1103/PhysRevB.74.064405}{{\sf Phys. Rev. B} {\sf
  {74} }{\sf (Aug, 2006) }{\sf 064405}}.
  \url{https://link.aps.org/doi/10.1103/PhysRevB.74.064405}.

\bibitem{Moradi2022}
H.~{Moradi}, S.~{Faroogh Moosavian}, and A.~{Tiwari}, {\it {Topological
  Holography: Towards a Unification of Landau and Beyond-Landau Physics}},
  \href{http://dx.doi.org/10.48550/arXiv.2207.10712}{{\sf arXiv e-prints} {\sf
  (July, 2022) }{\sf arXiv:2207.10712}}.

\bibitem{Chatterjee2023}
A.~{Chatterjee} and X.-G. {Wen}, {\it {Holographic theory for continuous phase
  transitions: Emergence and symmetry protection of gaplessness}},
  \href{http://dx.doi.org/10.1103/PhysRevB.108.075105}{{\sf Phys. Rev. B} {\sf
  {108} }{\sf no.~7, }{\sf (Aug., 2023) }{\sf 075105}}.

\bibitem{Zhang2023}
C.~Zhang and M.~Levin, {\it Exactly solvable model for a deconfined quantum
  critical point in 1d},
  \href{http://dx.doi.org/10.1103/PhysRevLett.130.026801}{{\sf Phys. Rev.
  Lett.} {\sf {130} }{\sf (Jan, 2023) }{\sf 026801}}.
  \url{https://link.aps.org/doi/10.1103/PhysRevLett.130.026801}.

\bibitem{Mudry19}
C.~Mudry, A.~Furusaki, T.~Morimoto, and T.~Hikihara, {\it Quantum phase
  transitions beyond landau-ginzburg theory in one-dimensional space
  revisited},  \href{http://dx.doi.org/10.1103/PhysRevB.99.205153}{{\sf Phys.
  Rev. B} {\sf {99} }{\sf (May, 2019) }{\sf 205153}}.
  \url{https://link.aps.org/doi/10.1103/PhysRevB.99.205153}.

\bibitem{Burnell2023}
F.~J. {Burnell}, S.~{Moudgalya}, and A.~{Prem}, {\it {Filling constraints on
  translation invariant dipole conserving systems}},
  \href{http://dx.doi.org/10.48550/arXiv.2308.16241}{{\sf arXiv e-prints} {\sf
  (Aug., 2023) }{\sf arXiv:2308.16241}}.

\bibitem{Bulmash2023}
D.~{Bulmash}, O.~{Hart}, and R.~{Nandkishore}, {\it {Multipole groups and
  fracton phenomena on arbitrary crystalline lattices}},
  \href{http://dx.doi.org/10.48550/arXiv.2301.10782}{{\sf arXiv e-prints} {\sf
  (Jan., 2023) }{\sf arXiv:2301.10782}}.

\bibitem{Han2023}
J.~H. {Han}, E.~{Lake}, H.~T. {Lam}, R.~{Verresen}, and Y.~{You}, {\it
  {Topological quantum chains protected by dipolar and other modulated
  symmetries}},  \href{http://dx.doi.org/10.48550/arXiv.2309.10036}{{\sf arXiv
  e-prints} {\sf (Sept., 2023) }{\sf arXiv:2309.10036}}.

\bibitem{Delfino2023}
G.~{Delfino} and Y.~{You}, {\it {Anyon Condensation Web and Multipartite
  Entanglement in 2D Fracton-like Theories}},
  \href{http://dx.doi.org/10.48550/arXiv.2310.09490}{{\sf arXiv e-prints} {\sf
  (Oct., 2023) }{\sf arXiv:2310.09490}}.

\bibitem{Lam2023}
H.~T. {Lam}, {\it {Classification of Dipolar Symmetry-Protected Topological
  Phases: Matrix Product States, Stabilizer Hamiltonians and Finite Tensor
  Gauge Theories}},  \href{http://dx.doi.org/10.48550/arXiv.2311.04962}{{\sf
  arXiv e-prints} {\sf (Nov., 2023) }{\sf arXiv:2311.04962}}.

\bibitem{Coleman75}
S.~Coleman, {\it Quantum sine-gordon equation as the massive thirring model},
  \href{http://dx.doi.org/10.1103/PhysRevD.11.2088}{{\sf Phys. Rev. D} {\sf
  {11} }{\sf (Apr, 1975) }{\sf 2088--2097}}.
  \url{https://link.aps.org/doi/10.1103/PhysRevD.11.2088}.

\bibitem{Mandelstam75}
S.~Mandelstam, {\it Soliton operators for the quantized sine-gordon equation},
  \href{http://dx.doi.org/10.1103/PhysRevD.11.3026}{{\sf Phys. Rev. D} {\sf
  {11} }{\sf (May, 1975) }{\sf 3026--3030}}.
  \url{https://link.aps.org/doi/10.1103/PhysRevD.11.3026}.

\bibitem{Witten84}
E.~Witten, {\it Non-abelian bosonization in two dimensions},
  \href{http://dx.doi.org/10.1007/BF01215276}{{\sf Communications in
  Mathematical Physics} {\sf {92} }{\sf no.~4, }{\sf (Dec, 1984) }{\sf
  455--472}}. \url{https://doi.org/10.1007/BF01215276}.

\bibitem{Schroer79}
B.~Schroer and T.~T. Truong, {\it Z2 duality algebra in d = 2 quantum field
  theory},  \href{http://dx.doi.org/10.1016/0550-3213(79)90375-4}{{\sf Nuclear
  Physics B} {\sf {154} }{\sf no.~1, }{\sf (1979) }{\sf 125--139}}.
  \url{https://www.sciencedirect.com/science/article/pii/0550321379903754}.

\bibitem{BenTov15}
Y.~{BenTov}, {\it {Fermion masses without symmetry breaking in two spacetime
  dimensions}},  \href{http://dx.doi.org/10.1007/JHEP07(2015)034}{{\sf Journal
  of High Energy Physics} {\sf {2015} }{\sf (July, 2015) }{\sf 34}}.

\bibitem{Froelich04}
J.~Fr\"ohlich, J.~Fuchs, I.~Runkel, and C.~Schweigert, {\it Kramers-wannier
  duality from conformal defects},
  \href{http://dx.doi.org/10.1103/PhysRevLett.93.070601}{{\sf Phys. Rev. Lett.}
  {\sf {93} }{\sf (Aug, 2004) }{\sf 070601}}.
  \url{https://link.aps.org/doi/10.1103/PhysRevLett.93.070601}.

\bibitem{Ruelle05}
P.~Ruelle, {\it Kramers-wannier dualities via symmetries},
  \href{http://dx.doi.org/10.1103/PhysRevLett.95.225701}{{\sf Phys. Rev. Lett.}
  {\sf {95} }{\sf (Nov, 2005) }{\sf 225701}}.
  \url{https://link.aps.org/doi/10.1103/PhysRevLett.95.225701}.

\bibitem{Brunner:2013ota}
I.~Brunner, N.~Carqueville, and D.~Plencner, {\it {Orbifolds and topological
  defects}},  \href{http://dx.doi.org/10.1007/s00220-014-2056-3}{{\sf Commun.
  Math. Phys.} {\sf {332} }{\sf (2014) }{\sf 669--712}},
  \href{http://arxiv.org/abs/1307.3141}{{\ttfamily arXiv:1307.3141 [hep-th]}}.

\bibitem{Carqueville:2012dk}
N.~Carqueville and I.~Runkel, {\it {Orbifold completion of defect
  bicategories}},  \href{http://dx.doi.org/10.4171/qt/76}{{\sf Quantum Topol.}
  {\sf {7} }{\sf no.~2, }{\sf (2016) }{\sf 203--279}}.

\bibitem{Lin21}
Y.-H. Lin and S.-H. Shao, {\it Duality defect of the monster cft},
  \href{http://dx.doi.org/10.1088/1751-8121/abd69e}{{\sf Journal of Physics A:
  Mathematical and Theoretical} {\sf {54} }{\sf no.~6, }{\sf (Jan, 2021) }{\sf
  065201}}. \url{https://dx.doi.org/10.1088/1751-8121/abd69e}.

\bibitem{Cobanera10}
E.~Cobanera, G.~Ortiz, and Z.~Nussinov, {\it Unified approach to quantum and
  classical dualities},
  \href{http://dx.doi.org/10.1103/PhysRevLett.104.020402}{{\sf Phys. Rev.
  Lett.} {\sf {104} }{\sf (Jan, 2010) }{\sf 020402}}.
  \url{https://link.aps.org/doi/10.1103/PhysRevLett.104.020402}.

\bibitem{Cobanera11}
E.~Cobanera, G.~Ortiz, and Z.~Nussinov, {\it The bond-algebraic approach to
  dualities},  \href{http://dx.doi.org/10.1080/00018732.2011.619814}{{\sf
  Advances in Physics} {\sf {60} }{\sf no.~5, }{\sf (2011) }{\sf 679--798}}.
  \url{https://doi.org/10.1080/00018732.2011.619814}.

\bibitem{Cobanera13}
E.~Cobanera, G.~Ortiz, and Z.~Nussinov, {\it Holographic symmetries and
  generalized order parameters for topological matter},
  \href{http://dx.doi.org/10.1103/PhysRevB.87.041105}{{\sf Phys. Rev. B} {\sf
  {87} }{\sf (Jan, 2013) }{\sf 041105}}.
  \url{https://link.aps.org/doi/10.1103/PhysRevB.87.041105}.

\bibitem{Santos15}
L.~H. Santos, {\it Rokhsar-kivelson models of bosonic symmetry-protected
  topological states},
  \href{http://dx.doi.org/10.1103/PhysRevB.91.155150}{{\sf Phys. Rev. B} {\sf
  {91} }{\sf (Apr, 2015) }{\sf 155150}}.
  \url{https://link.aps.org/doi/10.1103/PhysRevB.91.155150}.

\bibitem{Tantivasadakarn21}
N.~Tantivasadakarn, W.~Ji, and S.~Vijay, {\it Non-abelian hybrid fracton
  orders},  \href{http://dx.doi.org/10.1103/PhysRevB.104.115117}{{\sf Phys.
  Rev. B} {\sf {104} }{\sf (Sep, 2021) }{\sf 115117}}.
  \url{https://link.aps.org/doi/10.1103/PhysRevB.104.115117}.

\bibitem{Redner81}
S.~Redner, {\it One-dimensional ising chain with competing interactions - exact
  results and connection with other statistical-models},
  \href{http://dx.doi.org/10.1007/BF01008476}{{\sf Journal of Statistical
  Physics} {\sf {25} }{\sf no.~1, }{\sf (1981) }{\sf 15--23}}.

\bibitem{Harada83}
I.~Harada, {\it Longitudinal and transverse susceptibilities of the
  one-dimensional ising model with competing interactions},
  \href{http://dx.doi.org/10.1143/JPSJ.52.4099}{{\sf Journal of the Physical
  Society of Japan} {\sf {52} }{\sf no.~12, }{\sf (1983) }{\sf 4099--4106}}.
  \url{https://doi.org/10.1143/JPSJ.52.4099}.

\bibitem{Majumdar69a}
C.~K. Majumdar and D.~K. Ghosh, {\it {On Next‐Nearest‐Neighbor Interaction
  in Linear Chain. I}},  \href{http://dx.doi.org/10.1063/1.1664978}{{\sf
  Journal of Mathematical Physics} {\sf {10} }{\sf no.~8, }{\sf (11, 2003)
  }{\sf 1388--1398}}. \url{https://doi.org/10.1063/1.1664978}.

\bibitem{Majumdar69b}
C.~K. Majumdar and D.~K. Ghosh, {\it {On Next‐Nearest‐Neighbor Interaction
  in Linear Chain. II}},  \href{http://dx.doi.org/10.1063/1.1664979}{{\sf
  Journal of Mathematical Physics} {\sf {10} }{\sf no.~8, }{\sf (11, 2003)
  }{\sf 1399--1402}}. \url{https://doi.org/10.1063/1.1664979}.

\bibitem{Majumdar70}
C.~K. Majumdar, {\it Antiferromagnetic model with known ground state},
  \href{http://dx.doi.org/10.1088/0022-3719/3/4/019}{{\sf Journal of Physics C:
  Solid State Physics} {\sf {3} }{\sf no.~4, }{\sf (Apr, 1970) }{\sf 911}}.
  \url{https://dx.doi.org/10.1088/0022-3719/3/4/019}.

\bibitem{Shastry81}
B.~S. Shastry and B.~Sutherland, {\it Excitation spectrum of a dimerized
  next-neighbor antiferromagnetic chain},
  \href{http://dx.doi.org/10.1103/PhysRevLett.47.964}{{\sf Phys. Rev. Lett.}
  {\sf {47} }{\sf (Sep, 1981) }{\sf 964--967}}.
  \url{https://link.aps.org/doi/10.1103/PhysRevLett.47.964}.

\bibitem{Haldane82}
F.~D.~M. Haldane, {\it Spontaneous dimerization in the $s=\frac{1}{2}$
  heisenberg antiferromagnetic chain with competing interactions},
  \href{http://dx.doi.org/10.1103/PhysRevB.25.4925}{{\sf Phys. Rev. B} {\sf
  {25} }{\sf (Apr, 1982) }{\sf 4925--4928}}.
  \url{https://link.aps.org/doi/10.1103/PhysRevB.25.4925}.

\bibitem{Jiang19}
S.~Jiang and O.~Motrunich, {\it Ising ferromagnet to valence bond solid
  transition in a one-dimensional spin chain: Analogies to deconfined quantum
  critical points},  \href{http://dx.doi.org/10.1103/PhysRevB.99.075103}{{\sf
  Phys. Rev. B} {\sf {99} }{\sf (Feb, 2019) }{\sf 075103}}.
  \url{https://link.aps.org/doi/10.1103/PhysRevB.99.075103}.

\bibitem{Tonegawa88}
T.~Tonegawa and I.~Harada, {\it Ground-state properties of the one-dimensional
  spin-1/2 heisenberg-xy antiferromagnet with competing interactions},
  \href{http://dx.doi.org/10.1051/jphyscol:19888648}{{\sf Journal de Physique}
  {\sf {49} }{\sf no.~C-8, 2, }{\sf (Dec, 1988) }{\sf 1411--1412}}.

\bibitem{Tonegawa92}
T.~Tonegawa, I.~Harada, and M.~Kaburagi, {\it Spin-fluid to dimer and n\'eel to
  dimer phase transitions in a spin-1/2 heisenberg chain with competing
  interactions},  \href{http://dx.doi.org/10.1143/JPSJ.61.4665}{{\sf Journal of
  the Physical Society of Japan} {\sf {61} }{\sf no.~12, }{\sf (1992) }{\sf
  4665--4666}}. \url{https://doi.org/10.1143/JPSJ.61.4665}.

\bibitem{Nersesyan98}
A.~A. Nersesyan, A.~O. Gogolin, and F.~H.~L. E\ss{}ler, {\it Incommensurate
  spin correlations in spin- $1/2$ frustrated two-leg heisenberg ladders},
  \href{http://dx.doi.org/10.1103/PhysRevLett.81.910}{{\sf Phys. Rev. Lett.}
  {\sf {81} }{\sf (Jul, 1998) }{\sf 910--913}}.
  \url{https://link.aps.org/doi/10.1103/PhysRevLett.81.910}.

\bibitem{Hikihara01}
T.~Hikihara, M.~Kaburagi, and H.~Kawamura, {\it Ground-state phase diagrams of
  frustrated spin-s xxz chains: Chiral ordered phases},
  \href{http://dx.doi.org/10.1103/PhysRevB.63.174430}{{\sf Phys. Rev. B} {\sf
  {63} }{\sf (Apr, 2001) }{\sf 174430}}.
  \url{https://link.aps.org/doi/10.1103/PhysRevB.63.174430}.

\bibitem{Furukawa12}
S.~Furukawa, M.~Sato, S.~Onoda, and A.~Furusaki, {\it Ground-state phase
  diagram of a spin-$\frac{1}{2}$ frustrated ferromagnetic xxz chain: Haldane
  dimer phase and gapped/gapless chiral phases},
  \href{http://dx.doi.org/10.1103/PhysRevB.86.094417}{{\sf Phys. Rev. B} {\sf
  {86} }{\sf (Sep, 2012) }{\sf 094417}}.
  \url{https://link.aps.org/doi/10.1103/PhysRevB.86.094417}.

\bibitem{Suzuki71}
M.~Suzuki, {\it {Relationship among Exactly Soluble Models of Critical
  Phenomena. I*): 2D Ising Model, Dimer Problem and the Generalized XY-Model}},
   \href{http://dx.doi.org/10.1143/PTP.46.1337}{{\sf Progress of Theoretical
  Physics} {\sf {46} }{\sf no.~5, }{\sf (11, 1971) }{\sf 1337--1359}}.
  \url{https://doi.org/10.1143/PTP.46.1337}.

\bibitem{PerezGarcia2008}
D.~P\'erez-Garc\'{\i}a, M.~M. Wolf, M.~Sanz, F.~Verstraete, and J.~I. Cirac,
  {\it String order and symmetries in quantum spin lattices},
  \href{http://dx.doi.org/10.1103/PhysRevLett.100.167202}{{\sf Phys. Rev.
  Lett.} {\sf {100} }{\sf (Apr, 2008) }{\sf 167202}}.
  \url{https://link.aps.org/doi/10.1103/PhysRevLett.100.167202}.

\bibitem{Pollman2012}
F.~Pollmann, E.~Berg, A.~M. Turner, and M.~Oshikawa, {\it Symmetry protection
  of topological phases in one-dimensional quantum spin systems},
  \href{http://dx.doi.org/10.1103/PhysRevB.85.075125}{{\sf Phys. Rev. B} {\sf
  {85} }{\sf (Feb, 2012) }{\sf 075125}}.
  \url{https://link.aps.org/doi/10.1103/PhysRevB.85.075125}.

\bibitem{Kitaev01}
A.~Y. Kitaev, {\it Unpaired majorana fermions in quantum wires},
  \href{http://dx.doi.org/10.1070/1063-7869/44/10s/s29}{{\sf Physics-Uspekhi}
  {\sf {44} }{\sf no.~10S, }{\sf (Oct, 2001) }{\sf 131--136}}.
  \url{https://doi.org/10.1070%2F1063-7869%2F44%2F10s%2Fs29}.

\bibitem{Turzillo2019}
A.~Turzillo and M.~You, {\it Fermionic matrix product states and
  one-dimensional short-range entangled phases with antiunitary symmetries},
  \href{http://dx.doi.org/10.1103/PhysRevB.99.035103}{{\sf Phys. Rev. B} {\sf
  {99} }{\sf (Jan, 2019) }{\sf 035103}}.
  \url{https://link.aps.org/doi/10.1103/PhysRevB.99.035103}.

\bibitem{Bourne2021}
C.~Bourne and Y.~Ogata, {\it The classification of symmetry protected
  topological phases of one-dimensional fermion systems},
  \href{http://dx.doi.org/10.1017/fms.2021.19}{{\sf Forum of Mathematics,
  Sigma} {\sf {9} }{\sf (2021) }{\sf e25}}.

\bibitem{Aksoy2022}
{\"O}.~M. Aksoy and C.~Mudry, {\it Elementary derivation of the stacking rules
  of invertible fermionic topological phases in one dimension},
  \href{http://dx.doi.org/10.1103/PhysRevB.106.035117}{{\sf Phys. Rev. B} {\sf
  {106} }{\sf (Jul, 2022) }{\sf 035117}}.
  \url{https://link.aps.org/doi/10.1103/PhysRevB.106.035117}.

\bibitem{Weicheng2022}
W.~Ye, M.~Guo, Y.-C. He, C.~Wang, and L.~Zou, {\it {Topological
  characterization of Lieb-Schultz-Mattis constraints and applications to
  symmetry-enriched quantum criticality}},
  \href{http://dx.doi.org/10.21468/SciPostPhys.13.3.066}{{\sf SciPost Phys.}
  {\sf {13} }{\sf (2022) }{\sf 066}}.
  \url{https://scipost.org/10.21468/SciPostPhys.13.3.066}.

\end{thebibliography}\endgroup
\bibliographystyle{bibstyle2017}
\end{small}

\end{document}